\documentclass{article}
\usepackage{epubtk}
\usepackage{graphicx}
\usepackage{amsmath}
\usepackage{amssymb}
\usepackage{booktabs}
\usepackage{bm}
\usepackage{enumitem}

\def\vp{\varphi}
\newcommand{\rd}{\mathrm{d}}
\newcommand{\ti}{\tilde}
\newcommand{\tr}{\tilde{r}}
\newcommand{\de}{\mathrm{d}}
\newcommand{\GB}{\mathcal{G}}

\newcommand{\fR}{\textit{f}(\textit{R})}

\begin{document}

\title{\textit{f}(\textit{R}) Theories}

\author{%
\epubtkAuthorData{Antonio De Felice}
		 {Department of Physics, Faculty of Science, Tokyo University of Science,\\
		  1-3, Kagurazaka, Shinjuku-ku, Tokyo 162-8601, Japan}
		 {defelice@rs.kagu.tus.ac.jp}
		 {http://sites.google.com/site/adefelic/}
\\
\and \\
\epubtkAuthorData{Shinji Tsujikawa}
		 {Department of Physics, Faculty of Science, Tokyo University of Science,\\
		  1-3, Kagurazaka, Shinjuku-ku, Tokyo 162-8601, Japan}
		 {shinji@rs.kagu.tus.ac.jp}
		 {http://www.rs.kagu.tus.ac.jp/shinji/Tsujikawae.html}
}

\date{}
\maketitle

\begin{abstract}
Over the past decade, \fR\ theories have been extensively studied as
one of the simplest modifications to General Relativity. In this
article we review various applications of \fR\ theories to cosmology
and gravity -- such as inflation, dark energy, local gravity
constraints, cosmological perturbations, and spherically symmetric
solutions in weak and strong gravitational backgrounds. We present a
number of ways to distinguish those theories from General Relativity
observationally and experimentally.
We also discuss the extension 
to other modified gravity theories such as 
Brans--Dicke theory and Gauss--Bonnet gravity, and address models 
that can satisfy both cosmological and local gravity constraints.
\end{abstract}

\epubtkKeywords{f(R) gravity, inflation, dark energy, cosmological
  perturbations, modified gravity}

\newpage

\section{Introduction}
\setcounter{equation}{0}

General Relativity (GR)~\cite{Einstein0,Einstein} is widely accepted as a
fundamental theory to describe the geometric properties of
spacetime. In a homogeneous and isotropic spacetime the Einstein
field equations give rise to the Friedmann equations that
describe the evolution of the universe. In fact, the standard big-bang
cosmology based on radiation and matter dominated epochs can be well
described within the framework of General Relativity.

However, the rapid development of observational cosmology which
started from 1990s shows that the universe has undergone two phases of
cosmic acceleration. The first one is called inflation~\cite{Star80,
  Kazanas, Guth, Sato}, which is believed to have occurred prior to
the radiation domination (see~\cite{Lyth, LiddleLyth, Bassett} for
reviews). This phase is required not only to solve the flatness and
horizon problems plagued in big-bang cosmology, but also to explain a
nearly flat spectrum of temperature anisotropies observed in Cosmic
Microwave Background (CMB)~\cite{CMB1}. The second accelerating phase
has started after the matter domination. The unknown component giving
rise to this late-time cosmic acceleration is called dark
energy~\cite{HutTur} (see~\cite{Sahni00, Carroll01, Paddy03,
  Peebles03, CST06, ATbook} for reviews). The existence of dark energy
has been confirmed by a number of observations -- such as supernovae Ia
(SN~Ia)~\cite{SN1, SN2, SN2d}, large-scale structure
(LSS)~\cite{Teg04, Teg06}, baryon acoustic oscillations
(BAO)~\cite{BAO1, BAO2}, and CMB~\cite{WMAP1, WMAP3, WMAP5}.

These two phases of cosmic acceleration cannot be explained by the
presence of standard matter whose equation of state $w=P/\rho$
satisfies the condition $w \ge 0$ (here $P$ and $\rho$ are the
pressure and the energy density of matter, respectively). In fact, we 
further require some component of negative pressure, with
$w<-1/3$, to realize the acceleration of the universe.  
The cosmological constant $\Lambda$ is the simplest candidate
of dark energy, which corresponds to $w=-1$.
However, if the cosmological constant originates from a vacuum 
energy of particle physics, its energy scale is too large to be
compatible with the dark energy density \cite{Weinbergreview}.
Hence we need to find some mechanism to obtain a small value of
$\Lambda$ consistent with observations.
Since the accelerated expansion in the very early universe
needs to end to connect to the radiation-dominated universe, 
the pure cosmological constant is not responsible for inflation.
A scalar field $\phi$ with a slowly varying potential can be 
a  candidate for inflation as well as for dark energy.

Although many scalar-field potentials for inflation have been constructed
in the framework of string theory and supergravity, the CMB
observations still do not show particular evidence to 
favor one of such models.
This situation is also similar in the context of dark
energy---there is a degeneracy as for the potential of the scalar field
(``quintessence''~\cite{quin,quin2,quinold1,quinold2,quinold3,quinold4,quinold5,quinold6}) 
due to the observational degeneracy to the dark energy equation 
of state around $w=-1$.
Moreover it is generally difficult to construct viable quintessence potentials
motivated from particle physics because the field mass responsible for
cosmic acceleration today is very small 
($m_{\phi} \simeq 10^{-33}$\,eV)~\cite{Carroll98,Kolda}.

While scalar-field models of inflation and dark energy correspond to a
modification of the energy-momentum tensor in Einstein equations,
there is another approach to explain the acceleration of the universe.
This corresponds to the modified gravity in which the
gravitational theory is modified compared to GR. The Lagrangian
density for GR is given by $f(R)=R-2\Lambda$, where $R$ is the Ricci
scalar and $\Lambda$ is the cosmological constant (corresponding to
the equation of state $w=-1$).  The presence of $\Lambda$ gives rise
to an exponential expansion of the universe, but we cannot use it for
inflation because the inflationary period needs to connect to the
radiation era. It is possible to use the cosmological constant for
dark energy since the acceleration today does not need to end.
However, if the cosmological constant originates from a vacuum energy
of particle physics, its energy density would be enormously larger
than the today's dark energy density.  While the
$\Lambda$-Cold Dark Matter ($\Lambda$CDM) model ($f(R)=R-2\Lambda$)
fits a number of observational data well~\cite{WMAP5, Kowalski}, there
is also a possibility for the time-varying equation of state of dark
energy~\cite{Alam, Alam2, Nesseris, Nesseris2, Zhao10}.

One of the simplest modifications to GR is the \fR\
gravity in which the Lagrangian density $f$ is an arbitrary function
of $R$~\cite{Bergmann, Ruz, Brei, Buch}. There are two formalisms in
deriving field equations from the action in \fR\ gravity.  The first
is the standard metric formalism in which the field equations are
derived by the variation of the action with respect to the metric
tensor $g_{\mu \nu}$. In this formalism the affine connection
$\Gamma^{\alpha}_{\beta \gamma}$ depends on $g_{\mu
\nu}$. Note that we will consider here and in the remaining
sections only torsion-free theories. The second is the Palatini
formalism~\cite{Palatini1919} in which $g_{\mu \nu}$ and
$\Gamma^{\alpha}_{\beta \gamma}$ are treated as independent variables
when we vary the action. These two approaches give rise to different
field equations for a non-linear Lagrangian density in $R$, while for
the GR action they are identical with each other. In this article we
mainly review the former approach unless otherwise stated. In
Section~\ref{Palasec} we discuss the Palatini formalism in detail.

The model with $f(R)=R+\alpha R^2$ ($\alpha>0$) can lead to the
accelerated expansion of the Universe because of the presence of the
$\alpha R^2$ term.  In fact, this is the first model of inflation
proposed by Starobinsky in 1980~\cite{Star80}. As we will see in
Section~\ref{cosmoinf}, this model is well consistent with the
temperature anisotropies observed in CMB and thus it can be a viable
alternative to the scalar-field models of inflation. Reheating after
inflation proceeds by a gravitational particle production during the
oscillating phase of the Ricci scalar~\cite{Starreheating, Vilenkin,
  Suen}.

The discovery of dark energy in 1998 also stimulated the idea that
cosmic acceleration today may originate from some modification of
gravity to GR. Dark energy models based on \fR\ theories have been
extensively studied as the simplest modified gravity scenario to
realize the late-time acceleration.  The model with a Lagrangian
density $f(R)= R-\alpha /R^{n}$ ($\alpha>0, n>0$) was proposed for
dark energy in the metric formalism~\cite{fRearly1, fRearly2p,
  fRearly2, fRearly3, fRearly4}. However it was shown that this model
is plagued by a matter instability~\cite{Dolgov, Faramatter} as well
as by a difficulty to satisfy local gravity constraints~\cite{OlmoPRL,
  Olmo05, Fara06, Erick06, Chiba07, Navarro, CapoTsuji}. Moreover it
does not possess a standard matter-dominated epoch because of a large
coupling between dark energy and dark matter~\cite{APT, APT2}. These
results show how non-trivial it is to obtain a viable \fR\
model. Amendola et al.~\cite{AGPT} derived conditions for the
cosmological viability of \fR\ dark energy models. In local regions
whose densities are much larger than the homogeneous cosmological
density, the models need to be close to GR for consistency with local
gravity constraints. A number of viable \fR\ models that can satisfy
both cosmological and local gravity constraints have been proposed 
in~\cite{AGPT, LiBarrow, AmenTsuji07, Hu07, Star07, Appleby,
  Tsuji08, Natalie, Cognola07, Linder09}. Since the law of gravity
gets modified on large distances in \fR\ models, this leaves several
interesting observational signatures such as the modification to the
spectra of galaxy clustering~\cite{matterper1, matterper2, SongHu1,
  SongHu2, Teg, TsujiUddin, Pogosian}, CMB~\cite{Zhang05, SongHu1,
  LiBarrow, Peiris}, and weak lensing~\cite{TsujiTate, Schmidt}.  In
this review we will discuss these topics in detail, paying particular
attention to the construction of viable \fR\ models and resulting
observational consequences.

The \fR\ gravity in the metric formalism corresponds to 
generalized Brans--Dicke (BD) theory~\cite{BD} with a BD parameter 
$\omega_{\mathrm{BD}}=0$~\cite{ohanlon, teyssandier, Chiba}.
Unlike original BD theory~\cite{BD}, there
exists a potential for a scalar-field degree of freedom (called
``scalaron''~\cite{Star80}) with a gravitational origin.  If the mass
of the scalaron always remains as light as the present Hubble
parameter $H_0$, it is not possible to satisfy local gravity
constraints due to the appearance of a long-range fifth force with a
coupling of the order of unity. One can design the field potential of
\fR\ gravity such that the mass of the field is heavy in 
the region of high density.
The viable \fR\ models mentioned above have
been constructed to satisfy such a condition. Then the interaction
range of the fifth force becomes short in the region of high density,
which allows the possibility that the models are compatible with local
gravity tests.  More precisely the existence of a matter coupling,
in the Einstein frame, gives rise to an extremum of the
effective field potential around which the field can be stabilized.
As long as a spherically symmetric body has a ``thin-shell'' around
its surface, the field is nearly frozen in most regions inside the
body.  Then the effective coupling between the field and
non-relativistic matter outside the body can be strongly suppressed
through the chameleon mechanism~\cite{chame1, chame2}.
The experiments for the violation of equivalence principle as well as a number of
solar system experiments place tight constraints on dark energy models
based on \fR\ theories~\cite{Hu07, Teg, Tsuji08, CapoTsuji, Van}.

The spherically symmetric solutions mentioned above have been derived
under the weak gravity backgrounds where the background metric is
described by a Minkowski space-time. In strong gravitational
backgrounds such as neutron stars and white dwarfs, we need to take
into account the backreaction of gravitational potentials to the field
equation.  The structure of relativistic stars in \fR\ gravity has
been studied by a number of 
authors~\cite{KM08, KM09, TTT09, Babi1, Upadhye, Nzioki, Babi2, Cooney}.
Originally the difficulty of obtaining relativistic stars was pointed
out in~\cite{KM08} in connection to the singularity problem of \fR\
dark energy models in the high-curvature regime~\cite{Frolov}. For
constant density stars, however, a thin-shell field profile has been
analytically derived in~\cite{TTT09} for chameleon models in the
Einstein frame. The existence of relativistic stars in \fR\ gravity
has been also confirmed numerically for the stars with
constant~\cite{Babi1, Upadhye} and varying~\cite{Babi2} densities. In
this review we shall also discuss this issue.

It is possible to extend \fR\ gravity to generalized BD theory
with a field potential and an arbitrary BD parameter $\omega_{\mathrm{BD}}$.
If we make a conformal transformation to the Einstein 
frame~\cite{DickeCT, Wald, Maeda, Wands94, Faraoni99, FujiiMaeda}, 
we can show that BD theory with a field potential corresponds to the coupled
quintessence scenario~\cite{coupled} with a coupling $Q$ between the field and
non-relativistic matter.  This coupling is related to the BD parameter
via the relation $1/(2Q^2)=3+2\omega_{\mathrm{BD}}$~\cite{chame2, TUMTY}.
One can recover GR by
taking the limit $Q \to 0$, i.e., $\omega_{\mathrm{BD}} \to \infty$.
The \fR\ gravity in the metric formalism
corresponds to $Q=-1/\sqrt{6}$~\cite{APT}, i.e., $\omega_{\mathrm{BD}}=0$.  For large
coupling models with $|Q|={\cal O}(1)$ it is possible to design
scalar-field potentials such that the chameleon mechanism works to
reduce the effective matter coupling, while at the same time the field
is sufficiently light to be responsible for the late-time cosmic
acceleration.  This generalized BD theory also leaves a number of
interesting observational and experimental signatures~\cite{TUMTY}.

In addition to the Ricci scalar $R$, one can construct other scalar
quantities such as $R_{\mu \nu}R^{\mu \nu}$ and $R_{\mu \nu \rho
  \sigma}R^{\mu \nu \rho \sigma}$ from the Ricci tensor $R_{\mu \nu}$
and Riemann tensor $R_{\mu \nu \rho \sigma}$~\cite{CaDe}.
For the Gauss--Bonnet (GB) curvature invariant defined by 
$\GB \equiv R^2-4R_{\alpha\beta}\,R^{\alpha\beta}
+R_{\alpha\beta\gamma\delta}\,R^{\alpha\beta\gamma\delta}$, it is known
that one can avoid the appearance of spurious spin-2 
ghosts~\cite{ghost1, ghost1d, ghost1dd}
(see also~\cite{ghost1ddd, ghost2, ghost2d, ghost2dd, Cal05, DeFelice06, DeFelice06d}). 
In order to give rise to some contribution of the GB term to the Friedmann
equation, we require that (i) the GB term couples to a scalar field
$\phi$, i.e., $F(\phi)\,\GB$ or (ii) the Lagrangian density $f$ is a
function of $\GB$, i.e., $f(\GB)$. 
The GB coupling in the case (i) appears in low-energy string effective 
action~\cite{string} and cosmological solutions in
such a theory have been studied extensively (see~\cite{GBearly1, GBearly2, GBearly3, GBearly4, GBearly5, GBearly6, Ohta04}
for the construction of nonsingular cosmological solutions and~\cite{NO05, Koi, Koi2, TS, Sanyal, Neupane, Neupaned, Neupane2, Amendola}
for the application to dark energy).
In the case (ii) it is possible to construct viable models that are consistent 
with both the background cosmological evolution and local 
gravity constraints~\cite{fGO, DeTsuji1, DeTsuji2}
(see also~\cite{Cognola, DeHind, Davis, LiMota, Zhou, Uddin}).
However density perturbations in perfect fluids exhibit negative instabilities 
during both the radiation and the matter domination, irrespective of the 
form of $f(\GB)$~\cite{LiMota, Mota09}. This growth of perturbations 
gets stronger on smaller scales, which is difficult to be compatible with 
the observed galaxy spectrum unless the deviation from GR is very small. 
We shall review such theories as well as other modified gravity theories.

This review is organized as follows. In Section~\ref{fieldsec} we
present the field equations of \fR\ gravity in the metric formalism.
In Section~\ref{inflationsec} we apply \fR\ theories to the
inflationary universe.  Section~\ref{denergysec} is devoted to the
construction of cosmologically viable \fR\ dark energy models. 
In Section~\ref{lgcsec} local gravity constraints on viable \fR\ 
dark energy models will be discussed.
In Section~\ref{persec} we provide the equations of linear 
cosmological perturbations for general modified gravity theories 
including metric \fR\ gravity as a special case.
In Section~\ref{cosmoinf} we study the spectra of scalar and tensor metric 
perturbations generated during inflation based on \fR\ theories.
In Section~\ref{cosmodark} we discuss the evolution of matter density 
perturbations in \fR\ dark energy models and 
place constraints on model parameters from the observations 
of large-scale structure and CMB.
Section~\ref{Palasec} is devoted to the viability of
the Palatini variational approach in \fR\ gravity.
In Section~\ref{BDsec} we construct viable dark energy models based on 
BD theory with a potential as an extension of \fR\ theories. 
In Section~\ref{starsec} the structure of relativistic stars in
\fR\ theories will be discussed in detail.
In Section~\ref{GBsec} we provide a brief review of Gauss--Bonnet 
gravity and resulting observational and experimental consequences.
In Section~\ref{othersec} we discuss a number of other aspects of 
\fR\ gravity and modified gravity.
Section~\ref{consec} is devoted to conclusions.

There are other review articles on \fR\ 
gravity \cite{SotFaraoni, Sot09, Woodardreview} and modified gravity~\cite{CST06, NOreview, Caporeview,Loboreview,Durrer}.
Compared to those articles, we put more weights on observational and 
experimental aspects of \fR\ theories.
This is particularly useful to place constraints on inflation and 
dark energy models based on \fR\ theories.
The readers who are interested in the more detailed history of 
\fR\ theories and fourth-order gravity may have a look 
at the review articles by Schmidt \cite{Schreview} and 
Sotiriou and Faraoni \cite{SotFaraoni}.

In this review we use units such that $c =\hbar=k_{B} =1$, 
where $c$ is the speed of light, $\hbar$ is reduced Planck's 
constant, and $k_{B}$ is Boltzmann's constant.
We define $\kappa^2=8\pi G=8\pi/m_{\mathrm{pl}}^2=1/M_{\mathrm{pl}}^2$, 
where $G$ is the gravitational constant, $m_{\mathrm{pl}}=1.22 \times
10^{19}\mathrm{\ GeV}$ is the Planck mass with a reduced value 
$M_{\mathrm{pl}}=m_{\mathrm{pl}}/\sqrt{8\pi}=2.44 \times
10^{18}\mathrm{\ GeV}$.
Throughout this review, we use a dot for the derivative with 
respect to cosmic time $t$ and 
 ``${}_{,X}$'' for the partial derivative with respect to the variable $X$, e.g., 
$f_{,R} \equiv \partial f/\partial R$ and $f_{,RR} \equiv \partial^2 f/\partial R^2$. 
We use the metric signature $(-,+,+,+)$.
The Greek indices $\mu$ and $\nu$ run from 0 to 3, 
whereas the Latin indices $i$
and $j$ run from 1 to 3 (spatial components).

\newpage

\section{Field Equations in the Metric Formalism}
\label{fieldsec}
\setcounter{equation}{0}

We start with the 4-dimensional action in \fR\ gravity:
\begin{equation}
S=\frac{1}{2\kappa^{2}}\int \mathrm{d}^{4}x\sqrt{-g}\,f(R)
+\int \mathrm{d}^4 x {\cal L}_M (g_{\mu \nu}, \Psi_M)\,,
\label{fRaction}
\end{equation}
where $\kappa^2=8\pi G$, $g$ is
the determinant of the metric $g_{\mu \nu}$, and ${\cal L}_{M}$ is a matter
Lagrangian\epubtkFootnote{Note that we do not take into account a direct coupling 
between the Ricci scalar and matter (such as $f_1(R) {\cal L}_M$) considered
in ~\cite{Mukoh03, Bertolami1, Bertolami2, Bertolami3, Faracou}.}
that depends on $g_{\mu \nu}$ and matter 
fields $\Psi_{M}$.
The Ricci scalar $R$ is defined by $R=g^{\mu \nu}R_{\mu \nu}$, where
the Ricci tensor $R_{\mu \nu}$ is
\begin{equation}
R_{\mu \nu}=R^\alpha{}_{\mu\alpha\nu}=
\partial_{\lambda} \Gamma^{\lambda}_{\mu \nu}
-\partial_{\mu} \Gamma^{\lambda}_{\lambda \nu}
+\Gamma^{\lambda}_{\mu \nu} \Gamma^{\rho}_{\rho \lambda}
-\Gamma^{\lambda}_{\nu \rho}  \Gamma^{\rho}_{\mu \lambda}\,.
\label{Rmunu}
\end{equation}
In the case of the torsion-less metric formalism, the
connections $\Gamma^{\alpha}_{\beta \gamma}$ are the usual metric
connections defined in terms of the metric tensor $g_{\mu \nu}$, as
\begin{equation}
\Gamma^{\alpha}_{\beta \gamma}=
\frac12 g^{\alpha \lambda} \left( 
\frac{\partial g_{\gamma \lambda}}{\partial x^{\beta}}
+\frac{\partial g_{\lambda \beta}}{\partial x^{\gamma}}
-\frac{\partial g_{\beta \gamma}}{\partial x^{\lambda}}
\right)\,.
\label{Gamma}
\end{equation}
This follows from the metricity relation, 
$\nabla_{\lambda}g_{\mu \nu}=
\partial g_{\mu \nu}/\partial x^{\lambda}
-g_{\rho \nu} \Gamma^{\rho}_{\mu \lambda}
-g_{\mu \rho} \Gamma^{\rho}_{\nu \lambda}=0$.


\subsection{Equations of motion}

The field equation can be derived by varying the action~(\ref{fRaction}) 
with respect to $g_{\mu\nu}$: 
\begin{equation}
\Sigma_{\mu\nu}\equiv F(R)R_{\mu\nu}(g)-\frac{1}{2}f(R)g_{\mu\nu}
-\nabla_{\mu}\nabla_{\nu}F(R)+g_{\mu\nu}\square F(R)
=\kappa^{2}T_{\mu\nu}^{(M)}\,,
\label{fREin}
\end{equation}
where $F(R)\equiv\partial f/\partial R$.
$T_{\mu \nu}^{(M)}$ is the energy-momentum tensor 
of the matter fields defined by the variational derivative 
of ${\cal L}_M$ in terms of $g^{\mu \nu}$:
\begin{equation}
T_{\mu\nu}^{(M)}=-\frac{2}{\sqrt{-g}}
\frac{\delta {\cal L}_M}{\delta g^{\mu \nu}}\,.
\label{Tmunu}
\end{equation}
This satisfies the continuity equation 
\begin{equation}
\nabla^{\mu}T_{\mu\nu}^{(M)}=0\,,
\label{conori}
\end{equation}
as well as $\Sigma_{\mu\nu}$, i.e., $\nabla^\mu\Sigma_{\mu\nu}=0$.\epubtkFootnote{This result is a consequence of the action principle, but it can be derived also by a direct calculation, using the Bianchi identities.}
The trace of Eq.~(\ref{fREin}) gives
\begin{equation}
3\,\square F(R)+F(R)R-2f(R)=\kappa^{2}T\,,
\label{trace}
\end{equation}
where $T=g^{\mu\nu}T_{\mu\nu}^{(M)}$ and 
$\square F=(1/\sqrt{-g})\partial_{\mu}(\sqrt{-g}g^{\mu \nu} \partial_{\nu}F)$.

Einstein gravity, without the cosmological constant, corresponds 
to $f(R)=R$ and $F(R)=1$, so that the term
$\square F(R)$ in Eq.~(\ref{trace}) vanishes. In this case we have
$R=-\kappa^{2}T$ and hence the Ricci scalar $R$ is
directly determined by the matter (the trace $T$).  In modified
gravity the term $\square F(R)$ does not vanish in
Eq.~(\ref{trace}), which means that there is a propagating scalar
degree of freedom, $\varphi \equiv F(R)$. 
The trace equation~(\ref{trace}) determines the dynamics of 
the scalar field $\varphi$ (dubbed {}``scalaron''~\cite{Star80}).

The field equation~(\ref{fREin}) can be written 
in the following form~\cite{Star07}
\begin{equation}
G_{\mu \nu}=\kappa^2 \left( T_{\mu\nu}^{(M)}
+T_{\mu\nu}^{(D)}
\right)\,,
\label{Einmo}
\end{equation}
where $G_{\mu \nu} \equiv R_{\mu \nu}-(1/2)g_{\mu \nu}R$
and 
\begin{equation}
\label{TmunuD}
\kappa^2 T_{\mu \nu}^{(D)} \equiv 
g_{\mu \nu} (f-R)/2+\nabla_{\mu} \nabla_{\nu}F
-g_{\mu \nu}\square F+(1-F) R_{\mu \nu}\,.
\end{equation}
Since $\nabla^{\mu}G_{\mu \nu}=0$ and 
$\nabla^{\mu}T_{\mu \nu}^{(M)}=0$, it follows that 
\begin{equation}
\nabla^{\mu}T_{\mu \nu}^{(D)}=0\,.
\label{TmunuD2}
\end{equation}
Hence the continuity equation holds,
not only for $\Sigma_{\mu\nu}$, but also
for the effective energy-momentum tensor $T_{\mu \nu}^{(D)}$
defined in Eq.~(\ref{TmunuD}).
This is sometimes convenient when we study the 
dark energy equation of state~\cite{Hu07, Star07} as well as 
the equilibrium description of thermodynamics 
for the horizon entropy~\cite{BGT}.

There exists a de~Sitter point that corresponds to a vacuum 
solution ($T=0$) at which the Ricci scalar is constant.
Since $\square F(R)=0$ at this point, we obtain
\begin{equation}
F(R)R-2f(R)=0\,.
\label{fRdeSitter}
\end{equation}
The model $f(R)=\alpha R^{2}$ satisfies this condition, so that it
gives rise to the exact de~Sitter solution~\cite{Star80}.  In the model
$f(R)=R+\alpha R^{2}$, because of the linear term in $R$, the inflationary expansion
ends when the term $\alpha R^{2}$ becomes smaller than the linear term 
$R$ (as we will see in Section~\ref{inflationsec}).
This is followed by a reheating stage in which the oscillation of 
$R$ leads to the gravitational particle production.
It is also possible to use the de~Sitter point given 
by Eq.~(\ref{fRdeSitter}) for dark energy.

We consider the spatially flat Friedmann--Lema\^{\i}tre--Robertson--Walker
(FLRW) spacetime with a time-dependent scale factor $a(t)$ and a metric 
\begin{equation}
\rd s^{2}=g_{\mu \nu} \rd x^{\mu} \rd x^{\nu}
=-\rd t^{2}+a^{2}(t)\,\rd {\bm x}^{2}\,,
\label{FLRW}
\end{equation}
where $t$ is cosmic time. For this metric the Ricci scalar $R$ is given by 
\begin{equation}
R=6 (2H^{2}+\dot{H})\,,
\label{R}
\end{equation}
where $H\equiv\dot{a}/a$ is the Hubble parameter
and a dot stands for a derivative with respect to $t$.
The present value of $H$ is given by 
\begin{equation}
H_0=100\,h\mathrm{\ km\,sec^{-1}\,Mpc^{-1}}
=2.1332\,h \times 10^{-42}\mathrm{\ GeV}\,,
\end{equation}
where $h=0.72 \pm 0.08$ describes the uncertainty 
of $H_0$~\cite{HSTco}.

The energy-momentum tensor of matter is given by 
$T^{\mu}{}^{(M)}_{\nu}= \mathrm{diag}\,(-\rho_M,P_M,P_M,P_M)$,
where $\rho_M$ is the energy density and $P_M$ is 
the pressure.
The field equations~(\ref{fREin}) in the flat FLRW background give
\begin{eqnarray}
\label{fRinf1}
3FH^{2} &=& (FR-f)/2-3H\dot{F}+\kappa^{2} \rho_M\,,\\ 
\label{fRinf2}
-2F\dot{H} &=& \ddot{F}-H\dot{F}
+\kappa^2 (\rho_M+P_M)\,,
\end{eqnarray}
where the perfect fluid satisfies the continuity equation 
\begin{equation}
\label{continuity}
\dot{\rho}_M+3H(\rho_M+P_M)=0\,.
\end{equation}
We also introduce the equation of state of matter, 
$w_M \equiv P_M/\rho_M$. As long as $w_M$ is constant, 
the integration of Eq.~(\ref{continuity}) gives 
$\rho_M \propto a^{-3(1+w_M)}$.
In Section~\ref{denergysec} we shall take into account 
both non-relativistic matter ($w_m=0$)
and radiation ($w_r=1/3$) to discuss cosmological 
dynamics of \fR\ dark energy models.

Note that there are some works about the Einstein static universes in 
\fR\ gravity~\cite{Boehmer:2007tr, Seahra:2009ft}.
Although Einstein static solutions exist for a wide variety of 
\fR\ models in the presence of a barotropic perfect fluid, 
these solutions have been shown to be unstable against 
either homogeneous or inhomogeneous perturbations~\cite{Seahra:2009ft}.


\subsection{Equivalence with Brans--Dicke theory}

The \fR\ theory in the metric formalism can be cast 
in the form of Brans--Dicke (BD) theory~\cite{BD} with a potential for 
the effective scalar-field degree of freedom (scalaron).
Let us consider the following action with a new field $\chi$, 
\begin{equation}
S=\frac{1}{2\kappa^{2}}\int \mathrm{d}^{4}x\sqrt{-g}\,
\left[ f(\chi)+f_{,\chi}(\chi) (R-\chi) \right]
+\int \mathrm{d}^4 x {\cal L}_M (g_{\mu \nu}, \Psi_M)\,.
\label{BifR}
\end{equation}
Varying this action with respect to $\chi$, we obtain 
\begin{equation}
f_{,\chi\chi} (\chi) (R-\chi)=0\,.
\end{equation}
Provided $f_{,\chi\chi}(\chi) \neq 0$ it follows that $\chi=R$.
Hence the action~(\ref{BifR}) recovers the action 
(\ref{fRaction}) in \fR\ gravity.
If we define
\begin{equation}
\varphi \equiv f_{,\chi}(\chi)\,,
\label{varphidef}
\end{equation}
the action~(\ref{BifR}) can be expressed as
\begin{equation}
S=\int \mathrm{d}^{4}x \sqrt{-g}
\left[ \frac{1}{2\kappa^2} \varphi R-U(\varphi) \right]
+\int \mathrm{d}^4 x {\cal L}_M (g_{\mu \nu}, \Psi_M)\,,
\label{BifR2}
\end{equation}
where $U(\varphi)$ is a field potential given by 
\begin{equation}
U(\varphi)=\frac{\chi (\varphi)\,\varphi-f(\chi(\varphi))}
{2\kappa^2}\,.
\label{Uphidef}
\end{equation}

Meanwhile the action in BD theory~\cite{BD} with a potential 
$U(\varphi)$ is given by 
\begin{equation}
S=\int \mathrm{d}^4 x \sqrt{-g} \left[ \frac12 \varphi R
-\frac{\omega_{\mathrm{BD}}}{2\varphi} (\nabla \varphi)^2
-U(\varphi) \right]
+\int \mathrm{d}^4 x {\cal L}_M (g_{\mu \nu}, \Psi_M)\,,
\label{BDac}
\end{equation}
where $\omega_{\mathrm{BD}}$ is the BD 
parameter and $(\nabla \varphi)^2 \equiv g^{\mu \nu}
\partial_{\mu} \varphi \partial_{\nu} \varphi$.
Comparing Eq.~(\ref{BifR2}) with Eq.~(\ref{BDac}), it follows
that \fR\ theory in the metric formalism is equivalent to 
BD theory with the parameter $\omega_{\mathrm{BD}}=0$~\cite{ohanlon, teyssandier, Chiba}
(in the unit $\kappa^2=1$).
In Palatini \fR\ theory where the metric $g_{\mu \nu}$
and the connection $\Gamma^{\alpha}_{\beta \gamma}$ are
treated as independent variables, the Ricci scalar is different from 
that in metric \fR\ theory.
As we will see in Sections~\ref{fieldpala} and \ref{BDtheory}, 
\fR\ theory in the Palatini formalism is equivalent to BD theory 
with the parameter $\omega_{\mathrm{BD}}=-3/2$.


\subsection{Conformal transformation}
\label{secconformo}

The action~(\ref{fRaction}) in \fR\ gravity corresponds to 
a non-linear function $f$ in terms of $R$.
It is possible to derive an action in the Einstein frame 
under the conformal 
transformation~\cite{DickeCT, Wald, Maeda, Wands94, Faraoni99, FujiiMaeda, Magnano}:
\begin{equation}
\tilde{g}_{\mu\nu}=\Omega^{2}\,g_{\mu\nu}\,,
\label{ctrans}
\end{equation}
where $\Omega^2$ is the conformal factor
and a tilde represents quantities in the Einstein frame.
The Ricci scalars $R$ and $\tilde{R}$ in the two frames 
have the following relation
\begin{equation}
R=\Omega^{2}(\tilde{R}+6\tilde{\square}\omega-6\tilde{g}^{\mu\nu}
\partial_{\mu}\omega \partial_{\nu}\omega)\,,\label{Rex}
\end{equation}
where 
\begin{equation}
\omega \equiv \ln\,\Omega\,,\qquad
\partial_{\mu}\omega \equiv \frac{\partial \omega}{\partial \tilde{x}^{\mu}}\,,
\qquad
\tilde{\square}\omega\equiv\frac{1}{\sqrt{-\tilde{g}}}
\partial_{\mu}(\sqrt{-\tilde{g}}\,\tilde{g}^{\mu\nu}\partial_{\nu}\omega)\,.
\label{omega}
\end{equation}

We rewrite the action~(\ref{fRaction}) in the form
\begin{equation}
S=\int\rd^{4}x\sqrt{-g}\left( \frac{1}{2\kappa^{2}}FR-U \right)
+\int \mathrm{d}^4 x {\cal L}_M (g_{\mu \nu}, \Psi_M)\,,\label{fRac2}
\end{equation}
where 
\begin{eqnarray}
U=\frac{FR-f}{2\kappa^{2}}\,.
\label{Udef}
\end{eqnarray}
Using Eq.~(\ref{Rex}) and the relation $\sqrt{-g}=\Omega^{-4}\sqrt{-\tilde{g}}$, the action~(\ref{fRac2}) is transformed as
\begin{equation}
S=\int\rd^{4}x\sqrt{-\tilde{g}}\left[\frac{1}{2\kappa^{2}}F\Omega^{-2}
(\tilde{R}+6\tilde{\square}\omega-6\tilde{g}^{\mu\nu}
\partial_{\mu}\omega \partial_{\nu}\omega )-\Omega^{-4}U\right]
+\int \mathrm{d}^4 x {\cal L}_M (\Omega^{-2}\,\tilde g_{\mu\nu}, \Psi_M)\,.
\label{fRac3}
\end{equation}
We obtain the Einstein frame action (linear action in $\tilde{R}$) for the choice 
\begin{equation}
\Omega^{2}=F\,.
\end{equation}
This choice is consistent if $F>0$. We introduce a new scalar field $\phi$ defined by 
\begin{equation}
\kappa \phi \equiv \sqrt{3/2}\,\,\ln\, F\,.
\label{kappaphi}
\end{equation}
From the definition of $\omega$ in Eq.~(\ref{omega})
we have that $\omega=\kappa \phi/\sqrt{6}$.
Using Eq.~(\ref{omega}), the integral 
$\int\rd^{4}x\sqrt{-\tilde{g}}\,\tilde{\square}\omega$
vanishes on account of the Gauss's theorem. Then
the action in the Einstein frame is
\begin{equation}
S_{E}=\int\rd^{4}x\sqrt{-\tilde{g}}\left[\frac{1}{2\kappa^{2}}\tilde{R}
-\frac{1}{2}\tilde{g}^{\mu\nu}\partial_{\mu}\phi\partial_{\nu}\phi-V(\phi)\right]
+\int \mathrm{d}^4 x {\cal L}_M (F^{-1}(\phi)\tilde{g}_{\mu \nu}, \Psi_M)\,,
\label{Ein}
\end{equation}
where 
\begin{equation}
V(\phi)=\frac{U}{F^2}=\frac{FR-f}{2\kappa^{2}F^2}\,.
\label{potentialein}
\end{equation}
Hence the Lagrangian density of the field $\phi$ is given by 
${\cal L}_{\phi}=-\frac{1}{2}\tilde{g}^{\mu\nu}
\partial_{\mu}\phi\partial_{\nu}\phi-V(\phi)$ with 
the energy-momentum tensor 
\begin{equation}
\label{Tphi}
\tilde{T}_{\mu \nu}^{(\phi)}=
-\frac{2}{\sqrt{-\tilde{g}}} \frac{\delta (\sqrt{-\tilde{g}}{\cal L}_{\phi})}
{\delta \tilde{g}^{\mu \nu}}=
\partial_{\mu}\phi \partial_{\nu} \phi-\tilde{g}_{\mu \nu}
\left[ \frac12 \tilde{g}^{\alpha \beta} \partial_{\alpha} \phi
\partial_{\beta} \phi+V(\phi) \right]\,.
\end{equation}

The conformal factor $\Omega^2=F=\exp (\sqrt{2/3}\,\kappa \phi)$
is field-dependent. From the matter action~(\ref{Ein}) 
the scalar field $\phi$ is directly coupled to 
matter in the Einstein frame. 
In order to see this more explicitly, we take the variation of the action~(\ref{Ein}) with respect to the field $\phi$:
\begin{equation}
-\partial_{\mu} \left( \frac{\partial (\sqrt{-\tilde{g}}{\cal L}_\phi)}
{\partial (\partial_{\mu}\phi)} \right)
+\frac{\partial (\sqrt{-\tilde{g}}{\cal L}_\phi)}{\partial \phi}+
\frac{\partial{\cal L}_M}{\partial \phi}=0\,,
\end{equation}
that is 
\begin{equation}
\label{fieldeq1}
\tilde{\square} \phi-V_{,\phi}+\frac{1}{\sqrt{-\tilde{g}}}
\frac{\partial {\cal L}_M}{\partial \phi}=0\,,\qquad
\mathrm{where} \qquad
\tilde{\square} \phi  \equiv\frac{1}{\sqrt{-\tilde{g}}}
\partial_{\mu}(\sqrt{-\tilde{g}}\,\tilde{g}^{\mu\nu}\partial_{\nu}
\phi)\,.
\end{equation}
Using Eq.~(\ref{ctrans}) and the relations $\sqrt{-\tilde{g}}=F^2 \sqrt{-g}$
and $\tilde{g}^{\mu \nu}=F^{-1}g^{\mu \nu}$, 
the energy-momentum tensor of matter is transformed as 
\begin{equation}
\label{TM}
\tilde{T}_{\mu \nu}^{(M)}=-\frac{2}{\sqrt{-\tilde{g}}}
\frac{\delta {\cal L}_M}{\delta \tilde{g}^{\mu \nu}}=
\frac{T_{\mu \nu}^{(M)}}{F}\,.
\end{equation}
The energy-momentum tensor of perfect fluids in the 
Einstein frame is given by 
\begin{equation}
\tilde{T}^{\mu}{}^{(M)}_{\nu}= \mathrm{diag} (-\tilde{\rho}_M, \tilde{P}_M, 
\tilde{P}_M, \tilde{P}_M)= \mathrm{diag} (-\rho_M/F^2, P_M/F^2, 
P_M/F^2, P_M/F^2)\,.
\end{equation}
The derivative of the Lagrangian density ${\cal L}_M={\cal L}_M(g_{\mu \nu})
={\cal L}_M(F^{-1}(\phi) \tilde{g}_{\mu \nu})$ with respect to 
$\phi$ is 
\begin{equation}
\frac{\partial {\cal L}_{M}}{\partial \phi}=
\frac{\delta {\cal L}_M}{\delta g^{\mu \nu}} 
\frac{\partial g^{\mu \nu}}{\partial \phi}=
\frac{1}{F(\phi)} \frac{\delta {\cal L}_M}{\delta \ti{g}^{\mu \nu}}
\frac{\partial (F(\phi)\ti{g}^{\mu \nu})}{\partial \phi}=
-\sqrt{-\ti{g}}\,\frac{F_{,\phi}}{2F}\,
\ti{T}_{\mu \nu}^{(M)} \ti{g}^{\mu \nu}\,.
\end{equation}

The strength of the coupling between the field and matter can be 
quantified by the following quantity
\begin{equation}
Q \equiv -\frac{F_{,\phi}}{2\kappa F}
=-\frac{1}{\sqrt{6}}\,,
\label{Qdef}
\end{equation}
which is constant in \fR\ gravity~\cite{APT}.
It then follows that 
\begin{equation}
\frac{\partial {\cal L}_{M}}{\partial \phi}=
\sqrt{-\tilde{g}}\,\kappa Q\tilde{T}\,,
\label{parL}
\end{equation}
where $\tilde{T}=\tilde{g}_{\mu \nu} \tilde{T}^{\mu \nu(M)}=
-\ti{\rho}_M+3\ti{P}_M$.
Substituting Eq.~(\ref{parL}) into Eq.~(\ref{fieldeq1}), we obtain 
the field equation in the Einstein frame:
\begin{equation}
\label{fieldeq2}
\tilde{\square} \phi-V_{,\phi}+
\kappa Q \tilde{T}=0\,.
\end{equation}
This shows that the field $\phi$ is directly coupled  to 
matter apart from radiation ($\tilde{T}=0$).

Let us consider the flat FLRW spacetime with the 
metric (\ref{FLRW}) in the Jordan frame.
The metric in the Einstein frame is given by 
\begin{eqnarray}
\mathrm{d} \tilde{s}^2=\Omega^2 \mathrm{d}s^2
&=& F (-\rd t^{2}+a^{2}(t)\,\rd {\bm x}^{2})\,, \nonumber \\
&=& -\rd \tilde{t}^{2}+\tilde{a}^{2}(\tilde{t})\,\rd {\bm x}^{2}\,,
\end{eqnarray}
which leads to the following relations (for $F>0$)
\begin{equation}
\rd \tilde{t}=\sqrt{F} \rd t\,,\qquad
\tilde{a}=\sqrt{F}a\,,
\label{transre}
\end{equation}
where
\begin{equation}
F=e^{-2Q\kappa \phi}\,.
\label{Fexp}
\end{equation}
Note that Eq.~(\ref{Fexp}) comes from the integration of 
Eq.~(\ref{Qdef}) for constant $Q$. 
The field equation~(\ref{fieldeq2}) can be expressed as
\begin{equation}
\frac{\rd^2 \phi}{\rd \tilde{t}^2}
+3\tilde{H} \frac{\rd \phi}{\rd \tilde{t}}
+V_{,\phi}=-\kappa Q (\tilde{\rho}_M-3\tilde{P}_M)\,,
\label{field3}
\end{equation}
where 
\begin{equation}
\tilde{H} \equiv \frac{1}{\ti{a}} \frac{\mathrm{d}\tilde{a}}
{\mathrm{d} \ti{t}}=\frac{1}{\sqrt{F}} \left( H+\frac{\dot{F}}{2F}
 \right)\,.
\label{transre2}
\end{equation}
Defining the energy density $\tilde{\rho}_{\phi}=\tfrac12(\rd \phi/\rd \tilde{t})^2+V(\phi)$
and the pressure $\tilde{P}_{\phi}=\tfrac12(\rd \phi/\rd \tilde{t})^2-V(\phi)$, 
Eq.~(\ref{field3}) can be written as
\begin{equation}
\label{cou1}
\frac{\rd \tilde{\rho}_{\phi}}{\rd \tilde{t}}+3\tilde{H}
(\tilde{\rho}_{\phi}+\tilde{P}_{\phi})=
-\kappa Q (\tilde{\rho}_M-3\tilde{P}_M)
\frac{\rd \phi}{\rd \tilde{t}}\,.
\end{equation}
Under the transformation (\ref{transre}) together with 
$\rho_M=F^2 \tilde{\rho}_M$, $P_M=F^2 \tilde{P}_M$, 
and $H=F^{1/2} [\tilde{H}-(\rd F/\rd \tilde{t})/2F]$, 
the continuity equation~(\ref{continuity}) is 
transformed as 
\begin{equation}
\label{cou2}
\frac{\rd \tilde{\rho}_M}{\rd \tilde{t}}+3\tilde{H}
(\tilde{\rho}_M+\tilde{P}_M)=
\kappa Q (\tilde{\rho}_M-3\tilde{P}_M)
\frac{\rd \phi}{\rd \tilde{t}}\,.
\end{equation}

Equations (\ref{cou1}) and (\ref{cou2}) show that the field and 
matter interacts with each other, while 
the total energy density $\tilde{\rho}_T=\tilde{\rho}_{\phi}+
\tilde{\rho}_{M}$ and the pressure $\tilde{P}_T=\tilde{P}_{\phi}+
\tilde{P}_{M}$ satisfy the continuity equation 
$\rd \tilde{\rho}_T/\rd \tilde{t}+3\tilde{H} (\tilde{\rho}_T+\tilde{P}_T)=0$.
More generally, Eqs.~(\ref{cou1}) and (\ref{cou2}) can be expressed
in terms of the energy-momentum tensors 
defined in Eqs.~(\ref{Tphi}) and (\ref{TM}):
\begin{equation}
\tilde{\nabla}_{\mu} \tilde{T}^{\mu (\phi)}_{\nu}=
-Q \tilde{T} \tilde{\nabla}_{\nu}\phi\,,\qquad
\tilde{\nabla}_{\mu} \tilde{T}^{\mu (M)}_{\nu}=
Q \tilde{T} \tilde{\nabla}_{\nu}\phi\,,
\end{equation}
which correspond to the same equations in coupled quintessence
studied in~\cite{coupled} (see also~\cite{coupledpre}).

In the absence of a field potential $V(\phi)$ 
(i.e., massless field) the field mediates a long-range
fifth force with a large coupling ($|Q| \simeq 0.4$), 
which contradicts with experimental tests in the 
solar system.
In \fR\ gravity a field potential with gravitational
origin is present, which allows the possibility 
of compatibility with local gravity tests
through the chameleon mechanism~\cite{chame1, chame2}.

In \fR\ gravity the field $\phi$ is coupled to non-relativistic matter
(dark matter, baryons) with a universal coupling $Q=-1/\sqrt{6}$.
We consider the frame in which the baryons obey the standard
continuity equation $\rho_m \propto a^{-3}$, i.e., the Jordan 
frame, as the ``physical'' frame in which physical quantities
are compared with observations and experiments.
It is sometimes convenient to refer the Einstein frame
in which a canonical scalar field is coupled to non-relativistic matter.
In both frames we are treating the same physics, but using the 
different time and length scales gives rise to the apparent 
difference between the observables in two frames.
Our attitude throughout the review is to discuss observables
in the Jordan frame. When we transform to the Einstein 
frame for some convenience, we go back to the Jordan frame
to discuss physical quantities.

\newpage

\section{Inflation in \fR\ Theories}
\label{inflationsec}
\setcounter{equation}{0}

Most models of inflation in the early universe are 
based on scalar fields appearing in superstring and supergravity theories.
Meanwhile, the first inflation model proposed by Starobinsky~\cite{Star80}
is related to the conformal anomaly in quantum 
gravity\epubtkFootnote{There are some other works about theoretical 
constructions of \fR\ models based on quantum gravity, supergravity
and extra dimensional theories~\cite{Ketov:2009wc, Klinkhamer:2008ff, Shojai:2008er, Machado:2007ea,Codello,Gunther,Gunther2,Saidov1,Saidov2}.}.
Unlike the models such as ``old inflation''~\cite{Kazanas, Guth, Sato} 
this scenario is not plagued by the graceful exit 
problem -- the period of cosmic acceleration 
is followed by the radiation-dominated epoch with a transient
matter-dominated phase~\cite{Starreheating, Vilenkin, Suen}.
Moreover it predicts nearly scale-invariant spectra of gravitational 
waves and temperature anisotropies consistent with CMB
observations~\cite{Star79, Mukha81, Star83, Kofman87, Hwang01}.
In this section we review the dynamics of inflation and reheating.
In Section~\ref{cosmoinf} we will discuss the power spectra of scalar and
tensor perturbations generated in \fR\ inflation models.


\subsection{Inflationary dynamics}
\label{infdynamics}

We consider the models of the form 
\begin{equation}
\label{fRinf}
f(R)=R+\alpha R^n\,,\qquad
(\alpha>0, n>0)\,,
\end{equation}
which include the Starobinsky's model~\cite{Star80}
as a specific case ($n=2$).
In the absence of the matter fluid ($\rho_M=0$), 
Eq.~(\ref{fRinf1}) gives
\begin{equation}
3(1+n \alpha R^{n-1})H^2=\frac12 (n-1)\alpha R^n
-3n(n-1)\alpha H R^{n-2} \dot{R}\,.
\label{Heqinf}
\end{equation}
The cosmic acceleration can be realized 
in the regime $F=1+n \alpha R^{n-1} \gg 1$.
Under the approximation $F \simeq n \alpha R^{n-1}$,
we divide Eq.~(\ref{Heqinf}) by $3n \alpha R^{n-1}$ to give
\begin{equation}
H^2 \simeq \frac{n-1}{6n} \left( R-6nH \frac{\dot{R}}{R} \right)\,.
\label{Heqinf2}
\end{equation}

During inflation the Hubble parameter $H$ evolves slowly 
so that one can use the approximation $|\dot{H}/H^2| \ll 1$
and $|\ddot{H}/(H\dot{H})| \ll 1$. 
Then Eq.~(\ref{Heqinf2}) reduces to 
\begin{equation}
\frac{\dot{H}}{H^2} \simeq  -\epsilon_1, \qquad
\epsilon_1 = \frac{2-n}{(n-1)(2n-1)}\,.
\end{equation}
Integrating this equation for $\epsilon_1>0$, we obtain the solution 
\begin{equation}
H \simeq \frac{1}{\epsilon_1 t}\,, \qquad
a \propto t^{1/\epsilon_1}\,.
\end{equation}
The cosmic acceleration occurs for $\epsilon_1<1$, 
i.e., $n>(1+\sqrt{3})/2$.
When $n=2$ one has $\epsilon_1=0$, so that $H$ is constant 
in the regime $F \gg 1$. The models with $n>2$ lead to 
super inflation characterized by $\dot{H}>0$ and 
$a \propto |t_0-t|^{-1/|\epsilon_1|}$ ($t_0$ is a constant).
Hence the standard inflation with decreasing $H$ occurs for 
$(1+\sqrt{3})/2<n<2$. 

In the following let us focus on the Starobinsky's 
model given by 
\begin{equation}
f(R)=R+R^2/(6M^2)\,,
\label{stamodel}
\end{equation}
where the constant $M$ has a dimension of mass.
The presence of the linear term in $R$ eventually
causes inflation to end.
Without neglecting this linear term, the combination of 
Eqs.~(\ref{fRinf1}) and (\ref{fRinf2}) gives
\begin{eqnarray}
& & \ddot{H}-\frac{\dot{H}^2}{2H}
+\frac12 M^2 H=-3H\dot{H}\,,
\label{steq1}
\\
& & \ddot{R}+3H\dot{R}+M^2 R=0\,.
\label{steq2}
\end{eqnarray}
During inflation the first two terms in Eq.~(\ref{steq1})
can be neglected relative to others, which gives
$\dot{H} \simeq -M^2/6$.
We then obtain the solution 
\begin{eqnarray}
H &\simeq& H_i-(M^2/6)(t-t_i)\,,
\label{apeq1} \\
a &\simeq& a_i \exp \left[ H_i (t-t_i)-(M^2/12)(t-t_i)^2
\right]\,,
\label{apeq2} \\
R &\simeq& 12H^2-M^2\,,
\label{apeq3} 
\end{eqnarray}
where $H_i$ and $a_i$ are the Hubble parameter and 
the scale factor at the onset of inflation ($t=t_i)$,
respectively. 
This inflationary solution is a transient attractor of
the dynamical system~\cite{Maeda88}.
The accelerated expansion continues as long as the slow-roll parameter
\begin{equation}
\epsilon_1= -\frac{\dot{H}}{H^2} \simeq 
\frac{M^2}{6H^2}\,,
\label{ep1re}
\end{equation}
is smaller than the order of unity, i.e., $H^2 \gtrsim M^2$.
One can also check that the approximate relation 
$3H \dot{R}+M^2R \simeq 0$ holds in Eq.~(\ref{steq2})
by using $R \simeq 12H^2$.
The end of inflation (at time $t=t_f$) is characterized by the condition 
$\epsilon_f \simeq 1$, i.e., $H_f\simeq M/\sqrt{6}$.
From Eq.~(\ref{apeq3}) this corresponds to the epoch at which 
the Ricci scalar decreases to $R \simeq M^2$.
As we will see later, the WMAP normalization of the CMB 
temperature anisotropies constrains the mass scale
to be $M \simeq 10^{13}\mathrm{\ GeV}$.
Note that the phase space analysis for the model~(\ref{stamodel})
was carried out in~\cite{Maeda88, Amendola92, Capo93}.

We define the number of e-foldings from $t=t_i$ to $t=t_f$:
\begin{equation}
N \equiv \int_{t_i}^{t_f} H\,\mathrm{d}t
 \simeq H_i (t_f-t_i) -\frac{M^2}{12} (t_f-t_i)^2\,.
 \label{efold1}
\end{equation}
Since inflation ends at $t_f \simeq t_i +6H_i/M^2$, it follows that 
\begin{equation}
N \simeq \frac{3H_i^2}{M^2} \simeq \frac{1}{2\epsilon_1(t_i)}\,,
\label{Nstamodel}
\end{equation}
where we used Eq.~(\ref{ep1re}) in the last approximate equality.
In order to solve horizon and flatness problems of the big bang cosmology
we require that $N \gtrsim 70$~\cite{LiddleLyth}, 
i.e.,\ $\epsilon_1 (t_i) \lesssim 7 \times 10^{-3}$.
The CMB temperature anisotropies correspond to the perturbations 
whose wavelengths crossed the Hubble radius around $N=55\mbox{\,--\,}60$
before the end of inflation.


\subsection{Dynamics in the Einstein frame}

Let us consider inflationary dynamics in the Einstein frame for the 
model~(\ref{stamodel}) in the absence of matter fluids (${\cal L}_M=0$).
The action in the Einstein frame corresponds to (\ref{Ein})
with a field $\phi$ defined by 
\begin{equation}
\phi=\sqrt{\frac32} \frac{1}{\kappa} \ln F=
\sqrt{\frac32} \frac{1}{\kappa}\,
\ln \left( 1+\frac{R}{3M^2} \right)\,.
\end{equation}
Using this relation, the field potential (\ref{potentialein}) 
reads~\cite{Maeda, Barrow88, BarrowCo}
\begin{equation}
V(\phi)=\frac{3M^2}{4\kappa^2}
\left( 1-e^{-\sqrt{2/3}\kappa \phi} \right)^2\,.
\label{poein}
\end{equation}

\epubtkImage{potential.png}{%
  \begin{figure}[hptb]
    \centerline{\includegraphics[width=3.4in]{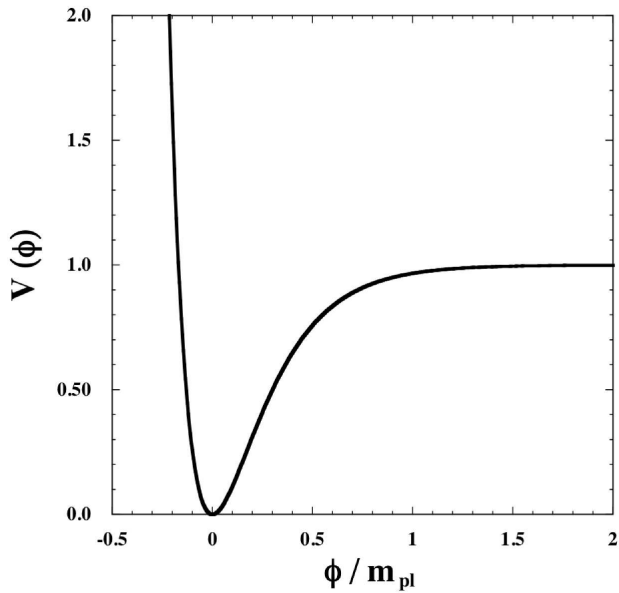}}
    \caption{The field potential~(\ref{poein}) in the Einstein frame
      corresponding to the model~(\ref{stamodel}). Inflation is
      realized in the regime $\kappa \phi \gg 1$.}
    \label{poein2} 
\end{figure}}

In Figure~\ref{poein2} we illustrate the potential (\ref{poein})
as a function of $\phi$. In the regime $\kappa \phi \gg 1$ the 
potential is nearly constant ($V(\phi) \simeq 3M^2/(4\kappa^2)$), 
which leads to slow-roll inflation.
The potential in the regime $\kappa \phi \ll 1$ is given by 
$V(\phi) \simeq (1/2)M^2 \phi^2$, so that the field oscillates 
around $\phi=0$ with a Hubble damping.
The second derivative of $V$ with respect to $\phi$ is 
\begin{equation}
V_{,\phi \phi}=-M^2 e^{-\sqrt{2/3}\kappa \phi}
\left( 1-2 e^{-\sqrt{2/3}\kappa \phi} \right)\,,
\label{Vphiphi}
\end{equation}
which changes from negative to positive at 
$\phi=\phi_1 \equiv \sqrt{3/2} (\ln 2)/\kappa \simeq 0.169m_{\mathrm{pl}}$.

Since $F \simeq 4H^2/M^2$ during inflation, the transformation 
(\ref{transre}) gives a relation between 
the cosmic time $\tilde{t}$ in the Einstein frame and 
that in the Jordan frame:
\begin{equation}
\tilde{t}=\int_{t_i}^{t} \sqrt{F}\,\mathrm{d}t
 \simeq \frac{2}{M} \left[ H_i (t-t_i)
 -\frac{M^2}{12} (t-t_i)^2 \right]\,,
 \label{tiltre}
\end{equation}
where $t=t_i$ corresponds to $\tilde{t}=0$.
The end of inflation ($t_f \simeq t_i +6H_i/M^2$) 
corresponds to $\tilde{t}_f=(2/M)N$ in the Einstein frame, 
where $N$ is given in Eq.~(\ref{efold1}).
On using Eqs.~(\ref{apeq2}) and (\ref{tiltre}), the scale factor 
$\tilde{a}=\sqrt{F}a$ in the Einstein frame evolves as 
\begin{equation}
\tilde{a}(\tilde{t}) \simeq
\left( 1-\frac{M^2}{12H_i^2} M\tilde{t}
\right)  \tilde{a}_i \,e^{M \tilde{t}/2}\,,
\label{aein}
\end{equation}
where $\tilde{a}_i=2H_ia_i/M$.
Similarly the evolution of the Hubble parameter 
$\tilde{H}=(H/\sqrt{F})[1+\dot{F}/(2HF)]$ 
is given by 
\begin{equation}
\tilde{H}(\tilde{t}) \simeq \frac{M}{2} \left[ 1-\frac{M^2}{6H_i^2}
\left( 1-\frac{M^2}{12H_i^2}M \tilde{t} \right)^{-2} \right]\,,
\label{Hein}
\end{equation}
which decreases with time.
Equations (\ref{aein}) and (\ref{Hein}) show that the universe 
expands quasi-exponentially in the Einstein frame as well.

The field equations for the action~(\ref{Ein})
are given by 
\begin{eqnarray}
& & 3\tilde{H}^2=\kappa^2 \left[ \frac12 
\left( \frac{\mathrm{d}\phi}
{\mathrm{d}\tilde{t}} \right)^2+V(\phi) \right]\,,
\label{eein1}\\
& & \frac{\rd^2 \phi}{\rd \tilde{t}^2}
+3\tilde{H} \frac{\rd \phi}{\rd \tilde{t}}
+V_{,\phi}=0
\label{eein2}\,.
\end{eqnarray}
Using the slow-roll approximations
$(\mathrm{d}\phi/\mathrm{d}  \tilde{t})^2 \ll V(\phi)$ and 
$|\mathrm{d}^2\phi/\mathrm{d}  \tilde{t}^2| \ll |\tilde{H} \mathrm{d}\phi/\mathrm{d} \tilde{t}|$
during inflation, one has $3 \tilde{H^2} \simeq \kappa^2 V(\phi)$ and 
$3\tilde{H} (\mathrm{d}\phi/\mathrm{d}\tilde{t})+V_{,\phi} \simeq 0$.
We define the slow-roll parameters
\begin{equation}
\tilde{\epsilon}_1 \equiv -\frac{\mathrm{d}\tilde{H}/\mathrm{d}\tilde{t}}
{\tilde{H}^2} \simeq \frac{1}{2\kappa^2} 
\left( \frac{V_{,\phi}}{V} \right)^2\,,\qquad
\tilde{\epsilon}_2 \equiv \frac{\mathrm{d}^2\phi/\mathrm{d}\tilde{t}^2}
{\tilde{H} (\mathrm{d}\phi/\mathrm{d}\tilde{t})}
\simeq \tilde{\epsilon}_1-\frac{V_{,\phi \phi}}{3 \tilde{H}^2}\,.
\end{equation}
For the potential (\ref{poein}) it follows that 
\begin{equation}
\tilde{\epsilon}_1 \simeq \frac43 (e^{\sqrt{2/3}\kappa \phi}-1)^{-2}\,,
\qquad
\tilde{\epsilon}_2 \simeq \tilde{\epsilon}_1+\frac{M^2}{3\tilde{H}^2}
e^{-\sqrt{2/3}\kappa \phi} (1-2e^{-\sqrt{2/3}\kappa \phi})\,,
\label{ep1ep2}
\end{equation}
which are much smaller than 1 during inflation ($\kappa \phi \gg 1$).
The end of inflation is characterized by the condition 
$\{\ti{\epsilon}_1,|\ti{\epsilon}_2| \}={\cal O}(1)$. 
Solving $\ti{\epsilon}_1=1$, we obtain the field value 
$\phi_f \simeq 0.19\,m_{\mathrm{pl}}$.

We define the number of e-foldings in the Einstein frame,
\begin{equation}
\tilde{N}=\int_{\tilde{t}_i}^{\tilde{t}_f} \tilde{H} \mathrm{d} \tilde{t}
\simeq \kappa^2 \int_{\phi_f}^{\phi_i} \frac{V}{V_{,\phi}} \mathrm{d}\phi\,, 
\end{equation}
where $\phi_i$ is the field value at the onset of inflation.
Since $\tilde{H}\mathrm{d}\tilde{t}=H\mathrm{d}t [1+\dot{F}/(2HF)]$, 
it follows that $\tilde{N}$ is identical to $N$ in the slow-roll limit:
$|\dot{F}/(2HF)| \simeq |\dot{H}/H^2| \ll 1$.
Under the condition $\kappa \phi_i \gg 1$ we have
\begin{equation}
\label{tilN}
\tilde{N} \simeq \frac34 e^{\sqrt{2/3}\kappa \phi_i}\,.
\end{equation}
This shows that $\phi_i \simeq 1.11m_{\mathrm{pl}}$ for 
$\tilde{N}=70$.
From Eqs.~(\ref{ep1ep2}) and (\ref{tilN}) together with 
the approximate relation $\tilde{H} \simeq M/2$, 
we obtain 
\begin{equation}
\tilde{\epsilon}_1 \simeq \frac{3}{4\tilde{N}^2}\,,\qquad
\tilde{\epsilon}_2 \simeq \frac{1}{\tilde{N}}\,,
\label{tiep}
\end{equation}
where, in the expression of $\tilde{\epsilon}_2$, 
we have dropped the terms of the order of $1/\tilde{N}^2$.
The results (\ref{tiep}) will be used to estimate the spectra 
of density perturbations in Section~\ref{cosmoinf}.


\subsection{Reheating after inflation}
\label{reheatingsec}

We discuss the dynamics of reheating and the resulting 
particle production in the Jordan frame for the model~(\ref{stamodel}).
The inflationary period is followed by a reheating phase in which 
the second derivative $\ddot{R}$ can no longer be neglected
in Eq.~(\ref{steq2}). Introducing $\hat{R}=a^{3/2}R$, we have

\begin{equation}
\label{ddotReq}
\ddot{\hat{R}}+\left(M^2-\frac34 H^2-\frac32 \dot{H} 
\right)\hat{R}=0\,.
\end{equation}
Since $M^2 \gg \{H^2, |\dot{H}| \}$ during reheating, the solution to
Eq.~(\ref{ddotReq}) is given by that of the harmonic oscillator with 
a frequency $M$. Hence the Ricci scalar exhibits a damped oscillation 
around $R=0$:
\begin{equation}
R \propto a^{-3/2} \sin (Mt)\,.
\end{equation}

Let us estimate the evolution of the Hubble parameter and the scale factor
during reheating in more detail.
If we neglect the r.h.s.\ of Eq.~(\ref{steq1}), we get 
the solution $H(t)= \mathrm{const} \times \cos^2 (Mt/2)$.
Setting $H(t)=f(t) \cos^2 (Mt/2)$ to derive 
the solution of Eq.~(\ref{steq1}), we obtain~\cite{Suen}
\begin{eqnarray}
\label{ddotH2}
f(t)=\frac{1}{C+(3/4)(t-t_{\mathrm{os}})+3/(4M) \sin [M (t-t_{\mathrm{os}})]}\,,
\end{eqnarray}
where $t_{\mathrm{os}}$ is the time at the onset of reheating.
The constant $C$ is determined by matching Eq.~(\ref{ddotH2})
with the slow-roll inflationary solution $\dot{H}=-M^2/6$ at $t=t_{\mathrm{os}}$.
Then we get $C=3/M$ and 
\begin{equation}
H(t)=\left[ \frac{3}{M}+\frac34 (t-t_{\mathrm{os}})+\frac{3}{4M}
\sin M(t-t_{\mathrm{os}}) \right]^{-1} \cos^2 \left[ \frac{M}{2}
(t-t_{\mathrm{os}}) \right]\,.
\end{equation}
Taking the time average of oscillations in the regime
$M(t-t_{\mathrm{os}}) \gg 1$, it follows that 
$\langle H \rangle \simeq (2/3)(t-t_{\mathrm{os}})^{-1}$.
This corresponds to the cosmic evolution during the matter-dominated
epoch, i.e., $\langle a \rangle \propto (t-t_{\mathrm{os}})^{2/3}$.
The gravitational effect of coherent oscillations of scalarons with mass $M$
is similar to that of a pressureless perfect fluid.
During reheating the Ricci scalar is approximately given by 
$R \simeq 6\dot{H}$, i.e.
\begin{equation}
R \simeq -3\left[ \frac{3}{M}+\frac34 (t-t_{\mathrm{os}})+\frac{3}{4M}
\sin M(t-t_{\mathrm{os}}) \right]^{-1}M \sin \left[ M(t-t_{\mathrm{os}})
\right]\,.
\end{equation}
In the regime $M(t-t_{\mathrm{os}}) \gg 1$ this behaves as
\begin{equation}
R \simeq -\frac{4M}{t-t_{\mathrm{os}}}
\sin \left[M(t-t_{\mathrm{os}}) \right]\,.
\label{Rap}
\end{equation}

In order to study particle production during reheating, 
we consider a scalar field $\chi$ with mass $m_{\chi}$.
We also introduce a nonminimal coupling $(1/2)\xi R \chi^2$
between the field $\chi$ and the Ricci scalar $R$~\cite{Birrell}.
Then the action is given by 
\begin{equation}
S=\int \mathrm{d}^4x \sqrt{-g} \left[ \frac{f(R)}{2\kappa^2}
-\frac12 g^{\mu \nu} \partial_{\mu} \chi \partial_{\nu} \chi
-\frac12 m_{\chi}^2 \chi^2-\frac12 \xi R \chi^2 \right]\,,
\end{equation}
where $f(R)=R+R^2/(6M^2)$. 
Taking the variation of this action with respect to $\chi$ gives
\begin{equation}
\square \chi - m_{\chi}^2 \chi -\xi R \chi=0\,.	
\end{equation}
We decompose the quantum field $\chi$ in terms of the 
Heisenberg representation:
\begin{equation}
\chi (t, {\bm x})=\frac{1}{(2\pi)^{3/2}} 
\int \mathrm{d}^3k \left( \hat{a}_k \chi_k (t) e^{-i {\bm k}\cdot {\bm x}}
+\hat{a}^{\dagger}_k \chi_k^* (t) e^{i {\bm k}\cdot {\bm x}}
\right)\,,
\end{equation}
where $\hat{a}_k$ and $\hat{a}^{\dagger}_k$ are annihilation and 
creation operators, respectively.
The field $\chi$ can be quantized in curved spacetime by generalizing
the basic formalism of quantum field theory in the flat spacetime.
See the book \cite{Birrell} for the detail of quantum field theory
in curved spacetime. 
Then each Fourier mode $\chi_k (t)$ obeys the following 
equation of motion 
\begin{equation}
\ddot{\chi}_k+3H \dot{\chi}_k+\left( \frac{k^2}{a^2}
+m_{\chi}^2+\xi R \right) \chi_k=0\,,
\label{chieq}
\end{equation}
where $k=|{\bm k}|$ is a comoving wavenumber.
Introducing a new field $u_k=a \chi_k$ and conformal time
$\eta=\int a^{-1} \mathrm{d}t$, we obtain 
\begin{equation}
\frac{\mathrm{d}^2 u_k}{\mathrm{d}\eta^2}+
\left[ k^2+m_{\chi}^2a^2+\left( \xi -\frac16 
\right)a^2 R \right]u_k=0\,,
\label{ukeq}
\end{equation}
where the conformal coupling correspond to 
$\xi=1/6$. This result states that, even though $\xi=0$ 
(that is, the field is minimally coupled to gravity), 
$R$ still gives a contribution to the effective mass of $u_k$.
In the following we first review the reheating scenario 
based on a minimally coupled massless field 
($\xi=0$ and $m_{\chi}=0$).
This corresponds to the gravitational particle production 
in the perturbative regime~\cite{Starreheating, Vilenkin, Suen}.
We then study the case in which the nonminimal coupling
$|\xi|$ is larger than the order of 1. 
In this case the non-adiabatic particle production 
\emph{preheating}~\cite{Tra90, KLS94, Yuri95, KLS97}
can occur via parametric resonance.

\subsubsection{Case: $\xi=0$ and $m_{\chi}=0$}

In this case there is no explicit coupling among the fields $\chi$ and $R$. 
Hence the $\chi$ particles are produced only gravitationally.
In fact, Eq.~(\ref{ukeq}) reduces to 
\begin{equation}
\frac{\mathrm{d}^2 u_k}{\mathrm{d}\eta^2}
+k^2 u_k=U u_k\,,
\label{ukeq2}
\end{equation}
where $U=a^2R/6$. 
Since $U$ is of the order of $(aH)^2$, 
one has $k^2 \gg U$ for the mode deep inside the Hubble radius.
Initially we choose the field in the vacuum state with the 
positive-frequency solution~\cite{Birrell}: $u_k^{(i)}=e^{-ik \eta}/\sqrt{2k}$.
The presence of the time-dependent term $U(\eta)$ leads to 
the creation of the particle $\chi$.
We can write the solution of Eq.~(\ref{ukeq2}) iteratively, 
as~\cite{Zeld}
\begin{equation}
u_k (\eta)=u_k^{(i)}+\frac1k
\int_{0}^{\eta} U(\eta') \sin [k(\eta-\eta')]
u_k (\eta')\mathrm{d}\eta'\,.
\end{equation}

After the universe enters the radiation-dominated epoch, 
the term $U$ becomes small so that the flat-space solution 
is recovered. The choice of decomposition of $\chi$ into $\hat a_k$ and $\hat a^\dagger_k$ 
is not unique. In curved spacetime it is possible to choose another decomposition
in term of new ladder operators $\hat {\cal A}_k$ and $\hat {\cal A}^\dagger_k$, 
which can be written in terms of $\hat a_k$ and $\hat a^\dagger_k$, such as 
$\hat{\cal A}_k=\alpha_k\hat a_k+\beta_k^*\hat a^\dagger_{-k}$. 
Provided that $\beta_k^*\neq0$, even though $\hat a_k\left\vert0\right>=0$, 
we have $\hat{\cal A}_k\left\vert0\right>\neq0$. 
Hence the vacuum in one basis is not the vacuum in the new basis, 
and according to the new basis, the particles are created.
The Bogoliubov coefficient describing the particle production is 
\begin{equation}
\beta_k=-\frac{i}{2k} \int_{0}^{\infty}
U(\eta') e^{-2ik \eta'} \mathrm{d}\eta'\,.
\end{equation}
The typical wavenumber in the $\eta$-coordinate is given by 
$k$, whereas in the $t$-coordinate it is $k/a$.
Then the energy density per unit comoving volume in the 
$\eta$-coordinate is~\cite{Suen}
\begin{eqnarray}
\rho_{\eta} &=&
\frac{1}{(2\pi)^3} \int_{0}^{\infty}
4\pi k^2 \mathrm{d}k \cdot k\,|\beta_k|^2 \nonumber \\
&=& \frac{1}{8\pi^2} \int_{0}^{\infty} \mathrm{d} \eta
\,U(\eta) \int_{0}^{\infty} \mathrm{d} \eta' U(\eta')
\int_{0}^{\infty} \mathrm{d}k \cdot k e^{2ik (\eta'-\eta)}
\nonumber \\
&=& \frac{1}{32\pi^2} \int_{0}^{\infty} \mathrm{d}\eta
\frac{\mathrm{d}U}{\mathrm{d}\eta} \int_0^{\infty}
\mathrm{d}\eta' \frac{U(\eta')}{\eta'-\eta}\,,
\label{rhoeta}
\end{eqnarray}
where in the last equality we have used the fact that the term
$U$ approaches 0 in the early and late times.

During the oscillating phase of the Ricci scalar the time-dependence
of $U$ is given by $U=I(\eta) \sin (\int_{0}^{\eta}\omega \mathrm{d}\bar{\eta})$, 
where $I(\eta)=ca(\eta)^{1/2}$ and $\omega= Ma$ ($c$ is a constant).
When we evaluate the term $\mathrm{d}U/\mathrm{d}\eta$ in Eq.~(\ref{rhoeta}), 
the time-dependence of $I(\eta)$ can be neglected.
Differentiating Eq.~(\ref{rhoeta}) in terms of $\eta$ 
and taking the limit $\int_{0}^{\eta}\omega \mathrm{d}\bar{\eta} \gg 1$, 
it follows that 
\begin{equation}
\frac{\mathrm{d}\rho_{\eta}}{\mathrm{d}\eta} \simeq
\frac{\omega}{32\pi}I^2(\eta) \cos^2 \left( 
\int_{0}^{\eta} \omega \mathrm{d} \bar{\eta}\right)\,,
\end{equation}
where we used the relation $\lim_{k \to \infty} \sin(kx)/x=\pi \delta (x)$.
Shifting the phase of the oscillating factor by $\pi/2$, we obtain 
\begin{equation}
\frac{\mathrm{d}\rho_{\eta}}{\mathrm{d}t} \simeq
\frac{MU^2}{32\pi}=
\frac{Ma^4 R^2}{1152\pi}\,.
\end{equation}
The proper energy density of the field $\chi$ is given by 
$\rho_{\chi}=(\rho_{\eta}/a)/a^3=\rho_{\eta}/a^4$.
Taking into account $g_*$ relativistic degrees of freedom, 
the total radiation density is 
\begin{equation}
\rho_M=\frac{g_*}{a^4}\rho_{\eta}
=\frac{g_*}{a^4} \int_{t_{\mathrm{os}}}^t 
\frac{Ma^4 R^2}{1152\pi} \mathrm{d}t\,,
\label{rhoMg}
\end{equation}
which obeys the following equation 
\begin{equation}
\dot{\rho}_M+4H\rho_M=\frac{g_*MR^2}{1152\pi}\,.
\end{equation}
Comparing this with the continuity equation~(\ref{continuity})
we obtain the pressure of the created particles, as 
\begin{equation}
P_M=\frac13 \rho_M-\frac{g_*MR^2}{3456\pi H}\,.
\label{PMg}
\end{equation}
Now the dynamical equations are given by Eqs.~(\ref{fRinf1})
and (\ref{fRinf2}) with the energy density (\ref{rhoMg}) 
and the pressure (\ref{PMg}).

In the regime $M (t-t_{\mathrm{os}}) \gg 1$ the evolution of 
the scale factor is given by $a \simeq a_0 (t-t_{\mathrm{os}})^{2/3}$, 
and hence
\begin{equation}
H^2 \simeq \frac{4}{9(t-t_{\mathrm{os}})^2}\,,
\end{equation}
where we have neglected the backreaction of created particles.
Meanwhile the integration of Eq.~(\ref{rhoMg}) gives
\begin{equation}
\rho_M \simeq \frac{g_* M^3}{240\pi} \frac{1}{t-t_{\mathrm{os}}}\,,
\end{equation}
where we have used the averaged relation $\langle R^2 \rangle \simeq 8M^2/(t-t_{\mathrm{os}})^2$ 
[which comes from Eq.~(\ref{Rap})].
The energy density $\rho_M$ evolves slowly compared to 
$H^2$ and finally it becomes a dominant contribution to the total 
energy density ($3H^2 \simeq 8\pi \rho_M/m_{\mathrm{pl}}^2$) at the
time $t_f \simeq t_{\mathrm{os}}+40m_{\mathrm{pl}}^2/(g_*M^3)$.
In~\cite{Suen} it was found that the transition from the
oscillating phase to the radiation-dominated epoch occurs slower
compared to the estimation given above.
Since the epoch of the transient matter-dominated era is about one order
of magnitude longer than the analytic estimation~\cite{Suen}, we take  
the value $t_f \simeq t_{\mathrm{os}}+400m_{\mathrm{pl}}^2/(g_*M^3)$
to estimate the reheating temperature $T_r$.
Since the particle energy density $\rho_M(t_f)$ is converted to 
the radiation energy density $\rho_r=g_* \pi^2 T_r^4/30$, 
the reheating temperature can be estimated 
as\epubtkFootnote{In~\cite{Suen} the reheating temperature is estimated by 
taking the maximum value of $\rho_M$ reached around the ten oscillations 
of $R$. Meanwhile we estimate $T_r$ at the epoch where $\rho_M$
becomes a dominant contribution to the total energy density (as in~\cite{Kolb}).} 
\begin{equation}
T_r \lesssim 3 \times 10^{17} g_*^{1/4}
\left( \frac{M}{m_{\mathrm{pl}}} \right)^{3/2} \mathrm{\ GeV}\,.
\end{equation}
As we will see in Section~\ref{cosmoinf}, the WMAP normalization of 
the CMB temperature anisotropies determines the mass scale 
to be $M \simeq 3 \times 10^{-6}m_{\mathrm{pl}}$.
Taking the value $g_*=100$, 
we have $T_r \lesssim 5 \times 10^{9} \mathrm{\ GeV}$.
For $t>t_{f}$ the universe enters the radiation-dominated epoch 
characterized by $a \propto t^{1/2}$, $R=0$, and $\rho_r \propto t^{-2}$.

\subsubsection{Case: $|\xi| \gtrsim 1$}

If $|\xi|$ is larger than the order of unity, one can expect the explosive particle 
production called preheating prior to the perturbative regime discussed above.
Originally the dynamics of such gravitational preheating was studied 
in~\cite{Bruce97, TMT} for a massive chaotic inflation model in Einstein gravity.
Later this was extended to the \fR\ model~(\ref{stamodel})~\cite{TMT2}.

Introducing a new field $X_k=a^{3/2}\chi_k$, Eq.~(\ref{chieq}) reads
\begin{equation}
\ddot{X}_k+\left( \frac{k^2}{a^2}+m_{\chi}^2+\xi R
-\frac94 H^2 -\frac32 \dot{H} \right)X_k=0\,.
\label{Xeq}
\end{equation}
As long as $|\xi|$ is larger than the order of unity, the last two terms 
in the bracket of Eq.~(\ref{Xeq}) can be neglected relative to $\xi R$.
Since the Ricci scalar is given by Eq.~(\ref{Rap}) in the regime 
$M (t-t_{\mathrm{os}}) \gg 1$, it follows that 
\begin{equation}
\ddot{X}_k+\left[ \frac{k^2}{a^2}+m_{\chi}^2
-\frac{4M \xi}{t-t_{\mathrm{os}}} \sin \{ M (t-t_{\mathrm{os}}) \}
\right]X_k \simeq 0\,.
\label{Xeq2}
\end{equation}

The oscillating term gives rise to parametric amplification of
the particle $\chi_k$. 
In order to see this we introduce the variable $z$ defined by 
$M(t-t_{\mathrm{os}})=2z \pm \pi/2$, where the plus and minus
signs correspond to the cases $\xi>0$ and $\xi<0$ respectively.
Then Eq.~(\ref{Xeq2}) reduces to the Mathieu equation 
\begin{equation}
\frac{\mathrm{d}^2}{\mathrm{d}z^2}X_k+
\left[ A_k-2q \cos (2z) \right]X_k \simeq 0\,,
\end{equation}
where 
\begin{equation}
A_k=\frac{4k^2}{a^2M^2}+\frac{4m_{\chi}^2}{M^2}\,,\qquad
q=\frac{8|\xi|}{M (t-t_{\mathrm{os}})}\,.
\end{equation}
The strength of parametric resonance depends on the 
parameters $A_k$ and $q$.
This can be described by a stability-instability chart of 
the Mathieu equation~\cite{Mathieu, KLS94, TMT2}.
In the Minkowski spacetime the parameters $A_k$ and $q$ 
are constant.
If $A_k$ and $q$ are in an instability band, then 
the perturbation $X_k$ grows exponentially 
with a growth index $\mu_k$, i.e., $X_k \propto e^{\mu_k z}$.
In the regime $q \ll 1$ the resonance occurs only in 
narrow bands around $A_k=\ell^2$, where $\ell=1,2,...$, 
with the maximum growth index $\mu_k=q/2$~\cite{KLS94}.
Meanwhile, for large $q~(\gg 1)$, a broad resonance can occur
for a wide range of parameter space and momentum 
modes~\cite{KLS97}.

In the expanding cosmological background both $A_k$
and $q$ vary in time. Initially the field $X_k$
is in the broad resonance regime ($q \gg 1$) for $|\xi|\gg 1$, but 
it gradually enters the narrow resonance regime ($q \lesssim 1$).
Since the field passes many instability and stability bands, 
the growth index $\mu_k$ stochastically changes with 
the cosmic expansion.
The non-adiabaticity of the change of the frequency 
$\omega_k^2=k^2/a^2+m_\chi^2
-4M\xi \sin\{ M(t-t_{\mathrm{os}}) \}/(t-t_{\mathrm{os}})$
can be estimated by the quantity
\begin{equation}
r_{\mathrm{na}} \equiv \left| \frac{\dot{\omega}_k}{\omega_k^2} \right|
=M\frac{|k^2/a^2+2M \xi \cos\{ M(t-t_{\mathrm{os}}) \}/(t-t_{\mathrm{os}})|}
{|k^2/a^2+m_\chi^2
-4M\xi \sin\{ M(t-t_{\mathrm{os}}) \}/(t-t_{\mathrm{os}})|^{3/2}}\,,
\label{dotomega}
\end{equation}
where the non-adiabatic regime corresponds to $r_{\mathrm{na}} \gtrsim 1$.
For small $k$ and $m_\chi$ we have 
$r_{\mathrm{na}} \gg 1$ around $M(t-t_{\mathrm{os}})=n\pi$, 
where $n$ are positive integers.
This corresponds to the time at which the Ricci scalar vanishes. 
Hence, each time $R$ crosses 0 during its 
oscillation, the non-adiabatic particle production occurs 
most efficiently. The presence of the mass term $m_{\chi}$ 
tends to suppress the non-adiabaticity parameter $r_{\mathrm{na}}$, but still it is 
possible to satisfy the condition $r_{\mathrm{na}} \gtrsim 1$
around $R=0$.

For the model~(\ref{stamodel}) it was shown in~\cite{TMT2}
that massless $\chi$ particles are resonantly amplified
for $|\xi| \gtrsim 3$. Massive particles with $m_{\chi}$ 
of the order of $M$ can be created for $|\xi| \gtrsim 10$.
Note that in the preheating scenario based on the model 
$V(\phi, \chi)=(1/2)m_{\phi}^2 \phi^2+(1/2)g^2\phi^2 \chi^2$
the parameter $q$ decreases more rapidly 
($q \propto 1/t^2$) than that in the model~(\ref{stamodel})~\cite{KLS97}.
Hence, in our geometric preheating scenario, we do not require 
very large initial values of $q$ [such as $q>{\cal O}(10^3)$]
to lead to the efficient parametric resonance.

While the above discussion is based on the linear analysis, 
non-linear effects (such as the mode-mode coupling of perturbations)
can be important at the late stage of preheating
(see, e.g., \cite{KLS97, Khle97}).
Also the energy density of created particles 
affects the background cosmological dynamics, 
which works as a backreaction to the Ricci scalar.
The process of the subsequent perturbative reheating 
stage can be affected by the explosive particle production 
during preheating. It will be of interest to take into account 
all these effects and study how the thermalization is reached
at the end of reheating. This certainly requires the detailed 
numerical investigation of lattice simulations, as developed 
in~\cite{Feld1, Feld2}. 

At the end of this section we should mention a number of interesting 
works about gravitational baryogenesis based on the interaction 
$(1/M_*^2) \int \mathrm{d}^4 x \sqrt{-g}\,J^{\mu}\,\partial_{\mu}R$
between the baryon number current $J^{\mu}$ and 
the Ricci scalar $R$
($M_*$ is the cut-off scale characterizing 
the effective theory)~\cite{Murayama, Lam06, Sadjadi07}.
This interaction can give rise to an equilibrium baryon asymmetry
which is observationally acceptable, even for the gravitational 
Lagrangian $f(R)=R^n$ with $n$ close to 1. 
It will be of interest to extend the analysis to 
more general \fR\ gravity models.

\newpage

\section{Dark Energy in \fR\ Theories}
\label{denergysec}

In this section we apply \fR\ theories to dark energy.
Our interest is to construct viable \fR\ models that can realize the 
sequence of radiation, matter, and accelerated epochs.
In this section we do not attempt to find unified models of inflation and 
dark energy based on \fR\ theories. 

Originally the model $f(R) = R-\alpha /R^{n}$ ($\alpha>0, n>0$) 
was proposed to explain the late-time cosmic 
acceleration~\cite{fRearly1, fRearly2p, fRearly2, fRearly3}
(see also~\cite{fRearly4, Soussa, Allem, Easson, Dick04, Allemandi:2004ca, Carloni, Barrow06}
for related works).
However, this model suffers from a number of problems 
such as matter instability~\cite{Dolgov, Faramatter}, 
the instability of cosmological 
perturbations~\cite{matterper1, matterper2, SongHu1, SongHu2, Teg}, 
the absence of the matter era~\cite{APT, APT2, Fairbairn07}, and
the inability to satisfy local gravity 
constraints~\cite{OlmoPRL, Olmo05, Fara06, Erick06, Chiba07, Navarro, CapoTsuji}.
The main reason why this model does not work is that 
the quantity $f_{,RR} \equiv \partial^2 f/\partial R^2$ is negative.
As we will see later, the violation of the condition $f_{,RR}>0$
gives rise to the negative mass squared $M^2$ for the 
scalaron field. Hence we require that $f_{,RR}>0$ to avoid
a tachyonic instability. 
The condition $f_{,R} \equiv \partial f/\partial R>0$ 
is also required to avoid the appearance of
ghosts (see Section~\ref{secperlag}).
Thus viable \fR\ dark energy models need to satisfy~\cite{Star07}
\begin{equation}
f_{,R}>0\,,\qquad f_{,RR}>0\,,\qquad \mathrm{for} \quad R \ge R_0~(>0)\,,
\label{fcon}
\end{equation}
where $R_0$ is the Ricci scalar today.

In the following we shall derive other conditions 
for the cosmological viability of \fR\ models.
This is based on the analysis of~\cite{AGPT}.
For the matter Lagrangian ${\cal L}_{M}$ in Eq.~(\ref{fRaction}) 
we take into account non-relativistic matter and radiation, whose
energy densities $\rho_m$ and $\rho_r$ satisfy 
\begin{eqnarray}
& & \dot{\rho}_{m}+3H\rho_{m}=0\,,\\
& & \dot{\rho}_{r}+4H\rho_{r}=0\,,
\end{eqnarray}
respectively.
From Eqs.~(\ref{fRinf1}) and (\ref{fRinf2}) it follows that 
\begin{eqnarray}
3FH^{2} &=& (FR-f)/2-3H\dot{F}+\kappa^{2}(\rho_{m}+\rho_{r})\,,
\label{FRWfR1}\\
-2F\dot{H} &=& \ddot{F}-H\dot{F}+
\kappa^{2}\left[\rho_{m}+(4/3)\rho_{r}\right]
\,.
\label{FRWfR2}
\end{eqnarray}


\subsection{Dynamical equations}

We introduce the following variables
\begin{equation}
x_{1}\equiv-\frac{\dot{F}}{HF}\,,\quad 
x_{2}\equiv-\frac{f}{6FH^{2}}\,,\quad 
x_{3}\equiv\frac{R}{6H^{2}}\,,\quad 
x_{4}\equiv\frac{\kappa^{2}\rho_{r}}{3FH^{2}}\,,
\end{equation}
together with the density parameters
\begin{equation}
\Omega_{m}\equiv\frac{\kappa^{2}\rho_{m}}{3FH^{2}}=1-x_{1}-x_{2}-x_{3}-x_{4},
\qquad\Omega_{r}\equiv x_{4}\,,\qquad\Omega_{\mathrm{DE}}\equiv x_{1}+x_{2}+x_{3}\,.
\label{fromedef}
\end{equation}
It is straightforward to derive the following equations 
\begin{eqnarray}
\frac{\rd x_{1}}{\rd N} &=&  -1-x_{3}-3x_{2}+x_{1}^{2}-x_{1}x_{3}+x_{4}~,\label{x1fR}\\
\frac{\rd x_{2}}{\rd N} &=&  \frac{x_{1}x_{3}}{m}-x_{2}(2x_{3}-4-x_{1})~,\label{x2fR}\\
\frac{\rd x_{3}}{\rd N} &=&  -\frac{x_{1}x_{3}}{m}-2x_{3}(x_{3}-2)~,\label{x3fR}\\
\frac{\rd x_{4}}{\rd N} &=&  -2x_{3}x_{4}+x_{1}\, x_{4}\,,\label{x4fR}
\end{eqnarray}
where $N=\ln a$ is the number of e-foldings, and
\begin{eqnarray}
m &\equiv&  \frac{\rd\ln F}{\rd\ln R}=\frac{Rf_{,RR}}{f_{,R}}\,,
\label{mdef}
\\
r &\equiv&  -\frac{\rd\ln f}{\rd\ln R}=-\frac{Rf_{,R}}{f}
=\frac{x_{3}}{x_{2}}\,.
\label{rdef}
\end{eqnarray}
From Eq.~(\ref{rdef}) the Ricci scalar $R$ can be expressed by 
$x_{3}/x_{2}$. Since $m$ depends on $R$, this means that $m$ is
a function of $r$, that is, $m=m(r)$. The $\Lambda$CDM model,
$f(R)=R-2\Lambda$, corresponds to $m=0$. Hence the quantity $m$
characterizes the deviation of the background dynamics from the
$\Lambda$CDM model. A number of authors studied cosmological dynamics
for specific \fR\ models~\cite{Clifton05, LiBarrow, Santiago, Fay07,
  AmenTsuji07, Souza, Goheer1, Bazeia07, Atazadeh07, Clifton:2008bn,
  Evans, Abdelwahab:2007jp, Goheer2, Paul, Ishakd1, Ishakd2}.

The effective equation of state of the system is defined by 
\begin{equation}
w_{\mathrm{eff}} \equiv -1-2\dot{H}/(3H^2)\,,
\label{ldef}
\end{equation}
which is equivalent to $w_{\mathrm{eff}}=-(2x_3-1)/3$.
In the absence of radiation ($x_{4}=0$) the fixed points for the
above dynamical system are
%
%
%
\begin{eqnarray}
 P_{1}:(x_{1},x_{2},x_{3})&=&(0,-1,2), \qquad\qquad \Omega_{m}=0, \qquad\qquad\qquad w_{\mathrm{eff}}=-1\,,\\
 P_{2}:(x_{1},x_{2},x_{3})&=&(-1,0,0), \qquad\qquad \Omega_{m}=2, \qquad\qquad\qquad w_{\mathrm{eff}}=1/3\,,\\
 P_{3}:(x_{1},x_{2},x_{3})&=&(1,0,0),  \qquad\qquad~~ \Omega_{m}=0, \qquad\qquad\qquad w_{\mathrm{eff}}=1/3\,,\\
 P_{4}:(x_{1},x_{2},x_{3})&=&(-4,5,0), \qquad\qquad \Omega_{m}=0, \qquad\qquad\qquad w_{\mathrm{eff}}=1/3\,,\\
 P_{5}:(x_{1},x_{2},x_{3})&=&\left(\frac{3m}{1+m},
-\frac{1+4m}{2(1+m)^{2}}\right., \left.\frac{1+4m}{2(1+m)}\right),\\
 && \qquad\qquad\qquad\qquad~ \Omega_{m}=1-\frac{m(7+10m)}{2(1+m)^{2}},  \qquad w_{\mathrm{eff}}=-\frac{m}{1+m},\\
 P_{6}:(x_{1},x_{2},x_{3})&=&\left(\frac{2(1-m)}{1+2m}\right.,\frac{1-4m}{m(1+2m)},
 -\left.\frac{(1-4m)(1+m)}{m(1+2m)}\right), \nonumber \\
 && \qquad\qquad\qquad\qquad~ \Omega_{m}=0,                    \qquad\qquad\qquad w_{\mathrm{eff}}=\frac{2-5m-6m^{2}}{3m(1+2m)}.
 \end{eqnarray}
The points $P_{5}$ and $P_{6}$ are on the line $m(r)=-r-1$ in the
$(r,m)$ plane.

The matter-dominated epoch ($\Omega_{m}\simeq1$ 
and $w_{\mathrm{eff}} \simeq 0$)
can be realized only by the point $P_{5}$ for $m$ close to 0. 
In the ($r,m$) plane this point exists around $(r,m)=(-1,0)$.
Either the point $P_1$ or $P_6$ can be responsible for
the late-time cosmic acceleration. The former is a de
Sitter point ($w_{\mathrm{eff}}=-1$) with $r=-2$, 
in which case the condition (\ref{fRdeSitter}) is satisfied.
The point $P_6$ can give rise to the accelerated expansion 
($w_{\mathrm{eff}}<-1/3$) provided that $m>(\sqrt{3}-1)/2$, 
or $-1/2<m<0$, or $m<-(1+\sqrt{3})/2$.

In order to analyze the stability of the above fixed points 
it is sufficient to consider only time-dependent
linear perturbations $\delta x_{i}(t)$ ($i=1,2,3$) 
around them (see~\cite{CLW, CST06} for the detail of such analysis).
For the point $P_{5}$ the eigenvalues for the $3\times3$ Jacobian
matrix of perturbations are
\begin{equation}
3(1+m_{5}'),\quad\frac{-3m_{5}\pm\sqrt{m_{5}(256m_{5}^{3}
+160m_{5}^{2}-31m_{5}-16)}}{4m_{5}(m_{5}+1)}\,,\label{P5eig}
\end{equation}
where $m_{5} \equiv m(r_5)$ and $m_{5}'\equiv\frac{\rd m}{\rd
  r}(r_{5})$ with $r_{5}\approx-1$.  In the limit that $|m_{5}|\ll1$ the
latter two eigenvalues reduce to $-3/4\pm\sqrt{-1/m_{5}}$.  
For the models with $m_5<0$, the solutions cannot 
remain for a long time around the point $P_{5}$
because of the divergent behavior of the eigenvalues as $m_5 \to -0$.
The model $f(R)=R-\alpha/R^{n}$ ($\alpha>0, n>0$)
falls into this category. On the other hand,
if $0<m_{5}<0.327$, the latter two eigenvalues in Eq.~(\ref{P5eig})
are complex with negative real parts. Then, provided that $m_{5}'>-1$,
the point $P_{5}$ corresponds to a saddle point 
with a damped oscillation.  
Hence the solutions can stay around this point for some time
and finally leave for the late-time acceleration.
Then the condition for the existence of the saddle 
matter era is
\begin{equation}
m(r)\simeq +0\,,\quad\frac{\rd m}{\rd r}>-1\,,
\quad \mathrm{at}\quad r=-1\,.
\label{m5con}
\end{equation}
The first condition implies that viable \fR\ models need to be close
to the $\Lambda$CDM model during the matter domination.
This is also required for consistency with local gravity constraints, 
as we will see in Section \ref{lgcsec}. 

The eigenvalues for the Jacobian matrix of perturbations about the
point $P_{1}$ are
\begin{equation}
-3,\quad-\frac{3}{2}\pm\frac{\sqrt{25-16/m_{1}}}{2}\,,
\end{equation}
where $m_{1}=m(r=-2)$. This shows that the condition for the stability
of the de~Sitter point $P_{1}$ is~\cite{Muller, Faraonista1, Faraonista2, AGPT}
\begin{equation}
0<m(r=-2) \le 1\,.
\label{m1con}
\end{equation}
The trajectories that start from the saddle matter point $P_{5}$ 
satisfying the condition (\ref{m5con}) and then approach the
stable de~Sitter point $P_{1}$ satisfying the condition (\ref{m1con})
are, in general, cosmologically viable.

One can also show that $P_6$ is stable and accelerated for
(a) $m_6'<-1$, $(\sqrt{3}-1)/2<m_6<1$, 
(b) $m_6'>-1$, $m_6<-(1+\sqrt{3})/2$,
(c) $m_6'>-1$, $-1/2<m_6<0$,
(d)  $m_6'>-1$, $m_6 \ge 1$.
Since both $P_5$ and $P_6$ are on the line $m=-r-1$, 
only the trajectories from $m_5'>-1$ to $m_6'<-1$ are allowed
(see Figure~\ref{mrplane}).
This means that only the case (a) is viable as a stable and 
accelerated fixed point $P_6$.
In this case the effective equation of state satisfies 
the condition $w_{\mathrm{eff}}>-1$.

From the above discussion the following two classes of
models are cosmologically viable.
\begin{itemize}
\item Class A: Models that connect $P_5$ ($r \simeq -1$, $m \simeq +0$) 
to $P_1$ ($r=-2, 0<m \le 1$)
\item Class B: Models that connect $P_5$ ($r \simeq -1$, $m \simeq +0$) 
to $P_6$ ($m=-r-1, (\sqrt{3}-1)/2<m<1$)
\end{itemize}
From Eq.~(\ref{fcon}) the viable \fR\ dark energy models
need to satisfy the condition $m>0$, which is consistent 
with the above argument.


\subsection{Viable \fR\ dark energy models}

We present a number of viable \fR\ models in the $(r,m)$
plane. First we note that the $\Lambda$CDM model corresponds to
$m=0$, in which case the trajectory is the straight line (i) in 
Figure~\ref{mrplane}.
The trajectory (ii) in Figure~\ref{mrplane} represents 
the model $f(R)=(R^{b}-\Lambda)^{c}$~\cite{AmenTsuji07}, which
corresponds to the straight line $m(r)=[(1-c)/c]r+b-1$ in the $(r,m)$
plane. The existence of a saddle matter epoch demands the condition $c
\ge 1$ and $bc \simeq 1$. The trajectory (iii) represents the
model~\cite{AGPT, LiBarrow}
\begin{equation}
f(R)=R-\alpha R^{n} \qquad (\alpha>0,~0<n<1)\,,
\label{powermodel}
\end{equation}
which corresponds to the curve $m=n(1+r)/r$.  The trajectory (iv) 
represents the model $m(r)=-C(r+1)(r^{2}+ar+b)$, in
which case the late-time accelerated attractor is the point $P_6$ with
$(\sqrt{3}-1)/2<m<1$.

\epubtkImage{mrplane.png}{%
  \begin{figure}[hptb]
    \centerline{\includegraphics[width=3.4in]{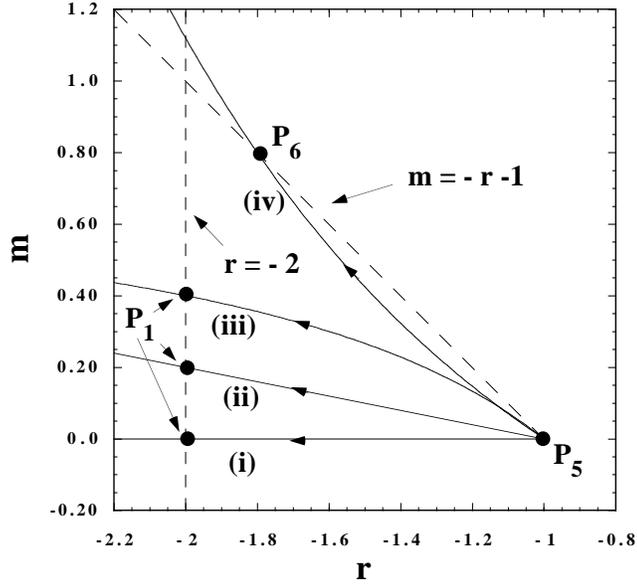}}
    \caption{Four trajectories in the $(r,m)$ plane. Each trajectory
      corresponds to the models: (i) $\Lambda$CDM, (ii)
      $f(R)=(R^{b}-\Lambda)^{c}$, (iii) $f(R)=R-\alpha R^{n}$ with
      $\alpha>0,0<n<1$, and (iv)
      $m(r)=-C(r+1)(r^{2}+ar+b)$. From~\cite{AmenTsuji07}.}
    \label{mrplane} 
\end{figure}}

In~\cite{AGPT} it was shown that $m$ needs to be
close to 0 during the radiation domination as well as the matter domination.
Hence the viable \fR\ models are close to the $\Lambda$CDM model
in the region $R\gg R_{0}$. The Ricci scalar 
remains positive from the radiation era up to 
the present epoch, as long as it does not oscillate
around $R=0$.
The model $f(R)=R-\alpha/R^n$ ($\alpha>0$, $n>0$) is not viable because
the condition $f_{,RR}>0$ is violated.

As we will see in Section \ref{lgcsec}, the local gravity constraints 
provide tight bounds on the deviation parameter $m$ in 
the region of high density ($R \gg R_0$), e.g., 
$m (R) \lesssim 10^{-15}$ 
for $R=10^5 R_0$~\cite{CapoTsuji,TUMTY}.
In order to realize a large deviation from the $\Lambda$CDM 
model such as $m(R)>{\cal O}(0.1)$ today ($R=R_0$)
we require that the variable $m$ changes rapidly from 
the past to the present. The \fR\ model given 
in Eq.~(\ref{powermodel}), for example, does not 
allow such a rapid variation, because $m$ evolves as
$m \simeq n(-r-1)$ in the region $R \gg R_0$.
Instead, if the deviation parameter has the dependence
\begin{equation}
m=C(-r-1)^p\,,\qquad p>1\,,~C>0\,,
\label{mpre}
\end{equation}
it is possible to lead to the rapid decrease of $m$
as we go back to the past.
The models that behave as Eq.~(\ref{mpre}) in the regime 
$R \gg R_0$ are
\begin{eqnarray}
 &  & \mathrm{(A)}~f(R)=R-\mu R_{c}\frac{(R/R_{c})^{2n}}{(R/R_{c})^{2n}+1}
 \qquad \mathrm{with}~~n,\mu,R_{c}>0\,,\label{Amodel}\\
 &  & \mathrm{(B)}~f(R)=R-\mu R_{c}\left[1-\left(1+R^{2}/R_{c}^{2}\right)^{-n}\right]
 \qquad \mathrm{with}~~n,\mu,R_{c}>0\,.
\label{Bmodel}
 \end{eqnarray}
The models (A) and (B) have been proposed by Hu and Sawicki~\cite{Hu07}
and Starobinsky~\cite{Star07}, respectively. 
Note that $R_c$ roughly corresponds to the order of $R_0$
for $\mu={\cal O}(1)$. 
This means that $p=2n+1$ for $R \gg R_0$.
In the next section we will show that both the models (A) and (B)
are consistent with local gravity constraints for $n \gtrsim 1$.

In the model (A) the following relation holds at the de~Sitter point: 
\begin{equation}
\mu=\frac{(1+x_d^{2n})^2}{x_d^{2n-1}(2+2x_d^{2n}-2n)}\,,
\label{lamb}
\end{equation}
where $x_d \equiv R_1/R_c$ and $R_1$ is the Ricci scalar
at the de~Sitter point.
The stability condition (\ref{m1con}) gives~\cite{Tsuji08}
\begin{equation}
2x_d^{4n}-(2n-1)(2n+4)x_d^{2n}+
(2n-1)(2n-2) \ge 0\,.
\label{Amodelcon}
\end{equation}
The parameter $\mu$ has a lower bound determined 
by the condition (\ref{Amodelcon}).
When $n=1$, for example, one has $x_d \ge \sqrt{3}$ and
$\mu \ge 8\sqrt{3}/9$.
Under Eq.~(\ref{Amodelcon}) one can show that 
the conditions (\ref{fcon}) are also satisfied.

Similarly the model (B) satisfies~\cite{Star07}
\begin{equation}
(1+x_d^2)^{n+2} \ge 1+(n+2)x_d^2+(n+1)(2n+1)x_d^4,
\label{Bmodelcon}
\end{equation}
with 
\begin{equation}
\mu=\frac{x_d(1+x_d^2)^{n+1}}
{2[(1+x_d^2)^{n+1}-1-(n+1)x_d^2]}\,.
\label{Bmodellam}
\end{equation}
When $n=1$ we have $x_d \ge \sqrt{3}$ and 
$\mu \ge 8\sqrt{3}/9$, which is the same as
in the model (A). For general $n$, however, the 
bounds on $\mu$ in the model (B) are not 
identical to those in the model (A).

Another model that leads to an even faster evolution of $m$
is given by~\cite{Tsuji08}
\begin{equation}
\mathrm{(C)}~f(R)=R-\mu R_{c} \tanh\,(R/R_{c})
\qquad \mathrm{with}~~\mu,R_{c}>0\,.
\label{tanh}
\end{equation}
A similar model was proposed by Appleby and Battye~\cite{Appleby}.
In the region $R \gg R_c$ the model~(\ref{tanh}) behaves as 
$f(R) \simeq R-\mu R_c \left[ 1-\exp (-2R/R_c) \right]$, which
may be regarded as a special case of (\ref{mpre}) in the 
limit that $p \gg 1$ \epubtkFootnote{The cosmological dynamics for the model 
$f(R)=R-\mu R_c \left[ 1-\exp (-2R/R_c) \right]$ was studied
  in~\cite{Linder09}.}. The Ricci scalar at the de~Sitter point is
determined by $\mu$, as 
\begin{equation}
\mu=\frac{x_d \cosh^2 (x_d)}{2\sinh (x_d) \cosh (x_d)-x_d}\,.
\end{equation}
From the stability condition~(\ref{m1con}) we obtain 
\begin{equation}
\mu>0.905\,,\qquad x_d>0.920\,.
\end{equation}

The models (A), (B) and (C) are close to the $\Lambda$CDM model
for $R \gg R_c$, but the deviation from it appears when $R$ decreases
to the order of $R_c$. This leaves a number of observational signatures
such as the phantom-like equation of state of dark energy and
the modified evolution of matter density perturbations.
In the following we discuss the dark energy equation of state in 
\fR\ models.  In Section~\ref{cosmodark} we study the evolution of 
density perturbations and resulting observational 
consequences in detail.


\subsection{Equation of state of dark energy}

In order to confront viable \fR\ models with SN~Ia observations, 
we rewrite Eqs.~(\ref{FRWfR1}) and (\ref{FRWfR2}) as follows:
\begin{eqnarray}
 3AH^{2}&=&\kappa^{2} \left(\rho_{m}+\rho_{r}+\rho_{\mathrm{DE}} 
 \right)\,,\label{mofR1}\\
 -2A\dot{H}&=&\kappa^{2}\left[\rho_{m}+(4/3)\rho_{r}
 +\rho_{\mathrm{DE}}+P_{\mathrm{DE}}\right]\,,
 \label{mofR2}
\end{eqnarray}
where $A$ is some constant and 
\begin{eqnarray}
\kappa^{2}\rho_{\mathrm{DE}} &\equiv& (1/2)(FR-f)-3H\dot{F}+3H^{2}
(A-F)\,,\label{fRrhode}\\
\kappa^{2}P_{\mathrm{DE}} &\equiv& \ddot{F}+2H\dot{F}
-(1/2)(FR-f)-(3H^{2}+2\dot{H})(A-F)\,.
\label{fRPde}
\end{eqnarray}
Defining $\rho_{\mathrm{DE}}$ and $P_{\mathrm{DE}}$ in the above way, 
we find that these satisfy the usual continuity equation
\begin{equation}
\dot{\rho}_{\mathrm{DE}}+3H(\rho_{\mathrm{DE}}+P_{\mathrm{DE}})=0\,.
\label{rhodecon}
\end{equation}
Note that this holds as a consequence of the Bianchi identities, 
as we have already mentioned in the discussion from 
Eq.~(\ref{Einmo}) to Eq.~(\ref{TmunuD2}).

The dark energy equation of state, $w_{\mathrm{DE}}\equiv P_{\mathrm{DE}}/\rho_{\mathrm{DE}}$,
is directly related to the one used in SN~Ia observations. 
From Eqs.~(\ref{mofR1}) and (\ref{mofR2}) it is given by 
\begin{equation}
w_{\mathrm{DE}}=-\frac{2A\dot{H}+3AH^2+\kappa^2 \rho_r/3}
{3AH^2-\kappa^2 (\rho_m+\rho_r)}
\simeq\frac{w_{\mathrm{eff}}}{1-(F/A)\Omega_m}\,,
\label{wDEfR}
\end{equation}
where the last approximate equality is valid in
the regime where the radiation density $\rho_r$ is negligible relative
to the matter density $\rho_m$.
The viable \fR\ models approach the $\Lambda$CDM model
in the past, i.e., $F \to 1$ as $R \to \infty$.  In
order to reproduce the standard matter era 
($3H^2 \simeq \kappa^2 \rho_m$) for $z \gg 1$, we can
choose $A=1$ in Eqs.~(\ref{mofR1}) and (\ref{mofR2}).  Another
possible choice is $A=F_0$, where $F_0$ is the present value of
$F$. This choice may be suitable if the deviation of $F_0$ from 1 is small
(as in scalar-tensor theory with a nearly massless scalar 
field~\cite{Torres,Boi00}). 
In both cases the equation of state $w_{\mathrm{DE}}$ can
be smaller than $-1$ before reaching the de~Sitter attractor
\cite{Hu07,AmenTsuji07,Tsuji08,Motohashi10}, 
while the effective equation of state 
$w_{\mathrm{eff}}$ is larger than $-1$.
This comes from the fact that the denominator in Eq.~(\ref{wDEfR})
becomes smaller than $1$ in the presence of the matter fluid.
Thus \fR\ gravity models give rise
to the phantom equation of state of dark energy 
without violating any stability conditions of the 
system. See~\cite{Dev,Mel09,Cardone09,Ali10} for observational constraints 
on the models (\ref{Amodel}) and (\ref{Bmodel})
by using the background expansion history of the universe.
Note that as long as the late-time attractor is the de~Sitter point 
the cosmological constant boundary crossing of $w_{\mathrm{eff}}$ reported in 
~\cite{Bamba08,Bamba09} does not typically occur, 
apart from small oscillations of $w_{\mathrm{eff}}$ 
around the de~Sitter point.

There are some works that try to reconstruct the forms 
of \fR\ by using some desired form for the evolution of 
the scale factor $a(t)$ or the observational 
data of SN~Ia~\cite{Caporecon,Capomatter,Mul06,Dobado06,Wu07,Fay07}.
We need to caution that the procedure of reconstruction 
does not in general guarantee the stability of solutions.
In scalar-tensor dark energy models, for example, it is known that 
a singular behavior sometimes arises at low-redshifts in such a
procedure~\cite{Esp01,Gan06}. 
In addition to the fact that the reconstruction method 
does not uniquely determine the forms of \fR, 
the observational data of the background expansion 
history alone is not yet sufficient to reconstruct \fR\ models 
in high precision. 

Finally we mention a number of 
works~\cite{Capoma1,Capoma2,Capoma3,Martins07,Iorio07,Sobouti1,Sobouti2,Boehmerma} 
about the use of metric 
\fR\ gravity as dark matter instead of dark energy.
In most of past works the power-law \fR\ model $f=R^n$
has been used to obtain spherically symmetric solutions for galaxy clustering.
In~\cite{Capoma2} it was shown that the theoretical rotation curves 
of spiral galaxies show good agreement with observational data
for $n=1.7$, while for broader samples the best-fit value of the power 
was found to be $n=2.2$~\cite{Martins07}.
However, these values are not compatible with the bound 
$|n-1|<7.2 \times 10^{-19}$
derived in~\cite{Barrow06,Clifton05} from a number of other observational 
constraints. Hence, it is unlikely that \fR\ gravity works as 
the main source for dark matter.

\newpage

\section{Local Gravity Constraints}
\label{lgcsec}
\setcounter{equation}{0}

In this section we discuss the compatibility of \fR\ models with
local gravity constraints~(see~\cite{OlmoPRL, Olmo05, Fara06, Erick06,Chiba07,Navarro,Teg} for early works, 
and~\cite{AmenTsuji07, Hu07,CapoTsuji} for experimental 
constraints on viable \fR\ dark energy models, and~\cite{Van, Dev,
Kainu07, Kainu08, Olmo07, PZhang, Cembranos, Zakharov, Jin, Baghram,
Ruggiero, Gerard:2006ia, Shao06, Stabile, Multamaki08, Hui,
Bisabr} for other related 
works). In an environment of high density such as Earth or Sun, the
Ricci scalar $R$ is much larger than the background cosmological
value $R_0$. If the outside of a spherically symmetric body is a
vacuum, the metric can be described by a Schwarzschild exterior
solution with $R=0$. In the presence of non-relativistic matter with
an energy density $\rho_m$, this gives rise to a contribution to the
Ricci scalar $R$ of the order $\kappa^2 \rho_m$.

If we consider local perturbations $\delta R$ on a background
characterized by the curvature $R_0$, the validity of the linear
approximation demands the condition $\delta R \ll R_0$. We first
derive the solutions of linear perturbations under the approximation
that the background metric $g_{\mu \nu}^{(0)}$ is described by the
Minkowski metric $\eta_{\mu \nu}$. In the case of Earth and Sun the
perturbation $\delta R$ is much larger than $R_0$, which means that
the linear theory is no longer valid. In such a non-linear regime the
effect of the chameleon mechanism~\cite{chame1,chame2} becomes
important, so that \fR\ models can be consistent with local gravity
tests.


\subsection{Linear expansions of perturbations in the spherically symmetric background}
\label{linearexpan}

First we decompose the quantities $R$, $F(R)$, and $T_{\mu\nu}$
into the background part and the perturbed part: 
$R=R_0+\delta R$, $F=F_{0}(1+\delta_{F})$, and 
$T_{\mu\nu}={}^{(0)}T_{\mu\nu}+\delta T_{\mu\nu}$
about the approximate Minkowski background 
($g_{\mu \nu}^{(0)} \approx \eta_{\mu \nu}$). 
In other words, although we consider $R$ close to a mean-field 
value $R_0$, the metric is still 
very close to the Minkowski case.
The linear expansion of Eq.~(\ref{trace}) in a time-independent 
background gives~\cite{Olmo05,Faraonista2,Chiba07,Navarro} 
\begin{equation}
\nabla^{2} \delta_{F}-M^{2} \delta_{F}
=\frac{\kappa^{2}}{3F_{0}}\delta T\,,\label{delpsi}
\end{equation}
where $\delta T\equiv\eta^{\mu\nu}\delta T_{\mu\nu}$ and 
\begin{equation}
M^{2}\equiv\frac{1}{3}\left[\frac{f_{,R}(R_{0})}
{f_{,RR}(R_{0})}-R_{0}\right]=\frac{R_{0}}{3}
\left[\frac{1}{m(R_{0})}-1\right]\,.
\label{Mpsi2}
\end{equation}
The variable $m$ is defined in Eq.~(\ref{mdef}).
Since $0<m(R_0)<1$ for viable \fR\ models, it follows 
that $M^2>0$ (recall that $R_0>0$).

We consider a spherically symmetric body with mass $M_{c}$, 
constant density $\rho~(=-\delta T)$, radius $r_{c}$, and
vanishing density outside the body. 
Since $\delta_{F}$ is a function of the distance $r$ from 
the center of the body, Eq.~(\ref{delpsi}) reduces to 
the following form inside the body ($r<r_c$):
\begin{equation}
\frac{\rd^{2}}{\rd r^{2}}\delta_{F}+\frac{2}{r}\frac{\rd}{\rd r}\delta_{F}
-M^{2}\,\delta_{F}=-\frac{\kappa^{2}}{3F_{0}}\rho\,,\label{rdeq}
\end{equation}
whereas the r.h.s.\ vanishes outside the body ($r>r_c$).
The solution of the perturbation $\delta_{F}$ 
for positive $M^{2}$ is given by 
\begin{eqnarray}
 (\delta_{F})_{r<r_{c}}&=&c_{1}\frac{e^{-Mr}}{r}+c_{2}\frac{e^{Mr}}{r}
+\frac{8\pi G\rho}{3F_{0}M^{2}}\,,\label{psiin}\\
 (\delta_{F})_{r>r_{c}}&=&c_{3}\frac{e^{-Mr}}{r}
+c_{4}\frac{e^{Mr}}{r}\,,\label{psiout}
\end{eqnarray}
where $c_i$ ($i=1,2,3,4$) are integration constants.  The requirement
that $(\delta_{F})_{r>r_{c}} \to 0$ as $r\to\infty$ gives $c_4=0$. The
regularity condition at $r=0$ requires that $c_{2}=-c_{1}$.  We match
two solutions (\ref{psiin}) and (\ref{psiout}) at $r=r_c$ by demanding
the regular behavior of $\delta_{F}(r)$ and $\delta_{F}'(r)$. 
Since $\delta_F\propto\delta R$, this implies that $R$ is also
continuous. If the mass $M$ satisfies the condition $M r_{c} \ll
1$, we obtain the following solutions
\begin{eqnarray}
 (\delta_F)_{r<r_{c}}&\simeq&\frac{4\pi G\rho}{3F_{0}}
\left(r_{c}^{2}-\frac{r^{2}}{3}\right)\,,
\label{depin}
\\
 (\delta_F)_{r>r_{c}}&\simeq&\frac{2GM_{c}}{3F_{0}r}e^{-Mr}\,.
\label{depout}
\end{eqnarray}

As we have seen in Section~\ref{secconformo}, the action 
(\ref{fRaction}) in \fR\ gravity can be transformed to the
Einstein frame action by a transformation of the metric.
The Einstein frame action is given by 
a linear action in $\tilde{R}$, where $\tilde{R}$ is 
a Ricci scalar in the new frame.
The first-order solution for the perturbation $h_{\mu\nu}$ of the
metric $\tilde{g}_{\mu\nu}=F_{0}\,(\eta_{\mu\nu}+h_{\mu\nu})$
follows from the first-order linearized Einstein equations
in the Einstein frame.
This leads to the solutions $h_{00}=2GM_{c}/(F_{0}r)$ and 
$h_{ij}=2GM_{c}/(F_{0}r)\,\delta_{ij}$.
Including the perturbation $\delta_F$ to the quantity $F$, 
the actual metric $g_{\mu\nu}$ is given by~\cite{Navarro}
\begin{equation}
g_{\mu\nu}=\frac{\tilde{g}_{\mu\nu}}{F}
\simeq \eta_{\mu\nu}+h_{\mu\nu}
-\delta_{F}\,\eta_{\mu\nu}\,.
\end{equation}
Using the solution (\ref{depout}) outside the body, 
the $(00)$ and $(ii)$ components of the metric $g_{\mu\nu}$ are
\begin{eqnarray}
g_{00}\simeq-1+\frac{2G_{\mathrm{eff}}^{(N)}M_{c}}{r}\,,
\qquad g_{ii}\simeq1+\frac{2G_{\mathrm{eff}}^{(N)}M_{c}}{r}\gamma\,,
\end{eqnarray}
where $G_{\mathrm{eff}}^{(N)}$ and $\gamma$ are the effective
gravitational coupling and the post-Newtonian parameter, 
respectively, defined by 
\begin{eqnarray}
G_{\mathrm{eff}}^{(N)}\equiv\frac{G}{F_{0}}
\left(1+\frac{1}{3}e^{-M r}\right)\,,\qquad
\gamma\equiv\frac{3-e^{-Mr}}{3+e^{-Mr}}\,.
\label{GeffN}
\end{eqnarray}

For the \fR\ models whose deviation from the $\Lambda$CDM model 
is small ($m \ll 1$), we have $M^{2} \simeq R_{0}/[3m(R_{0})]$
and $R \simeq 8\pi G\rho$. This gives the following estimate
\begin{equation}
(Mr_c)^2 \simeq 2\frac{\Phi_c}{m(R_0)}\,,
\end{equation}
where $\Phi_c=GM_c/(F_0r_c)=4\pi G \rho r_c^2/(3F_0)$ is 
the gravitational potential at the surface of the body.
The approximation $Mr_c \ll 1$ used to derive 
Eqs.~(\ref{depin}) and (\ref{depout}) corresponds 
to the condition 
\begin{equation}
m(R_0) \gg \Phi_c\,.
\label{mPhi}
\end{equation}

Since $F_{0}\delta_{F}=f_{,RR}(R_{0})\delta R$, it follows that
\begin{equation}
\delta R=\frac{f_{,R}(R_{0})}{f_{,RR}(R_{0})}
\delta_{F}\,.
\end{equation}
The validity of the linear expansion requires that 
$\delta R\ll R_{0}$, which translates into
$\delta_{F}\ll m(R_{0})$. 
Since $\delta_{F} \simeq 2GM_{c}/(3F_{0}r_{c})=2\Phi_c/3$
at $r=r_{c}$, one has $\delta_{F}\ll m(R_{0}) \ll 1$
under the condition (\ref{mPhi}).
Hence the linear analysis given above is valid for $m(R_0) \gg \Phi_c$.

For the distance $r$ close to $r_{c}$ the post Newtonian parameter 
in Eq.~(\ref{GeffN}) is given by $\gamma\simeq1/2$ 
(i.e., because $Mr \ll 1$).
The tightest experimental bound on $\gamma$ is 
given by~\cite{Will1,Bertotti,Will2}:
\begin{equation}
|\gamma-1|<2.3 \times 10^{-5}\,,
\label{gammacon}
\end{equation}
which comes from the time-delay effect of the Cassini tracking for Sun.
This means that \fR\ gravity models with the
light scalaron mass ($M r_{c} \ll 1$) do not satisfy local gravity
constraints~\cite{OlmoPRL,Olmo05,Fara06,Erick06,Chiba07,Navarro,Kainu07,Kainu08}. 
The mean density of Earth or Sun is of the order of
$\rho\simeq 1\mbox{\,--\,}10\mathrm{\ g/cm}^{3}$, which is much larger than the present
cosmological density $\rho_{c}^{(0)}\simeq 10^{-29}\mathrm{\ g/cm}^{3}$.
In such an environment the condition $\delta R \ll R_0$ is violated
and the field mass $M$ becomes large such that $M r_{c} \gg 1$. 
The effect of the chameleon mechanism~\cite{chame1,chame2} 
becomes important in this non-linear 
regime ($\delta R \gg R_0$)~\cite{Teg, Hu07, CapoTsuji, Van}. 
In Section \ref{chameleonsec} we will show that the \fR\ models can be 
consistent with local gravity constraints provided that the 
chameleon mechanism is at work.


\subsection{Chameleon mechanism in \fR\ gravity}
\label{chameleonsec}

Let us discuss the chameleon mechanism~\cite{chame1,chame2} 
in metric \fR\ gravity.
Unlike the linear expansion approach given in Section 
\ref{linearexpan}, this corresponds to a non-linear effect 
arising from a
large departure of the Ricci scalar from its background 
value $R_{0}$. The mass of an effective scalar field degree 
of freedom depends on the density of its environment.
If the matter density is sufficiently high, the
field acquires a heavy mass about the potential minimum.
Meanwhile the field has a lighter mass in a low-density 
cosmological environment
relevant to dark energy so that it can propagate freely.
As long as the spherically symmetric body has a thin-shell
around its surface, the effective coupling between the field and 
matter becomes much smaller than the bare coupling $|Q|$.
In the following we shall review the chameleon mechanism 
for general couplings $Q$ and then proceed to constrain 
\fR\ dark energy models from local gravity tests.

\subsubsection{Field profile of the chameleon field}

The action~(\ref{fRaction}) in \fR\ gravity can be transformed to 
the Einstein frame action~(\ref{Ein}) with the coupling $Q=-1/\sqrt{6}$
between the scalaron field $\phi=\sqrt{3/(2\kappa^2)}\ln \,F$ 
and non-relativistic matter.
Let us consider a spherically symmetric body with radius 
$\tilde{r}_c$ in the Einstein frame.
We approximate that the background geometry 
is described by the Minkowski space-time.
Varying the action~(\ref{Ein}) with respect to 
the field $\phi$, we obtain
\begin{equation}
\frac{\mathrm{d}^2 \phi}{\mathrm{d} \tr^2}+
\frac{2}{\tr} \frac{\mathrm{d}\phi}{\mathrm{d}\tr}-
\frac{\mathrm{d}V_{\mathrm{eff}}}{\mathrm{d}\phi}=0\,,
\label{dreq}
\end{equation}
where $\tr$ is a distance from the center of symmetry
that is related to the distance $r$ in the Jordan frame
via $\tilde{r}=\sqrt{F}r=e^{-Q\kappa \phi}r$.
The effective potential $V_{\mathrm{eff}}$ is defined by 
\begin{equation}
V_{\mathrm{eff}}(\phi)=V(\phi)+
e^{Q \kappa \phi}\rho^*\,,
\label{Veff}
\end{equation}
where $\rho^*$ is a conserved quantity in the Einstein 
frame~\cite{chame2}. Recall that the field potential $V(\phi)$
is given in Eq.~(\ref{potentialein}).
The energy density $\tilde{\rho}$ in the Einstein 
frame is related with the energy density $\rho$
in the Jordan frame via the relation 
$\tilde{\rho}=\rho/F^2=e^{4Q \kappa \phi}\rho$.
Since the conformal transformation gives rise to a coupling $Q$
between matter and the field, $\tilde{\rho}$ is not a 
conserved quantity.
Instead the quantity $\rho^*=e^{3Q\kappa \phi}\rho=e^{-Q\kappa \phi} \tilde{\rho}$ 
corresponds to a conserved quantity, which satisfies $\tilde{r}^3\rho^{*}=r^3\rho$.
Note that Eq.~(\ref{dreq}) is consistent with Eq.~(\ref{fieldeq2}).

In the following we assume that a spherically symmetric body has a constant density
$\rho^*=\rho_A$ inside the body ($\tr<\tr_c$) and that the energy
density outside the body ($\tr>\tr_c$) is $\rho^*=\rho_B$ ($\ll \rho_A$).
The mass $M_c$ of the body and the gravitational potential $\Phi_c$ at the
radius $\tr_c$ are given by $M_c=(4\pi/3)\tr_c^3 \rho_A$ and
$\Phi_c=GM_c/\tr_c$, respectively. 
The effective potential has minima at the field values
$\phi_A$ and $\phi_B$:
\begin{eqnarray}
V_{,\phi} (\phi_A)+\kappa Q e^{Q \kappa \phi_A}\rho_A &=& 0\,,\\
V_{,\phi} (\phi_B)+\kappa Q e^{Q \kappa \phi_B}\rho_B &=& 0\,.
\end{eqnarray}
The former corresponds to the region of
high density with a heavy mass squared 
$m_A^2 \equiv V_{\mathrm{eff},\phi \phi}(\phi_A)$, 
whereas the latter to a lower density region with a lighter 
mass squared $m_B^2 \equiv V_{\mathrm{eff},\phi \phi}(\phi_B)$.
In the case of Sun, for example, the field value $\phi_B$ is 
determined by the homogeneous dark matter/baryon 
density in our galaxy, i.e., $\rho_B \simeq 10^{-24}\mathrm{\ g/cm}^3$.

\epubtkImage{potz-potzeffrho.png}{%
  \begin{figure}[hptb]
    \centerline{\includegraphics[width=3.7in]{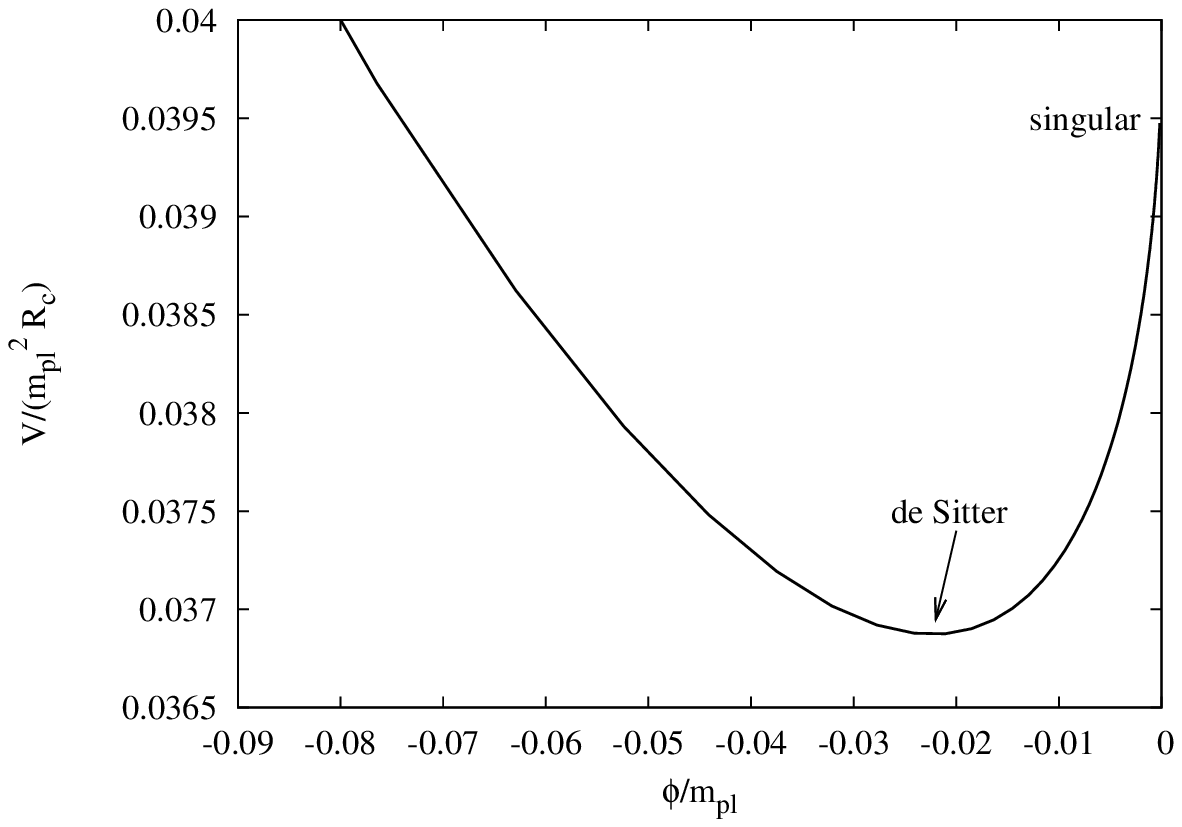}}
    \centerline{\includegraphics[width=3.7in]{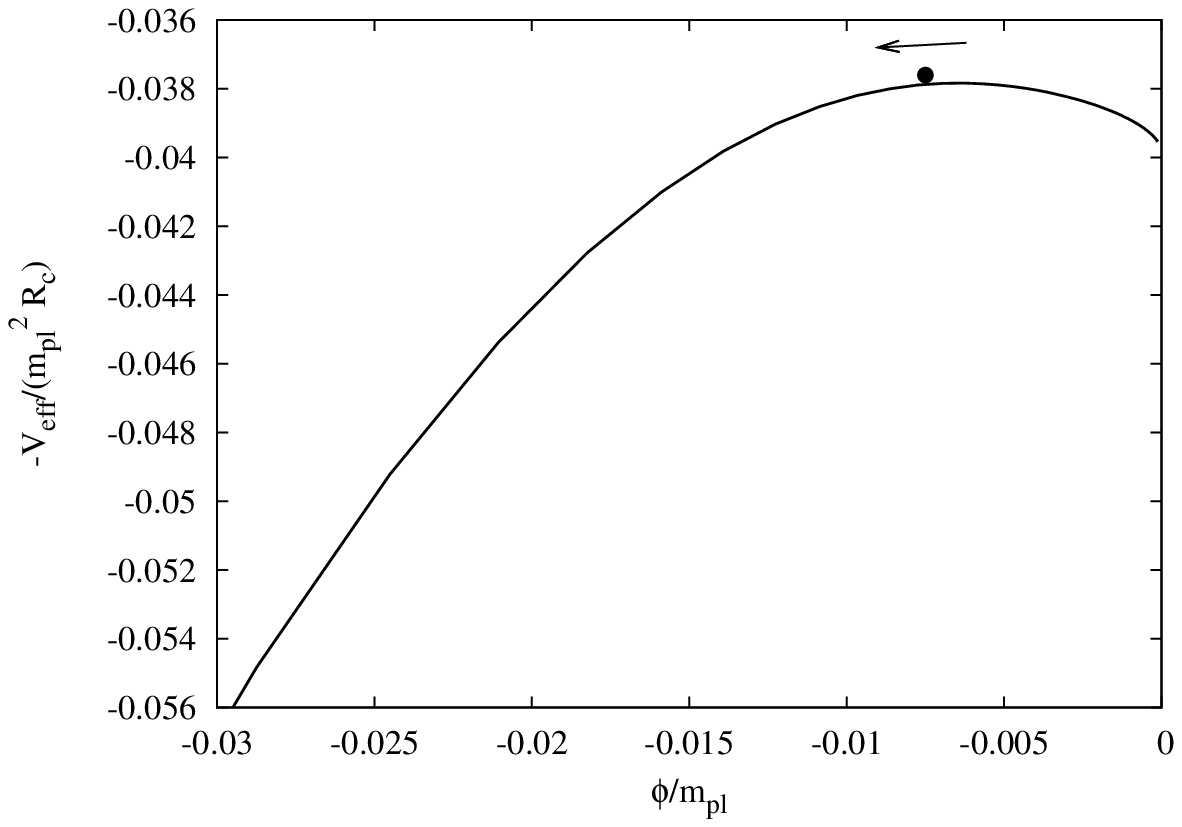}}
    \caption{(Top) The potential $V(\phi)=(FR-f)/(2\kappa^2F^2)$
      versus the field $\phi=\sqrt{3/(16\pi)}m_{\mathrm{pl}}\,\ln F$
      for the Starobinsky's dark energy model~(\ref{Bmodel}) with
      $n=1$ and $\mu=2$. (Bottom) The inverted effective potential
      $-V_{\mathrm{eff}}$ for the same model parameters as the top
      with $\rho^*=10R_cm_{\mathrm{pl}}^2$. The field value, at which
      the inverted effective potential has a maximum, is different
      depending on the density $\rho^*$, see Eq.~(\ref{kappaphiM}). In
      the upper panel ``de~Sitter'' corresponds to the minimum of the
      potential, whereas ``singular'' means that the curvature
      diverges at $\phi=0$.}
    \label{pofigfR} 
\end{figure}}

When $Q>0$ the effective potential has a minimum for the models 
with $V_{,\phi}<0$, which occurs, e.g., 
for the inverse power-law potential $V(\phi)=M^{4+n}\phi^{-n}$.
The \fR\ gravity corresponds to a negative coupling  ($Q=-1/\sqrt{6}$), 
in which case the effective potential has a minimum for $V_{,\phi}>0$.
As an example, let us consider the shape of the effective potential 
for the models (\ref{Amodel}) and (\ref{Bmodel}).
In the region $R \gg R_c$ both models behave as
\begin{equation}
f(R) \simeq R-\mu R_c \left[1-
\left( R_c/R  \right)^{2n} \right]\,.
\label{fRasy}
\end{equation}
For this functional form it follows that 
\begin{eqnarray}
F &=& e^{\frac{2}{\sqrt{6}}\kappa \phi}
=1-2n \mu(R/R_c)^{-(2n+1)}\,,
\label{Fun} \\
V(\phi) &=& \frac{\mu R_c}{2\kappa^2}
e^{-\frac{4}{\sqrt{6}}\kappa \phi} 
\left[1-(2n+1)\left( \frac{-\kappa \phi}{\sqrt{6}n \mu}
\right)^{\frac{2n}{2n+1}}\right]\,.
\label{Vein}
\end{eqnarray}
The r.h.s.\ of Eq.~(\ref{Fun}) is smaller than 1, so that $\phi<0$. 
The limit $R \to \infty$ corresponds to $\phi \to -0$.
In the limit $\phi \to -0$ one has $V \to \mu R_c/(2\kappa^2)$ and 
$V_{,\phi} \to \infty$. 
This property can be seen in the upper panel of Figure~\ref{pofigfR}, 
which shows the potential $V(\phi)$
for the model~(\ref{Bmodel}) with parameters $n=1$ and $\mu=2$.
Because of the existence of the coupling term $e^{-\kappa \phi/\sqrt{6}}\rho^*$,
the effective potential $V_{\mathrm{eff}}(\phi)$ has a minimum at 
\begin{equation}
\kappa \phi_M=-\sqrt{6}n \mu \left( \frac{R_c}{\kappa^2 \rho^*}
\right)^{2n+1}\,.
\label{kappaphiM}
\end{equation}
Since $R \sim \kappa^2 \rho^* \gg R_c$ in the region of high density, 
the condition $|\kappa \phi_M| \ll 1$ is in fact justified
(for $n$ and $\mu$ of the order of unity).
The field mass $m_\phi$ about the minimum of 
the effective potential is given by 
\begin{equation}
m_{\phi}^2=
\frac{1}{6n(n+1)\mu} R_c
\left( \frac{\kappa^2 \rho^*}{R_c} \right)^{2(n+1)}\,.
\end{equation}
This shows that, in the regime $R \sim \kappa^2 \rho^* \gg R_c$,
$m_{\phi}$ is much larger than the present Hubble
parameter $H_0$ ($\sim \sqrt{R_c}$).
Cosmologically the field evolves along the instantaneous minima
characterized by Eq.~(\ref{kappaphiM}) and then it approaches
a de~Sitter point which appears as a minimum of the potential 
in the upper panel of Figure~\ref{pofigfR}.

In order to solve the ``dynamics'' of
the field $\phi$ in Eq.~(\ref{dreq}), we need to consider the
inverted effective potential $(-V_{\mathrm{eff}})$.
See the lower panel of Figure~\ref{pofigfR} for illustration 
[which corresponds to the model~(\ref{Bmodel})].
We impose the following boundary conditions:
\begin{eqnarray}
\frac{\mathrm{d}\phi}{\mathrm{d}\tr}(\tr=0)&=&0\,, 
\label{bou1}\\
\phi(\tr \to\infty)&=&\phi_{B}\,.
\label{bou2}
\end{eqnarray}
The boundary condition (\ref{bou2}) can be also understood 
as $\lim_{\tr\to\infty}\de\phi/\de\tr=0$.
The field $\phi$ is at rest at $\tr=0$ and starts to roll down the
potential when the matter-coupling term 
$\kappa Q\rho_{A}e^{Q \kappa \phi}$
in Eq.~(\ref{dreq}) becomes important at a radius $\tr_{1}$.  
If the field value at $\tr=0$ is close to $\phi_A$, the field stays
around $\phi_A$ in the region $0<\tr<\tr_1$.
The body has a thin-shell if $\tr_1$ is close to the radius 
$\tr_c$ of the body. 

In the region $0 < \tr<\tr_1$ one can approximate the r.h.s.\ of 
Eq.~(\ref{dreq}) as $\rd V_{\mathrm{eff}}/\rd \phi \simeq m_A^2 (\phi-\phi_A)$ 
around $\phi=\phi_A$, where 
$m_A^2=R_c(\kappa^2 \rho_A/R_c)^{2(n+1)}/[6n(n+1)]$.
Hence the solution to Eq.~(\ref{dreq}) is given by 
$\phi(\tr)=\phi_A+A e^{-m_A \tr}/\tr+B e^{m_A \tr}/\tr$, where 
$A$ and $B$ are constants.
In order to avoid the divergence of $\phi$ at $\tr=0$
we demand the condition $B=-A$, 
in which case the solution is  
\begin{eqnarray}
\label{sol1}
\phi(\tr)=\phi_A+\frac{A (e^{-m_A \tr}-e^{m_A \tr})}{\tr} \qquad
(0 < \tr <\tr_1).
\end{eqnarray}
In fact, this satisfies the boundary condition (\ref{bou1}).

In the region $\tr_1<\tr<\tr_c$ the field $|\phi(\tr)|$ evolves
toward larger values with the increase of $\tr$.
In the lower panel of Figure~\ref{pofigfR} the field stays
around the potential maximum for $0<\tr<\tr_1$, 
but in the regime $\tr_1<\tr<\tr_c$
it moves toward the left (largely negative $\phi$ region).
Since $|V_{,\phi}| \ll |\kappa Q \rho_A e^{Q \kappa\phi}|$ 
in this regime we have that 
$\rd V_{\mathrm{eff}}/\rd \phi \simeq \kappa Q \rho_A$ 
in Eq.~(\ref{dreq}), where 
we used the condition $Q\kappa \phi \ll 1$.
Hence we obtain the following solution 
\begin{equation}
\label{sol2}
\phi(\tr)=\frac16 \kappa Q \rho_A \tr^2 -\frac{C}{\tr}+D
\qquad (\tr_1 < \tr < \tr_c),
\end{equation}
where $C$ and $D$ are constants.

Since the field acquires a sufficient kinetic energy in the region 
$\tr_1<\tr<\tr_c$, the field climbs up the potential hill 
toward the largely negative $\phi$ region 
outside the body ($\tr>\tr_c$).
The shape of the effective potential changes relative to that 
inside the body because the density drops from $\rho_A$
to $\rho_B$.
The kinetic energy of the field dominates over the
potential energy, which means that the term 
$\mathrm{d}V_{\mathrm{eff}}/\mathrm{d} \phi$
in Eq.~(\ref{dreq}) can be neglected.
Recall that one has $|\phi_B| \gg |\phi_A|$ under the 
condition $\rho_A \gg \rho_B$ [see Eq.~(\ref{kappaphiM})].
Taking into account the mass term 
$m_B^2=R_c(\kappa^2 \rho_B/R_c)^{2(n+1)}/[6n(n+1)]$, 
we have $\rd V_{\mathrm{eff}}/\rd \phi 
\simeq m_B^2 (\phi-\phi_B)$ on the r.h.s.\ of Eq.~(\ref{dreq}).
Hence we obtain the solution 
$\phi(\tr)=\phi_B+E e^{-m_B (\tr-\tr_c)}/\tr+
F e^{m_B (\tr-\tr_c)}/\tr$
with constants $E$ and $F$.
Using the boundary condition (\ref{bou2}), it follows that 
$F=0$ and hence
\begin{equation}
\label{sol3}
\phi(\tr)=\phi_B+E \frac{e^{-m_B (\tr-\tr_c)}}{\tr}
\qquad(\tr >\tr_c)\,.
\end{equation}

Three solutions (\ref{sol1}), (\ref{sol2}) and (\ref{sol3}) should be
matched at $\tr=\tr_1$ and $\tr=\tr_c$ by imposing continuous conditions
for $\phi$ and $\rd \phi/\rd \tr$.
The coefficients $A$, $C$, $D$ and $E$ 
are determined accordingly~\cite{TT08}:
\begin{eqnarray}
\label{re1}
C &=& \frac{s_1 s_2 [(\phi_B-\phi_A)+(\tr_1^2-\tr_c^2)\kappa Q \rho_A/6]
+[ s_2 \tr_1^2 (e^{-m_A \tr_1}-e^{m_A \tr_1})-s_1 \tr_c^2] \kappa Q\rho_A/3}
{m_A (e^{-m_A \tr_1}+e^{m_A \tr_1})s_2-m_B s_1}\,,\\
\label{re2}
A &=& -\frac{1}{s_1} (C+\kappa Q \rho_A \tr_1^3/3)\,,\\
\label{re3}
E &=& -\frac{1}{s_2}(C+\kappa Q \rho_A \tr_c^3/3)\,, \\
\label{re4}
D &=& \phi_B - \frac16 \kappa Q \rho_A \tr_c^2+
\frac{1}{\tr_c}(C+E)\,,
\end{eqnarray}
where 
\begin{eqnarray}
s_1 &\equiv& m_A \tr_1 (e^{-m_A \tr_1}+e^{m_A \tr_1})+
e^{-m_A \tr_1}-e^{m_A \tr_1}\,,\\
s_2 &\equiv& 1+m_B \tr_c\,.
\end{eqnarray}
If the mass $m_B$ outside the body is small to satisfy the condition 
$m_B \tr_c \ll 1$ and $m_A \gg m_B$, we can neglect the contribution 
of the $m_B$-dependent terms in Eqs.~(\ref{re1})\,--\,(\ref{re4}).
Then the field profile is given by~\cite{TT08}
\begin{eqnarray}
\label{ne1}
\phi(\tr) &=& \phi_A-\frac{1}{m_A (e^{-m_A \tr_1}+e^{m_A \tr_1})}
\left[ \phi_B-\phi_A+\frac12 \kappa Q \rho_A (\tr_1^2-\tr_c^2) \right]
\frac{e^{-m_A \tr}-e^{m_A \tr}}{\tr}\,, \nonumber \\
& & (0<\tr<\tr_1)\,, \\
\label{ne2}
\phi(\tr) &=& \phi_B+\frac16 \kappa Q \rho_A (\tr^2-3\tr_c^2)
+\frac{\kappa Q \rho_A \tr_1^3}{3\tr} \nonumber \\
& & -\left[ 1+\frac{e^{-m_A \tr_1}-e^{m_A \tr_1}}
{m_A \tr_1 (e^{-m_A \tr_1}+e^{m_A \tr_1})} \right]
\left[ \phi_B-\phi_A+\frac12 \kappa Q \rho_A (\tr_1^2-\tr_c^2) \right]
\frac{\tr_1}{\tr}\,, \nonumber \\
& & (\tr_1<\tr<\tr_c)\,,\\
\label{ne3}
\phi(\tr) &=& \phi_B - \Biggl[ \tr_1 (\phi_B -\phi_A)+
\frac16 \kappa Q \rho_A \tr_c^3
\left(2+\frac{\tr_1}{\tr_c} \right) 
\left(1-\frac{\tr_1}{\tr_c} \right)^2 \nonumber \\
& & \qquad +\frac{e^{-m_A \tr_1}-e^{m_A \tr_1}}
{m_A  (e^{-m_A \tr_1}+e^{m_A \tr_1})} 
\left\{ \phi_B-\phi_A+\frac12 \kappa Q \rho_A (\tr_1^2-\tr_c^2)
\right\} \Biggr] \frac{e^{-m_B (\tr-\tr_c)}}{\tr}\,, \nonumber \\
& & (\tr>\tr_c)\,.
\end{eqnarray}
Originally a similar field profile was derived in~\cite{chame1,chame2}
by assuming that the field is frozen at $\phi=\phi_A$ 
in the region $0<\tr<\tr_1$.

The radius $r_1$ is determined by the following condition
\begin{eqnarray}
m_A^2 \left[ \phi(\tr_1) -\phi_A \right]=\kappa Q \rho_A\,.
\label{macon}
\end{eqnarray}
This translates into
\begin{eqnarray}
\label{r1con}
 \phi_B -\phi_A +\frac12 \kappa Q \rho_A 
(\tr_1^2-\tr_c^2) =\frac{6Q \Phi_c}{\kappa (m_A \tr_c)^2}
\frac{m_A \tr_1 (e^{m_A \tr_1}+e^{-m_A \tr_1})}
{e^{m_A \tr_1}-e^{-m_A \tr_1}}\,,
\end{eqnarray}
where $\Phi_c=\kappa^2 M_c/(8\pi \tr_c)
=\kappa^2 \rho_A \tr_c^2/6$
is the gravitational potential at the surface of the body.
Using this relation, the field profile (\ref{ne3})
outside the body reduces to  
\begin{eqnarray}
\label{fieldout}
 \phi (\tr) &=& \phi_B -\frac{2Q\Phi_c}{\kappa} 
 \left[ 1 -\frac{\tr_1^3}{\tr_c^3} +3\frac{\tr_1}{\tr_c}
 \frac{1}{(m_A \tr_c)^2} \left\{
 \frac{m_A \tr_1 (e^{m_A \tr_1}+e^{-m_A \tr_1})}
 {e^{m_A \tr_1}-e^{-m_A \tr_1}}-1 \right\} \right]
 \frac{\tr_c e^{-m_B (\tr-\tr_c)}}{\tr}\,, \nonumber \\
 & & (\tr>\tr_c)\,.
\end{eqnarray}

If the field value at $\tr=0$ is away from $\phi_A$, 
the field rolls down the potential for $\tr>0$.
This corresponds to taking the limit $\tr_1 \to 0$ 
in Eq.~(\ref{fieldout}), in which case the field profile 
outside the body is given by 
\begin{equation}
\phi(\tr)=\phi_B- \frac{2Q}{\kappa} 
\frac{GM_c}{\tr} e^{-m_B(\tr-\tr_c)}\,.
\end{equation}
This shows that the effective coupling is of the order 
of $Q$ and hence for $|Q|={\cal O}(1)$ local gravity 
constraints are not satisfied.

\subsubsection{Thin-shell solutions}

Let us consider the case in which $\tr_1$ 
is close to $\tr_c$, i.e.
\begin{equation}
\Delta \tr_c \equiv \tr_c-\tr_1 \ll \tr_c\,.
\label{delrc}
\end{equation}
This corresponds to the thin-shell regime in which 
the field is stuck inside the star except 
around its surface. If the field is sufficiently 
massive inside the star to satisfy the condition 
$m_A \tilde{r}_c \gg 1$, Eq.~(\ref{r1con}) gives the 
following relation 
\begin{equation}
\epsilon_{\mathrm{th}} \equiv \frac{\kappa (\phi_B-\phi_A)}
{6Q\Phi_c} \simeq \frac{\Delta \tr_c}{\tr_c}+
\frac{1}{m_A \tr_c}\,,
\label{epth}
\end{equation}
where $\epsilon_{\mathrm{th}}$ is called the thin-shell 
parameter~\cite{chame1,chame2}.
Neglecting second-order terms with respect to 
$\Delta \tr_c/\tr_c$ and $1/(m_A \tr_c)$ in Eq.~(\ref{fieldout}), 
it follows that 
\begin{equation}
\label{phirthin}
\phi(\tr) \simeq \phi_B -\frac{2Q_{\mathrm{eff}}}{\kappa} 
\frac{GM_c}{\tr} e^{-m_B (\tr-\tr_c)}\,,
\end{equation}
where $Q_{\mathrm{eff}}$ is the effective coupling given by 
\begin{eqnarray}
\label{Qeff0}
Q_{\mathrm{eff}}=3Q \epsilon_{\mathrm{th}}\,.
\end{eqnarray}

Since $\epsilon_{\mathrm{th}} \ll 1$ under the conditions 
$\Delta \tr_c/\tr_c \ll 1$ and $1/(m_A \tr_c) \ll 1$, 
the amplitude of the effective coupling $Q_{\mathrm{eff}}$ 
becomes much smaller than 1.  
In the original papers of Khoury and Weltman~\cite{chame1,chame2}
the thin-shell solution was derived by assuming that the field is
frozen with the value $\phi=\phi_A$
in the region $0<\tr<\tr_1$. In this case the
thin-shell parameter is given by $\epsilon_{\mathrm{th}} \simeq \Delta
\tr_c/\tr_c$, which is different from Eq.~(\ref{epth}). 
However, this difference is not important
because the condition $ \Delta \tr_c/\tr_c \gg 1/(m_A
\tr_c)$ is satisfied for most of viable models~\cite{TT08}.

\subsubsection{Post Newtonian parameter}

We derive the bound on the thin-shell parameter from 
experimental tests of the post Newtonian parameter in the solar system.
The spherically symmetric metric in the Einstein frame is described 
by~\cite{Teg} 
\begin{equation}
\mathrm{d}\tilde{s}^{2}=\tilde{g}_{\mu\nu}\rd \tilde{x}^{\mu}
\rd \tilde{x}^{\nu}=-[1-2\tilde{{\cal A}}(\tilde{r})]\mathrm{d}t^{2}
+[1+2\tilde{{\cal B}}(\tilde{r})]\mathrm{d}\tilde{r}^{2}
+\tilde{r}^{2}\mathrm{d}\Omega^{2}\,,
\label{Einmet}
\end{equation}
where $\tilde{{\cal A}}(\tilde{r})$ and $\tilde{{\cal B}}(\tilde{r})$ are 
functions of $\tilde{r}$ and $\rd\Omega^{2}=
\rd\theta^{2}+(\sin^{2}\theta)\,\rd\phi^{2}$.
In the weak gravitational background 
($\tilde{{\cal A}}(\tilde{r}) \ll 1$ and $\tilde{{\cal B}}(\tilde{r}) \ll 1$)
the metric outside the spherically symmetric body 
with mass $M_{c}$ is given by $\tilde{{\cal A}}(\tr) 
\simeq \tilde{{\cal B}}(\tr) \simeq GM_{c}/\tr$.

Let us transform the metric (\ref{Einmet}) back to that in the Jordan frame
under the inverse of the conformal transformation, 
$g_{\mu \nu}=e^{2Q\kappa \phi}\tilde{g}_{\mu \nu}$. 
Then the metric in the Jordan frame, 
$\rd s^2=e^{2Q\kappa \phi} \rd \tilde{s}^2
=g_{\mu \nu} \rd x^{\mu} \rd x^{\nu}$, 
is given by 
\begin{equation}
\mathrm{d} s^{2}=-[1-2 {\cal A}(r)]\mathrm{d}t^{2}+
 [1+2{\cal B}(r)]\mathrm{d} r^{2}+r^{2}\mathrm{d}\Omega^{2}\,.
 \end{equation}
Under the condition $|Q \kappa \phi|\ll1$ we obtain
the following relations 
\begin{equation}
\tilde{r}=e^{Q\kappa \phi}r\,,\quad {\cal A}(r) \simeq 
\tilde{{\cal A}}(\tr)-Q\kappa \phi(\tr)\,,\quad
{\cal B}(r) \simeq \tilde{{\cal B}}(\tr)
-Q \kappa \tr \frac{\mathrm{d}\phi(\tr)}{\mathrm{d}\tr}\,.
\label{tilrAr}
\end{equation}
In the following we use the approximation $r \simeq \tr$,
which is valid for $|Q \kappa \phi|\ll1$.
Using the thin-shell solution (\ref{phirthin}), it follows that 
\begin{eqnarray}
{\cal A}(r)=\frac{GM_{c}}{r}\left[1+6Q^{2}
\epsilon_{\mathrm{th}}\left(1-r/r_{c}\right)\right]\,,\qquad
{\cal B}(r)
=\frac{GM_{c}}{r}\left(1-6Q^{2}\epsilon_{\mathrm{th}}\right)\,,
\end{eqnarray}
where we have used the approximation $|\phi_{B}|\gg|\phi_{A}|$ and
hence $\phi_{B} \simeq 6Q\Phi_{c}\epsilon_{\mathrm{th}}/\kappa$.

The term $Q \kappa \phi_{B}$ in Eq.~(\ref{tilrAr})
is smaller than ${\cal A}(r)=GM_{c}/r$ under
the condition $r/r_{c}<(6Q^{2}\epsilon_{\mathrm{th}})^{-1}$.
Provided that the field $\phi$ reaches the value $\phi_{B}$ with the
distance $r_{B}$ satisfying the condition 
$r_{B}/r_{c}<(6Q^{2}\epsilon_{\mathrm{th}})^{-1}$,
the metric ${\cal A}(r)$ does not change its sign for $r<r_{B}$.
The post-Newtonian parameter $\gamma$ is given by 
\begin{equation}
\gamma \equiv \frac{{\cal B}(r)}{{\cal A}(r)}
\simeq\frac{1-6Q^{2}\epsilon_{\mathrm{th}}}
{1+6Q^{2}\epsilon_{\mathrm{th}}(1-r/r_{c})}\,.
\label{gam}
\end{equation}
The experimental bound (\ref{gammacon}) can be satisfied as long as
the thin-shell parameter $\epsilon_{\mathrm{th}}$ is much smaller than 1.
If we take the distance $r=r_c$, the constraint (\ref{gammacon}) 
translates into 
\begin{equation}
\epsilon_{\mathrm{th},\odot}<3.8\times10^{-6}/Q^{2}\,,
\label{solarepbound}
\end{equation}
where $\epsilon_{\mathrm{th},\odot}$ is the thin-shell parameter
for Sun. In \fR\ gravity ($Q=-1/\sqrt{6}$) this corresponds to 
$\epsilon_{\mathrm{th},\odot}<2.3 \times 10^{-5}$.

\subsubsection{Experimental bounds from the violation of equivalence principle}

Let us next discuss constraints on the thin-shell
parameter from the possible violation of equivalence principle
(EP).  The tightest bound comes from the solar system tests of weak EP
using the free-fall acceleration of Earth ($a_{\oplus}$) and 
Moon ($a_{\mathrm{Moon}}$) toward Sun~\cite{chame2}.
The experimental bound on the difference of two accelerations 
is given by~\cite{Will1,Bertotti,Will2}
\begin{eqnarray}
\frac{|a_{\oplus}-a_{\mathrm{Moon}}|}
{|a_{\oplus}+a_{\mathrm{Moon}}|/2}<10^{-13}\,.
\label{etamoon}
\end{eqnarray}

Provided that Earth, Sun, and Moon have thin-shells, the
field profiles outside the bodies are given by Eq.~(\ref{phirthin})
with the replacement of corresponding quantities. 
The presence of the field $\phi (r)$ with an effective coupling 
$Q_{\mathrm{eff}}$ gives rise to an extra acceleration, 
$a^{\mathrm{fifth}}=|Q_{\mathrm{eff}}\nabla\phi(r)|$.
Then the accelerations $a_{\oplus}$ and $a_{\mathrm{Moon}}$ 
toward Sun (mass $M_{\odot}$) are~\cite{chame2}
\begin{eqnarray}
a_{\oplus} & \simeq & \frac{GM_{\odot}}{r^{2}}\left[1+18Q^{2}
\epsilon_{\mathrm{th},\oplus}^{2}\frac{\Phi_{\oplus}}{\Phi_{\odot}}
\right]\,,\\
a_{\mathrm{Moon}} & \simeq & \frac{GM_{\odot}}{r^{2}}
\left[1+18Q^{2}\epsilon_{\mathrm{th},\oplus}^{2}
\frac{\Phi_{\oplus}^{2}}{\Phi_{\odot}\Phi_{\mathrm{Moon}}}\right]\,,
\end{eqnarray}
where $\epsilon_{\mathrm{th},\oplus}$ is the thin-shell parameter of
Earth, and $\Phi_{\odot}\simeq2.1\times10^{-6}$, 
$\Phi_{\oplus}\simeq7.0\times10^{-10}$,
$\Phi_{\mathrm{Moon}} \simeq 3.1\times 10^{-11}$ are the gravitational
potentials of Sun, Earth and Moon, respectively. 
Hence the condition (\ref{etamoon}) translates into~\cite{CapoTsuji,TUMTY}
\begin{equation}
\epsilon_{\mathrm{th},\oplus}<8.8\times10^{-7}/|Q|\,,
\label{boep}
\end{equation}
which corresponds to $\epsilon_{\mathrm{th},\oplus}<2.2 \times 10^{-6}$
in \fR\ gravity.
This bound provides a tighter bound on model parameters 
compared to (\ref{solarepbound}).

Since the condition $|\phi_B| \gg |\phi_A|$ is satisfied for $\rho_A \gg \rho_B$, 
one has $\epsilon_{\mathrm{th},\oplus} \simeq \kappa \phi_B/(6Q \Phi_{\oplus})$ from 
Eq.~(\ref{epth}). Then the bound (\ref{boep}) translates into 
\begin{equation}
|\kappa \phi_{B,\oplus}|<3.7 \times 10^{-15}\,.
\label{boep2}
\end{equation}

\subsubsection{Constraints on model parameters in \fR\ gravity}

We place constraints on the \fR\ models given in Eqs.~(\ref{Amodel}) and (\ref{Bmodel})
by using the experimental bounds discussed above. 
In the region of high density where $R$ is much larger than $R_c$,
one can use the asymptotic form (\ref{fRasy}) to discuss local gravity constraints. 
Inside and outside the spherically symmetric body the effective potential 
$V_{\mathrm{eff}}$ for the model~(\ref{fRasy}) has two minima at 
\begin{eqnarray}
\kappa \phi_{A} \simeq -\sqrt{6}n\mu
\left( \frac{R_{c}}{\kappa^2 \rho_{A}} \right)^{2n+1}\,,
\qquad
\kappa \phi_{B} \simeq -\sqrt{6}n\mu
\left( \frac{R_{c}}{\kappa^2 \rho_{B}} \right)^{2n+1}\,.
\end{eqnarray}

The bound (\ref{boep2}) translates into 
\begin{equation}
\frac{n\mu}{x_{d}^{2n+1}}\left(\frac{R_{1}}
{\kappa^2 \rho_{B}} \right)^{2n+1}<1.5\times10^{-15}\,,
\label{consmo1}
\end{equation}
where $x_{d}\equiv R_{1}/R_{c}$
and $R_1$ is the Ricci scalar at the late-time de~Sitter point. 
In the following we consider the case in which the Lagrangian density
is given by (\ref{fRasy}) for $R\ge R_{1}$. 
If we use the models (\ref{Amodel}) and (\ref{Bmodel}),
then there are some modifications for the estimation of $R_{1}$. 
However this change should be insignificant when we place 
constraints on model parameters.

At the de~Sitter point the model~(\ref{fRasy}) satisfies 
the condition $\mu=x_{d}^{2n+1}/[2(x_{d}^{2n}-n-1)]$.
Substituting this relation into Eq.~(\ref{consmo1}), we find
\begin{equation}
\frac{n}{2(x_{d}^{2n}-n-1)}
\left(\frac{R_{1}}{\kappa^2 \rho_{B}} \right)^{2n+1}
<1.5\times10^{-15}\,.
\label{cons2}
\end{equation}
For the stability of the de~Sitter point we require that $m(R_{1})<1$,
which translates into the condition $x_{d}^{2n}>2n^{2}+3n+1$. Hence
the term $n/[2(x_{d}^{2n}-n-1)]$ in Eq.~(\ref{cons2}) is smaller
than 0.25 for $n>0$.

We use the approximation that $R_{1}$ and $\rho_{B}$ are of 
the orders of the present cosmological density $10^{-29}\mathrm{\ g/cm}^{3}$
and the baryonic/dark matter density $10^{-24}\mathrm{\ g/cm}^{3}$ 
in our galaxy, respectively.
From Eq.~(\ref{cons2}) we obtain the bound~\cite{CapoTsuji}
\begin{equation}
n>0.9\,.
\label{bound3}
\end{equation}
Under this condition one can see an appreciable deviation from 
the $\Lambda$CDM model cosmologically as $R$ decreases to 
the order of $R_{c}$.

If we consider the model~(\ref{powermodel}), it was shown in~\cite{CapoTsuji} that 
the bound (\ref{boep2}) gives the constraint $n<3 \times 10^{-10}$.
This means that the deviation from the $\Lambda$CDM model is very
small. Meanwhile, for the models (\ref{Amodel}) and (\ref{Bmodel}), 
the deviation from the $\Lambda$CDM model can be large
even for $n={\cal O}(1)$, while satisfying local gravity constraints.
We note that the model~(\ref{tanh}) is also consistent 
with local gravity constraints.

\newpage

\section{Cosmological Perturbations}
\label{persec}
\setcounter{equation}{0}

The \fR\ theories have one extra scalar degree of freedom compared
to the $\Lambda$CDM model. This feature results in more freedom for
the background. As we have seen previously, a viable cosmological sequence 
of radiation, matter, and accelerated epochs is possible provided some conditions
hold for \fR. In principle, however, one can specify any given $H=H(a)$ and
solve Eqs.~(\ref{fRinf1}) and (\ref{fRinf2}) for those $f(R(a))$ 
compatible with the given $H(a)$.

Therefore the background cosmological evolution is not in general 
enough to distinguish \fR\ theories from other theories. 
Even worse, for the same $H(a)$, there may be some
different forms of \fR\ which fulfill the Friedmann equations. 
Hence other observables are needed in order to distinguish between
different theories. In order to achieve this goal, perturbation theory turns 
out to be of fundamental importance. More than this, perturbations theory in
cosmology has become as important as in particle physics, since it
gives deep insight into these theories by providing information
regarding the number of independent degrees of freedom, their speed of
propagation, their time-evolution: all observables to be confronted
with different data sets.

The main result of the perturbation analysis in \fR\ gravity can be 
understood in the following way. Since it is possible to express this theory 
into a form of scalar-tensor theory, this should correspond to having 
a scalar-field degree of freedom which propagates with 
the speed of light. Therefore no extra vector or tensor modes come from 
the \fR\ gravitational sector. Introducing
matter fields will in general increase the number of degrees of
freedom, e.g., a perfect fluid will only add another propagating
scalar mode and a vector mode as well. 
In this section we shall provide perturbation equations for the general Lagrangian 
density $f(R, \phi)$ including metric \fR\ gravity 
as a special case.


\subsection{Perturbation equations}

We start with a general perturbed metric about 
the flat FLRW background~\cite{Bardeen,Kodama,Ellis,Ellis2,Mukha92}
\begin{equation}
\de s^2=-(1+2\alpha)\, \de t^2-2 a (t)\, (\partial_i \beta-S_i) \de t\, \de x^i
+a^2(t) (\delta_{ij}+2\psi \delta_{ij}+2\partial_i \partial_j \gamma
+2\partial_j F_i+h_{ij})\, \de x^i \,\de x^j\,,
\label{permetric}
\end{equation}
where $\alpha$, $\beta$, $\psi$, $\gamma$ are scalar perturbations, 
$S_i$, $F_i$ are vector perturbations, and $h_{ij}$ is the tensor perturbations, 
respectively. In this review we focus on scalar and tensor perturbations, because
vector perturbations are generally unimportant in cosmology~\cite{Bassett}.

For generality we consider the following action 
\begin{equation}
S=\int \mathrm{d}^{4}x\sqrt{-g}\,\left[ \frac{1}{2\kappa^{2}}f(R,\phi)
-\frac12 \omega(\phi) g^{\mu\nu} \partial_{\mu}\phi 
\partial_{\nu} \phi
-V(\phi) \right]
+S_{M}(g_{\mu\nu},\Psi_{M})\,,
\label{geneaction}
\end{equation}
where $f(R, \phi)$ is a function of the Ricci scalar $R$ and the scalar 
field $\phi$, $\omega(\phi)$ and $V(\phi)$ are functions of $\phi$, 
and $S_M$ is a matter action. We do not take into account an explicit 
coupling between the field $\phi$ and matter.
The action~(\ref{geneaction}) covers not only \fR\ gravity 
but also other modified gravity theories such as Brans--Dicke theory, 
scalar-tensor theories, and dilaton gravity.
We define the quantity $F(R,\phi) \equiv \partial f/\partial R$.
Varying the action~(\ref{geneaction}) with respect to $g_{\mu \nu}$
and $\phi$, we obtain the following field equations
\begin{eqnarray}
& & FR_{\mu\nu}-\frac{1}{2}f g_{\mu\nu}
-\nabla_{\mu}\nabla_{\nu}F+g_{\mu\nu}\square  F \nonumber \\
& & =\kappa^{2} \left[ \omega \left( \nabla_{\mu} \phi
\nabla_{\nu} \phi -\frac12 g_{\mu \nu} \nabla^{\lambda}\phi
\nabla_{\lambda}\phi \right)-Vg_{\mu \nu}+
T_{\mu\nu}^{(M)} \right],\\
& & \square \phi+\frac{1}{2\omega} \left(
\omega_{,\phi} \nabla^{\lambda}\phi \nabla_{\lambda}\phi
-2V_{,\phi}+\frac{f_{,\phi}}{\kappa^2} \right)=0\,,
\end{eqnarray}
where $T_{\mu\nu}^{(M)}$ is the energy-momentum tensor 
of matter.

We decompose $\phi$ and $F$ into homogeneous and perturbed parts,
$\phi=\bar{\phi}+\delta \phi$ and $F=\bar{F}+\delta F$, respectively.
In the following we omit the bar for simplicity.
The energy-momentum tensor of an ideal fluid with perturbations is 
\begin{equation}
T^0_0=-(\rho_M+\delta \rho_M)\,,\quad
T^{0}_i=-(\rho_M+P_M)\partial_i v\,,\quad
T^{i}_j=(P_M+\delta P_M) \delta^i_j\,,
\end{equation}
where $v$ characterizes the velocity potential of the fluid.
The conservation of the energy-momentum tensor ($\nabla^{\mu}T_{\mu \nu}=0$)
holds for the theories with the action~(\ref{geneaction})~\cite{Koivistopre}.

For the action~(\ref{geneaction}) the background equations 
(without metric perturbations) are given by 
\begin{eqnarray}
& &3FH^2=\frac12 (RF-f) -3H \dot{F}+\kappa^2
\left[ \frac12 \omega \dot{\phi}^2+V(\phi)+\rho_M \right]\,,
\label{fback1} 
\\
& &-2F\dot{H}=\ddot{F}-H\dot{F}+\kappa^2 \omega \dot{\phi}^2
+\kappa^2 (\rho_M+P_M)\,,
\label{fback2}
\\
& &\ddot{\phi}+3H\dot{\phi}+\frac{1}{2\omega}
\left( \omega_{,\phi} \dot{\phi}^2+2V_{,\phi}
-\frac{f_{,\phi}}{\kappa^2} \right)=0\,,
\label{fback3}
\\
& & \dot{\rho}_M+3H (\rho_M+P_M)=0\,,
\label{fback4}
\end{eqnarray}
where $R$ is given in Eq.~(\ref{R}).

For later convenience, we define the following perturbed quantities
\begin{equation}
\chi \equiv a(\beta+a {\dot \gamma})\,,\qquad
A \equiv 3(H \alpha -\dot{\psi})-\frac{\Delta}{a^2}\chi\,.
\end{equation}
Perturbing Einstein equations at linear order, we obtain 
the following equations~\cite{Hwang02,Hwang05}
(see also~\cite{Mukha81,Star83,Kofman87,MKP87,Hwang90,Hwang91,Hwang96,Perro04,Carloniper,Ananda,Mulnew,JiWang})
\begin{eqnarray}
& &\frac{\Delta}{a^2}\psi+HA=-\frac{1}{2F}
\biggl[ \left( 3H^2+3\dot{H}+\frac{\Delta}{a^2} \right)\delta F
-3H \dot{\delta F}+\frac12 \left( \kappa^2\omega_{,\phi} \dot{\phi}^2
+2\kappa^2V_{,\phi}-f_{,\phi} \right)\delta \phi 
\nonumber\\
& &~~~~~~~~~~~~~~~~~~~~~~~~~~~
+\kappa^2 \omega \dot{\phi} \dot{\delta \phi}
+(3H \dot{F}-\kappa^2 \omega \dot{\phi}^2)\alpha+\dot{F}A+
\kappa^2 \delta \rho_M \biggr]\,,
\label{per1} \\
& &H\alpha-\dot{\psi}=\frac{1}{2F}
\left[ \kappa^2 \omega \dot{\phi} \delta \phi+\dot{\delta F}
-H\delta F-\dot{F}\alpha+\kappa^2 (\rho_M+P_M) v \right]\,,
\label{per2} 
\\
& &\dot{\chi}+H\chi-\alpha-\psi=\frac{1}{F}
(\delta F-\dot{F} \chi)\,,
\label{per3} 
\\
& & \dot{A}+2HA+\left( 3H+\frac{\Delta}{a^2} \right)\alpha
=\frac{1}{2F} \biggl[ 3\ddot{\delta F}+3H\dot{\delta F}
-\left( 6H^2+\frac{\Delta}{a^2} \right)\delta F
+4\kappa^2 \omega \dot{\phi} \dot{\delta \phi} \nonumber \\
& &~~~~~~~~~~~~~~~~~~~~~~~~~~~~~~~~~~~~~~~~~~~
+(2\kappa^2 \omega_{,\phi} \dot{\phi}^2-2\kappa^2 V_{,\phi}
+f_{,\phi}) \delta \phi-3\dot{F}\dot{\alpha}-\dot{F}A \nonumber \\
& &~~~~~~~~~~~~~~~~~~~~~~~~~~~~~~~~~~~~~~~~~~~
-(4\kappa^2 \omega \dot{\phi}^2+3H \dot{F}+6 \ddot{F})\alpha
+\kappa^2 (\delta \rho_M+\delta P_M) \biggr]\,,
\label{per4} 
\end{eqnarray}
\begin{eqnarray}
& & \ddot{\delta F}+3H \dot{\delta F}-\left( \frac{\Delta}{a^2}+\frac{R}{3}
\right)\delta F+\frac23 \kappa^2 \dot{\phi} \dot{\delta \phi}
+\frac13 (\kappa^2 \omega_{,\phi} \dot{\phi}^2-4\kappa^2V_{,\phi}+2f_{,\phi})
\delta \phi \nonumber \\
& & =\frac13 \kappa^2 (\delta \rho_M-3\delta P_M)+\dot{F}(A+\dot{\alpha})
+\left( 2\ddot{F}+3H\dot{F}+\frac23 \kappa^2 \omega \dot{\phi}^2 \right)
\alpha-\frac13 F \delta R\,,
\label{per5} 
\\
& &\delta \ddot{\phi}+\left( 3H+
\frac{\omega_{,\phi}}{\omega} \dot{\phi} \right)
\delta \dot{\phi}+\left[ -\frac{\Delta}{a^2}
+\left( \frac{\omega_{,\phi}}{\omega} \right)_{,\phi}
\frac{\dot{\phi}^2}{2}+
\left(\frac{2V_{,\phi}-f_{,\phi}}{2\omega}
\right)_{,\phi} \right]\delta \phi
 \nonumber \\
& &=\dot{\phi}\dot{\alpha}+
\left( 2\ddot{\phi}+3H\dot{\phi}+
\frac{\omega_{,\phi}}{\omega} \dot{\phi}^2
\right) \alpha+\dot{\phi}A
+\frac{1}{2\omega} F_{,\phi} \delta R\,,
\label{per5d} 
\\
& & \dot{\delta \rho_M}+3H (\delta \rho_M+\delta P_M)
=(\rho_M+P_M) \left( A-3H \alpha+\frac{\Delta}{a^2}v \right)\,,
\label{per6} 
\\
& &\frac{1}{a^3 (\rho_M+P_M)}\,\frac{\rd}{\rd t}
[a^3 (\rho_M+P_M)v]=\alpha+
\frac{\delta P_M}{\rho_M+P_M}\,,
\label{per7} 
\end{eqnarray}
where $\delta R$ is given by 
\begin{equation}
\delta R=-2 \left[ \dot{A}+4HA+\left( \frac{\Delta}{a^2}+
3\dot{H} \right)\alpha+2\frac{\Delta}{a^2}\psi \right]\,.
\label{per8}
\end{equation}

We shall solve the above equations in two different contexts:
(i) inflation (Section~\ref{cosmoinf}), and (ii) the matter dominated 
epoch followed by the late-time cosmic acceleration (Section~\ref{cosmodark}).


\subsection{Gauge-invariant quantities}

Before discussing the detail for the evolution of cosmological 
perturbations, we construct a number of gauge-invariant 
quantities. This is required to avoid the appearance of
unphysical modes.
Let us consider the gauge transformation
\begin{equation}
\hat{t}=t+\delta t\,,\qquad 
\hat{x}^i=x^i+\delta^{ij}\partial_j \delta x\,,
\label{gauge}
\end{equation}
where $\delta t$ and $\delta x$
characterize the time slicing and the spatial threading, 
respectively. Then the scalar metric perturbations $\alpha$, 
$\beta$, $\psi$ and $E$ transform as~\cite{Bardeen,Bassett,MalikWands}
\begin{eqnarray}
& & \hat{\alpha}=\alpha-\dot{\delta t}  \,,\\
& & \hat{\beta}=\beta-a^{-1} \delta t+a \dot{\delta x}\,,\\
& & \hat{\psi}=\psi-H\delta t\,,\\
& & \hat{\gamma}=\gamma-\delta x\,.
\end{eqnarray}

Matter perturbations such as $\delta \phi$ and $\delta \rho$
obey the following transformation rule 
\begin{eqnarray}
& & \hat{\delta\phi}=\delta \phi-\dot{\phi}\,\delta t\,,\\
& & \hat{\delta\rho}=\delta \rho-\dot{\rho}\,\delta t\,.
\end{eqnarray}
Note that the quantity $\delta F$ is also subject to the same
transformation: $\hat{\delta F}=\delta F-\dot{F} \delta t$.
We express the scalar part of the 3-momentum energy-momentum 
tensor $\delta T^0_i$ as
\begin{equation}
\delta T^0_i=\partial_i \delta q\,.
\end{equation}
For the scalar field and the perfect fluid one has 
$\delta q=-\dot{\phi} \delta \phi$ and 
$\delta q=-(\rho_M+P_M)v$, respectively.
This quantity transforms as 
\begin{equation}
\hat{\delta q}=\delta q+(\rho+P)\delta t\,.
\end{equation}

One can construct a number of gauge-invariant quantities
unchanged under the transformation (\ref{gauge}):
\begin{eqnarray}
& & \Phi=\alpha-\frac{\mathrm{d}}{\mathrm{d}t} \left[ a^2
(\gamma+\beta/a) \right]\,,\qquad
\Psi=-\psi+a^2H (\dot{\gamma}+\beta/a)\,,\\
& & {\cal R}=\psi+\frac{H}{\rho+P}\delta q\,,
\qquad
{\cal R}_{\delta \phi}=\psi-\frac{H}{\dot{\phi}}\delta \phi\,, 
\qquad
{\cal R}_{\delta F}=\psi-\frac{H}{\dot{F}}\delta F\,,\\
& & \delta \rho_q=\delta \rho-3H \delta q\,. \label{delrho}
\end{eqnarray}
Since $\delta q=-\dot{\phi} \delta \phi$ for single-field inflation with 
a potential $V(\phi)$, ${\cal R}$ is identical to ${\cal R}_{\delta \phi}$
[where we used $\rho=\dot{\phi}^2/2+V(\phi)$ and 
 $P=\dot{\phi}^2/2-V(\phi)$].
In \fR\ gravity one can introduce a scalar field $\phi$
as in Eq.~(\ref{kappaphi}), so that 
${\cal R}_{\delta F}={\cal R}_{\delta \phi}$.
From the gauge-invariant quantity (\ref{delrho}) it is also 
possible to construct another gauge-invariant quantity for 
the matter perturbation of perfect fluids:
\begin{equation}
\delta_M=\frac{\delta \rho_M}{\rho_M}+3H(1+w_M)v\,,
\label{delM}
\end{equation}
where $w_M=P_M/\rho_M$.

We note that the tensor perturbation $h_{ij}$ is invariant 
under the gauge transformation~\cite{MalikWands}.

We can choose specific gauge conditions to fix the gauge degree of freedom.
After fixing a gauge, two scalar variables $\delta t$ and $\delta x$
are determined accordingly.
The Longitudinal gauge corresponds to the gauge choice $\hat{\beta}=0$
and $\hat{\gamma}=0$, under which $\delta t=a(\beta+a\dot{\gamma})$ 
and $\delta x=\gamma$. In this gauge one has 
$\hat{\Phi}=\hat{\alpha}$ and $\hat{\Psi}=-\hat{\psi}$, so that 
the line element (without vector and tensor perturbations) is given by 
\begin{equation}
\mathrm{d}s^2=-(1+2\Phi)\mathrm{d}t^2+a^2(t)(1-2\Psi)
\delta_{ij}\mathrm{d}x^i \mathrm{d}x^j\,,
\label{longauge}
\end{equation}
where we omitted the hat for perturbed quantities.

The uniform-field gauge corresponds to 
$\hat{\delta \phi}=0$ which fixes $\delta t=\delta \phi/\dot{\phi}$.
The spatial threading $\delta x$ is fixed by choosing 
either $\hat{\beta}=0$ or $\hat{\gamma}=0$ 
(up to an integration constant in the former case).
For this gauge choice one has $\hat{{\cal R}}_{\delta \phi}=\hat{\psi}$.
Since the spatial curvature ${}^{(3)}{\cal R}$ on the constant-time
hypersurface is related to $\psi$ via the relation 
${}^{(3)}{\cal R}=-4\nabla^2 \psi/a^2$, the quantity ${\cal R}$
is often called the curvature perturbation on the uniform-field hypersurface.
We can also choose the gauge condition $\hat{\delta q}=0$ or 
$\hat{\delta F}=0$.

\newpage

\section{Perturbations Generated During Inflation}
\label{cosmoinf}

Let us consider scalar and tensor perturbations generated during inflation
for the theories (\ref{geneaction})
without taking into account the perfect fluid ($S_M=0$).
In \fR\ gravity the contribution of the field $\phi$ such as $\delta \phi$
is absent in the perturbation equations (\ref{per1})\,--\,(\ref{per5d}).
One can choose the gauge condition $\delta F=0$, so that 
${\cal R}_{\delta F}=\psi$.
In scalar-tensor theory in which $F$ is the function of $\phi$ alone
(i.e., the coupling of the form $F(\phi)R$ without a non-linear term in $R$), 
the gauge choice $\delta \phi=0$ leads to ${\cal R}_{\delta \phi}=\psi$.
Since $\delta F=F_{,\phi}\delta \phi=0$ in this case, 
we have ${\cal R}_{\delta F}={\cal R}_{\delta \phi}=\psi$.

We focus on the effective single-field theory such as
\fR\ gravity and scalar-tensor theory with the coupling $F(\phi)R$, 
by choosing the gauge condition $\delta \phi=0$ and $\delta F=0$.
We caution that this analysis does not cover the theory such as 
${\cal L}=\xi(\phi)\,R+\alpha R^2$~\cite{Rador}, because the quantity 
$F$ depends on both $\phi$ and $R$ (in other words,
$\delta F=F_{,\phi}\delta \phi+F_{,R}\delta R$).
In the following we write the curvature perturbations ${\cal R}_{\delta F}$
and ${\cal R}_{\delta \phi}$ as ${\cal R}$.


\subsection{Curvature perturbations}
\label{cursec}

Since $\delta \phi=0$ and $\delta F=0$ in 
Eq.~(\ref{per2}) we obtain 
\begin{equation}
\alpha=\frac{\dot{\cal R}}{H+\dot{F}/(2F)}\,.
\label{alpha}
\end{equation}
Plugging Eq.~(\ref{alpha}) into Eq.~(\ref{per1}),
we have 
\begin{equation}
A=-\frac{1}{H+\dot{F}/(2F)} \left[ \frac{\Delta}{a^2}
{\cal R}+\frac{3H\dot{F}-\kappa^2 \omega \dot{\phi}^2}
{2F \{ H+\dot{F}/(2F) \}}\dot{{\cal R}} \right]\,.
\label{Are}
\end{equation}
Equation~(\ref{per4}) gives 
\begin{equation}
\dot{A}+\left( 2H +\frac{\dot{F}}{2F} \right)A
+\frac{3\dot{F}}{2F} \dot{\alpha}+
\left[ \frac{3\ddot{F}+6H\dot{F}+\kappa^2 \omega \dot{\phi}^2}
{2F}+\frac{\Delta}{a^2} \right] \alpha=0\,,
\label{dotAre}
\end{equation}
where we have used the background equation~(\ref{fback2}).
Plugging Eqs.~(\ref{alpha}) and (\ref{Are}) into Eq.~(\ref{dotAre}), 
we find that the curvature perturbation satisfies the following 
simple equation in Fourier space 
\begin{equation}
\ddot{{\cal R}}+\frac{(a^3 Q_s)^{\dot{}}}{a^3Q_s} 
\dot{{\cal R}}+\frac{k^2}{a^2}{\cal R}=0\,,
\label{ddotR}
\end{equation}
where $k$ is a comoving wavenumber and 
\begin{equation}
Q_s \equiv \frac{\omega \dot{\phi}^2+3\dot{F}^2/(2\kappa^2 F)}
{[H+\dot{F}/(2F)]^2}\,.
\label{Qs}
\end{equation}
Introducing the variables $z_s=a\sqrt{Q_s}$ and $u=z_s {\cal R}$, 
Eq.~(\ref{ddotR}) reduces to 
\begin{equation}
u''+\left( k^2 -\frac{z_s''}{z_s} \right) u=0\,,
\label{veq}
\end{equation}
where a prime represents a derivative with respect to 
the conformal time $\eta=\int a^{-1} \rd t$.

In General Relativity with a canonical scalar field $\phi$ one has 
$\omega=1$ and $F=1$, which corresponds to $Q_s=\dot{\phi}^2/H^2$.
Then the perturbation $u$ corresponds to $u=a[-\delta \phi+(\dot{\phi}/H)\psi]$. 
In the spatially flat gauge ($\psi=0$) this reduces to 
$u=-a\delta \phi$, which implies that the perturbation $u$
corresponds to a canonical scalar field $\delta \chi=a\delta \phi$.
In modified gravity theories it is not clear at this stage that the 
perturbation $u=a\sqrt{Q_s}{\cal R}$ corresponds a canonical 
field that should be quantized, because Eq.~(\ref{ddotR}) is 
unchanged by multiplying a constant term to the quantity 
$Q_s$ defined in Eq.~(\ref{Qs}).
As we will see in Section~\ref{secperlag}, this problem is overcome 
by considering a second-order perturbed action for 
the theory (\ref{geneaction}) from the beginning. 

In order to derive the spectrum of curvature perturbations generated
during inflation, we introduce the following variables~\cite{Hwang01}
\begin{equation}
\epsilon_1 \equiv -\frac{\dot{H}}{H^2}\,,\quad
\epsilon_2 \equiv \frac{\ddot{\phi}}{H \dot{\phi}}\,,\quad
\epsilon_3 \equiv \frac{\dot{F}}{2HF}\,,\quad
\epsilon_4 \equiv \frac{\dot{E}}{2HE}\,,
\end{equation}
where $E \equiv F[\omega+3\dot{F}^2/(2\kappa^2 \dot{\phi}^2 F)]$.
Then the quantity $Q_s$ can be expressed as
\begin{equation}
Q_s=\dot{\phi}^2 \frac{E}{F H^2(1+\epsilon_3)^2}\,.
\label{Qsre}
\end{equation}
If the parameter $\epsilon_1$ is constant, it follows that 
$\eta=-1/[(1-\epsilon_1)aH]$~\cite{Stewart93}.
If $\dot{\epsilon}_i=0$ ($i=1, 2, 3, 4$) one has 
\begin{equation}
\frac{z_s''}{z_s}=\frac{\nu_{\cal R}^2-1/4}{\eta^2}\,, 
\qquad
\mathrm{with} 
\qquad
\nu_{\cal R}^2=\frac14+\frac{(1+\epsilon_1+\epsilon_2
-\epsilon_3+\epsilon_4)(2+\epsilon_2-\epsilon_3+\epsilon_4)}
{(1-\epsilon_1)^2}\,.
\label{nuR}
\end{equation}
Then the solution to Eq.~(\ref{veq}) can be expressed as a linear
combination of Hankel functions,
\begin{equation}
u=\frac{\sqrt{\pi |\eta|}}{2} e^{i(1+2\nu_{\cal R})\pi/4}
\left[ c_1 H_{\nu_{\cal R}}^{(1)} (k|\eta|)+
c_2 H_{\nu_{\cal R}}^{(2)} (k|\eta|) \right]\,,
\end{equation}
where $c_1$ and $c_2$ are integration constants.

During inflation one has $|\epsilon_i| \ll 1$, so that 
$z_s''/z_s \approx (aH)^2$.
For the modes deep inside the Hubble radius ($k \gg aH$, i.e., $|k\eta| \gg 1$)
the perturbation $u$ satisfies the standard equation of a canonical field in the
Minkowski spacetime: $u''+k^2u \simeq 0$.
After the Hubble radius crossing ($k=aH$) during inflation, 
the effect of the gravitational term $z_s''/z_s$ becomes important.
In the super-Hubble limit ($k \ll aH$, i.e., $|k\eta| \ll 1$)
the last term on the l.h.s.\ of Eq.~(\ref{ddotR}) can be neglected, 
giving the following solution 
\begin{equation}
{\cal R}=c_1+c_2 \int \frac{\mathrm{d}t}{a^3 Q_s}\,,
\end{equation}
where $c_1$ and $c_2$ are integration constants.
The second term can be identified as a decaying mode, 
which rapidly decays during inflation
(unless the field potential has abrupt features).
Hence the curvature perturbation approaches a constant 
value $c_1$ after the Hubble radius crossing ($k<aH$).

In the asymptotic past ($k\eta \to -\infty$) the solution to Eq.~(\ref{veq})
is determined by a vacuum state in quantum field theory~\cite{Birrell}, as
$u \to e^{-ik \eta}/\sqrt{2k}$.
This fixes the coefficients to be $c_1=1$ and $c_2=0$, 
giving the following solution 
\begin{equation}
u=\frac{\sqrt{\pi |\eta|}}{2} e^{i(1+2\nu_{\cal R})\pi/4}
H_{\nu_{\cal R}}^{(1)} (k|\eta|)\,.
\label{vso}
\end{equation}

We define the power spectrum of curvature perturbations,
\begin{equation}
{\cal P}_{{\cal R}} \equiv \frac{4\pi k^3}{(2\pi)^3}
\left| {\cal R} \right|^2\,.
\end{equation}
Using the solution (\ref{vso}), we obtain the 
power spectrum~\cite{Hwang05}
\begin{equation}
{\cal P}_{{\cal R}}=\frac{1}{Q_s} \left( (1-\epsilon_1)
\frac{\Gamma(\nu_{\cal R})}{\Gamma (3/2)}
\frac{H}{2\pi} \right)^2 
\left(\frac{|k\eta|}{2}\right)^{3-2\nu_{\cal R}}\,,
\label{PR}
\end{equation}
where we have used the relations $H_{\nu}^{(1)}(k|\eta|) \to -(i/\pi)
\Gamma (\nu) (k|\eta|/2)^{-\nu}$ for $k \eta \to 0$
and $\Gamma (3/2)=\sqrt{\pi}/2$.
Since the curvature perturbation is frozen after the Hubble 
radius crossing, the spectrum (\ref{PR}) should be evaluated 
at $k=aH$. The spectral index of ${\cal R}$, which is defined by 
$n_{\cal R}-1=\mathrm{d}\ln {\cal P}_{\cal R}/\mathrm{d}\ln k|_{k=aH}$, is
\begin{equation}
n_{\cal R}-1=3-2\nu_{\cal R}\,,
\label{nR}
\end{equation}
where $\nu_{\cal R}$ is given in Eq.~(\ref{nuR}).
As long as $|\epsilon_i|$ ($i=1,2,3,4$) are much smaller than 1 during 
inflation, the spectral index reduces to 
\begin{equation}
n_{\cal R}-1 \simeq -4\epsilon_1-2\epsilon_2+2\epsilon_3
-2\epsilon_4\,,
\label{nuR2}
\end{equation}
where we have ignored those terms higher than the order of 
$\epsilon_i$'s. Provided that $|\epsilon_i| \ll 1$ the spectrum 
is close to scale-invariant ($n_{\cal R} \simeq 1$).
From Eq.~(\ref{PR}) the power spectrum of curvature
perturbations can be estimated as 
\begin{equation}
{\cal P}_{\cal R} \simeq \frac{1}{Q_s} 
\left( \frac{H}{2\pi} \right)^2\,.
\label{PRf}
\end{equation}
A minimally coupled scalar field $\phi$ in Einstein gravity
corresponds to $\epsilon_3=0$, $\epsilon_4=0$ and $Q_s=\dot{\phi}^2/H^2$, 
in which case we obtain the standard results 
$n_{\cal R}-1 \simeq -4\epsilon_1-2\epsilon_2$
and ${\cal P}_{\cal R} \simeq H^4/(4\pi^2 \dot{\phi}^2)$
in slow-roll inflation~\cite{Stewart93,Liddle92}.


\subsection{Tensor perturbations}

Tensor perturbations $h_{ij}$ have two polarization states, which are 
generally written as $\lambda=+, \times$~\cite{LiddleLyth}.
In terms of polarization tensors $e_{ij}^{+}$ and $e_{ij}^{\times}$
they are given by 
\begin{equation}
h_{ij}=h_{+} e_{ij}^{+} + h_{\times} e_{ij}^{\times}\,.
\end{equation}
If the direction of a momentum ${\bm k}$ is along the $z$-axis, 
the non-zero components of polarization tensors are given by 
$e_{xx}^{+}=-e_{yy}^{+}=1$ and 
$e_{xy}^{\times}=e_{yx}^{\times}=1$.

For the action~(\ref{geneaction}) the Fourier components 
$h_{\lambda}$ ($\lambda=+, \times$) obey the following 
equation~\cite{Hwang96}
\begin{equation}
\ddot{h_{\lambda}}+\frac{(a^3F)^{\dot{}}}{a^3F}\dot{h_{\lambda}}
+\frac{k^2}{a^2}h_{\lambda}=0\,.
\end{equation}
This is similar to Eq.~(\ref{ddotR}) of curvature perturbations, 
apart from the difference of the factor $F$ instead of $Q_s$.
Defining new variables $z_t=a\sqrt{F}$ and 
$u_{\lambda}=z_th_{\lambda}/\sqrt{16\pi G}$, it follows that 
\begin{equation}
u_{\lambda}''+\left( k^2-\frac{z_t''}{z_t}
\right) u_{\lambda}=0\,.
\end{equation}
We have introduced the factor $16\pi G$ to relate a dimensionless 
massless field $h_{\lambda}$ with a massless scalar field $u_{\lambda}$
having a unit of mass.

If $\dot{\epsilon}_i=0$, we obtain
\begin{equation}
\frac{z_t''}{z_t}=\frac{\nu_{t}^2-1/4}{\eta^2}\,, 
\qquad
\mathrm{with} 
\qquad
\nu_{t}^2=\frac14+\frac{(1+\epsilon_3)(2-\epsilon_1+\epsilon_3)}
{(1-\epsilon_1)^2}\,.
\label{nut}
\end{equation}
We follow the similar procedure to the one given in Section~\ref{cursec}.
Taking into account polarization states, the spectrum of tensor 
perturbations after the Hubble radius crossing is given by 
\begin{equation}
{\cal P}_T=4 \times \frac{16\pi G}{a^2F}
\frac{4\pi k^3}{(2\pi)^3}|u_{\lambda}|^2 \simeq
\frac{16}{\pi} \left( \frac{H}{m_{\mathrm{pl}}} \right)^2 
\frac{1}{F} \left( (1-\epsilon_1) \frac{\Gamma (\nu_t)}
{\Gamma (3/2)} \right)^2
\left( \frac{|k\eta|}{2} \right)^{3-2\nu_t}\,,
\end{equation}
which should be evaluated at the Hubble radius crossing
($k=aH$). The spectral index of ${\cal P}_T$ is 
\begin{equation}
n_T=3-2\nu_t\,,
\label{nT}
\end{equation}
where $\nu_t$ is given in Eq.~(\ref{nut}).
If $|\epsilon_i| \ll 1$, this reduces to
\begin{equation}
n_T \simeq -2\epsilon_1-2\epsilon_3\,.
\label{nT2}
\end{equation}
Then the amplitude of tensor perturbations is 
given by 
\begin{equation}
{\cal P}_T \simeq \frac{16}{\pi}
\left( \frac{H}{m_{\mathrm{pl}}} \right)^2 \frac{1}{F}\,.
\label{PTf}
\end{equation}

We define the tensor-to-scalar ratio
\begin{equation}
r \equiv \frac{{\cal P}_T}{{\cal P}_{\cal R}}
 \simeq \frac{64\pi}{m_{\mathrm{pl}}^2}
\frac{Q_s}{F}\,.
\label{rfR}
\end{equation}
For a minimally coupled scalar field $\phi$ in Einstein gravity,
it follows that $n_T \simeq -2\epsilon_1$, 
${\cal P}_T \simeq 16H^2/(\pi m_{\mathrm{pl}}^2)$, and 
$r \simeq 16 \epsilon_1$.


\subsection{The spectra of perturbations in inflation based on \fR\ gravity}

Let us study the spectra of scalar and tensor perturbations
generated during inflation in metric \fR\ gravity. 
Introducing the quantity $E=3\dot{F}^2/(2\kappa^2)$, 
we have $\epsilon_4=\ddot{F}/(H\dot{F})$ and 
\begin{equation}
Q_s=\frac{6F\epsilon_3^2}{\kappa^2 (1+\epsilon_3)^2}
=\frac{E}{F H^2(1+\epsilon_3)^2}\,.
\label{Qsf}
\end{equation}
Since the field kinetic term $\dot{\phi}^2$ is absent, 
one has $\epsilon_2=0$ in Eqs.~(\ref{nuR}) and (\ref{nuR2}).
Under the conditions $|\epsilon_i| \ll 1$ ($i=1,3,4$),
the spectral index of curvature perturbations 
is given by $n_{\cal R}-1 \simeq -4\epsilon_1
+2\epsilon_3-2\epsilon_4$.

In the absence of the matter fluid, Eq.~(\ref{fRinf2})
translates into 
\begin{equation}
\epsilon_1=-\epsilon_3 (1-\epsilon_4)\,,
\end{equation}
which gives $\epsilon_1 \simeq -\epsilon_3$
for $|\epsilon_4| \ll 1$.
Hence we obtain~\cite{Hwang01}
\begin{equation}
n_{\cal R}-1 \simeq -6\epsilon_1-2\epsilon_4\,.
\label{ncalR1}
\end{equation}
From Eqs.~(\ref{PRf}) and (\ref{Qsf}), the amplitude of ${\cal R}$
is estimated as 
\begin{equation}
{\cal P}_{\cal R} \simeq \frac{1}{3\pi F}
\left( \frac{H}{m_{\mathrm{pl}}} \right)^2
\frac{1}{\epsilon_3^2}\,.
\label{PRfR}
\end{equation}

Using the relation $\epsilon_1 \simeq -\epsilon_3$, 
the spectral index (\ref{nT2}) of tensor perturbations is given by 
\begin{equation}
n_T \simeq 0\,,
\end{equation}
which vanishes at first-order of slow-roll approximations.
From Eqs.~(\ref{PTf}) and (\ref{PRfR}) we obtain the 
tensor-to-scalar ratio
\begin{equation}
r \simeq 48 \epsilon_3^2 \simeq 48 \epsilon_1^2\,.
\label{rcon}
\end{equation}

\subsubsection{The model $f(R)=\alpha R^n$ ($n>0$)}

Let us consider the inflation model: $f(R)=\alpha R^n$ ($n>0$).
From the discussion given in Section~\ref{infdynamics} the slow-roll parameters 
$\epsilon_i$ ($i=1,3, 4$) are constants:
\begin{equation}
\epsilon_1=\frac{2-n}{(n-1)(2n-1)}\,,\qquad
\epsilon_3=-(n-1)\epsilon_1\,,\qquad
\epsilon_4=\frac{n-2}{n-1}\,.
\end{equation}
In this case one can use the exact results (\ref{nR}) and (\ref{nT})
with $\nu_{\cal R}$ and $\nu_{t}$ given in Eqs.~(\ref{nuR}) and 
(\ref{nut}) (with $\epsilon_2=0$).
Then the spectral indices are 
\begin{equation}
n_{\cal R}-1=n_T=-\frac{2(n-2)^2}{2n^2-2n-1}\,.
\end{equation}

If $n=2$ we obtain the scale-invariant spectra with 
$n_{\cal R}=1$ and $n_T=0$.
Even the slight deviation from $n=2$ leads to a 
rather large deviation from the scale-invariance.
If $n=1.7$, for example, one has $n_{\cal R}-1=n_T=-0.13$, 
which does not match with the WMAP 5-year constraint:
$n_{\cal R}=0.960 \pm 0.013$~\cite{WMAP5}.

\subsubsection{The model $f(R)=R+R^2/(6M^2)$}

For the model $f(R)=R+R^2/(6M^2)$, the spectrum 
of the curvature perturbation ${\cal R}$ shows some deviation 
from the scale-invariance.
Since inflation occurs in the regime $R \gg M^2$ and $|\dot{H}| \ll H^2$, 
one can approximate $F\simeq R/(3M^2) \simeq 4H^2/M^2$.
Then the power spectra (\ref{PRfR}) and (\ref{PTf}) yield
\begin{equation}
{\cal P}_{\cal R} \simeq \frac{1}{12\pi} 
\left( \frac{M}{m_{\mathrm{pl}}} \right)^2 \frac{1}{\epsilon_1^2}\,,
\qquad
{\cal P}_{T} \simeq \frac{4}{\pi} \left( \frac{M}{m_{\mathrm{pl}}} 
\right)^2\,,
\end{equation}
where we have employed the relation $\epsilon_3 \simeq -\epsilon_1$.

Recall that the evolution of the Hubble parameter during inflation 
is given by Eq.~(\ref{apeq1}).
As long as the time $t_k$ at the Hubble radius crossing ($k=aH$)
satisfies the condition $(M^2/6)(t_k-t_i) \ll H_i$, one 
can approximate $H(t_k) \simeq H_i$.
Using Eq.~(\ref{apeq1}), the number of e-foldings from 
$t=t_k$ to the end of inflation can be estimated as 
\begin{equation}
N_k \simeq \frac{1}{2\epsilon_1 (t_k)}\,.
\end{equation}
Then the amplitude of the curvature perturbation is given by 
\begin{equation}
{\cal P}_{\cal R} \simeq \frac{N_k^2}{3\pi} 
\left( \frac{M}{m_{\mathrm{pl}}} \right)^2\,.
\end{equation}
The WMAP 5-year normalization corresponds to 
${\cal P}_{\cal R}=(2.445 \pm 0.096) \times 10^{-9}$
at the scale $k=0.002\mathrm{\ Mpc}^{-1}$~\cite{WMAP5}.
Taking the typical value $N_k=55$, the mass $M$ 
is constrained to be
\begin{equation}
M \simeq 3 \times 10^{-6}m_{\mathrm{pl}}\,.
\end{equation}
Using the relation $F \simeq 4H^2/M^2$, it follows that $\epsilon_4 \simeq -\epsilon_1$.
Hence the spectral index (\ref{ncalR1}) reduces to 
\begin{equation}
n_{\cal R}-1 \simeq -4\epsilon_1 \simeq -\frac{2}{N_k}
=-3.6 \times 10^{-2} \left( \frac{N_k}{55} \right)^{-1}\,.
\label{ncalR}
\end{equation}
For $N_k=55$ we have $n_{\cal R} \simeq 0.964$, which is in the 
allowed region of the WMAP 5-year constraint 
($n_{\cal R}=0.960 \pm 0.013$ at the 68\% confidence level~\cite{WMAP5}). 
The tensor-to-scalar ratio (\ref{rcon}) can be estimated as
\begin{equation}
r \simeq \frac{12}{N_k^2} \simeq 4.0 \times 10^{-3}
\left( \frac{N_k}{55} \right)^{-2}\,,
\label{rJor}
\end{equation}
which satisfies the current observational bound $r<0.22$~\cite{WMAP5}.
We note that a minimally coupled field with the potential $V(\phi)=m^2\phi^2/2$
in Einstein gravity (chaotic inflation model~\cite{Linde83})
gives rise to a larger tensor-to-scalar ratio 
of the order of $0.1$.
Since future observations such as the Planck satellite 
are expected to reach the level of $r={\cal O}(10^{-2})$, 
they will be able to discriminate between the chaotic inflation 
model and the Starobinsky's \fR\ model.

\subsubsection{The power spectra in the Einstein frame}

Let us consider the power spectra in the Einstein frame.
Under the conformal transformation $\tilde{g}_{\mu \nu}=Fg_{\mu \nu}$, 
the perturbed metric (\ref{permetric}) is transformed as 
\begin{eqnarray}
\de \ti{s}^2 &=& F \mathrm{d}s^2 \nonumber \\
&=&-(1+2\ti{\alpha})\, \de \ti{t}^2-2 \ti{a} (\ti{t})\, (\partial_i \ti{\beta}-\ti{S}_i) 
\de \ti{t}\, \de \ti{x}^i \nonumber \\
& &+\ti{a}^2(\ti{t}) (\delta_{ij}+2\ti{\psi} \delta_{ij}+2\partial_i \partial_j 
\ti{\gamma}+2\partial_j \ti{F}_i+\ti{h}_{ij})\, \de \ti{x}^i \,\de \ti{x}^j\,.
\label{permetric2}
\end{eqnarray}
We decompose the conformal factor into the background and 
perturbed parts, as 
\begin{equation}
F(t, {\bm x})=\bar{F} (t) \left( 1+\frac{\delta F (t, {\bm x})}
{\bar{F} (t)} \right)\,.
\end{equation}
In what follows we omit a bar from $F$.
We recall that the background quantities are transformed as
Eqs.~(\ref{transre}) and (\ref{transre2}).
The transformation of scalar metric perturbations is given by 
\begin{equation}
\ti{\alpha}=\alpha+\frac{\delta F}{2F}\,,\qquad
\ti{\beta}=\beta\,,\qquad
\ti{\psi}=\psi+\frac{\delta F}{2F}\,,\qquad
\ti{\gamma}=\gamma\,.
\end{equation}
Meanwhile vector and tensor perturbations are invariant 
under the conformal transformation 
($\ti{S}_i=S_i$, $\ti{F}_i=F_i$, $\ti{h}_{ij}=h_{ij}$).

Using the above transformation law, one can easily show that the curvature perturbation 
${\cal R}=\psi-H\delta F/\dot{F}$ in \fR\ gravity is invariant
under the conformal transformation:
\begin{equation}
\ti{\cal R}={\cal R}\,.
\end{equation}
Since the tensor perturbation is also invariant, the tensor-to-scalar 
ratio $\ti{r}$ in the Einstein frame is identical to that in the Jordan frame.
For example, let us consider the model $f(R)=R+R^2/(6M^2)$.
Since the action in the Einstein frame is given by Eq.~(\ref{Ein}), 
the slow-roll parameters $\ti{\epsilon}_3$ and $\ti{\epsilon}_4$
vanish in this frame. Using Eqs.~(\ref{nuR2}) and (\ref{tiep}), the spectral 
index of curvature perturbations is given by 
\begin{equation}
\ti{n}_{\cal R}-1 \simeq -4\ti{\epsilon}_1-2\ti{\epsilon}_2
\simeq -\frac{2}{\ti{N}_k}\,,
\label{tilnR}
\end{equation}
where we have ignored the term of the order of
$1/\ti{N}_k^2$.
Since $\ti{N}_k \simeq N_k$ in the slow-roll limit ($|\dot{F}/(2HF)| \ll 1$), 
Eq.~(\ref{tilnR}) agrees with the result (\ref{ncalR}) in the Jordan frame.
Since $Q_s=(\mathrm{d}\phi/\mathrm{d} \ti{t})^2/\ti{H}^2$ in the Einstein frame, 
Eq.~(\ref{rfR}) gives the tensor-to-scalar ratio
\begin{equation}
\ti{r}=\frac{64\pi}{m_{\mathrm{pl}}^2} \left(
\frac{\mathrm{d}\phi}{\mathrm{d} \ti{t}} \right)^2 
\frac{1}{\ti{H}^2} \simeq 16\ti{\epsilon}_1 \simeq \frac{12}{\ti{N}_k^2}\,,
\label{rEin}
\end{equation}
where the background equations (\ref{eein1}) and (\ref{eein2})
are used with slow-roll approximations.
Equation~(\ref{rEin}) is consistent with the result (\ref{rJor})
in the Jordan frame.

The equivalence of the curvature perturbation between the Jordan 
and Einstein frames also holds for scalar-tensor theory 
with the Lagrangian ${\cal L}=F(\phi)R/(2\kappa^2)
-(1/2)\omega(\phi)g^{\mu \nu} \partial_{\mu}\phi 
\partial_{\nu}\phi-V(\phi)$~\cite{Makino,Fakirper}.
For the non-minimally coupled scalar field with 
$F(\phi)=1-\xi \kappa^2 \phi^2$~\cite{Futamase,Fakir}
the spectral indices of scalar and tensor perturbations have 
been derived by using such equivalence~\cite{Komatsuper,Gumjudpai}.


\subsection{The Lagrangian for cosmological perturbations}
\label{secperlag}

In Section~\ref{cursec} we used the fact that the field which should be
quantized corresponds to $u=a\sqrt{Q}_s {\cal R}$. 
This can be justified by writing down the action~(\ref{permetric})
expanded at second-order in the perturbations~\cite{Mukha92}.
We recall again that we are considering an effective single-field 
theory such as \fR\ gravity and scalar-tensor theory 
with the coupling $F(\phi) R$.
Carrying out the expansion of the action~(\ref{geneaction}) in second order, 
we find that the action for the curvature perturbation ${\cal R}$
(either ${\cal R}_{\delta F}$ or ${\cal R}_{\delta \phi}$) 
is given by~\cite{Hwang97}
\begin{equation}
\label{eq:actPhid}
\delta S^{(2)}=\int \de t\,\de^3x\,a^3\,Q_s\left[\frac12\,\dot{\cal R}^2
 -\frac12\,\frac{1}{a^2}(\nabla{\cal R})^2\right]\, ,
\end{equation}
where $Q_s$ is given in Eq.~(\ref{Qs}).
In fact, the variation of this action in terms of the field ${\cal R}$
gives rise to Eq.~(\ref{ddotR}) in Fourier space.
We note that there is another approach called the Hamiltonian formalism
which is also useful for the quantization of cosmological perturbations.
See~\cite{Ha1,Ha2,Ha3,Ha4} for this approach in the context of \fR\ gravity and
modified gravitational theories.

Introducing the quantities $u=z_s {\cal R}$ and $z_s=a\sqrt{Q}_s$, 
the action~(\ref{eq:actPhid}) can be written as 
\begin{equation}
\label{eq:actPhid2}
\delta S^{(2)}=\int \de \eta\,\de^3x
\left[\frac12\,u'^2
 -\frac12 (\nabla u)^2+
 \frac12 \frac{z_s''}{z_s}u^2 \right]\,,
\end{equation}
where a prime represents a derivative with respect to 
the conformal time $\eta=\int a^{-1} \mathrm{d}t$.
The action~(\ref{eq:actPhid2}) leads to Eq.~(\ref{veq}) in Fourier space.
The transformation of the action~(\ref{eq:actPhid}) to (\ref{eq:actPhid2})
gives rise to the effective mass\epubtkFootnote{If we define 
$X=\sqrt{Q_s}{\cal R}$
and plugging it into Eq.~(\ref{eq:actPhid}), we obtain the perturbed action 
for the field $X$ after the partial integration:
\begin{equation}
\delta S^{(2)}=\int \rd t \,\rd^3 x\,\sqrt{-g^{(0)}} 
\left[ \frac12 \dot{X}^2
-\frac12 \frac{1}{a^2}(
\nabla X )^2-\frac12 M_s^2 X^2 \right]\,,
\nonumber 
\end{equation}
where $\sqrt{-g^{(0)}}=a^3$ and $M_s$ is defined in Eq.~(\ref{Ms2}).
Then, for the field $X$, we obtain the Klein--Gordon equation
$\square X=M_s^2 X$ in the large-scale limit ($k \to 0$), 
which defines the mass $M_s$ in an invariant way 
in the FLRW background.}
\begin{equation}
M_s^2 \equiv -\frac{1}{a^2}\frac{z_s''}{z_s}
=\frac{\dot Q_s^2}{4Q_s^2}-\frac{\ddot Q_s}{2Q_s}
-\frac{3H\dot Q_s}{2Q_s}\,.
\label{Ms2}
\end{equation}

We have seen in Eq.~(\ref{nuR}) that during inflation 
the quantity $z_s''/z_s$ can be 
estimated as $z_s''/z_s \simeq 2(aH)^2$ in the slow-roll limit, 
so that $M_s^2 \simeq -2H^2$.
For the modes deep inside the Hubble radius ($k \gg aH$)
the action~(\ref{eq:actPhid2}) reduces to the one for a canonical 
scalar field $u$ in the flat spacetime.
Hence the quantization should be done for the field $u=a\sqrt{Q_s}{\cal R}$, 
as we have done in Section~\ref{cursec}.

From the action~(\ref{eq:actPhid2}) we understand a number of 
physical properties in \fR\ theories and scalar-tensor theories
with the coupling $F(\phi) R$ listed below.
\begin{enumerate}
\item Having a standard d'Alambertian operator, the mode has speed of
  propagation equal to the speed of light. This leads to a standard
  dispersion relation $\omega=k/a$ for the high-$k$ modes in
  Fourier space.
\item The sign of $Q_s$ corresponds to the sign of the kinetic energy of ${\cal R}$. 
  The negative sign corresponds to a ghost (phantom) scalar field.  
  In \fR\ gravity (with $\dot{\phi}=0$) the ghost appears for $F<0$.
  In Brans--Dicke theory with $F(\phi)=\kappa^2 \phi$ and 
  $\omega(\phi)=\omega_{\mathrm{BD}}/\phi$~\cite{BD} (where $\phi>0$) the 
  condition for the appearance of the ghost ($\omega \dot{\phi}^2+3\dot{F}^2/(2\kappa^2 F)<0$)
  translates into $\omega_{\mathrm{BD}}<-3/2$.
  In these cases one would encounter serious problems related to 
  vacuum instability~\cite{Trodden03,Cline03}.
\item The field $u$ has the effective mass squared given in Eq.~(\ref{Ms2}).
In \fR\ gravity it can be written as
\begin{equation}
\label{eq:massa}
M_s^2=-\frac{72 F^2 H^4}{(2 F H+f_{,RR} \dot R)^2}+\frac{1}{3} F \left(\frac{288 H^3-12 H R}
{2 F H+f_{,RR} \dot R}+\frac{1}{f_{,RR}}\right)+\frac{f_{,RR}^2 \dot R^2}{4 F^2}
-24 H^2+\frac76 R\,,
\end{equation}
where  we used the background equation~(\ref{fRinf2}) 
to write $\dot H$ in terms of $R$ and $H^2$.
In Fourier space the perturbation $u$ obeys the equation of motion 
\begin{equation}
u''+\left( k^2+M_s^2 a^2 \right)u=0\,.
\end{equation}
For $k^2/a^2 \gg M_s^2$, the field $u$ propagates with speed of light. 
For small $k$ satisfying $k^2/a^2 \ll M_s^2$, 
we require a positive $M_s^2$ to 
avoid the tachyonic instability of perturbations.
Recall that the viable dark energy models based on \fR\ theories need to satisfy 
$Rf_{,RR}\ll F$ (i.e., $m=Rf_{,RR}/f_{,R} \ll 1$) at early times, 
in order to have successful cosmological evolution from radiation 
domination till matter domination. 
At these epochs the mass squared is approximately given by 
\begin{equation}
\label{eq:massfR}
M_s^2 \simeq \frac{F}{3f_{,RR}}\,,
\end{equation}
which is consistent with the result (\ref{Mpsi2}) derived by the 
linear analysis about the Minkowski background.
Together with the ghost condition $F>0$, this leads to $f_{,RR}>0$. 
Recall that these correspond to the conditions presented in 
Eq.~(\ref{fcon}).
\end{enumerate}

\newpage

\section{Observational Signatures of Dark Energy Models in \fR\ Theories}
\label{cosmodark}

In this section we discuss a number of observational signatures of 
dark energy models based on metric \fR\ gravity.
Our main interest is to distinguish these models from the
$\Lambda$CDM model observationally. 
In particular we study the evolution of matter density perturbations 
as well as the gravitational potential to confront \fR\ models 
with the observations of large-scale structure (LSS) and Cosmic 
Microwave Background (CMB).
The effect on weak lensing will be discussed in Section~\ref{wlensingsec}
in more general modified gravity theories including \fR\ gravity.


\subsection{Matter density perturbations}
\label{mattersec}

Let us consider the perturbations of non-relativistic matter with 
the background energy density $\rho_m$ and the negligible 
pressure ($P_m=0$).
In Fourier space Eqs.~(\ref{per6}) and (\ref{per7}) give 
\begin{eqnarray}
& & \dot{\delta \rho_m}+3H \delta \rho_m=
\rho_m \left( A-3H\alpha -\frac{k^2}{a^2}v \right)\,,
\label{matter1} \\
& &  \dot{v}=\alpha\,,
\label{matter2}
\end{eqnarray}
where in the second line we have used the continuity equation, 
$\dot{\rho}_m+3H\rho_m=0$.
The density contrast defined in Eq.~(\ref{delM}), i.e.
\begin{equation}
\delta_m=\frac{\delta \rho_m}{\rho_m}+3Hv\,,
\label{delmdefi}
\end{equation}
obeys the following equation from 
Eqs.~(\ref{matter1}) and (\ref{matter2}):
\begin{equation}
\ddot{\delta}_m+2H \dot{\delta}_m+\frac{k^2}{a^2}
(\alpha -\dot{\chi})=3\ddot{B}+6H\dot{B}\,,
\label{delmeq}
\end{equation}
where $B \equiv Hv-\psi$ and we used the relation 
$A=3(H\alpha-\dot{\psi})+(k^2/a^2)\chi$.

In the following we consider the evolution of perturbations 
in \fR\ gravity in the Longitudinal gauge (\ref{longauge}).
Since $\chi=0$, $\alpha=\Phi$, $\psi=-\Psi$, and $A=3(H\Phi+\dot{\Psi})$
in this case, Eqs.~(\ref{per1}), (\ref{per3}),
(\ref{per5}), and (\ref{delmeq}) give
\begin{eqnarray}
& & \frac{k^2}{a^2}\Psi+3H (H \Phi+\dot{\Psi})
=-\frac{1}{2F} \biggl[ \left( 3H^2+3\dot{H}-\frac{k^2}{a^2}
\right) \delta F-3H \dot{\delta F} \nonumber \\
& &~~~~~~~~~~~~~~~~~~~~~~~~~~~~~~~~~~~~~~~~
+3H \dot{F} \Phi+
3\dot{F} (H \Phi+\dot{\Psi})+\kappa^2 \delta \rho_m \biggr]\,, 
\label{perdark1} \\
& &  \Psi-\Phi=\frac{\delta F}{F}\,,
\label{perdark2} \\
& & \ddot{\delta F}+3H \dot{\delta F}+\left( \frac{k^2}{a^2}
+M^2 \right) \delta F=\frac{\kappa^2}{3}\delta \rho_m
+\dot{F}(3H\Phi+3\dot{\Psi}+\dot{\Phi}) 
+(2\ddot{F}+3H\dot{F})\Phi\,,
\label{perdark3} \\
& & \ddot{\delta}_m+2H \dot{\delta}_m+\frac{k^2}{a^2}\Phi
=3\ddot{B}+6H\dot{B}\,,
\label{perdark4}
\end{eqnarray}
where $B=Hv+\Psi$.
In order to derive Eq.~(\ref{perdark3}), we have used
the mass squared $M^2=(F/F_{,R}-R)/3$ introduced in 
Eq.~(\ref{Mpsi2}) together with the relation 
$\delta R=\delta F/F_{,R}$.

Let us consider the wavenumber $k$ deep inside the Hubble radius 
($k \gg aH$).
In order to derive the equation of matter perturbations approximately, 
we use the quasi-static approximation under which the dominant terms in 
Eqs.~(\ref{perdark1})\,--\,(\ref{perdark4}) correspond to 
those including $k^2/a^2$, $\delta \rho_m$ (or $\delta_m$) and $M^2$.
In General Relativity this approximation was first used by Starobinsky 
in the presence of a minimally coupled scalar field~\cite{Star98}, 
which was numerically confirmed in~\cite{Ma99}.
This was further extended to scalar-tensor theories~\cite{Boi00,CST06,Tsujimatter07}
and \fR\ gravity~\cite{Tsujimatter07,TsujiUddin}.
Precisely speaking, in \fR\ gravity, this approximation corresponds to 
\begin{equation}
\label{loncon1}
\left\{ \frac{k^2}{a^2}|\Phi|, \frac{k^2}{a^2}|\Psi|,
\frac{k^2}{a^2}|\delta F|, M^2|\delta F| \right\} \gg 
\left\{ H^2 |\Phi|, H^2 |\Psi|, H^2 |B|, H^2 |\delta F| \right\}\,,
\end{equation}
and 
\begin{equation}
\label{loncon2}
|\dot{X}| \lesssim |HX|\,,\quad \mathrm{where} \quad
X=\Phi, \Psi, F, \dot{F}, \delta F, \dot{\delta F}\,.
\end{equation}
From Eqs.~(\ref{perdark1}) and (\ref{perdark2}) it then follows that 
\begin{equation}
\Psi \simeq \frac{1}{2F} \left( \delta F-\frac{a^2}{k^2}  
\kappa^2 \delta \rho_m \right)\,,\qquad
\Phi \simeq -\frac{1}{2F} \left( \delta F +\frac{a^2}{k^2}
\kappa^2 \delta \rho_m \right)\,.
\end{equation}
Since $(k^2/a^2+M^2)\delta F \simeq \kappa^2\delta \rho_m/3$ from 
Eq.~(\ref{perdark3}), we obtain
\begin{equation}
\frac{k^2}{a^2}\Psi \simeq -\frac{\kappa^2 \delta \rho_m}{2F} 
\frac{2+3M^2a^2/k^2}{3(1+M^2a^2/k^2)}\,,
\qquad
\frac{k^2}{a^2}\Phi \simeq
-\frac{\kappa^2 \delta \rho_m}{2F}
\frac{4+3M^2a^2/k^2}{3(1+M^2a^2/k^2)}\,.
\label{Phiex}
\end{equation}
We also define the effective gravitational potential 
\begin{equation}
\Phi_{\mathrm{eff}} \equiv (\Phi+\Psi)/2\,.
\label{Phieffdef}
\end{equation}
This quantity characterizes the deviation of light rays, which 
is linked with the Integrated Sachs--Wolfe (ISW) effect 
in CMB~\cite{SongHu1} and weak lensing observations~\cite{Sapone08}.
From Eq.~(\ref{Phiex}) we have 
\begin{equation}
\Phi_{\mathrm{eff}} \simeq -\frac{\kappa^2}{2F}
\frac{a^2}{k^2} \delta \rho_m\,.
\label{Phieff}
\end{equation}

From Eq.~(\ref{per2}) the term $Hv$ is of the 
order of $H^2\Phi/(\kappa^2\rho_m)$ provided that 
the deviation from the $\Lambda$CDM model is not significant.
Using Eq.~(\ref{Phiex}) we find that the ratio 
$3Hv/(\delta \rho_m/\rho_m)$ is of the order of 
$(aH/k)^2$, which is much smaller than unity 
for sub-horizon modes. Then the gauge-invariant perturbation
$\delta_m$ given in Eq.~(\ref{delmdefi}) can be approximated 
as $\delta_m \simeq \delta \rho_m/\rho_m$.
Neglecting the r.h.s.\ of Eq.~(\ref{perdark4}) relative to the l.h.s.\ 
and using Eq.~(\ref{Phiex}) with $\delta \rho_m \simeq \rho_m \delta_m$, 
we get the equation for matter perturbations:
\begin{equation}
\ddot{\delta}_m+2H \dot{\delta}_m-4\pi G_{\mathrm{eff}} 
\rho_m \delta_m \simeq 0\,,
\label{delmfi}
\end{equation}
where $G_{\mathrm{eff}}$ is the effective (cosmological)
gravitational coupling defined by~\cite{Tsujimatter07,TsujiUddin}
\begin{equation}
G_{\mathrm{eff}} \equiv \frac{G}{F}
\frac{4+3M^2a^2/k^2}{3(1+M^2a^2/k^2)}\,.
\end{equation}

We recall that viable \fR\ dark energy models are constructed to 
have a large mass $M$ in the region of high density ($R \gg R_0$).
During the radiation and deep matter eras the deviation parameter 
$m=Rf_{,RR}/f_{,R}$ is much smaller than 1, so that the mass
squared satisfies
\begin{equation}
M^2=\frac{R}{3} \left( \frac{1}{m}-1 \right) \gg R\,.
\end{equation}
If $m$ grows to the order of 0.1 by the present epoch, 
then the mass $M$ today can be of the order of $H_0$.
In the regimes $M^2 \gg k^2/a^2$ and $M^2 \ll k^2/a^2$
the effective gravitational coupling has the asymptotic 
forms $G_{\mathrm{eff}} \simeq G/F$ and 
$G_{\mathrm{eff}} \simeq 4G/(3F)$, respectively. 
The former corresponds to the ``General Relativistic (GR) regime''
in which the evolution of $\delta_m$ mimics that in GR, 
whereas the latter corresponds to the ``scalar-tensor regime''
in which the evolution of $\delta_m$ is non-standard.
For the \fR\ models (\ref{Amodel}) and (\ref{Bmodel}) 
the transition from the former regime to 
the latter regime, which is characterized by the condition $M^2=k^2/a^2$, 
can occur during the matter domination for the wavenumbers
relevant to the matter power spectrum~\cite{Hu07,Star07,Tsuji08,Moraes08,Moraes09}.

In order to derive Eq.~(\ref{delmfi}) we used the approximation 
that the time-derivative terms of $\delta F$ on the l.h.s.\ of 
Eq.~(\ref{perdark3}) is neglected. In the regime $M^2 \gg k^2/a^2$, 
however, the large mass $M$ can induce rapid oscillations 
of $\delta F$.
In the following we shall study the evolution of 
the oscillating mode~\cite{Star07}.
For sub-horizon perturbations Eq.~(\ref{perdark3}) 
is approximately given by 
\begin{equation}
\ddot{\delta F}+3H \dot{\delta F}+\left(\frac{k^2}
{a^2} +M^2 \right)\delta F \simeq
\frac{\kappa^2}{3} \delta \rho_m\,.
\end{equation}
The solution of this equation is the sum of the matter induce mode
$\delta F_{\mathrm{ind}} \simeq (\kappa^2/3)\delta \rho_m/(k^2/a^2+M^2)$ 
and the oscillating mode $\delta F_{\mathrm{osc}}$ satisfying 
\begin{eqnarray}
\ddot{\delta F}_{\mathrm{osc}}+3H \dot{\delta F}_{\mathrm{osc}}+
\left(\frac{k^2}{a^2} +M^2 \right)\delta F_{\mathrm{osc}}=0\,.
\label{ddotdelF}
\end{eqnarray}

As long as the frequency $\omega=\sqrt{k^2/a^2+M^2}$
satisfies the adiabatic condition $|\dot{\omega}| \ll \omega^2$, 
we obtain the solution of Eq.~(\ref{ddotdelF})
under the WKB approximation: 
\begin{eqnarray}
\delta F_{\mathrm{osc}} \simeq ca^{-3/2}\frac{1}{\sqrt{2\omega}}
\cos \left( \int  \omega \mathrm{d}t \right)\,,
\end{eqnarray}
where $c$ is a constant.
Hence the solution of the perturbation $\delta R$ is 
expressed by~\cite{Star07,Tsuji08}
\begin{eqnarray}
\label{delrso}
\delta R \simeq 
\frac{1}{3f_{,RR}} \frac{\kappa^2 \delta \rho_m}
{k^2/a^2+M^2}
+c a^{-3/2}\frac{1}{f_{,RR}\sqrt{2\omega}}
\cos \left( \int  \omega \mathrm{d}t \right)\,.
\end{eqnarray}

For viable \fR\ models, the scale factor $a$ and the background Ricci scalar
$R^{(0)}$ evolve as $a \propto t^{2/3}$ and
$R^{(0)} \simeq 4/(3t^2)$ during the matter era.
Then the amplitude of $\delta R_{\mathrm{osc}}$ relative to $R^{(0)}$
has the time-dependence 
\begin{eqnarray}
\frac{|\delta R_{\mathrm{osc}}|}{R^{(0)}} \propto
\frac{M^2t}{(k^2/a^2+M^2)^{1/4}}\,.
\end{eqnarray}
The \fR\ models (\ref{Amodel}) and (\ref{Bmodel}) 
behave as $m(r)=C(-r-1)^p$ with $p=2n+1$ in the regime $R \gg R_c$.
During the matter-dominated epoch the mass $M$ evolves 
as $M \propto t^{-(p+1)}$.
In the regime $M^2 \gg k^2/a^2$ one has 
$|\delta R_{\mathrm{osc}}|/R^{(0)} \propto t^{-(3p+1)/2}$
and hence the amplitude of the oscillating mode 
decreases faster than $R^{(0)}$. However the contribution of 
the oscillating mode tends to be more important as we go back to the past.
In fact, this behavior was confirmed in the numerical simulations 
of~\cite{Tsuji08,Appleby08}.
This property persists in the radiation-dominated epoch as well.
If the condition $|\delta R| <R^{(0)}$ is violated, then $R$ can 
be negative such that the condition $f_{,R}>0$ or $f_{,RR}>0$
is violated for the models (\ref{Amodel}) and (\ref{Bmodel}).
Thus we require that $|\delta R|$ is smaller 
than $R^{(0)}$ at the beginning of the radiation era.
This can be achieved by choosing the constant $c$ in Eq.~(\ref{delrso}) 
to be sufficiently small, which amounts to a fine tuning 
for these models.

For the models (\ref{Amodel}) and (\ref{Bmodel}) one has 
$F=1-2n \mu (R/R_c)^{-2n-1}$ in the regime $R \gg R_c$.
Then the field $\phi$ defined in Eq.~(\ref{kappaphi}) rapidly 
approaches $0$ as we go back to the past.
Recall that in the Einstein frame the effective potential of the field 
has a potential minimum around $\phi=0$ because of 
the presence of the matter coupling.
Unless the oscillating mode of the field perturbation $\delta \phi$
is strongly suppressed relative to the background field $\phi^{(0)}$, 
the system can access the curvature singularity at $\phi=0$~\cite{Frolov}.
This is associated with the condition 
$|\delta R| <R^{(0)}$ discussed above.
This curvature singularity appears in the past, 
which is not related to the future singularities studied in 
~\cite{Nojirisin1,Nojirisin2}.
The past singularity can be cured by taking into account 
the $R^2$ term~\cite{Appleby09}, as we will see in Section~\ref{curesinsec}.
We note that the \fR\ models proposed in~\cite{Miranda}
[e.g., $f(R)=R-\alpha R_c \ln (1+R/R_c)$] to cure the singularity 
problem satisfy neither the local gravity 
constraints~\cite{Thongkool} nor observational constraints 
of large-scale structure~\cite{delaCruz}.

As long as the oscillating mode $\delta R_{\mathrm{osc}}$ is negligible 
relative to the matter-induced mode $\delta R_{\mathrm{ind}}$, we can estimate
the evolution of matter perturbations $\delta_m$ as well as the effective
gravitational potential $\Phi_{\mathrm{eff}}$.
Note that in~\cite{delaCruz08,Motohashi} the perturbation equations have been 
derived without neglecting the oscillating mode.
As long as the condition $|\delta R_{\mathrm{osc}}|<|\delta R_{\mathrm{ind}}|$ 
is satisfied initially, the approximate equation~(\ref{delmfi}) is accurate 
to reproduce the numerical solutions~\cite{delaCruz08,Moraes09}.
Equation~(\ref{delmfi}) can be written as
\begin{equation}
\frac{\mathrm{d}^2 \delta_m}{\mathrm{d}N^2}+\left( \frac12
-\frac32w_{\mathrm{eff}} \right) \frac{\mathrm{d}\delta_m}{\mathrm{d}N}
-\frac32 \Omega_m \frac{4+3M^2a^2/k^2}
{3(1+M^2a^2/k^2)}=0\,,
\end{equation}
where $N=\ln a$, $w_{\mathrm{eff}}=-1-2\dot{H}/(3H^2)$, and 
$\Omega_m=8\pi G \rho_m/(3FH^2)$.
The matter-dominated epoch corresponds to $w_{\mathrm{eff}}=0$
and $\Omega_m=1$.
In the regime $M^2 \gg k^2/a^2$ the evolution of 
$\delta_m$ and $\Phi_{\mathrm{eff}}$ during the matter 
dominance is given by 
\begin{equation}
\delta_m \propto t^{2/3}\,,\qquad 
\Phi_{\mathrm{eff}}= \mathrm{constant}\,,
\label{Phiso1}
\end{equation}
where we used Eq.~(\ref{Phieff}).
The matter-induced mode $\delta R_{\mathrm{ind}}$ relative to the background 
Ricci scalar $R^{(0)}$ evolves as $|\delta R_{\mathrm{ind}}|/R^{(0)} \propto t^{2/3} 
\propto \delta_m$.
At late times the perturbations can enter the regime $M^2 \ll k^2/a^2$, 
depending on the wavenumber $k$ and the mass $M$. 
When $M^2 \ll k^2/a^2$, the evolution of $\delta_m$ and 
$\Phi_{\mathrm{eff}}$ during the matter era is~\cite{Star07}
\begin{equation}
\delta_m \propto t^{(\sqrt{33}-1)/6}\,,\qquad 
\Phi_{\mathrm{eff}} \propto t^{(\sqrt{33}-5)/6}\,.
\label{Phiso2}
\end{equation}
For the model $m(r)=C(-r-1)^p$, the evolution of the matter-induced
mode in the region $M^2 \ll k^2/a^2$ is given by 
$|\delta R_{\mathrm{ind}}|/R^{(0)} \propto t^{-2p+(\sqrt{33}-5)/6}$.
This decreases more slowly relative to the ratio 
$|\delta R_{\mathrm{osc}}|/R^{(0)}$~\cite{Tsuji08}, 
so the oscillating mode tends to be unimportant with time.


\subsection{The impact on large-scale structure}

We have shown that the evolution of matter perturbations during 
the matter dominance is given by $\delta_m \propto t^{2/3}$
for $M^2 \gg k^2/a^2$ (GR regime) and 
$\delta_m \propto t^{(\sqrt{33}-1)/6}$
for $M^2 \ll k^2/a^2$ (scalar-tensor regime), respectively.
The existence of the latter phase gives rise to the modification 
to the matter power spectrum~\cite{matterper1,matterper2,SongHu1,SongHu2,Teg}
(see also~\cite{TsujiUddin,Pogosian,Pola08,Borisov08,Narikawa,Girones,Motohashi10}
for related works).
The transition from the GR regime to the scalar-tensor regime
occurs at $M^2=k^2/a^2$. 
If it occurs during the matter dominance ($R \simeq 3H^2$), 
the condition $M^2=k^2/a^2$ translates into~\cite{Moraes09}
\begin{equation}
m \simeq (aH/k)^2\,,
\label{mkcon}
\end{equation}
where we have used the relation 
$M^2 \simeq R/(3m)$ (valid for $m \ll 1$).

We are interested in the wavenumbers $k$ relevant to 
the linear regime of the galaxy power spectrum~\cite{Teg04,Teg06}:
\begin{equation}
0.01\,h\mathrm{\ Mpc}^{-1} \lesssim k \lesssim
0.,h\mathrm{\ Mpc}^{-1}\,,
\label{kscale}
\end{equation}
where $h = 0.72 \pm 0.08$ corresponds to the uncertainty 
of the Hubble parameter today.
Non-linear effects are important for 
$k \gtrsim 0.2\,h\mathrm{\ Mpc}^{-1}$.
The current observations on large scales around 
$k \sim 0.01\,h\mathrm{\ Mpc}^{-1}$
are not so accurate but can be improved in future.
The upper bound $k=0.2\,h\mathrm{\ Mpc}^{-1}$ corresponds to 
$k\simeq 600 a_0H_0$, where the subscript ``0'' represents 
quantities today. If the transition from the GR regime to 
the scalar-tensor regime occurred by the present 
epoch (the redshift $z = 0$) for the mode $k=600 a_0H_0$, 
then the parameter $m$ today is constrained to be 
\begin{equation}
m(z=0) \gtrsim 3 \times 10^{-6}\,.
\label{mbound}
\end{equation}
When $m(z=0) \lesssim 3 \times 10^{-6}$ the linear perturbations 
have been always in the GR regime by today, in which case 
the models are not distinguished from the $\Lambda$CDM model.
The bound (\ref{mbound}) is relaxed for non-linear
perturbations with $k \gtrsim 0.2\,h\mathrm{\ Mpc}^{-1}$, 
but the linear analysis is not valid in such cases.

If the transition characterized by the condition (\ref{mkcon})
occurs during the deep matter era ($z \gg 1$), we can
estimate the critical redshift $z_k$ at the transition point.
In the following let us consider the models (\ref{Amodel}) 
and (\ref{Bmodel}). In addition to the approximations
$H^2 \simeq H_0^2 \Omega_m^{(0)} (1+z)^3$ and 
$R \simeq 3H^2$ during the matter dominance, 
we use the the asymptotic forms
$m \simeq C(-r-1)^{2n+1}$ and $r \simeq -1-\mu R_c/R$
with $C=2n(2n+1)/{\mu}^{2n}$.
Since the dark energy density today can be approximated 
as $\rho_{\mathrm{DE}}^{(0)} \approx \mu R_c/2$, 
it follows that $\mu R_c \approx 6H_0^2 
\Omega_{\mathrm{DE}}^{(0)}$.
Then the condition (\ref{mkcon}) translates into the 
critical redshift~\cite{Moraes09}
\begin{equation}
z_k=\left[ \left( \frac{k}{a_0H_0} \right)^2 
\frac{2n(2n+1)}{\mu^{2n}} 
\frac{(2\Omega_{\mathrm{DE}}^{(0)})^{2n+1}}
{(\Omega_m^{0})^{2(n+1)}} \right]^{1/(6n+4)}-1\,.
\label{zk}
\end{equation}
For $n=1$, $\mu=3$, $\Omega_m^{(0)}=0.28$, and 
$k=300a_0H_0$ the numerical value of the critical redshift 
is $z_k=4.5$, which is in good agreement with the 
analytic value estimated by (\ref{zk}).

The estimation (\ref{zk}) shows that, for larger $k$,
the transition occurs earlier. The time $t_k$ at the transition 
has a $k$-dependence: $t_k \propto k^{-3/(6n+4)}$.
For $t>t_k$ the matter perturbation evolves as 
$\delta_m \propto t^{(\sqrt{33}-1)/6}$ by the time 
$t=t_{\Lambda}$ corresponding to the onset of cosmic 
acceleration ($\ddot{a}=0$).
The matter power spectrum $P_{\delta_m}=|\delta_m|^2$
at the time $t_{\Lambda}$ shows a difference compared 
to the case of the $\Lambda$CDM model~\cite{Star07}:
\begin{equation}
\frac{P_{\delta_m}(t_\Lambda)}
{P_{\delta_m}{}^{\Lambda\mathrm{CDM}}(t_\Lambda)}
=\left(\frac{t_\Lambda}{t_k}\right)
^{2\left(\frac{\sqrt{33}-1}{6}-\frac23\right)}
\propto k^{\frac{\sqrt{33}-5}{6n+4}}\,.
\label{Pratio}
\end{equation}
We caution that, when $z_k$ is close to 
$z_{\Lambda}$ (the redshift at $t=t_{\Lambda}$), the
estimation~(\ref{Pratio}) begins to lose its accuracy. The ratio of
the two power spectra today,
i.e., $P_{\delta_m}(t_0)/P_{\delta_m}{}^{\Lambda\mathrm{CDM}} (t_0)$ is
in general different from Eq.~(\ref{Pratio}). However, numerical
simulations in~\cite{Tsuji08} show that the difference is small
for $n$ of the order of unity.

The modified evolution~(\ref{Phiso2}) of the effective gravitational 
potential for $z<z_k$ leads to the integrated Sachs--Wolfe (ISW) effect
in CMB anisotropies~\cite{SongHu1,LiBarrow,Peiris}. However this is limited to 
very large scales (low multipoles) in the CMB spectrum.
Meanwhile the galaxy power spectrum is directly affected by 
the non-standard evolution of matter perturbations.
From Eq.~(\ref{Pratio}) there should be a difference between
the spectral indices of the CMB spectrum and 
the galaxy power spectrum
on the scale (\ref{kscale})~\cite{Star07}:
\begin{equation}
\Delta n_s =\frac{\sqrt{33}-5}{6n+4}\,.
\label{DelnfR}
\end{equation}
Observationally we do not find any strong signature for
the difference of slopes of the two spectra.
If we take the mild bound $\Delta n_s<0.05$, 
we obtain the constraint $n>2$.
Note that in this case the local gravity constraint 
(\ref{bound3}) is also satisfied.

In order to estimate the growth rate of matter perturbations, 
we introduce the growth index $\gamma$ 
defined by~\cite{Peebles} 
\begin{equation}
f_\delta \equiv \frac{\dot{\delta}_m}{H \delta_m}=
(\tilde{\Omega}_m)^{\gamma}\,,
\end{equation}
where $\tilde{\Omega}_m=\kappa^2 \rho_m/(3H^2)=F\Omega_m$.
This choice of $\tilde{\Omega}_m$ comes from writing Eq.~(\ref{FRWfR1})
in the form $3H^2=\rho_{\mathrm{DE}}+\kappa^2 \rho_m$, 
where $\rho_{\mathrm{DE}} \equiv (FR-f)/2-3H\dot{F}+3H^2(1-F)$ and 
we have ignored the contribution of radiation.
Since the viable \fR\ models are close to the $\Lambda$CDM model 
in the region of high density, the quantity $F$ approaches 1 
in the asymptotic past. Defining $\rho_{\mathrm{DE}}$ and $\tilde{\Omega}_m$
in the above way, the Friedmann equation can be cast in the usual 
GR form with non-relativistic matter and dark energy~\cite{Star07,Moraes08,Moraes09}.

The growth index in the $\Lambda$CDM model corresponds to 
$\gamma \simeq 0.55$~\cite{Wang98,Linder05}, 
which is nearly constant for $0<z<1$. 
In \fR\ gravity, if the perturbations are in the GR regime
($M^2 \gg k^2/a^2$) today, $\gamma$ is close to the GR 
value. Meanwhile, if the transition to the scalar-tensor regime occurred
at the redshift $z_k$ larger than 1, the growth index becomes 
smaller than 0.55~\cite{Moraes08}.
Since $0<\tilde{\Omega}_m<1$, the smaller $\gamma$ implies 
a larger growth rate.

\epubtkImage{gammaevo.png}{%
  \begin{figure}[hptb]
    \centerline{\includegraphics[width=3.3in]{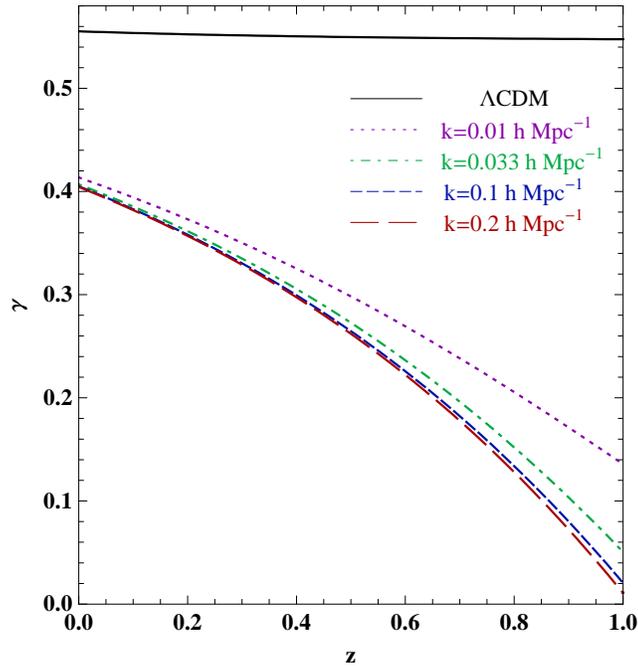}}
    \caption{Evolution of $\gamma$ versus the redshift $z$ in the
      model~(\ref{Amodel}) with $n=1$ and $\mu=1.55$ for four
      different values of $k$. For these model parameters the
      dispersion of $\gamma$ with respect to $k$ is very small. All
      the perturbation modes shown in the figure have reached the
      scalar-tensor regime ($M^2 \ll k^2/a^2$) by
      today. From~\cite{Moraes09}.}
    \label{gammaevo} 
\end{figure}}

\epubtkImage{ngammas.png}{%
  \begin{figure}[hptb]
    \centerline{\includegraphics[width=3.3in]{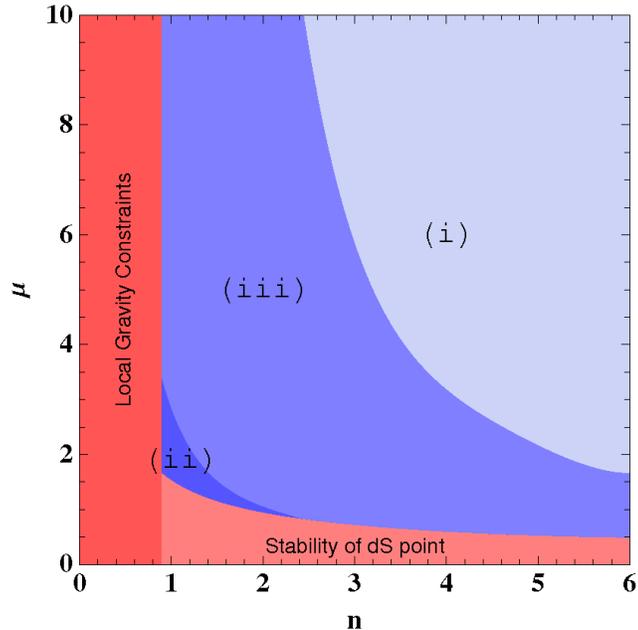}}
    \caption{The regions (i), (ii) and (iii) for the
      model~(\ref{Bmodel}). We also show the bound $n>0.9$ coming from
      the local gravity constraints as well as the condition
      (\ref{Bmodelcon}) coming from the stability of the de~Sitter
      point. From~\cite{Moraes09}.}
    \label{ngammaphasek} 
\end{figure}}

In Figure~\ref{gammaevo} we plot the evolution of the growth index 
$\gamma$ in the model~(\ref{Amodel}) with $n=1$ 
and $\mu=1.55$ for a number of different wavenumbers.
In this case the present value of $\gamma$
is degenerate around $\gamma_0 \simeq 0.41$
independent of the scales of our interest.
For the wavenumbers $k=0.1\,h\mathrm{\ Mpc}^{-1}$ and 
$k=0.01\,h\mathrm{\ Mpc}^{-1}$ the transition redshifts 
correspond to $z_k=5.2$ and $z_k=2.7$, respectively.
Hence these modes have already entered the scalar-tensor 
regime by today.

From Eq.~(\ref{zk}) we find that $z_k$ gets smaller for larger
$n$ and $\mu$. If the mode $k=0.2\,h\mathrm{\ Mpc}^{-1}$ crossed the transition point 
at $z_k>{\cal O}(1)$ and the mode $k=0.01\,h\mathrm{\ Mpc}^{-1}$
has marginally entered (or has not entered) the scalar-tensor regime 
by today, then the growth indices should be strongly dispersed.
For sufficiently large values of $n$ and $\mu$ one can expect 
that the transition to the regime $M^2 \ll k^2/a^2$ has not occurred
by today. The following three cases appear depending on the values 
of $n$ and $\mu$~\cite{Moraes09}:
\begin{enumerate}[label=(\roman{*})]
\item All modes have the values of $\gamma_0$ close to the 
$\Lambda$CDM value: $\gamma_0= 0.55$, 
i.e., $0.53 \lesssim \gamma_0 \lesssim 0.55$. 
\item All modes have the values of $\gamma_0$ close to the value
in the range $0.40 \lesssim \gamma_0 \lesssim 0.43$. 
\item The values of $\gamma_0$ are dispersed in the range $0.40
\lesssim \gamma_0 \lesssim 0.55$.
\end{enumerate}
The region (i) corresponds to the opposite of the inequality (\ref{mbound}), 
i.e., $m(z=0) \lesssim 3 \times 10^{-6}$, 
in which case $n$ and $\mu$ take large values.
The border between (i) and (iii) is characterized by the condition 
$m(z=0) \approx 3 \times 10^{-6}$.
The region (ii) corresponds to small values of $n$ and $\mu$ 
(as in the numerical simulation of Figure~\ref{gammaevo}), in which 
case the mode $k=0.01\,h\mathrm{\ Mpc}^{-1}$ 
entered the scalar-tensor regime for $z_k>{\cal O}(1)$.

The regions (i), (ii), (iii) can be found numerically by solving the perturbation
equations. In Figure~\ref{ngammaphasek} we plot those regions
for the model~(\ref{Bmodel}) together with the bounds coming from 
the local gravity constraints as well as the stability of the late-time 
de~Sitter point. Note that the result in the model~(\ref{Amodel}) is 
also similar to that in the model~(\ref{Bmodel}).
The parameter space for $n \lesssim 3$ and 
$\mu={\cal O}(1)$ is dominated by either the region (ii) 
or the region (iii). While the present observational constraint 
on $\gamma$ is quite weak, the unusual converged or dispersed 
spectra found above can be useful to distinguish metric \fR\ 
gravity from the $\Lambda$CDM model in future observations.
We also note that for other viable \fR\ models such as 
(\ref{tanh}) the growth index today can be as small as 
$\gamma_0 \simeq 0.4$~\cite{Moraes09}. 
If future observations detect such unusually small values 
of $\gamma_0$, this can be a smoking gun for \fR\ models.


\subsection{Non-linear matter perturbations}

So far we have discussed the evolution of linear perturbations
relevant to the matter spectrum for the scale 
$k \lesssim 0.01\mbox{\,--\,}0.2\,h\mathrm{\ Mpc}^{-1}$.
For smaller scale perturbations the effect of non-linearity 
becomes important. In GR there are some mapping 
formulas from the linear power spectrum to the non-linear 
power spectrum such as the halo fitting 
by Smith et al.~\cite{Smith}.
In the halo model the non-linear power spectrum $P(k)$
is defined by the sum of two pieces~\cite{Cooray}:
\begin{equation}
P(k)=I_1(k)+I_2(k)^2 P_L(k)\,,
\label{PkI}
\end{equation}
where $P_L(k)$ is a linear power spectrum and 
\begin{equation}
I_1(k)=\int \frac{\mathrm{d}M}{M} \left( \frac{M}{\rho_m^{(0)}}
\right)^2 \frac{\mathrm{d}n}{\mathrm{d} \ln M}\,y^2 (M,k)\,,\qquad
I_2(k)=\int \frac{\mathrm{d}M}{M} \left( \frac{M}{\rho_m^{(0)}}
\right)^2 \frac{\mathrm{d}n}{\mathrm{d} \ln M}\,b(M)y (M,k)\,.
\end{equation}
Here $M$ is the mass of dark matter halos, 
$\rho_m^{(0)}$ is the dark matter density today, 
$\mathrm{d}n/\mathrm{d} \ln M$ is the mass
function describing the comoving number density of halos, 
$y(M, k)$ is the Fourier transform of the halo density profile, 
and $b(M)$ is the halo bias.

In modified gravity theories, Hu and Sawicki (HS)~\cite{Huparametrization} 
provided a fitting formula to describe a non-linear power spectrum 
based on the halo model.
The mass function $\mathrm{d}n/\mathrm{d} \ln M$ and the halo profile
$\rho$ depend on the root-mean-square $\sigma(M)$ of
a linear density field.
The Sheth-Torman mass function~\cite{Sheth} and the 
Navarro-Frenk-White halo profile~\cite{NFW} are usually employed in GR.
Replacing $\sigma$ for $\sigma_{\mathrm{GR}}$ obtained in the GR
dark energy model that follows the same expansion history as
the modified gravity model, we obtain a non-linear power spectrum 
$P(k)$ according to Eq.~(\ref{PkI}).
In~\cite{Huparametrization} this non-linear spectrum is 
called $P_{\infty}(k)$.
It is also possible to obtain a non-linear spectrum $P_0(k)$ by 
applying a usual (halo) mapping formula in GR to modified gravity. 
This approach is based on the assumption that the 
growth rate in the linear regime determines
the non-linear spectrum.
Hu and Sawicki proposed a parametrized 
non-linear spectrum that interpolates between two spectra 
$P_{\infty}(k)$ and $P_0(k)$~\cite{Huparametrization}:
\begin{equation}
P(k)=\frac{P_0(k)+c_{\mathrm{nl}}\Sigma^2 (k) P_{\infty}(k)}
{1+c_{\mathrm{nl}}\Sigma^2 (k)}\,,
\label{Hupara}
\end{equation}
where $c_{\mathrm{nl}}$ is a parameter which controls whether 
$P(k)$ is close to $P_0(k)$ or $P_{\infty}(k)$.
In~\cite{Huparametrization} they have taken the form 
$\Sigma^2(k)=k^3 P_L(k)/(2\pi^2)$.

\epubtkImage{nonlinearfig.png}{%
  \begin{figure}[hptb]
    \centerline{\includegraphics[width=5.8in]{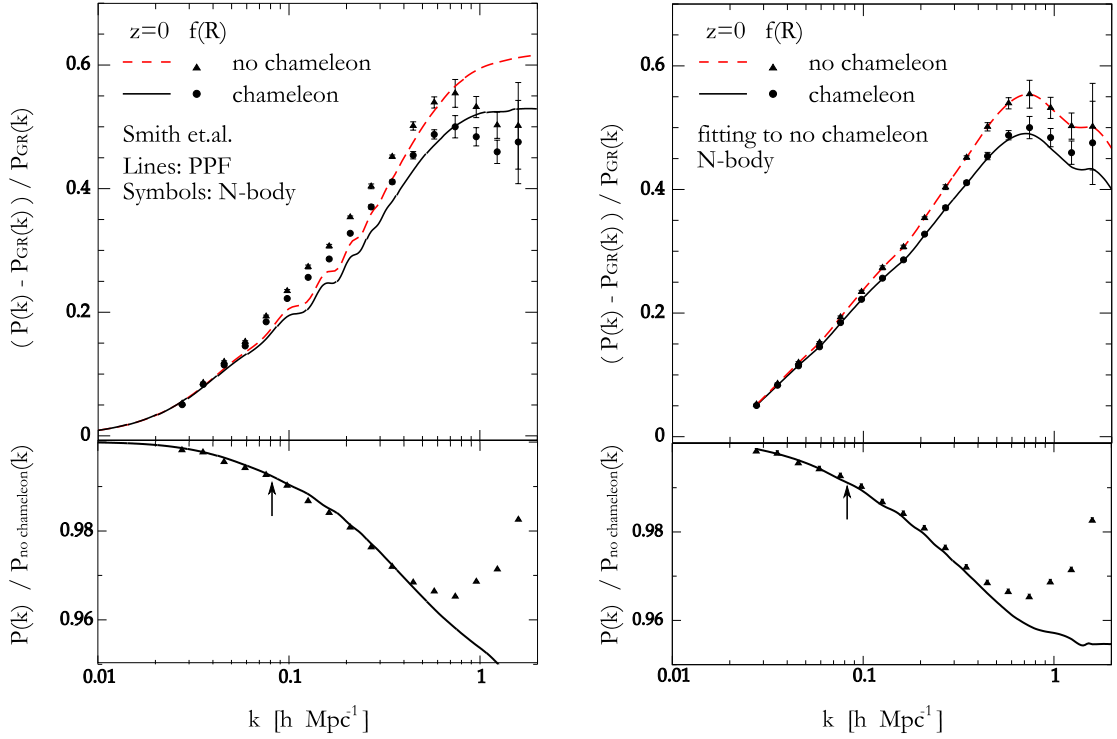}}
    \caption{Comparison between $N$-body simulations and the two
      fitting formulas in the \fR\ model~(\ref{Amodel}) with
      $n=1/2$. The circles and triangles show the results of $N$-body
      simulations with and without the chameleon mechanism,
      respectively. The arrow represents the maximum value of
      $k\,(=0.08\,h\mathrm{\ Mpc}^{-1})$ by which the perturbation theory is
      valid. (Left) The fitting formula by Smith et al.~\cite{Smith}
      is used to predict $P_{\mathrm{non-GR}}$ and
      $P_{\mathrm{GR}}$. The solid and dashed lines correspond to the
      power spectra with and without the chameleon mechanism,
      respectively. For the chameleon case $c_{\mathrm{nl}}(z)$ is
      determined by the perturbation theory with
      $c_{\mathrm{nl}}(z=0)=0.085$. (Right) The $N$-body results
      in~\cite{Oyaizu2} are interpolated to derive
      $P_{\mathrm{non-GR}}$ without the chameleon mechanism. The
      obtained $P_{\mathrm{non-GR}}$ is used for the HS fitting
      formula to derive the power spectrum $P$ in the chameleon
      case. From~\cite{Koyama09}.}
    \label{nonlinearfig} 
\end{figure}}

The validity of the HS fitting formula (\ref{Hupara}) should be checked
with $N$-body simulations in modified gravity models.
In~\cite{Oyaizu1,Oyaizu2,Oyaizu3} $N$-body simulations
were carried out for the \fR\ model~(\ref{Amodel}) 
with $n=1/2$ (see also~\cite{Stabenau,Laszlo} 
for $N$-body simulations in other modified gravity models).
The chameleon mechanism should be 
at work on small scales (solar-system scales)
for the consistency with local gravity constraints.
In~\cite{Oyaizu2} it was found that the chameleon mechanism
tends to suppress the enhancement of the power spectrum in 
the non-linear regime that corresponds to the recovery of GR.
On the other hand, in the post Newtonian intermediate regime, 
the power spectrum is enhanced compared to the GR case 
at the measurable level. 

Koyama et al.~\cite{Koyama09} studied the validity of 
the HS fitting formula by comparing it with 
the results of $N$-body simulations.
Note that in this paper the parametrization (\ref{Hupara})
was used as a fitting formula without employing the halo model
explicitly. In their notation $P_0$ corresponds to ``$P_{\mathrm{non-GR}}$''
derived without non-linear interactions responsible for the recovery 
of GR (i.e., gravity is modified down to small scales in the same manner
as in the linear regime), whereas $P_{\infty}$ corresponds to 
``$P_{\mathrm{GR}}$'' obtained in the GR dark energy model
following the same expansion history as that in the modified
gravity model. Note that $c_{\mathrm{nl}}$ characterizes how the theory 
approaches GR by the chameleon mechanism.
Choosing $\Sigma$ as
\begin{equation}
\Sigma^2(k, z)=\left( \frac{k^3}{2\pi^2} P_L(k, z) \right)^{1/3}\,,
\end{equation}
where $P_L$ is the linear power spectrum in the modified gravity model, 
they showed that, in the \fR\ model~(\ref{Amodel}) with $n=1/2$, 
the formula~(\ref{Hupara}) can fit the solutions in perturbation theory 
very well by allowing the time-dependence of the parameter $c_{\mathrm{nl}}$ 
in terms of the redshift $z$.
In the regime $0<z<1$ the parameter $c_{\mathrm{nl}}$ is approximately given by 
$c_{\mathrm{nl}}(z=0)=0.085$.

In the left panel of Figure~\ref{nonlinearfig} the relative 
difference of the non-linear power spectrum $P(k)$ from the 
GR spectrum $P_{\mathrm{GR}}(k)$ is plotted as a dashed curve
(``no chameleon'' case with $c_{\mathrm{nl}}=0$)  and 
as a solid curve (``chameleon'' case with non-zero $c_{\mathrm{nl}}$
derived in the perturbative regime).
Note that in this simulation the fitting formula by 
Smith et al.~\cite{Smith} is used to obtain 
the non-linear power spectrum from the linear one.
The agreement with $N$-body simulations is not very good 
in the non-linear regime ($k>0.1\,h\mathrm{\ Mpc}^{-1}$).
In~\cite{Koyama09} the power spectrum $P_{\mathrm{non-GR}}$ 
in the no chameleon case (i.e., $c_{\mathrm{nl}}=0$) was 
derived by interpolating the $N$-body 
results in~\cite{Oyaizu2}.
This is plotted as the dashed line in the right panel 
of Figure~\ref{nonlinearfig}.
Using this spectrum $P_{\mathrm{non-GR}}$ for $c_{\mathrm{nl}} \neq 0$,
the power spectrum in $N$-body simulations in the chameleon case
can be well reproduced by the fitting formula (\ref{Hupara}) for the scale
$k<0.5\,h\mathrm{\ Mpc}^{-1}$ (see the solid line in Figure~\ref{nonlinearfig}).
Although there is some deviation in the regime $k>0.5\,h\mathrm{\ Mpc}^{-1}$, 
we caution that $N$-body simulations have large errors in this regime.
See~\cite{ClusterHu} for clustered abundance constraints on 
the \fR\ model~(\ref{Amodel}) derived by the calibration of 
$N$-body simulations.

In the quasi non-linear regime a normalized skewness, 
$S_3=\langle \delta_m^3 \rangle/\langle \delta_m^2 \rangle^2$, 
of matter perturbations can provide a good test for the picture 
of gravitational instability from Gaussian initial conditions~\cite{Bernar}.
If large-scale structure grows via gravitational instability from 
Gaussian initial perturbations, the skewness 
in a universe dominated by pressureless matter is 
known to be $S_3 = 34/7$ in GR~\cite{Peebles}.
In the $\Lambda$CDM model the skewness depends weakly on 
the expansion history of the universe (less than a few percent)~\cite{Kamion}.
In \fR\ dark energy models the difference of the skewness from 
the $\Lambda$CDM model is only less than a few percent~\cite{Tateskewness}, 
even if the growth rate of matter perturbations is significantly different.
This is related to the fact that in the Einstein frame dark energy 
has a universal coupling $Q=-1/\sqrt{6}$ with all non-relativistic
matter, unlike the coupled quintessence scenario with different 
couplings between dark energy and matter species (dark matter, 
baryons)~\cite{Amenskew}.


\subsection{Cosmic Microwave Background}
\label{CMBsec}

The effective gravitational potential (\ref{Phieffdef}) is directly 
related to the ISW effect in CMB anisotropies.
This contributes to the temperature anisotropies today 
as an integral~\cite{Sugiyama95,Dodelsonbook}
\begin{equation}
\Theta_{\mathrm{ISW}} \equiv 
\int_0^{\eta_0} \mathrm{d}\eta e^{-\tau}
\frac{\mathrm{d}\Phi_{\mathrm{eff}}}{\mathrm{d}\eta} 
j_{\ell} [k(\eta_0-\eta)]\,,
\end{equation}
where $\tau$ is the optical depth, 
$\eta=\int a^{-1} \mathrm{d}t$ is the conformal time with the present 
value $\eta_0$, and $j_{\ell} [k(\eta_0-\eta)]$ is the spherical Bessel function 
for CMB multipoles $\ell$ and the wavenumber $k$.
In the limit $\ell \gg 1$ (i.e., small-scale limit) the spherical Bessel 
function has a dependence $j_{\ell}(x) \simeq (1/\ell) (x/\ell)^{\ell-1/2}$, 
which is suppressed for large $\ell$.
Hence the dominant contribution to the ISW effect comes from  
the low $\ell$ modes ($\ell={\cal O}(1)$).

\epubtkImage{iswfig.png}{%
  \begin{figure}[hptb]
    \centerline{
      \includegraphics[width=2.8in]{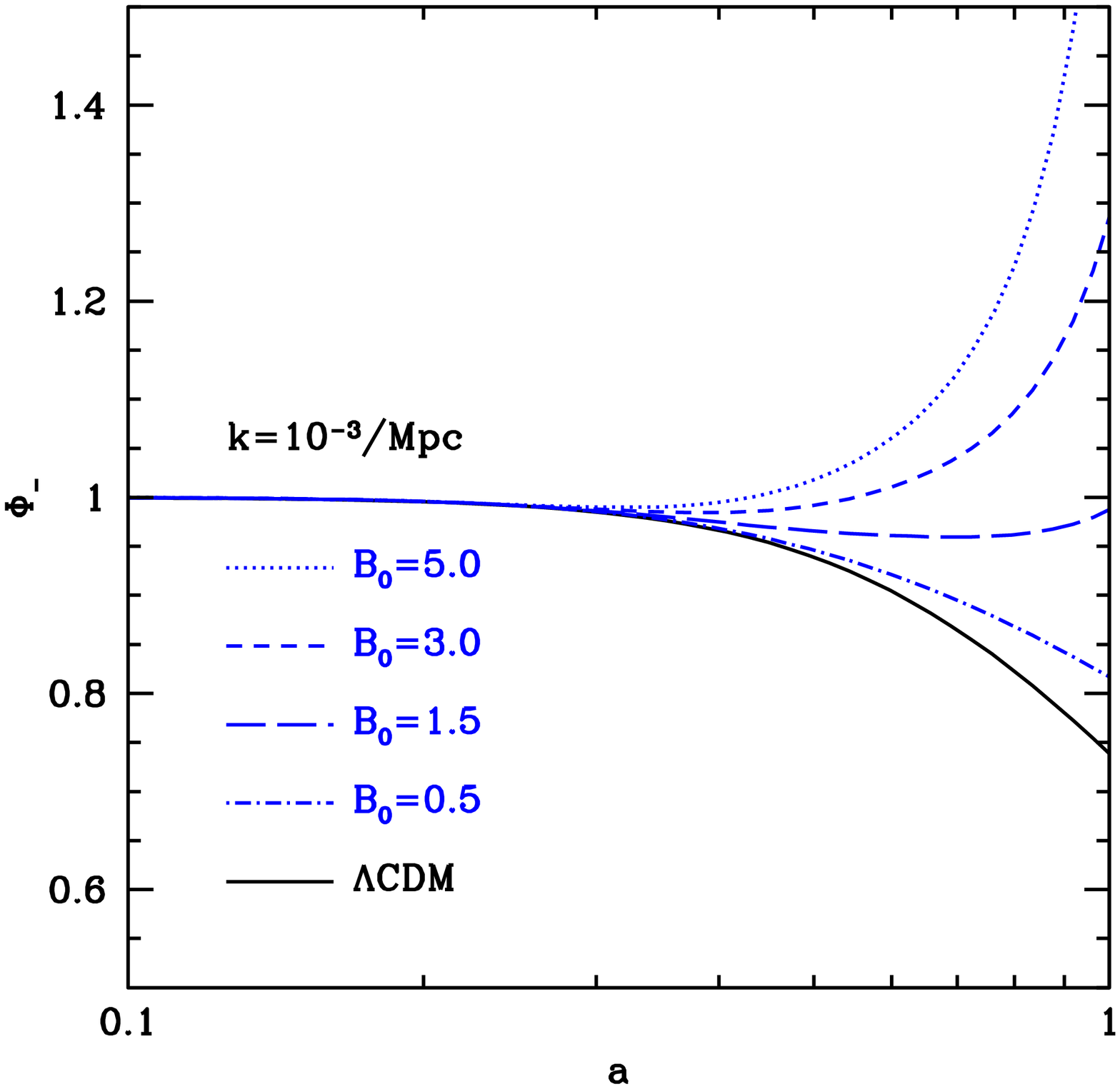}
      \includegraphics[width=2.8in]{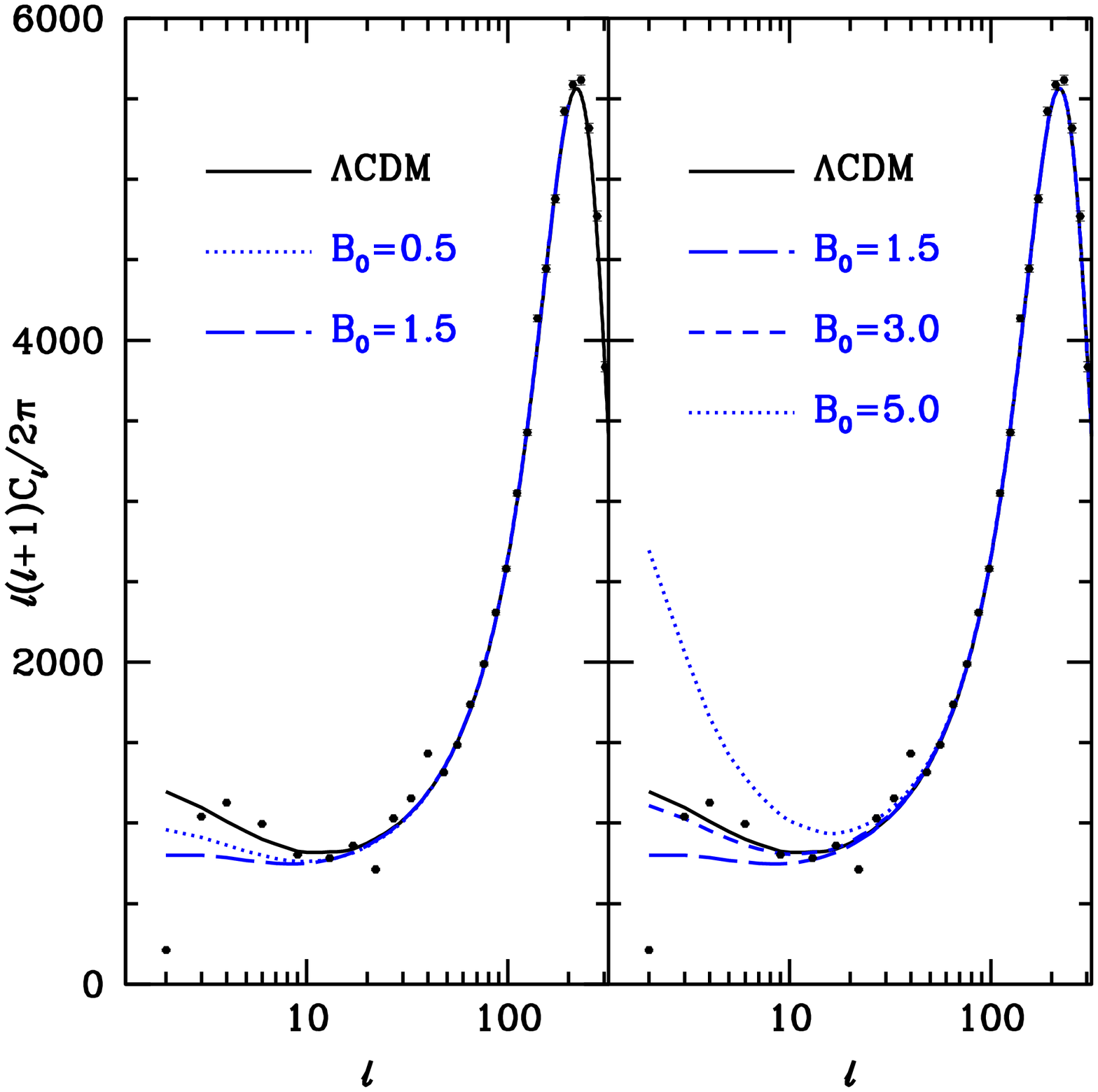}
    }
    \caption{(Left) Evolution of the effective gravitational potential
      $\Phi_{\mathrm{eff}}$ (denoted as $\Phi_{-}$ in the figure)
      versus the scale factor $a$ (with the present value $a=1$) on
      the scale $k^{-1}=10^{3}\mathrm{\ Mpc}$ for the $\Lambda$CDM model and
      \fR\ models with $B_0$ = 0.5, 1.5, 3.0, 5.0. As the parameter
      $B_0$ increases, the decay of $\Phi_{\mathrm{eff}}$ decreases
      and then turns into growth for $B_0 \gtrsim 1.5$. (Right) The
      CMB power spectrum $\ell (\ell+1)C_{\ell}/(2\pi)$ for the
      $\Lambda$CDM model and \fR\ models with $B_0$ = 0.5, 1.5, 3.0,
      5.0. As $B_0$ increases, the ISW contributions to low multipoles
      decrease, reach the minimum around $B_0$ = 1.5, and then
      increase. The black points correspond to the WMAP 3-year
      data~\cite{WMAP3}. From~\cite{Peiris}.}
    \label{ISWfig}
\end{figure}}

In the $\Lambda$CDM model the effective gravitational potential
is constant during the matter dominance, 
but it begins to decay after the Universe 
enters the epoch of cosmic acceleration (see the left panel
of Figure~\ref{ISWfig}).
This late-time variation of $\Phi_{\mathrm{eff}}$ leads to the contribution 
to $\Theta_{\mathrm{ISW}}$, which works as the ISW effect.

For viable \fR\ dark energy models the evolution of $\Phi_{\mathrm{eff}}$
during the early stage of the matter era is constant as 
in the $\Lambda$CDM model.
After the transition to the scalar-tensor regime, the effective 
gravitational potential evolves as $\Phi_{\mathrm{eff}} \propto 
t^{(\sqrt{33}-5)/6}$ during the matter dominance
[as we have shown in Eq.~(\ref{Phiso2})].
The evolution of $\Phi_{\mathrm{eff}}$ during the accelerated epoch 
is also subject to change compared to the $\Lambda$CDM model.
In the left panel of Figure~\ref{ISWfig} we show the evolution of 
$\Phi_{\mathrm{eff}}$ versus the scale factor $a$ 
for the wavenumber $k=10^{-3}\mathrm{\ Mpc}^{-1}$ in several different 
cases. In this simulation the background cosmological evolution is 
fixed to be the same as that in the $\Lambda$CDM model.
In order to quantify the difference from the $\Lambda$CDM model
at the level of perturbations, \cite{PZhang,SongHu1,Peiris}
defined the following quantity
\begin{equation}
B\equiv m\,\frac{\dot{R}}{R}\,\frac{H}{\dot{H}}\,,
\end{equation}
where $m=Rf_{,RR}/f_{,R}$.
If the effective equation of state
$w_{\mathrm{eff}}$ defined in Eq.~(\ref{ldef}) is constant, it then 
follows that $R=3H^2(1-3w_{\mathrm{eff}})$ and hence 
$B=2m$. The stability of cosmological perturbations requires 
the condition $B>0$~\cite{SongHu1,SongHu2}.
The left panel of Figure~\ref{ISWfig} shows that, as we 
increase the values of $B$ today ($=B_0$), the evolution of
$\Phi_{\mathrm{eff}}$ at late times tends to be significantly 
different from that in the $\Lambda$CDM model.
This comes from the fact that, for increasing $B$, the transition 
to the scalar-tensor regime occurs earlier.

From the right panel of Figure~\ref{ISWfig} we find that, as 
$B_0$ increases, the CMB spectrum for low multipoles first decreases 
and then reaches the minimum around $B_0=1.5$.
This comes from the reduction in the decay rate of $\Phi_{\mathrm{eff}}$
relative to the $\Lambda$CDM model, see the left panel of 
Figure~\ref{ISWfig}.
Around $B_0=1.5$ the effective gravitational potential is nearly 
constant, so that the ISW effect is almost absent 
(i.e., $\Theta_{\mathrm{ISW}} \approx 0$).
For $B_0 \gtrsim 1.5$ the evolution of $\Phi_{\mathrm{eff}}$ turns into 
growth. This leads to the increase of the large-scale CMB spectrum, 
as $B_0$ increases.
The spectrum in the case $B_0=3.0$ is similar to that in 
the $\Lambda$CDM model.
The WMAP 3-year data rule out $B_0 >4.3$ at the 95\% confidence
level because of the excessive ISW effect~\cite{Peiris}.

There is another observational constraint coming from the
angular correlation between the CMB temperature field and 
the galaxy number density field induced by the ISW 
effect~\cite{SongHu1}.
The \fR\ models predict that, for $B_0 \gtrsim 1$, the
galaxies are anticorrelated with the CMB because of the sign 
change of the ISW effect.
Since the anticorrelation has not been observed in the 
observational data of CMB and LSS, this places an upper bound 
of $B_0 \lesssim 1$~\cite{Peiris}.
This is tighter than the bound $B_0<4.3$ coming from 
the CMB angular spectrum discussed above.

Finally we briefly mention stochastic gravitational waves 
produced in the early 
universe~\cite{Mendoza,Corda07,Capogra1,Capogra2,Corda09,Corda09d,DeLaurentis:2009kz,Alves:2009eg}. 
For the inflation model $f(R)=R+R^{2}/(6M^{2})$
the primordial gravitational waves are generated
with the tensor-to-scalar ratio $r$ of the order of $10^{-3}$, 
see Eq.~(\ref{rJor}).
It is also possible to generate stochastic gravitational waves
after inflation under the modification of gravity.
Capozziello et al.~\cite{Capogra1,Capogra2} studied the evolution of tensor 
perturbations for a toy model $f=R^{1+\epsilon}$ in the 
FLRW universe with the power-law evolution of the 
scale factor. Since the parameter $\epsilon$ 
is constrained to be very small 
($|\epsilon|<7.2 \times 10^{-19}$)~\cite{Barrow06,Clifton05}, 
it is very difficult to detect the signature of \fR\ gravity in the 
stochastic gravitational wave background.
This property should hold for viable \fR\ dark energy models 
in general, because the deviation from GR during the 
radiation and the deep matter era is very small.

\newpage

\section{Palatini Formalism}
\label{Palasec}
\setcounter{equation}{0}

In this section we discuss \fR\ theory in the Palatini 
formalism~\cite{Palatini1919}.
In this approach the action~(\ref{fRaction}) is varied with respect to both 
the metric $g_{\mu \nu}$ and the connection 
$\Gamma^{\alpha}_{\beta \gamma}$.
Unlike the metric approach, $g_{\mu \nu}$ and 
$\Gamma^{\alpha}_{\beta \gamma}$ are treated as
independent variables.
Variations using the Palatini approach~\cite{Ferraris,Vollick,Vollick2,Flanagan0,Flanagan,Flanagan2} 
lead to second-order field equations which are free from the instability
associated with negative signs of $f_{,RR}$~\cite{Meng,Meng2}.
We note that even in 1930's Lanczos~\cite{Lanczos} proposed a specific combination of 
curvature-squared terms that lead to a second-order and divergence-free
modified Einstein equation.

The background cosmological dynamics of Palatini \fR\ gravity
has been investigated in~\cite{Sot,Sotinf,Motapala,FayTavakol,Popwalski}, 
which shows that the sequence of radiation, matter, and 
accelerated epochs can be realized
even for the model $f(R)=R-\alpha/R^n$ with $n>0$
(see also~\cite{Meng3,NOpala,Popwalski}).
The equations for matter density perturbations were 
derived in~\cite{KoivistoPala}. Because of a large 
coupling $Q$ between dark energy and non-relativistic matter
dark energy models based on Palatini \fR\ gravity
are not compatible with the observations of large-scale structure, 
unless the deviation from the $\Lambda$CDM model is very 
small~\cite{KoivistoPala2,LiPala0,LiPala,TsujiUddin}.
Such a large coupling also gives rise to non-perturbative corrections 
to the matter action, which leads to a conflict with the Standard
Model of particle physics~\cite{Flanagan0,Flanagan,Flanagan2}
(see also~\cite{Kaloper,OlmoPRL2,Olmo08,Olmo09,Barausse1}).

There are also a number of works \cite{Olmo05,Olmo07,Dominguez,sotnewton} about the Newtonian limit 
in the Palatini formalism (see also~\cite{Allemandi,Allemandi2,Bustelo,Kainu,Ruggiero,Ruggiero2}).
In particular it was shown in~\cite{Barausse1,Barausse2} that the non-dynamical 
nature of the scalar-field degree of freedom can lead to a divergence 
of non-vacuum static spherically symmetric solutions at the surface
of a compact object for commonly-used polytropic equations of state.
Hence Palatini \fR\ theory is difficult to be 
compatible with a number of observations and experiments, 
as long as the models are constructed to explain the late-time 
cosmic acceleration. Moreover it is also known that in Palatini gravity 
the Cauchy problem~\cite{Wald} is not well-formulated due to the presence of 
higher derivatives of matter fields in field equations~\cite{Lanahan}
(see also~\cite{Salgado,CapoCauchy} for related works).
We also note that the matter Lagrangian
(such as the Lagrangian of Dirac particles) cannot be simply 
assumed to be independent of connections. 
Even in the presence of above mentioned problems 
it will be useful to review this theory
because we can learn the way of modifications of gravity from GR
to be consistent with observations and experiments.


\subsection{Field equations}
\label{fieldpala}

Let us derive field equations by treating $g_{\mu \nu}$ and
$\Gamma^{\alpha}_{\beta \gamma}$ as
independent variables. Varying the action~(\ref{fRaction}) with respect to 
$g_{\mu \nu}$, we obtain 
\begin{equation}
\label{Pala1}
F(R)R_{\mu \nu}(\Gamma) -\frac12 f(R)g_{\mu \nu}
=\kappa^2 T_{\mu \nu}^{(M)}\,,
\end{equation}
where $F(R)=\partial f/\partial R$, 
$R_{\mu \nu}(\Gamma)$ is the Ricci tensor 
corresponding to the connections $\Gamma^{\alpha}_{\beta \gamma}$, 
and $T_{\mu \nu}^{(M)}$ is defined in Eq.~(\ref{Tmunu}).
Note that $R_{\mu \nu}(\Gamma)$ is in general different from 
the Ricci tensor calculated in terms of metric connections $R_{\mu \nu}(g)$.
The trace of Eq.~(\ref{Pala1}) gives
\begin{eqnarray}
\label{Pala2}
F(R)R-2f(R)=\kappa^2 T\,,
\end{eqnarray}
where $T=g^{\mu \nu}T_{\mu \nu}^{(M)}$.
Here the Ricci scalar $R(T)$ is directly 
related to $T$ and it is different from the Ricci scalar
$R(g)=g^{\mu \nu}R_{\mu \nu}(g)$ in the metric formalism.
More explicitly we have the following relation~\cite{SotFaraoni}
\begin{equation}
R(T)=R(g)+\frac{3}{2(f'(R(T)))^2}(\nabla_{\mu}f'(R(T)))
(\nabla^{\mu}f'(R(T)))+\frac{3}{f'(R(T))} \square f'(R(T))\,,
\label{Rrela}
\end{equation}
where a prime represents a derivative in terms of $R(T)$.
The variation of the action~(\ref{fRaction}) with respect to 
the connection leads to the following equation 
\begin{eqnarray}
\label{Pala3}
&&R_{\mu \nu}(g) -\frac12 g_{\mu \nu} R(g)
=\frac{\kappa^2 T_{\mu \nu}}{F}
-\frac{FR(T)-f}{2F}g_{\mu \nu}
+\frac{1}{F}(\nabla_{\mu} \nabla_{\nu}F
-g_{\mu \nu} \square F) \nonumber \\
& &~~~~~~~~~~~~~~~~~~~~~~~~~~~~~
-\frac{3}{2F^2} \left[ \partial_{\mu}F \partial_{\nu}F
-\frac12 g_{\mu \nu} (\nabla F)^2 \right]\,.
\end{eqnarray}

In Einstein gravity ($f(R)=R-2\Lambda$ and $F(R)=1$)
the field equations (\ref{Pala2}) and (\ref{Pala3})
are identical to the equations (\ref{trace}) and (\ref{fREin}), respectively.
However, the difference appears for the \fR\ models 
which include non-linear terms in $R$.
While the kinetic term $\square F$ is present in Eq.~(\ref{trace}), 
such a term is absent in Palatini \fR\ gravity.
This has the important consequence that the oscillatory 
mode, which appears in the metric formalism, does not exist
in the Palatini formalism. 
As we will see later on, Palatini \fR\ theory corresponds to Brans--Dicke (BD) 
theory~\cite{BD} with a parameter $\omega_{\mathrm{BD}}=-3/2$ 
in the presence of a field potential.
Such a theory should be treated separately, compared to 
BD theory with $\omega_{\mathrm{BD}} \neq -3/2$ in 
which the field kinetic term is present.

As we have derived the action~(\ref{BifR2}) from (\ref{BifR}),
the action in Palatini \fR\ gravity is equivalent to 
\begin{equation}
S=\int \mathrm{d}^{4}x \sqrt{-g}
\left[ \frac{1}{2\kappa^2} \varphi R(T)-U(\varphi) \right]
+\int \mathrm{d}^4 x {\cal L}_M (g_{\mu \nu}, \Psi_M)\,,
\label{BifRPa}
\end{equation}
where 
\begin{equation}
\varphi=f'(R(T))\,,\qquad
U=\frac{R(T)f'(R(T))-f(R(T))}{2\kappa^2}\,.
\label{Palare}
\end{equation}
Since the derivative of $U$ in terms of $\varphi$ is 
$U_{,\varphi}=R/(2\kappa^2)$, 
we obtain the following relation from Eq.~(\ref{Pala2}):
\begin{equation}
4U-2\varphi U_{,\varphi}=T\,.
\label{BifRPa3}
\end{equation}

Using the relation (\ref{Rrela}), the action~(\ref{BifRPa}) can 
be written as 
\begin{equation}
S=\int \mathrm{d}^{4}x \sqrt{-g}
\left[ \frac{1}{2\kappa^2} \varphi R(g)
+\frac{3}{4\kappa^2}\frac{1}{\varphi} (\nabla \varphi)^2
-U(\varphi) \right]+\int \mathrm{d}^4 x {\cal L}_M (g_{\mu \nu}, \Psi_M)\,.
\label{BifRPa2}
\end{equation}
Comparing this with Eq.~(\ref{BDac}) in the unit $\kappa^2=1$, 
we find that Palatini \fR\ gravity is equivalent to BD theory with the parameter 
$\omega_{\mathrm{BD}}=-3/2$~\cite{Flanagan,Olmo05,Sotirioueq}.
As we will see in Section~\ref{BDtheory}, this equivalence can be also seen 
by comparing Eqs.~(\ref{Pala1}) and (\ref{Pala3}) with those obtained
by varying the action~(\ref{BDac}) in BD theory. 
In the above discussion we have implicitly assumed that ${\cal L}_M$ 
does not explicitly depend on the Christoffel connections $\Gamma^{\lambda}_{\mu \nu}$. 
This is true for a scalar field or a perfect fluid, but it is not necessarily 
so for other matter Lagrangians such as those describing vector fields.

There is another way for taking the variation of the action, known
as the metric-affine formalism~\cite{Hehl,Liberati,Liberati2,Capome}.
In this formalism the matter action $S_M$ depends 
not only on the metric $g_{\mu \nu}$ but also on the connection 
$\Gamma^{\lambda}_{\mu \nu}$.
Since the connection is independent of the metric in this approach, 
one can define the quantity called hypermomentum~\cite{Hehl}, as
$\Delta_{\lambda}^{\mu \nu} \equiv (-2/\sqrt{-g})\delta {\cal L}_M
/\delta \Gamma^{\lambda}_{\mu \nu}$.
The usual assumption that the connection is symmetric is also 
dropped, so that the antisymmetric quantity called the
Cartan torsion tensor, $S_{\mu \nu}^{\lambda} \equiv \Gamma^{\lambda}_{[\mu \nu]}$, 
is defined. The non-vanishing property of $S_{\mu \nu}^{\lambda}$
allows the presence of torsion in this theory.
If the condition $\Delta_{\lambda}^{[\mu \nu]}=0$ holds, 
it follows that the Cartan torsion tensor vanishes
($S_{\mu \nu}^{\lambda}=0$)~\cite{Liberati}.
Hence the torsion is induced by matter fields with the anti-symmetric
hypermomentum. The \fR\ Palatini gravity
belongs to \fR\ theories in the metric-affine formalism with 
$\Delta_{\lambda}^{\mu \nu}=0$. In the following we 
do not discuss further \fR\ theory in the metric-affine formalism.
Readers who are interested in those theories may refer to 
the papers~\cite{Liberati2,SotFaraoni}.


\subsection{Background cosmological dynamics}

We discuss the background cosmological evolution of dark energy 
models based on Palatini \fR\ gravity.
We shall carry out general analysis without specifying
the forms of \fR.
We take into account non-relativistic matter and radiation whose energy 
densities are $\rho_m$ and $\rho_r$, respectively.
In the flat FLRW background (\ref{FLRW}) 
we obtain the following equations
\begin{eqnarray}
 FR-2f&=&-\kappa^{2}\rho_{m}\,,\label{Palaeq1} \\
  6F \left( H+\frac{\dot{F}}{2F} \right)^2-f&=&
 \kappa^2 (\rho_m+2\rho_r)\,,\label{Palaeq2}
\end{eqnarray}
together with the continuity equations, 
$\dot{\rho}_m+3H \rho_m=0$ and 
$\dot{\rho}_r+4H \rho_r=0$.
Combing Eqs.~(\ref{Palaeq1}) and (\ref{Palaeq2}) together with 
continuity equations, it follows that 
\begin{eqnarray}
\dot{R}&=&\frac{3\kappa^{2}H\rho_{m}}{F_{,R}R-F}
=-3H\frac{FR-2f}{F_{,R}R-F}\,,
\label{Pala4} \\
 H^{2}&=&\frac{2\kappa^{2}(\rho_{m}+\rho_{r})+FR-f}
{6F\xi}\,, \label{Pala5}
\end{eqnarray}
where 
\begin{equation}
\xi \equiv \left[1-\frac{3}{2}\frac{F_{,R}(FR-2f)}{F(F_{,R}R-F)}\right]^{2}\,.
\end{equation}

In order to discuss cosmological dynamics it is convenient to introduce
the dimensionless variables:
\begin{equation}
y_{1}\equiv\frac{FR-f}{6F\xi H^{2}}\,,\qquad 
y_{2}\equiv\frac{\kappa^{2}\rho_{r}}{3F\xi H^{2}}\,,
\label{y1y2}
\end{equation}
by which Eq.~(\ref{Pala5}) can be written as 
\begin{equation}
\frac{\kappa^2 \rho_m}{3F \xi H^2}=1-y_1-y_2\,.
\end{equation}
Differentiating $y_1$ and $y_2$ with respect to $N=\ln\,a$, 
we obtain~\cite{FayTavakol}
\begin{eqnarray}
 \frac{\mathrm{d}y_{1}}{\mathrm{d}N}&=&y_{1}\left[3-3y_{1}+y_{2}
 +C(R)(1-y_{1})\right]\,,\label{auto1}\\
 \frac{\mathrm{d}y_{2}}{\mathrm{d}N}&=&
 y_{2}\left[-1-3y_{1}+y_{2}-C(R)y_{1}\right]\,,
 \label{auto2}
\end{eqnarray}
where 
\begin{equation}
C(R)\equiv\frac{R\dot{F}}{H(FR-f)}=
-3\frac{(FR-2f)F_{,R}R}{(FR-f)(F_{,R}R-F)}\,.
\label{CR}
\end{equation}
The following constraint equation also holds
\begin{equation}
\frac{1-y_{1}-y_{2}}{2y_{1}}=-\frac{FR-2f}{FR-f}\,.
\label{y1y2re}
\end{equation}
Hence the Ricci scalar $R$ can be expressed in terms of $y_1$ and $y_2$.

Differentiating Eq.~(\ref{Pala4}) with respect to $t$, 
it follows that 
\begin{equation}
\frac{\dot{H}}{H^{2}}=-\frac32+\frac32y_{1}-\frac12 y_{2}
-\frac{\dot{F}}{2HF}-\frac{\dot{\xi}}{2H\xi}
+\frac{\dot{F}R}{12F\xi H^{3}}\,,
\end{equation}
from which we get the effective equation of state:
\begin{equation}
w_{\mathrm{eff}}=-1-\frac23 \frac{\dot{H}}{H^2}=
-y_{1}+\frac{1}{3}y_{2}+\frac{\dot{F}}{3HF}
+\frac{\dot{\xi}}{3H\xi}-\frac{\dot{F}R}{18F\xi H^{3}}\,.
\end{equation}
The cosmological dynamics is known by solving Eqs.~(\ref{auto1})
and (\ref{auto2}) with Eq.~(\ref{CR}). If $C(R)$ is not constant, 
then one can use Eq.~(\ref{y1y2re}) to express $R$ and $C(R)$
in terms of $y_1$ and $y_2$.

The fixed points of Eqs.~(\ref{auto1}) and (\ref{auto2}) can be 
found by setting $\mathrm{d}y_1/\mathrm{d}N=0$ and $\mathrm{d}y_2/\mathrm{d}N=0$.
Even when $C(R)$ is not constant, except for the cases $C(R)=-3$ and 
$C(R)=-4$, we obtain the following fixed points~\cite{FayTavakol}:
\begin{enumerate}
\item $P_r$: $(y_1, y_2)=(0, 1)$\,,
\item $P_m$: $(y_1, y_2)=(0, 0)$\,,
\item $P_d$: $(y_1, y_2)=(1, 0)$\,.
\end{enumerate}
The stability of the fixed points can be analyzed by considering 
linear perturbations about them.
As long as $\mathrm{d}C/\mathrm{d}y_1$ and $\mathrm{d}C/\mathrm{d}y_2$
are bounded, the eigenvalues $\lambda_1$ and $\lambda_2$ of 
the Jacobian matrix of linear perturbations are given by 
\begin{enumerate}
\item $P_r$: $(\lambda_1, \lambda_2)=(4+C(R), 1)$\,,
\item $P_m$: $(\lambda_1, \lambda_2)=(3+C(R), -1)$\,,
\item $P_d$: $(\lambda_1, \lambda_2)=(-3-C(R), -4-C(R))$\,.
\end{enumerate}

In the $\Lambda$CDM model ($f(R)=R-2\Lambda$) one has
$w_{\mathrm{eff}}=-y_1+y_2/3$ and $C(R)=0$. Then the points $P_r$,
$P_m$, and $P_d$ correspond to $w_{\mathrm{eff}}=1/3$, $(\lambda_1,
\lambda_2)=(4, 1)$ (radiation domination, unstable),
$w_{\mathrm{eff}}=0$, $(\lambda_1, \lambda_2)=(3, -1)$ (matter
domination, saddle), and $w_{\mathrm{eff}}=-1$, $(\lambda_1,
\lambda_2)=(-3, -4)$ (de~Sitter epoch, stable), respectively. Hence
the sequence of radiation, matter, and de~Sitter epochs is in fact
realized.

\epubtkImage{palatini.png}{%
  \begin{figure}[hptb]
    \centerline{\includegraphics[width=3.4in]{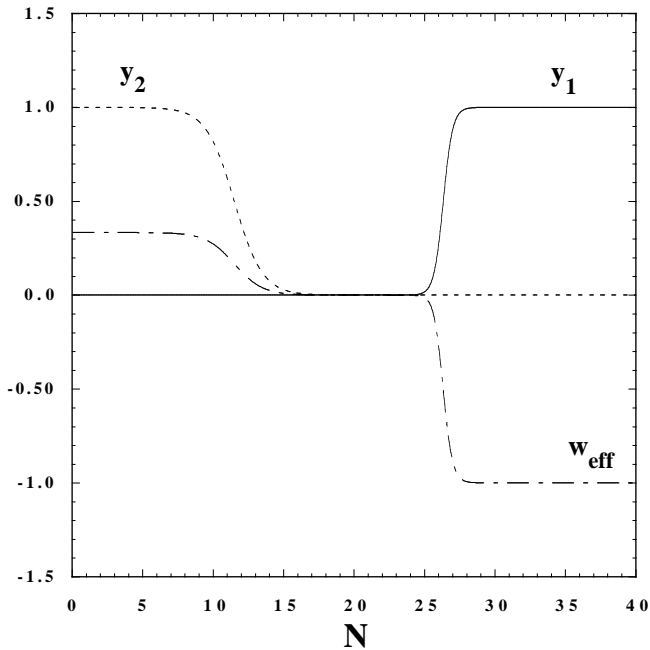}}
    \caption{The evolution of the variables $y_1$ and $y_2$ for the
      model $f(R)=R-\beta/R^n$ with $n=0.02$, together with the
      effective equation of state $w_{\mathrm{eff}}$. Initial
      conditions are chosen to be $y_1=10^{-40}$ and
      $y_{2}=1.0-10^{-5}$. From~\cite{FayTavakol}.}
    \label{palatini} 
\end{figure}}

Let us next consider the model $f(R)=R-\beta/R^{n}$ with $\beta>0$
and $n>-1$. In this case the quantity $C(R)$ is 
\begin{equation}
C(R)=3n\frac{R^{1+n}-(2+n)\beta}{R^{1+n}+n(2+n)\beta}\,.
\end{equation}
The constraint equation~(\ref{y1y2re}) gives 
\begin{equation}
\frac{\beta}{R^{1+n}}=\frac{2y_1}
{3y_{1}+n(y_{1}-y_{2}+1)-y_{2}+1}\,.
\end{equation}
The late-time de~Sitter point corresponds to $R^{1+n}=(2+n) \beta$, 
which exists for $n>-2$. Since $C(R)=0$ in this case, the de~Sitter 
point $P_d$ is stable with the eigenvalues 
$(\lambda_1, \lambda_2)=(-3, -4)$.
During the radiation and matter domination we have 
$\beta/R^{1+n} \ll 1$ (i.e., $f(R) \simeq R$) and hence $C(R)=3n$.
$P_r$ corresponds to the radiation point ($w_{\mathrm{eff}}=1/3$)
with the eigenvalues $(\lambda_1, \lambda_2)=(4+3n, 1)$, 
whereas $P_m$ to the matter point ($w_{\mathrm{eff}}=0$)
with the eigenvalues $(\lambda_1, \lambda_2)=(3+3n, -1)$.
Provided that $n>-1$, $P_r$ and $P_m$ correspond to 
unstable and saddle points respectively, in which case 
the sequence of radiation, matter, and de~Sitter eras 
can be realized.
For the models $f(R)=R+\alpha R^m-\beta/R^n$, 
it was shown in~\cite{FayTavakol}
that unified models of inflation and dark energy 
with radiation and matter eras  
are difficult to be realized.

In Figure~\ref{palatini} we plot the evolution of $w_{\mathrm{eff}}$
as well as $y_1$ and $y_2$ for the model 
$f(R)=R-\beta/R^n$ with $n=0.02$.
This shows that the sequence of ($P_r$) radiation domination
($w_{\mathrm{eff}}=1/3$), ($P_m$) matter domination ($w_{\mathrm{eff}}=0$), 
and de~Sitter acceleration ($w_{\mathrm{eff}}=-1$) is realized.
Recall that in metric \fR\ gravity the model 
$f(R)=R-\beta/R^n$ ($\beta>0$, $n>0$) is not viable because 
$f_{,RR}$ is negative. In Palatini \fR\ gravity
the sign of $f_{,RR}$ does not matter because there is no 
propagating degree of freedom with a mass $M$
associated with the second derivative $f_{,RR}$~\cite{Sotiriou:2006sf}.

In~\cite{Motapala, FayTavakol} the dark energy model $f(R)=R-\beta/R^n$ 
was constrained by the combined analysis of independent observational data.
From the joint analysis of Super-Nova Legacy Survey~\cite{SNLS}, BAO~\cite{BAO1}
and the CMB shift parameter~\cite{WMAP3},
the constraints on two parameters $n$ and $\beta$ 
are $n\in\left[-0.23,0.42\right]$ and 
$\beta\in\left[2.73,10.6\right]$ at the 95\% 
confidence level (in the unit of $H_0=1$)~\cite{FayTavakol}.
Since the allowed values of $n$ are close to 0, the 
above model is not particularly favored over the 
$\Lambda$CDM model.
See also~\cite{CapoPala,Santos,Santos08,Palaback1,Palaback2} for observational 
constraints on \fR\ dark energy models based on the Palatini formalism.


\subsection{Matter perturbations}
\label{matterperPala}

We have shown that \fR\ theory in the Palatini formalism
can give rise to the late-time cosmic acceleration 
preceded by radiation and matter eras.
In this section we study the evolution of matter density perturbations
to confront Palatini \fR\ gravity with the observations of 
large-scale 
structure~\cite{KoivistoPala,KoivistoPala2,Koivistopre,Uddin07,Lee08,TsujiUddin}.
Let us consider the perturbation $\delta \rho_m$ of non-relativistic matter 
with a homogeneous energy density $\rho_m$.
Koivisto and Kurki-Suonio~\cite{KoivistoPala} derived perturbation 
equations in Palatini \fR\ gravity.
Using the perturbed metric~(\ref{permetric}) with the same variables as 
those introduced in Section~\ref{persec}, the perturbation
equations are given by 
\begin{eqnarray}
\label{pa1}
 \frac{\Delta}{a^2}\psi&+&\left( H+\frac{\dot{F}}
{2F} \right)A+\frac{1}{2F}
\left(\frac{3\dot{F}^2}{2F}+3H\dot{F} \right)\alpha
\nonumber \\
&=&\frac{1}{2F} \left[
\left( 3H^2-\frac{3\dot{F}^2}{4F^2}-\frac{R}{2}
-\frac{\Delta}{a^2} \right) \delta F+
\left( \frac{3\dot{F}}{2F}+3H \right) \dot{\delta F}
-\kappa^2 \delta \rho_m \right]\,,\\
\label{pa2}
 H\alpha&-&\dot{\psi}=\frac{1}{2F}
\left[ \dot{\delta F}- \left(H+\frac{3\dot{F}}{2F}
\right) \delta F-\dot{F} \alpha+\kappa^2 \rho_m v
\right]\,, \\
\label{pa3}
 \dot{\chi}&+&H \chi-\alpha-\psi=
\frac{1}{F} ( \delta F-\dot{F}\chi)\,, \\
\label{pa4}
 \dot{A}&+&\left(2H+\frac{\dot{F}}{2F}\right)
A+\left( 3\dot{H}+\frac{3\ddot{F}}{F}+
\frac{3H\dot{F}}{2F}-\frac{3\dot{F}^2}{F^2}
+\frac{\Delta}{a^2} \right) \alpha+\frac32 \frac{\dot{F}}
{F}\dot{\alpha} \nonumber \\
&=&\frac{1}{2F} \left[ \kappa^2 \delta \rho_m+
\left(6H^2+6\dot{H}+\frac{3\dot{F}^2}{F^2}-R
-\frac{\Delta}{a^2}\right)\delta F+
\left( 3H-\frac{6\dot{F}}{F}
\right) \dot{\delta F}+3 \ddot{\delta F}
 \right], \\
\label{pa5}
 R\delta F&-&F \delta R=-\kappa^2\delta \rho_m\,,
\end{eqnarray}
where the Ricci scalar $R$ can be understood as $R(T)$.

From Eq.~(\ref{pa5}) the perturbation $\delta F$ can be 
expressed by the matter perturbation $\delta \rho_m$, as 
\begin{equation}
\delta F=\frac{F_{,R}}{R} \frac{\kappa^2 \delta \rho_m}{1-m}\,,
\end{equation}
where $m=RF_{,R}/F$.
This equation clearly shows that the perturbation $\delta F$
is sourced by the matter perturbation only, unlike 
metric \fR\ gravity
in which the oscillating mode of $\delta F$ is present.
The matter perturbation $\delta \rho_m$ and the velocity 
potential $v$ obey the same equations as given in 
Eqs.~(\ref{matter1}) and (\ref{matter2}), which results in 
Eq.~(\ref{delmeq}) in Fourier space.

Let us consider the perturbation equations in Fourier space.
We choose the Longitudinal gauge ($\chi=0$)
with $\alpha=\Phi$ and $\psi=-\Psi$. 
In this case Eq.~(\ref{pa3}) gives
\begin{equation}
\Psi-\Phi=\frac{\delta F}{F}\,.
\label{PsiPhi}
\end{equation}
Under the quasi-static approximation on sub-horizon scales used in 
Section~\ref{mattersec}, Eqs.~(\ref{pa1}) and (\ref{delmeq}) reduce to  
\begin{eqnarray}
\frac{k^2}{a^2}\Psi &\simeq& \frac{1}{2F} \left( \frac{k^2}{a^2}
\delta F-\kappa^2 \delta \rho_m \right)\,,\label{Psire} \\
\ddot{\delta}_m&+&2H \dot{\delta}_m+\frac{k^2}{a^2}\Phi 
\simeq 0\,.
\end{eqnarray}
Combining Eq.~(\ref{PsiPhi}) with Eq.~(\ref{Psire}), we obtain 
\begin{equation}
\frac{k^2}{a^2}\Psi=-\frac{\kappa^2 \delta \rho_m}{2F}
\left(1-\frac{\zeta}{1-m} \right)\,,\qquad
\frac{k^2}{a^2}\Phi=-\frac{\kappa^2 \delta \rho_m}{2F}
\left(1+\frac{\zeta}{1-m} \right)\,,
\end{equation}
where 
\begin{equation}
\zeta \equiv \frac{k^2}{a^2} \frac{F_{,R}}{F}
=\frac{k^2}{a^2 R}m\,.
\end{equation}
Then the matter perturbation satisfies the following 
Eq.~\cite{TsujiUddin}
\begin{equation}
\ddot{\delta}_m+2H \dot{\delta}_m-\frac{\kappa^2 \rho_m}
{2F} \left( 1+\frac{\zeta}{1-m} \right) \delta_m \simeq 0\,.
\label{delmeqPala}
\end{equation}
The effective gravitational potential defined 
in Eq.~(\ref{Phieffdef}) obeys
\begin{equation}
\Phi_{\mathrm{eff}} \simeq -\frac{\kappa^2 \rho_m}{2F}
\frac{a^2}{k^2} \delta_m\,.
\end{equation}
In the above approximation we do not need to worry about
the dominance of the oscillating mode of perturbations
in the past.
Note also that the same approximate equation of $\delta_m$
as Eq.~(\ref{delmeqPala}) can be derived for different 
gauge choices~\cite{TsujiUddin}.

The parameter $\zeta$ is a crucial quantity 
to characterize the evolution of perturbations.
This quantity can be estimated as $\zeta \approx (k/aH)^2m$, 
which is much larger than $m$ for sub-horizon modes ($k \gg aH$).
In the regime $\zeta \ll 1$ the matter perturbation evolves 
as $\delta_m \propto t^{2/3}$.
Meanwhile the evolution of $\delta_m$ in the regime $\zeta \gg 1$
is completely different from that in GR.
If the transition characterized by $\zeta=1$ occurs before today, 
this gives rise to the modification to the matter spectrum
compared to the GR case.

In the regime $\zeta \gg 1$, let us study the evolution of matter perturbations
during the matter dominance. We shall consider the case in which 
the parameter $m$ (with $|m| \ll 1)$ evolves as
\begin{equation}
m \propto t^p\,,
\end{equation}
where $p$ is a constant.
For the model $f(R)=R-\mu R_c (R/R_c)^{n}$ ($n<1$) the power $p$ corresponds to 
$p=1+n$, whereas for the models (\ref{Amodel}) and (\ref{Bmodel}) with $n>0$
one has $p=1+2n$. 
During the matter dominance the parameter $\zeta$ evolves as 
$\zeta=\pm (t/t_k)^{2p+2/3}$, where the subscript ``$k$'' denotes the value
at which the perturbation crosses $\zeta=\pm 1$.
Here $+$ and $-$ signs correspond to the cases $m>0$
and $m<0$, respectively.
Then the matter perturbation equation~(\ref{delmeqPala}) reduces to 
\begin{equation}
\frac{\mathrm{d}^2 \delta_m}{\mathrm{d}N^2}+
\frac12 \frac{\mathrm{d} \delta_m}{\mathrm{d}N}
-\frac32 \left[ 1\pm e^{(3p+1)(N-N_k)} \right]
\delta_m=0\,.
\label{delPa}
\end{equation}

When $m>0$, the growing mode solution to Eq.~(\ref{delPa})
is given by 
\begin{equation}
\delta_m \propto
\exp \left( \frac{\sqrt{6}
e^{(3p+1)(N-N_k)/2}}{3p+1}\right)\,,
\qquad
f_\delta \equiv \frac{\dot{\delta}_m}{H \delta_m}=
\frac{\sqrt{6}}{2} e^{(3p+1)(N-N_k)/2}\,.
\end{equation}
This shows that the perturbations exhibit violent growth for 
$p>-1/3$, which is not compatible with observations of 
large-scale structure. In metric \fR\ gravity the growth 
of matter perturbations is much milder.

When $m<0$, the perturbations show a damped oscillation:
\begin{equation}
\delta_m \propto
e^{-(3p+2)(N-N_k)/4}
\,\cos (x+\theta)\,,\qquad
f_{\delta}= -\frac14 (3p+2) -\frac{3p+1}{2}
x \tan (x+\theta)\,,
\end{equation}
where $x=\sqrt{6}e^{(3p+1)(N-N_k)/2}/(3p+1)$, and 
$\theta$ is a constant.
The averaged value of the growth rate $f_{\delta}$ is given by
$\bar{f_{\delta}}=-(3p+2)/4$, but it shows a divergence every time
$x$ changes by $\pi$.
These negative values of $f_{\delta}$ are also difficult to be 
compatible with observations.

The \fR\ models can be consistent with observations
of large-scale structure if the universe does not 
enter the regime $|\zeta|>1$ by today. 
This translates into the condition~\cite{TsujiUddin} 
\begin{equation}
\left| m(z=0) \right| \lesssim (a_0H_0/k)^2\,.
\label{mcon}
\end{equation}
Let us consider the wavenumbers 
$0.01\,h\mathrm{\ Mpc}^{-1} \lesssim k \lesssim 0.2\,h\mathrm{\ Mpc}^{-1}$
that corresponds to the  linear regime of the matter power spectrum.
Since the wavenumber $k=0.2\,h\mathrm{\ Mpc}^{-1}$ corresponds to 
$k \approx 600a_0H_0$ (where ``0'' represents present quantities), 
the condition (\ref{mcon}) gives the bound $\left|m(z=0) \right| 
\lesssim 3 \times 10^{-6}$.

If we use the observational constraint of the growth rate, 
$f_\delta \lesssim 1.5$~\cite{Mcdonald,Viel,DiPorto}, then the deviation 
parameter $m$ today is constrained to be $\left| m (z=0) \right| \lesssim
10^{-5}$-$10^{-4}$ for the model $f(R)=R-\lambda R_c (R/R_c)^{n}$
($n<1$) as well as for the models (\ref{Amodel}) and 
(\ref{Bmodel})~\cite{TsujiUddin}.
Recall that, in metric \fR\ gravity, the deviation 
parameter $m$ can grow to the order of 0.1 by today.
Meanwhile \fR\ dark energy models based on 
the Palatini formalism are hardly distinguishable from 
the $\Lambda$CDM model~\cite{KoivistoPala2,LiPala0,LiPala,TsujiUddin}.
Note that the bound on $m(z=0)$ becomes even severer by 
considering perturbations in non-linear regime.
The above peculiar evolution of matter perturbations is associated
with the fact that the coupling between non-relativistic matter 
and a scalar-field degree of freedom is very strong (as 
we will see in Section~\ref{BDtheory}).

The above results are based on the fact that dark matter 
is described by a cold and perfect fluid with no pressure.
In~\cite{Koivistogene} it was suggested that 
the tight bound on the parameter $m$ can be relaxed 
by considering imperfect dark matter with a shear stress.
Although the approach taken in~\cite{Koivistogene} did not 
aim to explain the origin of a dark matter stress $\Pi$ that cancels
the $k$-dependent term in Eq.~(\ref{delmeqPala}), 
it will be of interest to further study whether some theoretically 
motivated choice of $\Pi$ really allows the possibility that 
Palatini \fR\ dark energy models can be distinguished
from the $\Lambda$CDM model.


\subsection{Shortcomings of Palatini \fR\ gravity}
\label{palaprosec}

In addition to the fact that Palatini \fR\ dark energy models
are hardly distinguished from the $\Lambda$CDM model from 
observations of large-scale structure, there are a number of 
problems in Palatini \fR\ gravity associated with non-dynamical 
nature of the scalar-field degree of freedom.

The dark energy model $f=R-\mu^4/R$ based on the 
Palatini formalism was shown to be
in conflict with the Standard 
Model of particle physics~\cite{Flanagan0,Flanagan,Flanagan2,Kaloper,Barausse1}
because of large non-perturbative corrections to the matter Lagrangian 
[here we use $R$ for the meaning of $R(T)$].
Let us consider this issue for a more general model $f=R-\mu^{2(n+1)}/R^{n}$.
From the definition of $\varphi$ in Eq.~(\ref{Palare}) the field potential $U(\varphi)$
is given by 
\begin{equation}
U(\varphi)=\frac{n+1}{2n^{n/(n+1)}}\frac{\mu^2}{\kappa^2}
(\varphi-1)^{n/(n+1)}\,,
\end{equation}
where $\varphi=1+n \mu^{2(n+1)}R^{-n-1}$.
Using Eq.~(\ref{BifRPa3}) for the vacuum ($T=0$), we obtain 
the solution 
\begin{equation}
\varphi (T=0)=\frac{2(n+1)}{n+2}\,.
\end{equation}

In the presence of matter we expand the field $\varphi$ as 
$\varphi=\varphi (T=0)+\delta \varphi$.
Substituting this into Eq.~(\ref{BifRPa3}), we obtain 
\begin{equation}
\delta \varphi \simeq \frac{n}{(n+2)^{\frac{n+2}{n+1}}}
\frac{\kappa^2 T}{\mu^2}\,.
\label{delvarphi}
\end{equation}
For $n={\cal O}(1)$ we have  
$\delta \varphi \approx \kappa^2 T/\mu^2=T/(\mu^2 M_{\mathrm{pl}}^2)$
with $\varphi(T=0) \approx 1$.
Let us consider a matter action of a Higgs scalar field $\phi$ with 
mass $m_{\phi}$:
\begin{equation}
S_M=\int \mathrm{d}^4x \sqrt{-g}
\left[ -\frac12 g^{\mu \nu} \partial_{\mu}\phi
\partial_{\nu}\phi-\frac12 m_\phi^2 \phi^2 \right]\,.
\end{equation}
Since $T \approx m_\phi^2 \delta \phi^2$ it follows that 
$\delta \varphi \approx m_\phi^2 \delta \phi^2/(\mu^2 M_{\mathrm{pl}}^2)$.
Perturbing the Jordan-frame action~(\ref{BifRPa2}) [which is equivalent to 
the action in Palatini \fR\ gravity] to second-order
and using the solution $\varphi \approx 1
+m_{\phi}^2\delta \phi^2/(\mu^2 M_{\mathrm{pl}}^2)$, 
we find that the effective action of the Higgs field
$\phi$ for an energy scale $E$  much lower than 
$m_{\phi}$ (=~100\,--\,1000~GeV) is given by~\cite{Barausse1}
\begin{equation}
\delta S_M \simeq \int \mathrm{d}^4x \sqrt{-g}
\left[ -\frac12 g^{\mu \nu} \partial_{\mu}\delta\phi
\partial_{\nu}\delta \phi-\frac12 m_\phi^2 \delta\phi^2 \right]
\left( 1+\frac{m_\phi^2 \delta \phi^2}
{\mu^2 M_{\mathrm{pl}}^2}+\cdots \right)\,.
\end{equation}
Since $\delta \phi \approx m_{\phi}$ for $E \ll m_{\phi}$, 
the correction term can be estimated as 
\begin{equation}
\delta \varphi \approx \frac{m_\phi^2 \delta \phi^2}
{\mu^2 M_{\mathrm{pl}}^2} \approx 
\left( \frac{m_\phi}{\mu} \right)^2 
\left( \frac{m_\phi}{M_{\mathrm{pl}}} \right)^2\,.
\end{equation}
In order to give rise to the late-time acceleration we require that 
$\mu \approx H_0 \approx 10^{-42}\mathrm{\ GeV}$.
For the Higgs mass $m_{\phi}=100\mathrm{\ GeV}$ it follows that 
$\delta \varphi \approx 10^{56} \gg 1$.
This correction is too large to be compatible with the 
Standard Model of particle physics.

The above result is based on the models 
$f(R)=R-\mu^{2(n+1)}/R^n$ with $n={\cal O}(1)$.
Having a look at Eq.~(\ref{delvarphi}), the only way to make the 
perturbation $\delta \varphi$ small is to choose $n$ 
very close to 0. This means that the deviation from the 
$\Lambda$CDM model is extremely small (see~\cite{Li2008}
for a related work).
In fact, this property was already found by the analysis of 
matter density perturbations in Section \ref{matterperPala}.
While the above analysis is based on the calculation in 
the Jordan frame in which test particles follow 
geodesics~\cite{Barausse1}, the same result was also 
obtained by the analysis in the Einstein 
frame~\cite{Flanagan0,Flanagan,Flanagan2,Kaloper}.

Another unusual property of Palatini \fR\ gravity is that 
a singularity with the divergent Ricci scalar
can appear at the surface of a static 
spherically symmetric star with a polytropic equation of 
state $P=c\rho_0^{\Gamma}$ with $3/2<\Gamma<2$
(where $P$ is the pressure and $\rho_0$ is the 
rest-mass density)~\cite{Barausse2,Barausse1}
(see also~\cite{Bustelo,Kainu}).
Again this problem is intimately related with the particular 
algebraic dependence (\ref{Pala2}) in Palatini \fR\ gravity. 
In~\cite{Barausse2} it was claimed that the appearance of 
the singularity does not very much depend on the functional forms 
of \fR\ and that the result is not specific to the choice 
of the polytropic equation of state.

The Palatini gravity has a close relation with 
an effective action which reproduces the dynamics of 
loop quantum cosmology~\cite{Singh}. 
~\cite{Olmore} showed that the model $f(R)=R+R^2/(6M^2)$, 
where $M$ is of the order of the Planck mass, is not plagued by 
a singularity problem mentioned above, while the singularity 
typically arises for the \fR\ models constructed to explain 
the late-time cosmic acceleration
(see also~\cite{Rei09} for a related work).
Since Planck-scale corrected Palatini \fR\ models may cure 
the singularity problem, it will be of interest to understand
the connection with quantum gravity around the cosmological 
singularity (or the black hole singularity).
In fact, it was shown in~\cite{Barra09} that non-singular bouncing solutions
can be obtained for power-law \fR\ Lagrangians
with a finite number of terms.

Finally we note that the extension of Palatini \fR\ gravity to 
more general theories including Ricci and Riemann tensors
was carried out in~\cite{LiPalatini,Shaw08,Borunda,Exiri,Li2008,Eran,GOlmo09}.
While such theories are more involved than Palatini \fR\ gravity,
it may be possible to construct viable modified gravity 
models of inflation or dark energy.

\newpage

\section{Extension to Brans--Dicke Theory}
\label{BDsec}
\setcounter{equation}{0}

So far we have discussed \fR\ gravity theories in the metric and 
Palatini formalisms.
In this section we will see that these theories 
are equivalent to Brans--Dicke (BD) theory~\cite{BD} in the presence of 
a scalar-field potential, by comparing field equations in \fR\ theories 
with those in BD theory.
It is possible to construct viable dark energy models based on 
BD theory with a constant parameter $\omega_{\mathrm{BD}}$.
We will discuss cosmological dynamics, local gravity constraints, and 
observational signatures of such generalized theory.


\subsection{Brans--Dicke theory and the equivalence with \fR\ theories}
\label{BDtheory}

Let us start with the following 4-dimensional action in BD theory
\begin{equation}
S=\int \mathrm{d}^{4}x\sqrt{-g}\left[\frac{1}{2}\vp R
-\frac{\omega_{\mathrm{BD}}}{2\vp}(\nabla\vp)^{2}
-U(\vp)\right]+S_{M}
(g_{\mu\nu},\Psi_{M})\,,
\label{BDaction}
\end{equation}
where $\omega_{\mathrm{BD}}$ is the BD parameter, 
$U(\vp)$ is a potential of the scalar field $\vp$, 
and $S_{M}$ is a matter action that depends on the 
metric $g_{\mu\nu}$ and matter fields $\Psi_{M}$. 
In this section we use the unit $\kappa^2=8\pi G=1/M_{\mathrm{pl}}^2=1$, 
but we recover the gravitational constant $G$ and the reduced 
Planck mass $M_{\mathrm{pl}}$ when the discussion becomes transparent.
The original BD theory~\cite{BD} does not 
possess the field potential $U(\vp)$.

Taking the variation of the action~(\ref{BDaction}) with respect to 
$g_{\mu \nu}$ and $\vp$, we obtain the following field equations
\begin{eqnarray}
R_{\mu\nu}(g)-\frac{1}{2}g_{\mu\nu}R(g)&=& \frac{1}{\vp}T_{\mu\nu}
-\frac{1}{\vp}g_{\mu\nu}U(\vp)+\frac{1}{\vp}(\nabla_{\mu}\nabla_{\nu}\vp
-g_{\mu\nu}\square\vp)\nonumber \\
&& +\frac{\omega_{\mathrm{BD}}}{\vp^{2}}
 \left[\partial_{\mu}\vp\partial_{\nu}\vp-\frac{1}{2}g_{\mu\nu}(\nabla \vp)^{2}\right]\,,\label{BDeq1}\\
(3+2\omega_{\mathrm{BD}})\Box\vp+4U(\vp)-2\vp U_{,\vp}&=& T\,,
\label{BDeq2}
\end{eqnarray}
where $R(g)$ is the Ricci scalar in metric \fR\ gravity, and 
$T_{\mu \nu}$ is the energy-momentum tensor of matter.
In order to find the relation with \fR\ theories in the metric and 
Palatini formalisms, we consider the following correspondence
\begin{equation}
\vp=F(R)\,,\qquad U(\vp)=\frac{RF-f}{2}\,.
\end{equation}
Recall that this potential (which is the gravitational origin) already 
appeared in Eq.~(\ref{Udef}).
We then find that Eqs.~(\ref{fREin}) and (\ref{trace}) 
in metric \fR\ gravity are equivalent 
to Eqs.~(\ref{BDeq1}) and (\ref{BDeq2})
with the BD parameter $\omega_{\mathrm{BD}}=0$.
Hence \fR\ theory in the metric formalism corresponds to 
BD theory with $\omega_{\mathrm{BD}}=0$~\cite{ohanlon,teyssandier,Chiba,Faraonieq,Capone}.
In fact we already showed this by rewriting the action~(\ref{fRaction})
in the form (\ref{BifR2}).
We also notice that Eqs.~(\ref{Pala3}) and (\ref{Pala2}) 
in Palatini \fR\ gravity are equivalent to Eqs.~(\ref{fREin}) 
and (\ref{trace}) with the BD parameter 
$\omega_{\mathrm{BD}}=-3/2$.
Then \fR\ theory in the Palatini formalism corresponds to BD theory 
with $\omega_{\mathrm{BD}}=-3/2$~\cite{Flanagan,Olmo05,Sotirioueq}.
Recall that we also showed this by rewriting the action~(\ref{fRaction})
in the form (\ref{BifRPa2}).
 
One can consider more general theories called scalar-tensor 
theories~\cite{FujiiMaeda} in which the Ricci scalar $R$ is 
coupled to a scalar field $\vp$.
The general 4-dimensional action for scalar-tensor theories 
can be written as
\begin{equation}
S=\int \mathrm{d}^{4}x\sqrt{-g}\left[\frac{1}{2}F(\vp)R
-\frac{1}{2}\omega(\vp)(\nabla\vp)^{2}-U(\vp) \right]
+S_{M}(g_{\mu\nu},\Psi_{M})\,,
\label{stensoraction}
\end{equation}
where $F(\vp)$ and $U(\vp)$ are functions of $\vp$. 
Under the conformal transformation $\tilde{g}_{\mu \nu}=Fg_{\mu \nu}$, 
we obtain the action in the Einstein frame~\cite{Maeda,Wands94}
\begin{equation}
S_{E}=\int \mathrm{d}^{4}x\sqrt{-\tilde{g}}\left[\frac{1}{2}\tilde{R}
-\frac{1}{2}(\tilde{\nabla}\phi)^{2}-V(\phi)\right]+
S_{M}(F^{-1}\tilde{g}_{\mu\nu},\Psi_{M})\,,
\label{SEframe}
\end{equation}
where $V=U/F^{2}$. We have introduced a new scalar field $\phi$ 
to make the kinetic term canonical:
\begin{equation}
\phi \equiv \int \mathrm{d}\vp\,\sqrt{\frac{3}{2}
\left(\frac{F_{,\vp}}{F}\right)^{2}
+\frac{\omega}{F}}\,.
\label{phire}
\end{equation}

We define a quantity $Q$ that characterizes the coupling 
between the field $\phi$ and non-relativistic matter 
in the Einstein frame:
\begin{equation}
Q\equiv-\frac{F_{,\phi}}{2F}=-\frac{F_{,\vp}}{F}
\left[\frac{3}{2}\left(\frac{F_{,\vp}}{F}\right)^{2}+
\frac{\omega}{F}\right]^{-1/2}\,.
\label{Q}
\end{equation}
Recall that, in metric \fR\ gravity,
we introduced the same quantity $Q$ in Eq.~(\ref{Qdef}), 
which is constant ($Q=-1/\sqrt{6}$). 
For theories with $Q=$constant, we obtain the following relations from 
Eqs.~(\ref{phire}) and (\ref{Q}):
\begin{equation}
F=e^{-2Q\phi}\,,\qquad
\omega=(1-6Q^{2})F\left(\frac{\mathrm{d}\phi}
{\mathrm{d}\vp}\right)^{2}\,.
\label{confactor}
\end{equation}
In this case the action~(\ref{stensoraction}) in the Jordan frame 
reduces to~\cite{TUMTY} 
\begin{equation}
S=\int \mathrm{d}^{4}x\sqrt{-g}\Bigg[\frac{1}{2}F(\phi)R-
\frac{1}{2}(1-6Q^{2})F(\phi)(\nabla\phi)^{2}-U(\phi)\Bigg]
+S_{M}(g_{\mu\nu},\Psi_{M})\,,
\quad \mathrm{with} \quad F(\phi)=e^{-2Q\phi}\,.
\label{action2}
\end{equation}
In the limit that $Q \to 0$ we have $F(\phi) \to 1$, so that
Eq.~(\ref{action2}) recovers the action of a minimally 
coupled scalar field in GR.

Let us compare the action~(\ref{action2}) with the action 
(\ref{BDaction}) in BD theory.
Setting $\varphi=F=e^{-2Q\phi}$, the former is
equivalent to the latter if the parameter $\omega_{\mathrm{BD}}$ 
is related to $Q$ via the relation~\cite{chame2,TUMTY}
\begin{equation}
3+2\omega_{\mathrm{BD}}=\frac{1}{2Q^{2}}\,.
\label{BD}
\end{equation}
This shows that the GR limit ($\omega_{\mathrm{BD}}\to\infty$)
corresponds to the vanishing coupling ($Q \to 0$).
Since $Q=-1/\sqrt{6}$ in metric \fR\ gravity one has
$\omega_{\mathrm{BD}}=0$, as expected.
The Palatini \fR\ gravity corresponds to $\omega_{\mathrm{BD}}=-3/2$, 
which corresponds to the infinite coupling ($Q^2 \to \infty$). 
In fact, Palatini gravity can be regarded as an isolated
``fixed point'' of a transformation involving a special 
conformal rescaling of the metric~\cite{Palatinifixed}.
In the Einstein frame of the Palatini formalism, the scalar field $\phi$ does not 
have a kinetic term and it can be integrated out. In general, 
this leads to a matter action which is non-linear, depending 
on the potential $U(\phi)$.
This large coupling poses a number of problems such as
the strong amplification of matter density perturbations and 
the conflict with the Standard Model of particle physics,
as we have discussed in Section \ref{Palasec}.

Note that BD theory is one of the examples in scalar-tensor theories
and there are some theories that give rise to non-constant values of $Q$.
For example, the action of a nonminimally coupled scalar field with a coupling $\xi$
corresponds to $F(\vp)=1-\xi \vp^2$ and $\omega (\vp)=1$, which
gives the field-dependent coupling 
$Q(\vp)=\xi \vp/[1-\xi \vp^2 (1-6\xi)]^{1/2}$.
In fact the dynamics of dark energy in such a theory has been 
studied by a number of authors~\cite{coupledpre,Uzan,Chiba99,Bartolo,Mata,Bacci,Riazuelo}.
In the following we shall focus on the constant coupling models 
with the action~(\ref{action2}).
We stress that this is equivalent to the action~(\ref{BDaction})
in BD theory.


\subsection{Cosmological dynamics of dark energy models based on Brans--Dicke theory}

The first attempt to apply BD theory to cosmic acceleration is the 
extended inflation scenario in which the BD field $\varphi$
is identified as an inflaton field~\cite{LaStein,SteinAceetta}.
The first version of the inflation model, which considered a
first-order phase transition in BD theory, 
resulted in failure due to the graceful exit problem~\cite{La2,Weinberg89,BarrowMaeda}.
This triggered further study of the possibility of realizing inflation 
in the presence of another scalar field~\cite{Linde90,Berkin}. 
In general the dynamics of such a multi-field system is more involved
than that in the single-field case~\cite{Bassett}. 
The resulting power spectrum of density perturbations 
generated during multi-field inflation in 
BD theory was studied by a number of 
authors~\cite{StaYoko,Garcia,CNY,StaYoko2}.

In the context of dark energy it is possible to construct 
viable single-field models based on BD theory.
In what follows we discuss cosmological dynamics of dark energy models 
based on the action~(\ref{action2}) in the flat FLRW background
given by (\ref{FLRW}) (see, e.g., \cite{TUMTY,coupledpre,Billyard,Gunzig,Agarwal,Jarv,Leach,Cooper} 
for dynamical analysis in scalar-tensor theories).
Our interest is to find conditions under which a sequence of radiation, 
matter, and accelerated epochs can be realized.
This depends upon the form of the field potential $U(\phi)$.
We first carry out general analysis without specifying the forms 
of the potential. We take into account non-relativistic matter with energy density 
$\rho_{m}$ and radiation with energy density $\rho_{r}$.
The Jordan frame is regarded as a physical frame due to the
usual conservation of non-relativistic matter ($\rho_{m}\propto a^{-3}$).
Varying the action~(\ref{action2}) with respect to $g_{\mu\nu}$ 
and $\phi$, we obtain the following equations
\begin{eqnarray}
 3FH^{2}&=&(1-6Q^{2})F\dot{\phi}^{2}/2+U-3H\dot{F}+\rho_{m}+\rho_{r}\,,\label{scabe1}\\
 2F\dot{H}&=&-(1-6Q^{2})F\dot{\phi}^{2}-\ddot{F}+H\dot{F}-\rho_{m}- 4\rho_{r}/3\,,\label{scabe2}\\
 (1-6Q^{2})&F&\left[ \ddot{\phi}+3H\dot{\phi}+\dot{F}/(2F)
 \dot{\phi} \right]+U_{,\phi}+QFR=0\,,\label{scabe3}
\end{eqnarray}
where $F=e^{-2Q\phi}$.

We introduce the following dimensionless variables 
\begin{equation}
x_{1}\equiv\frac{\dot{\phi}}{\sqrt{6}H}\,,\qquad 
x_{2}\equiv\frac{1}{H}\sqrt{\frac{U}{3F}}\,,\qquad 
x_{3}\equiv\frac{1}{H}\sqrt{\frac{\rho_{r}}{3F}}\,,
\end{equation}
and also the density parameters
\begin{equation}
\Omega_{m}\equiv\frac{\rho_{m}}{3FH^{2}}\,,\qquad\Omega_{r}\equiv x_{3}^{2}\,,
\qquad\Omega_{\mathrm{DE}}\equiv(1-6Q^{2})x_{1}^{2}+x_{2}^{2}+2\sqrt{6}Qx_{1}\,.
\end{equation}
These satisfy the relation $\Omega_{m}+\Omega_{r}+\Omega_{\mathrm{DE}}=1$
from Eq.~(\ref{scabe1}).
From Eq.~(\ref{scabe2}) it follows that 
\begin{equation}
\frac{\dot{H}}{H^{2}}=-\frac{1-6Q^{2}}{2}\left(3+3x_{1}^{2}-3x_{2}^{2}+x_{3}^{2}-6Q^{2}x_{1}^{2}
+2\sqrt{6}Qx_{1}\right)
+3Q(\lambda x_{2}^{2}-4Q)\,.
\label{dotHsca}
\end{equation}
Taking the derivatives of $x_{1}$, $x_{2}$ and $x_{3}$
with respect to $N=\ln\,a$, we find
\begin{eqnarray}
\frac{\mathrm{d}x_{1}}{\mathrm{d}N} &=& \frac{\sqrt{6}}{2}(\lambda x_{2}^{2}-\sqrt{6}x_{1}) 
\nonumber  \\
&& +\frac{\sqrt{6}Q}{2}\left[(5-6Q^{2})x_{1}^{2}+2\sqrt{6}Qx_{1}-3x_{2}^{2}+x_{3}^{2}-1\right]
-x_{1}\frac{\dot{H}}{H^{2}}\,,\label{scaau1}\\
\frac{\mathrm{d}x_{2}}{\mathrm{d}N} &=& \frac{\sqrt{6}}{2}(2Q-\lambda)x_{1}x_{2}-x_{2}\frac{\dot{H}}{H^{2}}\,,\label{scaau2}\\
\frac{\mathrm{d}x_{3}}{\mathrm{d}N} &=& \sqrt{6}Qx_{1}x_{3}-2x_{3}-x_{3}\frac{\dot{H}}{H^{2}}\,,\label{scaau3}
\end{eqnarray}
where $\lambda\equiv-U_{,\phi}/U$.

\begin{table}[t]
  \caption[The critical points of dark energy models.]{The critical
    points of dark energy models based on the action~(\ref{action2})
    in BD theory with constant $\lambda=-U_{,\phi}/U$ in the absence
    of radiation ($x_3=0$). The effective equation of state
    $w_{\mathrm{eff}}=-1-2\dot{H}/(3H^2)$ is known from
    Eq.~(\ref{dotHsca}).}
  \label{tablecri}
  \centering
  \begin{tabular}{lcccc}
    \toprule
    Name &  $x_1$ & $x_2$ & $\Omega_m$ & $w_{\mathrm{eff}}$  \\
    \midrule
    (a) $\phi$MDE & $\frac{\sqrt{6}Q}{3(2Q^{2}-1)}$ & 0 & $\frac{3-2Q^{2}}{3(1-2Q^{2})^{2}}$ 
    & $\frac{4Q^{2}}{3(1-2Q^{2})}$ \\
    \\
    (b1) Kinetic 1& $\frac{1}{\sqrt{6}Q+1}$ & 0 & 0 & $\frac{3-\sqrt{6}Q}{3(1+\sqrt{6}Q)}$
    \\
    \\
    (b2) Kinetic 2 & $\frac{1}{\sqrt{6}Q-1}$ & 0 & 0 & $\frac{3+\sqrt{6}Q}{3(1-\sqrt{6}Q)}$
    \\
    \\
    (c) Field dominated & $\frac{\sqrt{6}(4Q-\lambda)}{6(4Q^{2}-Q\lambda-1)}$  &
    $\left[\frac{6-\lambda^{2}+8Q\lambda-16Q^{2}}{6(4Q^{2}-Q\lambda-1)^{2}}\right]^{1/2}$
    & 0 & $-\frac{20Q^{2}-9Q\lambda-3+\lambda^{2}}{3(4Q^{2}-Q\lambda-1)}$\\
    \\
    (d) Scaling solution & $\frac{\sqrt{6}}{2\lambda}$  &
    $\sqrt{\frac{3+2Q\lambda-6Q^{2}}{2\lambda^{2}}}$
    & $1-\frac{3-12Q^{2}+7Q\lambda}{\lambda^{2}}$ & $-\frac{2Q}{\lambda}$\\
    \\
    (e) de~Sitter & 0  & 1 & 0 & $-1$\\
    \bottomrule
  \end{tabular}
\end{table}

If $\lambda$ is a constant, i.e., for the exponential potential 
$U=U_0 e^{-\lambda \phi}$, one can derive fixed points
for Eqs.~(\ref{scaau1})\,--\,(\ref{scaau3})
by setting $\mathrm{d}x_i/\mathrm{d}N=0$ ($i=1,2,3$).
In Table~\ref{tablecri} we list the fixed points of the system 
in the absence of radiation ($x_3=0$).
Note that the radiation point corresponds to 
$(x_1, x_2, x_3)=(0,0,1)$.
The point (a) is the $\phi$-matter-dominated epoch ($\phi$MDE)
during which the density of non-relativistic matter is a non-zero constant. 
Provided that $Q^2 \ll 1$ this can be used for the matter-dominated 
epoch. The kinetic points (b1) and (b2) are responsible
neither for the matter era nor for the accelerated epoch (for $|Q| \lesssim 1$).
The point (c) is the scalar-field dominated solution, which can be 
used for the late-time acceleration for $w_{\mathrm{eff}}<-1/3$.
When $Q^2 \ll 1$ this point yields the cosmic acceleration 
for $-\sqrt{ 2}+4Q<\lambda<\sqrt{2}+4Q$.
The scaling solution (d) can be responsible for the matter era
for $|Q| \ll |\lambda|$, but in this case the condition 
$w_{\mathrm{eff}}<-1/3$ for the
point (c) leads to $\lambda^{2} \lesssim 2$. 
Then the energy fraction of the pressureless matter for the point 
(d) does not satisfy the condition $\Omega_{m}\simeq1$. 
The point (e) gives rise to the de~Sitter expansion, which exists
for the special case with $\lambda=4Q$
[which can be also regarded as the special case of the point (c)].
From the above discussion the viable cosmological trajectory 
for constant $\lambda$ is the sequence from the point (a) to 
the scalar-field dominated point (c)
under the conditions $Q^{2}\ll1$ and $-\sqrt{2}+4Q<\lambda<\sqrt{2}+4Q$.
The analysis based on the Einstein frame action~(\ref{SEframe}) 
also gives rise to the $\phi$MDE followed by the 
scalar-field dominated solution~\cite{coupled,coupledpre}.

Let us consider the case of non-constant $\lambda$.
The fixed points derived above may be regarded as ``instantaneous'' 
points\epubtkFootnote{In strict mathematical sense the instantaneous
fixed point is not formally defined because it varies with time.}~\cite{Macorra,Ng} 
varying with the time-scale smaller than $H^{-1}$.
As in metric \fR\ gravity ($Q=-1/\sqrt{6}$) we are interested 
in large coupling models with $|Q|$ of the order of unity.
In order for the potential $U(\phi)$ to satisfy local gravity constraints, 
the field needs to be heavy 
in the region $R \gg R_0 \sim H_0^2$ 
such that $|\lambda| \gg 1$. Then it is possible to realize the matter era
by the point (d) with $|Q|\ll|\lambda|$.
Moreover the solutions can finally approach the de~Sitter solution (e) 
with $\lambda=4Q$ or the field-dominated solution (c). 
The stability of the point (e) was analyzed in 
~\cite{TUMTY,Faraonista2,Faraonista3}
by considering linear perturbations
$\delta x_1$, $\delta x_2$ and $\delta F$.
One can easily show that the point (e) is stable for 
\begin{equation}
Q\frac{\mathrm{d}\lambda}{\mathrm{d}F}(F_{1}) > 0 \quad \to \quad
\frac{\mathrm{d}\lambda}{\mathrm{d}\phi}(\phi_{1})<0\,,
\label{stadesitter}
\end{equation}
where $F_1=e^{-2Q \phi_1}$ with $\phi_1$ being the field
value at the de~Sitter point.
In metric \fR\ gravity ($Q=-1/\sqrt{6}$) this 
condition is equivalent to $m=Rf_{,RR}/f_{,R}<1$.

For the \fR\ model~(\ref{fRasy}) the field $\phi$ is related to 
the Ricci scalar $R$ via the relation 
$e^{2\phi/\sqrt{6}}=1-2n \mu (R/R_c)^{-(2n+1)}$. 
Then the potential $U=(FR-f)/2$
in the Jordan frame can be expressed as
\begin{equation}
U(\phi)=\frac{\mu R_c}{2} \left[ 1-\frac{2n+1}
{(2n \mu)^{2n/(2n+1)}} \left( 1-e^{2\phi/\sqrt{6}}
\right)^{2n/(2n+1)} \right]\,.
\label{UphifR}
\end{equation}
For theories with general couplings $Q$ we consider the 
following potential~\cite{TUMTY}
\begin{equation}
U(\phi)=U_{0}\left[1-C(1-e^{-2Q\phi})^{p}\right]
\qquad (U_{0}>0,~C>0,~0<p<1)\,,
\label{modelscalar}
\end{equation}
which includes the potential (\ref{UphifR}) in \fR\ gravity
as a specific case with the correspondence
$U_0=\mu R_c/2$ and $C=(2n+1)/(2n \mu)^{2n/(2n+1)}$, 
$Q=-1/\sqrt{6}$, and $p=2n/(2n+1)$.
The potential behaves as $U(\phi) \to U_0$ for $\phi \to 0$ and 
$U(\phi)\to U_{0}(1-C)$ in the limits $\phi\to\infty$ 
(for $Q>0$) and $\phi\to-\infty$ (for $Q<0$).
This potential has a curvature singularity at $\phi=0$ as in 
the models (\ref{Amodel}) and (\ref{Bmodel}) of \fR\ gravity, 
but the appearance of the singularity can be avoided by 
extending the potential to the regions $\phi>0$ ($Q<0$) 
or $\phi<0$ ($Q>0$) with a field mass bounded from above.
The slope $\lambda=-U_{,\phi}/U$ is given by 
\begin{equation}
\lambda=\frac{2Cp\, Qe^{-2Q\phi}(1-e^{-2Q\phi})^{p-1}}{1-C(1-e^{-2Q\phi})^{p}}\,.
\label{lambdascalar}
\end{equation}

During the radiation and deep matter eras one has 
$R=6(2H^2+\dot{H}) \simeq\rho_{m}/F$ from Eqs.~(\ref{scabe1})\,--\,(\ref{scabe2})
by noting that $U_{0}$ is negligibly small relative to the background fluid density. 
From Eq.~(\ref{scabe3}) the field is nearly frozen at a value 
satisfying the condition $U_{,\phi}+Q\rho_{m}\simeq0$. Then 
the field $\phi$ evolves along the instantaneous minima given by
\begin{equation}
\phi_{m}\simeq\frac{1}{2Q}\left(\frac{2U_{0}pC}{\rho_{m}}
\right)^{1/(1-p)}\,.
\label{phim}
\end{equation}
As long as $\rho_{m}\gg2U_{0}pC$ we have
that $|\phi_{m}|\ll1$.
In this regime the slope $\lambda$ in Eq.~(\ref{lambdascalar})
is much larger than 1.
The field value $|\phi_{m}|$ increases for decreasing $\rho_{m}$
and hence the slope $\lambda$ decreases with time.

Since $\lambda \gg 1$ around $\phi=0$, the instantaneous fixed point (d) can
be responsible for the matter-dominated epoch provided that $|Q|\ll \lambda$.
The variable $F=e^{-2Q\phi}$ decreases in time irrespective of the
sign of the coupling $Q$ and hence $0<F<1$. 
The de~Sitter point is characterized by $\lambda=4Q$, i.e., 
\begin{equation}
C=\frac{2(1-F_{1})^{1-p}}{2+(p-2)F_{1}}\,.
\label{Cdef}
\end{equation}
The de~Sitter solution is present as long as the solution of 
this equation exists in the region $0<F_{1}<1$.
From Eq.~(\ref{lambdascalar}) the derivative of $\lambda$
in terms of $\phi$ is given by 
\begin{equation}
\frac{\mathrm{d}\lambda}{\mathrm{d}\phi}=-\frac{4CpQ^{2}F(1-F)^{p-2}
[1-pF-C(1-F)^{p}]}{[1-C(1-F)^{p}]^{2}}\,.
\label{dlamphi}
\end{equation}
When $0<C<1$, we can show that the function 
$g(F) \equiv 1-pF-C(1-F)^{p}$ is positive and hence 
the condition $\mathrm{d}\lambda/\mathrm{d}\phi<0$ is satisfied.
This means that the de~Sitter point (e) is a stable attractor.
When $C>1$, the function $g(F)$ can be negative.
Plugging Eq.~(\ref{Cdef}) into Eq.~(\ref{dlamphi}), we find that 
the de~Sitter point is stable for 
\begin{equation}
F_1>\frac{1}{2-p}\,.
\label{F1con}
\end{equation}
If this condition is violated, the solutions choose 
another stable fixed point  [such as the point (c)] as an attractor.

The above discussion shows that for the model~(\ref{modelscalar}) the matter point (d) 
can be followed by the stable de~Sitter solution (e) for $0<C<1$.
In fact numerical simulations in~\cite{TUMTY} show that the
sequence of radiation, matter and de~Sitter epochs can be in fact realized.
Introducing the energy density $\rho_{\mathrm{DE}}$
and the pressure $P_{\mathrm{DE}}$ of dark energy as we have done for metric 
\fR\ gravity, the dark energy equation of state $w_{\mathrm{DE}}=P_{\mathrm{DE}}/\rho_{\mathrm{DE}}$
is given by the same form as Eq.~(\ref{wDEfR}).
Since for the model~(\ref{modelscalar}) $F$ increases toward the past, 
the phantom equation of state ($w_{\mathrm{DE}}<-1$) as well as the cosmological constant
boundary crossing ($w_{\mathrm{DE}}=-1$) occurs 
as in the case of metric \fR\ gravity~\cite{TUMTY}.

As we will see in Section \ref{lgcmaless}, for a light scalar field, 
it is possible to satisfy local gravity constraints for $|Q| \lesssim 10^{-3}$.
In those cases the potential does not need to be steep such 
that $\lambda \gg 1$ in the region $R \gg R_0$.
The cosmological dynamics for such nearly flat potentials 
have been discussed by a number of authors in several classes of 
scalar-tensor theories~\cite{Peri1,Nesseris2,Martin,Gan06}.
It is also possible to realize the condition $w_{\mathrm{DE}}<-1$, while
avoiding the appearance of a ghost~\cite{Martin,Gan06}.


\subsection{Local gravity constraints}
\label{lgcmaless}

We study local gravity constraints (LGC) for BD theory 
given by the action~(\ref{action2}).
In the absence of the potential $U(\phi)$ the BD parameter 
$\omega_{\mathrm{BD}}$ is constrained to be $\omega_{\mathrm{BD}}>4 \times 10^4$
from solar-system experiments~\cite{Will1,Bertotti,Will2}.
This bound also applies to the case of a nearly massless 
field with the potential $U(\phi)$ in which the Yukawa correction 
$e^{-Mr}$ is close to unity (where $M$ is a scalar-field mass 
and $r$ is an interaction length).
Using the bound $\omega_{\mathrm{BD}}>4 \times 10^4$
in Eq.~(\ref{BD}), we find that 
\begin{equation}
|Q|<2.5 \times 10^{-3}\,.
\end{equation}
This is a strong constraint under which the
cosmological evolution for such theories is difficult
to be distinguished from the $\Lambda$CDM model ($Q=0$).

If the field potential is present, the models with large couplings
($|Q|={\cal O}(1)$) can be consistent with local gravity constraints
as long as the mass $M$ of the field $\phi$ is sufficiently large in the
region of high density. For example, the potential (\ref{modelscalar})
is designed to have a large mass in the high-density region so that
it can be compatible with experimental tests for the violation of
equivalence principle through the chameleon mechanism~\cite{TUMTY}.
In the following we study conditions under which 
local gravity constraints can be satisfied 
for the model~(\ref{modelscalar}). 

As in the case of metric \fR\ gravity, let us consider a configuration in which 
a spherically symmetric body has a constant density $\rho_{A}$ 
inside the body with a constant density $\rho=\rho_{B}~(\ll\rho_{A})$
outside the body.
For the potential $V=U/F^{2}$ in the Einstein frame one has
$V_{,\phi}\simeq-2U_{0}QpC(2Q\phi)^{p-1}$ under the 
condition $|Q\phi| \ll 1$. Then the field
values at the potential minima inside and outside the body are 
\begin{equation}
\phi_{i}\simeq\frac{1}{2Q}\left(\frac{2U_{0}\, p\, C}{\rho_{i}}\right)^{1/(1-p)}\,,
\qquad i=A, B\,.
\label{phiAB}
\end{equation}
The field mass squared $m_{i}^{2}\equiv V_{,\phi\phi}$
at $\phi=\phi_{i}$ ($i=A, B$) is approximately given by 
\begin{equation}
m_{i}^{2}\simeq\frac{1-p}{(2^{p}\, pC)^{1/(1-p)}}Q^{2}
\left(\frac{\rho_{i}}{U_{0}}\right)^{(2-p)/(1-p)}U_{0}\,.
\label{Mphi}
\end{equation}

Recall that $U_{0}$ is roughly the same order as the present 
cosmological density $\rho_{0} \simeq 10^{-29}\mathrm{\ g/cm}^{3}$.
The baryonic/dark matter density in our galaxy corresponds to 
$\rho_{B}\simeq10^{-24}\mathrm{\ g/cm}^{3}$.
The mean density of Sun or Earth is about $\rho_A={\cal O}(1)\mathrm{\ g/cm}^{3}$.
Hence $m_A$ and $m_B$ are in general much larger than $H_{0}$
for local gravity experiments in our environment.
For $m_A \tilde{r}_c \gg 1$ the chameleon mechanism we discussed 
in Section~\ref{chameleonsec} can be directly applied to BD theory 
whose Einstein frame action is given by Eq.~(\ref{SEframe})
with $F=e^{-2Q\phi}$.

The bound (\ref{boep2}) coming from the possible violation
of equivalence principle in the solar system translates into 
\begin{equation}
\left(2U_{0}pC/\rho_{B}\right)^{1/(1-p)}<7.4\times10^{-15}\,|Q|\,.
\label{delrcon2}
\end{equation}
Let us consider the case in which the solutions finally approach 
the de~Sitter point (e) in Table \ref{tablecri}.
At this de~Sitter point we have $3F_{1}H_{1}^{2}=U_{0}[1-C(1-F_{1})^{p}]$
with $C$ given in Eq.~(\ref{Cdef}). 
Then the following relation holds
\begin{equation}
U_{0}=3H_{1}^{2}\left[2+(p-2)F_{1}\right]/p\,.
\end{equation}
Substituting this into Eq.~(\ref{delrcon2}) we obtain
\begin{equation}
\left(R_{1}/\rho_{B}\right)^{1/(1-p)}(1-F_{1})<7.4\times10^{-15}|Q|\,,
\end{equation}
where $R_{1}=12H_{1}^{2}$ is the Ricci scalar at the de~Sitter point.
Since $(1-F_{1})$ is smaller than 1/2 from Eq.~(\ref{F1con}), 
it follows that $(R_{1}/\rho_{B})^{1/(1-p)}<1.5\times10^{-14}|Q|$.
Using the values $R_{1}=10^{-29}\mathrm{\ g/cm}^{3}$ and
$\rho_{B}=10^{-24}\mathrm{\ g/cm}^{3}$, we get the bound for
$p$~\cite{TUMTY}:
\begin{equation}
p>1-\frac{5}{13.8-\log_{10}\,|Q|}\,.
\label{EPsca}
\end{equation}
When $|Q|=10^{-1}$ and $|Q|=1$ we have $p>0.66$ and $p>0.64$,
respectively. Hence the model can be compatible with local gravity experiments
even for $|Q|={\cal O}(1)$.


\subsection{Evolution of matter density perturbations}

Let us next study the evolution of perturbations in non-relativistic 
matter for the action~(\ref{action2}) with the potential $U(\phi)$ and 
the coupling $F(\phi)=e^{-2Q\phi}$. 
As in metric \fR\ gravity, the matter perturbation $\delta_{m}$ 
satisfies Eq.~(\ref{perdark4}) in the Longitudinal gauge.
We define the field mass squared as $M^2 \equiv U_{,\phi \phi}$.
For the potential consistent with local gravity constraints [such as 
(\ref{modelscalar})], the mass $M$ is much larger than the present 
Hubble parameter $H_0$ during the radiation and deep matter eras.
Note that the condition $M^2 \gg R$ is satisfied 
in most of the cosmological epoch as in the case of metric \fR\ gravity.

The perturbation equations for the action~(\ref{action2}) are given in 
Eqs.~(\ref{per1})\,--\,(\ref{per7}) with $f=F(\phi)R$, $\omega=(1-6Q^2)F$, 
and $V=U$. We use the unit $\kappa^2=1$, 
but we restore $\kappa^2$ when it is necessary.
In the Longitudinal gauge one has $\chi=0$, $\alpha=\Phi$, $\psi=-\Psi$, 
and $A=3(H\Phi+\dot{\Psi})$ in these equations.
Since we are interested in sub-horizon modes, we use the approximation 
that the terms containing $k^2/a^2$, $\delta \rho_m$, $\delta R$, and 
$M^2$ are the dominant contributions in Eqs.~(\ref{per1})\,--\,(\ref{per8}).
We shall neglect the contribution of the time-derivative terms 
of $\delta \phi$ in Eq.~(\ref{per5d}). As we have discussed for metric 
\fR\ gravity in Section~\ref{mattersec}, this amounts to neglecting 
the oscillating mode of perturbations. 
The initial conditions of the field perturbation in the radiation 
era need to be chosen so that the oscillating mode $\delta {\phi}_{\mathrm{osc}}$ 
is smaller than the matter-induced mode $\delta {\phi}_{\mathrm{ind}}$.
In Fourier space Eq.~(\ref{per5d}) gives
\begin{equation}
\left( \frac{k^2}{a^2}+\frac{M^2}{\omega} \right) 
\delta \phi_{\mathrm{ind}} \simeq \frac{1}{2\omega}F_{,\phi} \delta R\,.
\end{equation}
Using this relation together with Eqs.~(\ref{per3}) and (\ref{per7}), 
it follows that 
\begin{equation}
\delta \phi_{\mathrm{ind}} \simeq \frac{2QF}{(k^{2}/a^{2})(1-2Q^{2})F+M^{2}}
\frac{k^{2}}{a^{2}}\Psi\,.
\label{delphieq}
\end{equation}
Combing this equation with Eqs.~(\ref{per1}) and (\ref{per3}), 
we obtain~\cite{TUMTY,Song10} (see also~\cite{Berts,Zhao09,ZhaoPRL})
\begin{eqnarray}
\frac{k^{2}}{a^{2}}\Psi&\simeq&-\frac{\kappa^2\delta\rho_{m}}{2F}
\frac{(k^{2}/a^{2})(1-2Q^{2})F+M^{2}}{(k^{2}/a^{2})F+M^{2}}\,,
\label{PsiPhisca0}
\\
\frac{k^{2}}{a^{2}}\Phi&\simeq&-\frac{\kappa^2\delta\rho_{m}}{2F}
\frac{(k^{2}/a^{2})(1+2Q^{2})F+M^{2}}{(k^{2}/a^{2})F+M^{2}}\,,
\label{PsiPhisca}
\end{eqnarray}
where we have recovered $\kappa^2$.
Defining the effective gravitational potential $\Phi_{\mathrm{eff}}=(\Phi+\Psi)/2$, 
we find that $\Phi_{\mathrm{eff}}$ satisfies the same form of equation as (\ref{Phieff}).

Substituting Eq.~(\ref{PsiPhisca}) into Eq.~(\ref{perdark4}), 
we obtain the equation of matter perturbations 
on sub-horizon scales [with the neglect of the r.h.s.\ of Eq.~(\ref{perdark4})] 
\begin{equation}
\ddot{\delta}_m+2H\dot{\delta}_m-
4\pi G_{\mathrm{eff}}\rho_{m}\delta_{m} \simeq 0\,,
\label{mattereqsca2a}
\end{equation}
where the effective gravitational coupling is 
\begin{equation}
G_{\mathrm{eff}}=\frac{G}{F}\frac{(k^{2}/a^{2})(1+2Q^{2})F
+M^{2}}{(k^{2}/a^{2})F+M^{2}}\,.
\label{eq:jfg}
\end{equation}
In the regime $M^2/F \gg k^2/a^2$ (``GR regime'') this reduces to 
$G_{\mathrm{eff}}=G/F$, so that the evolution of $\delta_m$ and $\Phi_{\mathrm{eff}}$
during the matter domination ($\Omega_m=\rho_m/(3FH^2) \simeq 1$)
is standard: $\delta_m \propto t^{2/3}$ and $\Phi_{\mathrm{eff}} \propto \mathrm{constant}$.

In the regime $M^2/F \ll k^2/a^2$ (``scalar-tensor regime'') we have
\begin{equation}
G_{\mathrm{eff}}\simeq\frac{G}{F}(1+2Q^{2})=\frac{G}{F}
\frac{4+2\omega_{\mathrm{BD}}}{3+2\omega_{\mathrm{BD}}}\,,
\label{Geffmassless}
\end{equation}
where we used the relation (\ref{BD}) between the coupling $Q$ 
and the BD parameter $\omega_{\mathrm{BD}}$.
Since $\omega_{\mathrm{BD}}=0$ in \fR\ gravity, it follows 
that $G_{\mathrm{eff}}=4G/(3F)$.
Note that the result~(\ref{Geffmassless}) agrees with the effective 
Newtonian gravitational coupling between two test 
masses~\cite{Boi00,Damour}.
The evolution of $\delta_m$ and $\Phi_{\mathrm{eff}}$ 
during the matter dominance 
in the regime $M^2/F \ll k^2/a^2$ is   
\begin{equation}
\delta_{m}\propto t^{(\sqrt{25+48Q^{2}}-1)/6}\,,\qquad
\Phi_{\mathrm{eff}} \propto t^{(\sqrt{25+48Q^{2}}-5)/6}\,.
\label{persol2}
\end{equation}
Hence the growth rate of $\delta_m$ for $Q \neq 0$
is larger than that for $Q=0$.

As an example, let us consider the potential (\ref{modelscalar}). 
During the matter era the field mass squared around 
the potential minimum (induced by the matter coupling) 
is approximately given by 
\begin{equation}
M^2 \simeq \frac{1-p}{(2^p pC)^{1/(1-p)}} Q^2
\left( \frac{\rho_m}{U_0} \right)^{(2-p)/(1-p)}U_0\,,
\end{equation}
which decreases with time.
The perturbations cross the point $M^{2}/F=k^{2}/a^{2}$
at time $t=t_{k}$, which depends on the wavenumber $k$.
Since the evolution of the mass during the matter domination 
is given by $M\propto t^{-\frac{2-p}{1-p}}$, 
the time $t_{k}$ has a scale-dependence: $t_{k}\propto k^{-\frac{3(1-p)}{4-p}}$.
More precisely the critical redshift $z_{k}$ at time $t_{k}$ can be
estimated as~\cite{TUMTY}
\begin{equation}
z_{k}\simeq\left[\left(\frac{k}{a_{0}H_{0}}\frac{1}{|Q|}\right)^{2(1-p)}
\frac{2^{p}pC}{(1-p)^{1-p}}\frac{1}{(3F_{0}\Omega_{m}^{(0)})^{2-p}}
\frac{U_{0}}{H_{0}^{2}}\right]^{\frac{1}{4-p}}-1\,,
\label{zk2}
\end{equation}
where the subscript ``0'' represents present quantities.
For the scales $30a_{0}H_{0} \lesssim k \lesssim 600a_{0}H_{0}$, 
which correspond to the linear regime of the matter power spectrum, 
the critical redshift can be in the region $z_{k}>1$.
Note that, for larger $p$,  $z_{k}$ decreases.

When $t<t_k$ and $t>t_k$ the matter perturbation evolves as 
$\delta_m \propto t^{2/3}$ and 
$\delta_{m}\propto t^{(\sqrt{25+48Q^{2}}-1)/6}$, respectively
(apart from the epoch of the late-time cosmic acceleration).
The matter power spectrum $P_{\delta_{m}}$
at time $t=t_{\Lambda}$ (at which $\ddot{a}=0$) shows a difference
compared to the $\Lambda$CDM model, which is given by 
\begin{equation}
\frac{P_{\delta_{m}}(t_{\Lambda})}{P_{\delta_{m}}^{\Lambda\mathrm{CDM}}(t_{\Lambda})}=\left(\frac{t_{\Lambda}}{t_{k}}\right)^{2\left(\frac{\sqrt{25+48Q^{2}}-1}{6}-\frac{2}{3}\right)}\propto
k^{\frac{(1-p)(\sqrt{25+48Q^{2}}-5)}{4-p}}\,.
\end{equation}

The CMB power spectrum is also modified by the non-standard evolution of
the effective gravitational potential $\Phi_{\mathrm{eff}}$ for $t>t_k$.
This mainly affects the low multipoles of CMB anisotropies 
through of the ISW effect.
Hence there is a difference between the spectral indices 
of the matter power spectrum and of the CMB spectrum 
on the scales ($0.01\,h\mathrm{\ Mpc}^{-1} \lesssim k \lesssim
0.2\,h\mathrm{\ Mpc}^{-1}$)~\cite{TUMTY}:
\begin{equation}
\Delta n_s(t_{\Lambda})=\frac{(1-p)(\sqrt{25+48Q^{2}}-5)}{4-p}\,.
\label{deln}
\end{equation}
Note that this covers the result (\ref{DelnfR}) in \fR\ gravity 
($Q=-1/\sqrt{6}$ and $p=2n/(2n+1)$) as a special case.
Under the criterion $\Delta n_s(t_{\Lambda})<0.05$
we obtain the bounds $p>0.957$
for $Q=1$ and $p>0.855$ for $Q=0.5$. 
As long as $p$ is close to 1, the model can be consistent with both 
cosmological and local gravity constraints.
The allowed region coming from the bounds $\Delta n_s(t_{\Lambda})<0.05$
and (\ref{EPsca}) are illustrated in Figure~\ref{constraintfig}.

\epubtkImage{constraint.png}{%
  \begin{figure}[hptb]
    \centerline{\includegraphics[width=3.6in]{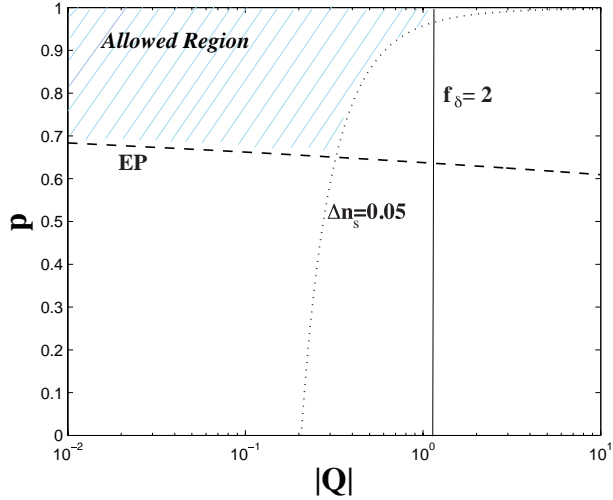}}
    \caption{The allowed region of the parameter space in the $(Q, p)$
      plane for BD theory with the potential (\ref{modelscalar}). We
      show the allowed region coming from the bounds $\Delta
      n_s(t_{\Lambda})<0.05$ and  $f_\delta<2$ as well as the the
      equivalence principle (EP) constraint (\ref{EPsca}).}
    \label{constraintfig} 
\end{figure}}

The growth rate of $\delta_m$ for $t>t_k$ is given by 
$f_\delta=\dot{\delta}_m/(H\delta_m)=(\sqrt{25+48Q^{2}}-1)/4$.
As we mentioned in Section~\ref{cosmodark}, 
the observational bound on $f_\delta$ is still weak in current observations.
If we use the criterion $f_\delta<2$ for the analytic estimation 
$f_\delta=(\sqrt{25+48Q^{2}}-1)/4$, we obtain the bound $Q<1.08$
(see Figure~\ref{constraintfig}).
The current observational data on the growth rate $f_\delta$ as well as its growth 
index $\gamma$ is not enough to place tight bounds on $Q$ and $p$, 
but this will be improved in future observations.

\newpage

\section{Relativistic Stars in \fR\ Gravity and Chameleon Theories}
\label{starsec}
\setcounter{equation}{0}

In Section~\ref{lgcsec} we discussed the existence of thin-shell 
solutions in metric \fR\ gravity in the Minkowski background, i.e., 
without the backreaction of metric perturbations. 
For the \fR\ dark energy models (\ref{Amodel}) and (\ref{Bmodel}), 
Frolov~\cite{Frolov} anticipated that the curvature singularity at $\phi=0$ 
(shown in Figure~\ref{pofigfR}) can be accessed in a strong 
gravitational background such as neutron stars.
Kobayashi and Maeda~\cite{KM08,KM09} studied spherically symmetric 
solutions for a constant density star with a vacuum exterior and
claimed the difficulty of obtaining thin-shell solutions
in the presence of the backreaction of metric perturbations.
In~\cite{TTT09} thin-shell solutions were derived analytically in 
the Einstein frame of BD theory (including \fR\ gravity)
under the linear expansion of the gravitational potential $\Phi_c$
at the surface of the body (valid for $\Phi_c<0.3$). 
In fact, the existence of such 
solutions was numerically confirmed for the inverse power-law
potential $V(\phi)=M^{4+n}\phi^{-n}$~\cite{TTT09}.

For the \fR\ models (\ref{Amodel}) and (\ref{Bmodel}), it was 
numerically shown that thin-shell solutions exist for $\Phi_c \lesssim 0.3$
by the analysis in the Jordan frame~\cite{Babi1,Upadhye,Babi2}
(see also~\cite{Cooney}).
In particular Babichev and Langlois~\cite{Babi1,Babi2} constructed static relativistic 
stars both for constant energy density configurations and for a polytropic
equation of state, provided that the pressure does not exceed one third of 
the energy density. 
Since the relativistic pressure tends to be stronger 
around the center of the spherically symmetric body for larger $\Phi_c$, 
the boundary conditions at the center of the body need to be carefully 
chosen to obtain thin-shell solutions numerically.
In this sense the analytic estimation of thin-shell solutions carried 
out in~\cite{TTT09} can be useful to show the existence of 
static star configurations, although such analytic solutions have 
been so far derived only for a constant density star.

In the following we shall discuss spherically symmetric solutions 
in a strong gravitational background with $\Phi_c \lesssim 0.3$
for BD theory with the action~(\ref{action2}).
This analysis covers metric \fR\ gravity as a special case 
(the scalar-field degree of freedom $\phi$ defined 
in Eq.~(\ref{kappaphi}) with $Q=-1/\sqrt{6}$).
While field equations will be derived in the Einstein frame, we can 
transform back to the Jordan frame to find the 
corresponding equations 
(as in the analysis of Babichev and Langlois~\cite{Babi2}).
In addition to the papers mentioned above, there are also 
a number of works about spherically symmetric solutions
for some equation state of matter~\cite{Kainu07,Kainu08,Multa06,Multa07,Multa08,Seifert}.


\subsection{Field equations}

We already showed that under the conformal transformation 
$\tilde{g}_{\mu \nu}=e^{-2Q \kappa \phi} g_{\mu \nu}$ the action 
(\ref{action2}) is transformed to the Einstein frame action:
\begin{equation}
\label{actionstar}
S_E=\int \mathrm{d}^4 x\sqrt{-\tilde{g}} 
\left[ \frac{1}{2\kappa^2} \tilde{R}
-\frac12 (\tilde{\nabla} \phi)^2-V(\phi) \right]
+\int \mathrm{d}^4x\,{\cal L}_M 
(e^{2Q \kappa \phi} \tilde{g}_{\mu \nu},\Psi_M)\,.
\end{equation}
Recall that in the Einstein frame this gives rise to a constant coupling $Q$
between non-relativistic matter and the field $\phi$.
We use the unit $\kappa^2=8\pi G=1$, but we restore the gravitational 
constant $G$ when it is required.

Let us consider a spherically symmetric static metric in the Einstein frame:
\begin{equation}
\label{smetric}
\rd \tilde{s}^2=-e^{2\Psi (\tr)} \rd t^2+e^{2\Phi (\tr)} \rd \tr^2+
\tr^2 \left( \rd \theta^2+\sin^2 \theta \rd \phi^2 \right)\,,
\end{equation}
where $\Psi (\tr)$ and $\Phi (\tr)$ are functions of the
distance $\tr$ from the center of symmetry.
For the action~(\ref{actionstar}) the energy-momentum 
tensors for the scalar field $\phi$ and the matter are given,
respectively, by 
\begin{eqnarray}
\tilde{T}_{\mu \nu}^{(\phi)} &=&
\partial_{\mu} \phi \partial_{\nu} \phi
-\tilde{g}_{\mu \nu}
\left[ \frac12 \tilde{g}^{\alpha \beta} 
\partial_{\alpha} \phi
\partial_{\beta} \phi +V(\phi) \right]\,, \\
\tilde{T}_{\mu \nu}^{(M)} &=& -\frac{2}{\sqrt{-\tilde{g}}}
\frac{\delta {\cal L}_M}{\delta \tilde{g}^{\mu \nu}}\,.
\end{eqnarray}

For the metric~(\ref{smetric}) the $(00)$ and $(11)$ 
components for the energy-momentum tensor of the field are 
\begin{equation}
\tilde{T}^{0(\phi)}_0=-\frac12 e^{-2\Phi} \phi'^2
-V(\phi)\,,\qquad
\tilde{T}^{1(\phi)}_1 = \frac12 e^{-2\Phi} \phi'^2
-V(\phi)\,,
\end{equation}
where a prime represents a derivative with respect to $\tr$.
The energy-momentum tensor of matter in the Einstein frame is given by 
$\tilde{T}^{\mu}_{\nu}= \mathrm{diag}\,(-\tilde{\rho}_M,\tilde{P}_M,\tilde{P}_M,\tilde{P}_M)$, 
which is related to $T^{\mu (M)}_{\nu}$ in the Jordan frame via 
$\tilde{T}^{\mu(M)}_{\nu}=e^{4Q\phi}\,T^{\mu (M)}_{\nu}$.
Hence it follows that $\tilde{\rho}_M=e^{4Q \phi}\rho_M$ and
$\tilde{P}_M=e^{4Q \phi}P_M$.

Variation of the action~(\ref{actionstar}) with respect to $\phi$ gives
\begin{equation}
-\partial_{\mu} \left( \frac{\partial (\sqrt{-\tilde{g}}{\cal L}_\phi)}
{\partial (\partial_{\mu}\phi)} \right)+\frac{\partial (\sqrt{-\tilde{g}}{\cal L}_\phi)}
{\partial \phi}+\frac{\partial {\cal L}_M}{\partial \phi}=0\,,
\end{equation}
where ${\cal L}_{\phi}=-(\tilde{\nabla}\phi)^2/2-V(\phi)$ is the field Lagrangian 
density. Since the derivative of ${\cal L}_M$ in terms of $\phi$ 
is given by Eq.~(\ref{parL}), i.e., $\partial {\cal L}_M/\partial \phi
=\sqrt{-\tilde{g}}Q (-\tilde{\rho}_M+3\tilde{P}_M)$, we obtain the 
equation of the field $\phi$~\cite{TTT09,Babi2}: 
\begin{equation}
\label{be1}
\phi''+\left( \frac{2}{\tr}+\Psi'-\Phi' \right) \phi'=
e^{2\Phi} \left[ V_{,\phi}+Q(\tilde{\rho}_M-3 \tilde{P}_M) \right]\,,
\end{equation}
where a tilde represents a derivative with respect to $\tr$.
From the Einstein equations it follows that 
\begin{eqnarray}
\label{be2}
\Phi' &=& \frac{1-e^{2\Phi}}{2\tr}+4\pi G \tr 
\left[ \frac12 \phi'^{2}+e^{2\Phi}V(\phi)
+e^{2\Phi}\tilde{\rho}_M \right]\,, \\
\label{be3}
\Psi' &=& \frac{e^{2\Phi}-1}{2 \tr}+4\pi G \tr
\left[ \frac12 \phi'^{2}-e^{2\Phi}V(\phi)
+e^{2\Phi} \tilde{P}_M \right]\,, \\
\label{be4}
\Psi''&+&\Psi'^{2}-\Psi' \Phi'+\frac{\Psi'-\Phi'}{\tr}
= -8\pi G \left[ \frac12 \phi'^{2}+
e^{2\Phi}V(\phi)-e^{2\Phi} \tilde{P}_M \right]\,.
\end{eqnarray}
Using the continuity equation $\nabla_{\mu} T^{\mu}_1=0$
in the Jordan frame, we obtain
\begin{equation}
\label{be5}
\tilde{P}_M'+(\tilde{\rho}_M+\tilde{P}_M)\Psi'
+Q\phi' (\tilde{\rho}_M-3\tilde{P}_M)=0\,.
\end{equation}
In the absence of the coupling $Q$ this reduces to 
the Tolman--Oppenheimer--Volkoff equation, 
$\tilde{P}_M'+(\tilde{\rho}_M+\tilde{P}_M)\Psi'=0$.

If the field potential $V(\phi)$ is responsible for dark energy, 
we can neglect both $V(\phi)$ and $\phi'^2$  
relative to $\tilde{\rho}_M$ in the local region
whose density is much larger than the cosmological density
($\rho_0 \sim 10^{-29}\mathrm{\ g/cm}^3$).
In this case Eq.~(\ref{be2}) is integrated to give
\begin{equation}
\label{Phim}
e^{2\Phi (\tr)}=\left[ 1-\frac{2Gm(\tr)}{\tr} \right]^{-1}\,,\quad
m(\tr)=\int_0^{\tr} 4\pi \bar{r}^2 \tilde{\rho}_M\,\mathrm{d} \bar{r}\,.
\end{equation}
Substituting Eqs.~(\ref{be2}) and (\ref{be3}) into
Eq.~(\ref{be1}), it follows that 
\begin{equation}
\label{fieldeq}
\phi''+\left[ \frac{1+e^{2\Phi}}{\tr}-4\pi G \tr e^{2\Phi}
(\tilde{\rho}_M-\tilde{P}_M) \right] \phi'= e^{2\Phi}
\left[ V_{,\phi} +Q (\tilde{\rho}_M-3\tilde{P}_M) \right]\,.
\end{equation}

The gravitational potential $\Phi$ around the surface of a compact 
object can be estimated as $\Phi \approx G \tilde{\rho}_M \tr_c^2$, where 
$\tilde{\rho}_M$ is the mean density of the star and $\tilde{r}_c$ is its radius.
Provided that $\Phi \ll 1$, Eq.~(\ref{fieldeq}) reduces to 
Eq.~(\ref{dreq}) in the Minkowski background 
(note that the pressure $\tilde{P}_M$ is also much smaller than 
the density $\tilde{\rho}_M$ for non-relativistic matter).


\subsection{Constant density star}
\label{constarsec}

Let us consider a constant density star with 
$\tilde{\rho}_M=\tilde{\rho}_A$.
We also assume that the density outside the star is constant, 
$\tilde{\rho}_M=\tilde{\rho}_B$.
We caution that the conserved density $\tilde{\rho}_M^{(c)}$
in the Einstein frame is given by 
$\tilde{\rho}_M^{(c)}=e^{-Q \phi} \tilde{\rho}_M$~\cite{chame2}.
However, since the condition $Q\phi \ll 1$ holds 
in most cases of our interest, we do not distinguish between 
$\tilde{\rho}_M^{(c)}$ and $\tilde{\rho}_M$ 
in the following discussion.

Inside the spherically symmetric body ($0<\tr<\tr_c$), 
Eq.~(\ref{Phim}) gives
\begin{equation}
\label{ePhi}
e^{2\Phi (\tr)}=\left( 1-\frac{8\pi G}{3} 
\tilde{\rho}_A \tr^2 \right)^{-1}\,.
\end{equation}
Neglecting the field contributions in 
Eqs.~(\ref{be2})\,--\,(\ref{be5}), the gravitational background 
for $0<\tr<\tr_c$ is characterized by the Schwarzschild 
interior solution.
Then the pressure $\tilde{P}_M(\tr)$ inside the body relative to 
the density $\tilde{\rho}_A$ can be analytically expressed as
\begin{equation}
\label{pm}
\frac{\tilde{P}_M (\tr)}{\tilde{\rho}_A}=
\frac{\sqrt{1-2(\tr^2/\tr_c^2)\Phi_c}-\sqrt{1-2\Phi_c}}
{3\sqrt{1-2\Phi_c}-\sqrt{1-2(\tr^2/\tr_c^2)\Phi_c }}
\quad \quad (0<\tr<\tr_c)\,,
\end{equation}
where $\Phi_c$ is the gravitational potential at the
surface of the body:
\begin{equation}
\label{Phic}
\Phi_c \equiv \frac{GM_c}{\tr_c}
=\frac16 \tilde{\rho}_A \tr_c^2\,.
\end{equation}
Here $M_c=4\pi \tilde{r}_c^3 \tilde{\rho}_A/3$ is 
the mass of the spherically symmetric body.
The density $\tilde{\rho}_B$ is much smaller than 
$\tilde{\rho}_A$, so that the metric outside the body can be
approximated by the Schwarzschild exterior solution 
\begin{equation}
\label{Phiout}
\Phi (\tr) \simeq \frac{GM_c}{\tr}=
\Phi_c \frac{\tr_c}{\tr}\,,
\qquad \tilde{P}_M (\tr) \simeq 0
\qquad (\tr>\tr_c)\,.
\end{equation}

In the following we shall derive the analytic field profile 
by using the linear expansion in terms of the gravitational potential
$\Phi_c$. This approximation is expected to be reliable for 
$\Phi_c<{\cal O}(0.1)$.
From Eqs.~(\ref{ePhi})\,--\,(\ref{Phic}) it follows that 
\begin{equation}
\Phi (\tr) \simeq \Phi_c \frac{\tr^2}{\tr_c^2}\,,\qquad
\frac{\tilde{P}_M (\tr)}{\tilde{\rho}_A} \simeq \frac{\Phi_c}{2}
\left(1-\frac{\tr^2}{\tr_c^2} \right)
\qquad (0<\tr<\tr_c)\,.
\end{equation}
Substituting these relations into Eq.~(\ref{fieldeq}), the field
equation inside the body is approximately given by 
\begin{equation}
\label{fieldinside}
\phi''+\frac{2}{\tr} \left( 1-\frac{\tr^2}{2\tr_c^2} \Phi_c \right)
\phi'-(V_{,\phi}+Q \tilde{\rho}_A ) \left( 1+2\Phi_c \frac{\tr^2}{\tr_c^2}
\right)+\frac32 Q \tilde{\rho}_A \Phi_c \left( 1-\frac{\tr^2}{\tr_c^2}
\right)=0\,.
\end{equation}
If $\phi$ is close to $\phi_A$ at $\tr=0$, the field stays
around $\phi_A$ in the region $0<\tr<\tr_1$.
The body has a thin-shell if $\tr_1$ is close to the radius 
$\tr_c$ of the body. 

In the region $0 <\tr<\tr_1$ the field derivative of the effective 
potential around $\phi=\phi_A$ can be approximated by
$\rd V_{\mathrm{eff}}/\rd \phi =V_{,\phi}+
Q\tilde{\rho}_A \simeq m_A^2 (\phi-\phi_A)$.
The solution to Eq.~(\ref{fieldinside}) can be obtained
by writing the field as $\phi=\phi_0+\delta \phi$, where $\phi_0$
is the solution in the Minkowski background  
and $\delta \phi$ is the perturbation induced by $\Phi_c$.
At linear order in $\delta \phi$ and $\Phi_c$ we obtain
\begin{equation}
\label{delphi}
\delta \phi''+\frac{2}{\tr}\delta \phi'-m_A^2 \delta \phi=
\Phi_c \left[ \frac{2m_A^2 \tr^2}{\tr_c^2} (\phi_0-\phi_A)
+\frac{\tr}{\tr_c^2}\phi_0' -\frac32 Q \tilde{\rho}_A 
\left( 1-\frac{\tr^2}{\tr_c^2} \right) \right]\,,
\end{equation}
where $\phi_0$ satisfies the equation 
$\phi_0''+(2/\tr)\phi_0'-m_A^2 (\phi_0-\phi_A)=0$.
The solution of $\phi_0$ with the boundary conditions 
$\mathrm{d}\phi_0/\mathrm{d}\tr=0$ at $\tr=0$ is given by 
$\phi_0 (\tr)=\phi_A+A(e^{-m_A \tr}-e^{m_A \tr})/\tr$, 
where $A$ is a constant. Plugging this into Eq.~(\ref{delphi}), 
we get the following solution for $\phi (\tr)$~\cite{TTT09}:
\begin{eqnarray}
\label{phiso1}
\phi (\tr)&=&\phi_A+\frac{A(e^{-m_A \tr}-e^{m_A \tr})}{\tr} \nonumber \\
&-&\frac{A \Phi_c}{m_A \tr_c^2} \left[ \left(\frac13 m_A^2 \tr^2
-\frac14 m_A \tr-\frac14 +\frac{1}{8m_A \tr} \right)e^{m_A \tr}+
\left(\frac13 m_A^2 \tr^2 +\frac14 m_A \tr-\frac14 -\frac{1}{8m_A \tr} 
\right)e^{-m_A \tr} \right] \nonumber \\
&-&\frac{3Q \tilde{\rho}_A \Phi_c}{2m_A^4 \tr_c^2}
\left[ m_A^2 (\tr^2-\tr_c^2)+6 \right]\,.
\end{eqnarray}

In the region $\tr_1<\tr<\tr_c$ the field $|\phi(\tr)|$ evolves 
towards larger values with increasing $\tr$.
Since the matter coupling term $Q \tilde{\rho}_A$ dominates 
over $V_{,\phi}$ in this regime, it follows that 
$\rd V_{\mathrm{eff}}/\rd \phi \simeq Q \tilde{\rho}_A$.
Hence the field perturbation $\delta \phi$ satisfies
\begin{equation}
\label{delphi2}
\delta \phi''+\frac{2}{\tr}\delta \phi'
=\Phi_c \left[ \frac{\tr}{\tr_c^2} \phi_0'
-\frac12 Q \tilde{\rho}_A \left( 3-7\,\frac{\tr^2}{\tr_c^2} 
\right) \right]\,,
\end{equation}
where $\phi_0$ obeys the equation 
$\phi_0''+(2/\tr)\phi_0'-Q \tilde{\rho}_A=0$.
Hence we obtain the solution 
\begin{equation}
\label{phiso2}
\phi(\tr)=-\frac{B}{\tr} \left(1-\Phi_c \frac{\tr^2}{2\tr_c^2}
\right)+C+\frac16 Q \rho_A \tr^2 \left( 1-\frac32 \Phi_c
+\frac{23}{20} \Phi_c \frac{\tr^2}{\tr_c^2} \right)\,,
\end{equation}
where $B$ and $C$ are constants.

In the region outside the body ($\tr>\tr_c$) the field $\phi$ 
climbs up the potential hill after it acquires sufficient kinetic 
energy in the regime $\tr_1<\tr<\tr_c$.
Provided that the field kinetic energy dominates over its
potential energy, the r.h.s.\ of Eq.~(\ref{fieldeq}) 
can be neglected relative to its l.h.s. of it.
Moreover the terms that include $\tilde{\rho}_M$ and $\tilde{P}_M$
in the square bracket on the l.h.s. of Eq.~(\ref{fieldeq})
is much smaller than the term $(1+e^{2\Phi})/\tr$.
Using Eq.~(\ref{Phiout}), it follows that 
\begin{eqnarray}
\label{phioutside}
\phi''+\frac{2}{\tr} \left( 1+
\frac{GM_c}{\tr} \right)\phi' \simeq 0\,,
\end{eqnarray}
whose solution satisfying the boundary condition 
$\phi (\tr \to \infty)=\phi_B$ is
\begin{eqnarray}
\label{phiso3}
\phi (\tr)=\phi_B+\frac{D}{\tr} 
\left(1+\frac{GM_c}{\tr} \right)\,,
\end{eqnarray}
where $D$ is a constant. 

The coefficients $A, B, C, D$ are known by matching the solutions
(\ref{phiso1}), (\ref{phiso2}), (\ref{phiso3}) and their derivatives
at $\tr=\tr_1$ and $\tr=\tr_c$.
If the body has a thin-shell, then the condition 
$\Delta \tr_c=\tr_c-\tr_1 \ll \tr_c$ is satisfied.
Under the linear expansion in terms of the three parameters 
$\Delta \tr_c/\tr_c$, $\Phi_c$, and $1/(m_A \tr_c)$
we obtain the following field profile~\cite{TTT09}:
\begin{eqnarray}
\label{phithin1}
\phi (\tr) &=& \phi_A+\frac{Q \tilde{\rho}_A}{m_A^2e^{m_A \tr_1}}
\frac{\tr_1}{\tr} \left( 1+\frac{m_A \tr_1^3 \Phi_c}{3\tr_c^2}
-\frac{\Phi_c \tr_1^2}{4\tr_c^2}  \right)^{-1}
(e^{m_A \tr}-e^{-m_A \tr}) \nonumber \\
&& +\frac{3Q \tilde{\rho}_A \Phi_c}{2m_A^2} 
\left[ 1-\frac{\tr^2}{\tr_c^2}-\frac{6}{(m_A \tr_c)^2} \right]
+\frac{\Phi_c \tr_1}{m_A \tr_c^2} 
\frac{Q \tilde{\rho}_A}{m_A^2e^{m_A \tr_1}}
 \left( 1+\frac{m_A \tr_1^3 \Phi_c}{3\tr_c^2}
-\frac{\Phi_c \tr_1^2}{4\tr_c^2} \right)^{-1} \nonumber 
\\
&& \times \biggl[ \left(\frac13 m_A^2 \tr^2
-\frac14 m_A \tr-\frac14 +\frac{1}{8m_A \tr} \right)e^{m_A \tr}+
\left(\frac13 m_A^2 \tr^2 +\frac14 m_A \tr-\frac14 -\frac{1}{8m_A \tr} 
\right)e^{-m_A \tr} \biggr] \nonumber \\
&& (0<\tr<\tr_1), \\
\label{phithin2}
\phi (\tr) &=& \phi_A+\frac{Q\tilde{\rho}_A \tr_c^2}{6}
\left[ 6\epsilon_{\mathrm{th}}+6C_1 \frac{\tr_1}{\tr}
\left( 1-\frac{\Phi_c \tr^2}{2\tr_c^2} \right)
-3\left(1-\frac{\Phi_c}{4}\right)+
\left( \frac{\tr}{\tr_c} \right)^2
\left( 1-\frac32 \Phi_c+\frac{23\Phi_c \tr^2}{20\tr_c^2}
\right) \right] \nonumber \\
&&  (\tr_1<\tr<\tr_c), \\
\label{phithin3}
\phi (\tr) &=& \phi_A+Q\tilde{\rho}_A \tr_c^2 \left[ 
\epsilon_{\mathrm{th}} -C_2 \frac{\tr_c}{\tr}
\left( 1+\Phi_c \frac{\tr_c}{\tr} \right) \right]
\qquad (\tr>\tr_c),
\end{eqnarray}
where $\epsilon_{\mathrm{th}}=(\phi_B-\phi_A)/(6Q\Phi_c)$ is 
the thin-shell parameter, and 
\begin{eqnarray}
C_1 &\equiv& (1-\alpha)\left[ -\epsilon_{\mathrm{th}}  \left(1+
\frac{\Phi_c \tr_1^2}{2\tr_c^2} \right)
+\frac12 \left(1 -\frac{\Phi_c}{4}
+\frac{\Phi_c \tr_1^2}{2\tr_c^2} \right)
- \frac{\tr_1^2}{2 \tr_c^2} \left(1-\frac32 \Phi_c
+\frac{7 \Phi_c \tr_1^2}{4 \tr_c^2} \right) \right] \nonumber \\
&& +\frac{\tr_1^2}{3 \tr_c^2} 
\left(1-\frac32 \Phi_c+\frac{9\Phi_c \tr_1^2}{5 \tr_c^2} \right)\,,\\
C_2 &\equiv&
(1-\alpha)\left[ \epsilon_{\mathrm{th}}  \frac{\tr_1}{\tr_c}
\left(1+\frac{\Phi_c \tr_1^2}{2\tr_c^2}-\frac{3\Phi_c}{2} \right)
-\frac{\tr_1}{2\tr_c}\left(1-\frac{7}{4}\Phi_c+\frac{\Phi_c \tr_1^2}{2\tr_c^2} \right)
+\frac{\tr_1^3}{2\tr_c^3} \left( 1-3\Phi_c+\frac{7\Phi_c \tr_1^2}
{4\tr_c^2} \right) \right] \nonumber \\
&& +\frac13 \left( 1-\frac65 \Phi_c \right)
-\frac{\tr_1^3}{3\tr_c^3}
\left( 1-3\Phi_c+\frac{9\Phi_c \tr_1^2}{5\tr_c^2} \right)\,,
\end{eqnarray}
where
\begin{equation}
\alpha\equiv \frac{(\tr_1^2/3\tr_c^2)\Phi_c+1/(m_A \tr_1)}
{1+(\tr_1^2/4\tr_c^2)\Phi_c+(m_A \tr_1^3 \Phi_c/3\tr_c^2)
[1-(\tr_1^2/2\tr_c^2)\Phi_c]}\,.
\end{equation}
As long as $m_A \tr_1 \Phi_c \gg 1$, the parameter
$\alpha$ is much smaller than 1.

In order to derive the above field profile we have used the fact that 
the radius $\tr_1$ is determined by the condition $m_A^2 
\left[ \phi (\tr_1)-\phi_A \right]=Q \tilde{\rho}_A$,
and hence
\begin{equation}
\label{phiAB2}
\phi_A-\phi_B=-Q \tilde{\rho}_A \tr_c^2
\left[ \frac{\Delta \tr_c}{\tr_c} \left(1+\Phi_c
-\frac12 \frac{\Delta \tr_c}{\tr_c} \right)+
\frac{1}{m_A \tr_c} \left( 1-\frac{\Delta \tr_c}
{\tr_c} \right)(1-\beta)\right]\,,
\end{equation}
where $\beta$ is defined by 
\begin{equation}
\label{beta}
\beta \equiv \frac{(m_A \tr_1^3 \Phi_c/3\tr_c^2)
(\tr_1^2/\tr_c^2)\Phi_c}
{1+(m_A \tr_1^3 \Phi_c/3\tr_c^2)
-(\tr_1^2/4\tr_c^2)\Phi_c}\,,
\end{equation}
which is much smaller than 1.
Using Eq.~(\ref{phiAB2}) we obtain the thin-shell parameter
\begin{equation}
\epsilon_{\mathrm{th}}=\frac{\Delta \tr_c}{\tr_c} \left(1+\Phi_c
-\frac12 \frac{\Delta \tr_c}{\tr_c} \right)+
\frac{1}{m_A \tr_c} \left( 1-\frac{\Delta \tr_c}{\tr_c} 
\right)(1-\beta)\,.
\end{equation}

In terms of a linear expansion of $\alpha, \beta, \Delta \tr_c/\tr_c,
\Phi_c$, the field profile (\ref{phithin3}) outside the body is 
\begin{equation}
\label{phiso}
\phi(\tr) \simeq \phi_B-2Q_{\mathrm{eff}}\frac{GM_c}{\tr}
\left(1+\frac{GM_c}{\tr} \right)\,,
\end{equation}
where the effective coupling is
\begin{equation}
\label{Qeff2}
Q_{\mathrm{eff}} =
3Q \left[ \frac{\Delta \tr_c}{\tr_c}
\left( 1-\frac{\Delta \tr_c}{\tr_c} \right)
+\frac{1}{m_A \tr_c} \left(1-
2\frac{\Delta \tr_c}{\tr_c}-\Phi_c-\alpha-\beta
 \right) \right]\,.
\end{equation}
To leading-order this gives 
$Q_{\mathrm{eff}}=3Q \left[ \Delta \tr_c/\tr_c+1/(m_A \tr_c) \right]
=3Q \epsilon_{\mathrm{th}}$, which agrees with the result (\ref{Qeff0}) 
in the Minkowski background.
As long as $\Delta \tr_c/\tr_c \ll 1$ and $1/(m_A \tr_c) \ll 1$, 
the effective coupling $|Q_{\mathrm{eff}}|$ can be much smaller than 
the bare coupling $|Q|$, even in a strong gravitational background.

From Eq.~(\ref{phithin1}) the field value and its derivative 
around the center of the body with radius $\tr \ll 1/m_A$
are given by 
\begin{eqnarray}
\label{phiv1}
\phi(\tr) &\simeq& \phi_A+\frac{2Q\tilde{\rho}_A \tr_1}
{m_A e^{m_A \tr_1}}
\left( 1+\frac{m_A \tr_1^3 \Phi_c}{3\tr_c^2}
-\frac{\Phi_c \tr_1^2}{4\tr_c^2}  \right)^{-1}
\left[ 1+\frac16 (m_A \tr)^2+\frac{\Phi_c}{2(m_A 
\tr_c)^2} \right]
\nonumber \\
&& +\frac{3Q\tilde{\rho}_A \Phi_c}
{2m_A^2} \left[1-\frac{\tr^2}{\tr_c^2}-
\frac{6}{(m_A \tr_c)^2} \right],\\
\label{phiv2}
\phi'(\tr) &\simeq& Q \tilde{\rho}_A \tr_c^2
\left[ \frac{2m_A \tr_1}{3e^{m_A \tr_1}}
\left( 1+\frac{m_A \tr_1^3 \Phi_c}{3\tr_c^2}
-\frac{\Phi_c \tr_1^2}{4 \tr_c^2}  \right)^{-1}
-\frac{3\Phi_c}{(m_A \tr_c)^2} \right]
\frac{\tr}{\tr_c^2}\,.
\end{eqnarray}
In the Minkowski background ($\Phi_c=0$), 
Eq.~(\ref{phiv2}) gives $\phi'(\tr)>0$ for $Q>0$
(or $\phi'(\tr)<0$ for $Q<0$).
In the strong gravitational background ($\Phi_c \neq 0$)
the second term in the square bracket of Eq.~(\ref{phiv2}) 
can lead to negative $\phi'(\tr)$ for $Q>0$ (or 
positive $\phi'(\tr)$ for $Q<0$), which leads to 
the evolution of $|\phi(\tr)|$ toward 0.
This effects comes from the presence of the strong relativistic 
pressure around the center of the body.
Unless the boundary conditions at $\tr=0$ are 
appropriately chosen the field tends to evolve 
toward $|\phi(\tr)|=0$, as seen in
numerical simulations in~\cite{KM08,KM09}
for the \fR\ model (\ref{Bmodel}).
However there exists a thin-shell field profile 
even for $\phi' (\tr)>0$ (and $Q=-1/\sqrt{6}$) 
around the center of the body.
In fact, the derivative $\phi'(\tr)$ can change its sign 
in the regime $1/m_A<\tr<\tr_1$ for thin-shell 
solutions, so that the field does not reach 
the curvature singularity at $\phi=0$~\cite{TTT09}.

\epubtkImage{strong.png}{%
  \begin{figure}[hptb]
    \centerline{
      \includegraphics[width=2.8in]{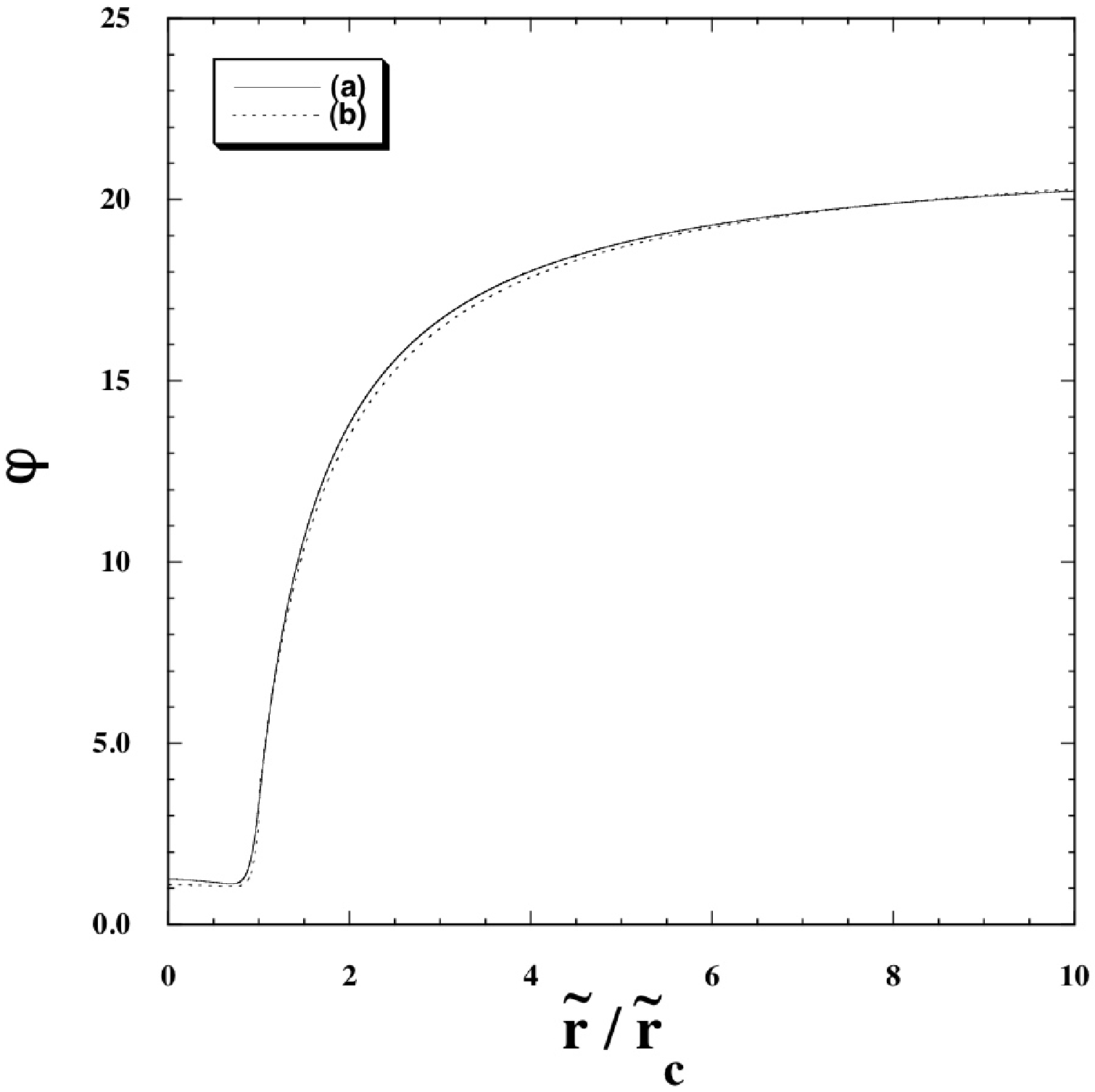}
      \includegraphics[width=2.8in]{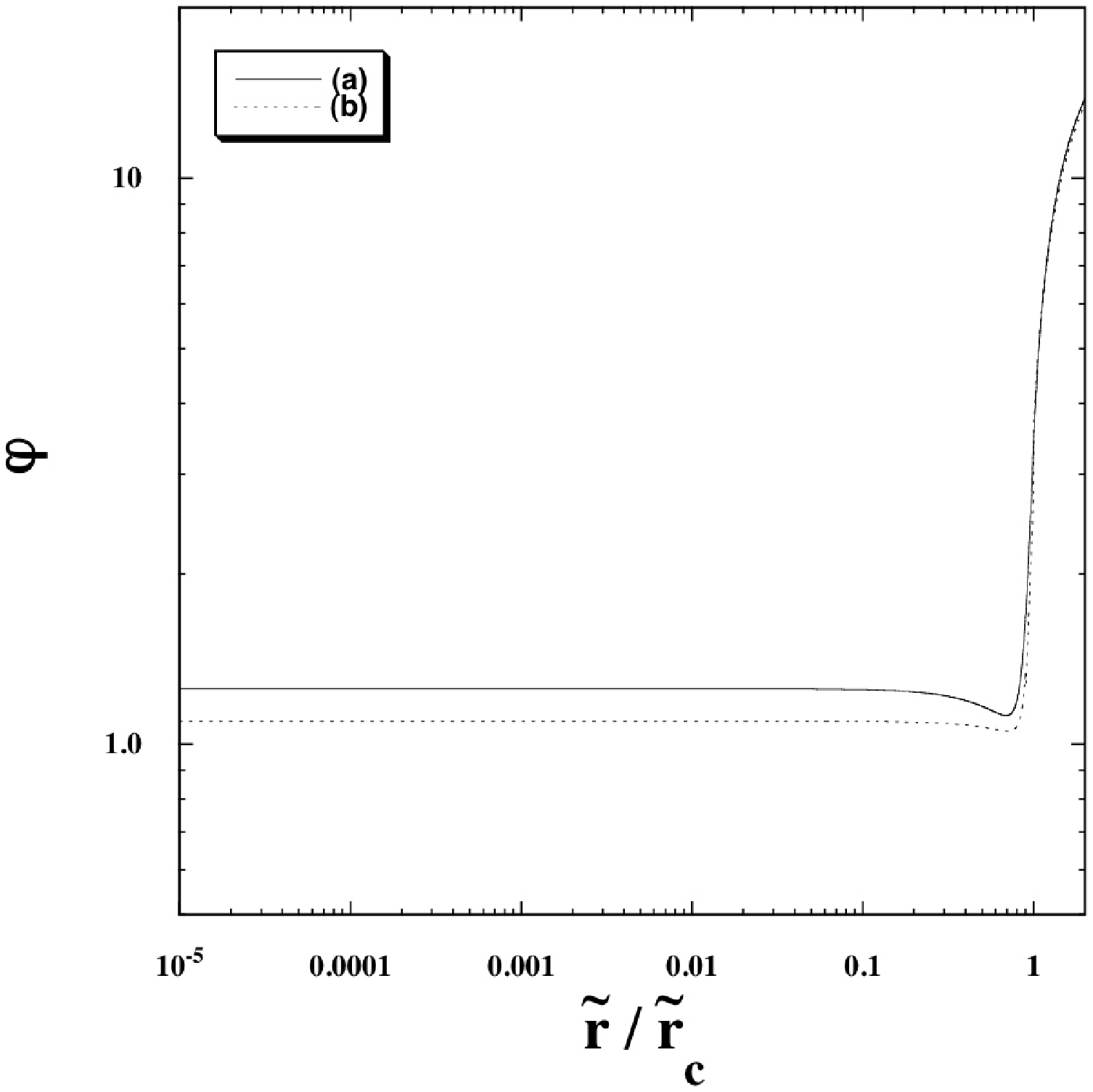}
    }
    \caption{The thin-shell field profile for the model
      $V=M^6\phi^{-2}$ with $\Phi_c=0.2$, $\Delta \tr_c/\tr_c=0.1$,
      $m_A \tr_c=20$, and $Q=1$. This case corresponds to
      $\tilde{\rho}_A/\tilde{\rho}_B=1.04 \times 10^4$, $\phi_A=8.99
      \times 10^{-3}$, $\phi_B=1.97 \times 10^{-1}$ and
      $\epsilon_{\mathrm{th}}=1.56 \times 10^{-1}$. The boundary
      condition of $\varphi=\phi/\phi_A$ at $x_i=\tr_i/\tr_c=10^{-5}$
      is $\varphi(x_i)=1.2539010$, which is larger than the analytic
      value $\varphi(x_i)=1.09850009$. The derivative $\varphi'(x_i)$
      is the same as the analytic value. The left and right panels
      show $\varphi(\tr)$ for $0<\tr/\tr_c<10$ and $0<\tr/\tr_c<2$,
      respectively. The black and dotted curves correspond to the
      numerically integrated solution and the analytic field profile
      (\ref{phithin1})\,--\,(\ref{phithin3}),
      respectively. From~\cite{TTT09}.}
    \label{strongfig}
\end{figure}}

For the inverse power-law potential $V(\phi)=M^{4+n}\phi^{-n}$,
the existence of thin-shell solutions was numerically confirmed in 
~\cite{TTT09} for $\Phi_c<0.3$.
Note that the analytic field profile (\ref{phithin1}) was used to set 
boundary conditions around the center of the body.
In Figure~\ref{strongfig} we show the normalized field $\varphi=\phi/\phi_A$ versus $\tr/\tr_c$ 
for the model $V(\phi)=M^{6}\phi^{-2}$ with $\Phi_c=0.2$, 
$\Delta \tr_c/\tr_c=0.1$, $m_A \tilde{r}_c=20$, and $Q=1$.
While we have neglected the term $V_{,\phi}$ relative to $Q \tilde{\rho}_A$
to estimate the solution in the region $\tr_1<\tr<\tr_c$ analytically, 
we find that this leads to some overestimation for the field value outside
the body ($\tr>\tr_c$). In order to obtain a numerical field profile similar
to the analytic one in the region $\tr>\tr_c$, we need to choose a field
value slightly larger than the analytic value around the center of the body.
The numerical simulation in Figure~\ref{strongfig} corresponds to the choice of 
such a boundary condition, which explicitly shows the presence of 
thin-shell solutions even for a strong gravitational background.


\subsection{Relativistic stars in metric \fR\ gravity}

The results presented above are valid for BD theory 
including metric \fR\ gravity with the coupling $Q=-1/\sqrt{6}$.
While the analysis was carried out in the Einstein frame, 
thin-shell solutions were numerically found in the Jordan frame
of metric \fR\ gravity for the models (\ref{Amodel}) and 
(\ref{Bmodel})~\cite{Babi1,Upadhye,Babi2}.
In these models the field $\phi=\sqrt{3/2}\,\ln\,F$
in the region of high density ($R \gg R_c$)
is very close to the curvature singularity at $\phi=0$.
Originally it was claimed in~\cite{Frolov,KM08} that
relativistic stars are absent because of the presence of 
this accessible singularity.
However, as we have discussed in Section \ref{constarsec},
the crucial point for obtaining thin-shell solutions is not the existence 
of the curvature singularity but the choice of appropriate boundary 
conditions around the center of the star.
For the correct choice of boundary conditions the field 
does not reach the singularity and thin-shell field profiles
can be instead realized.
In the Starobinsky's model~(\ref{Bmodel}), static 
configurations of a constant density star have been 
found for the gravitational potential $\Phi_c$ 
smaller than 0.345~\cite{Upadhye}.

\epubtkImage{phiv.png}{%
  \begin{figure}[htbp]
    \centerline{\includegraphics[width=3.3in]{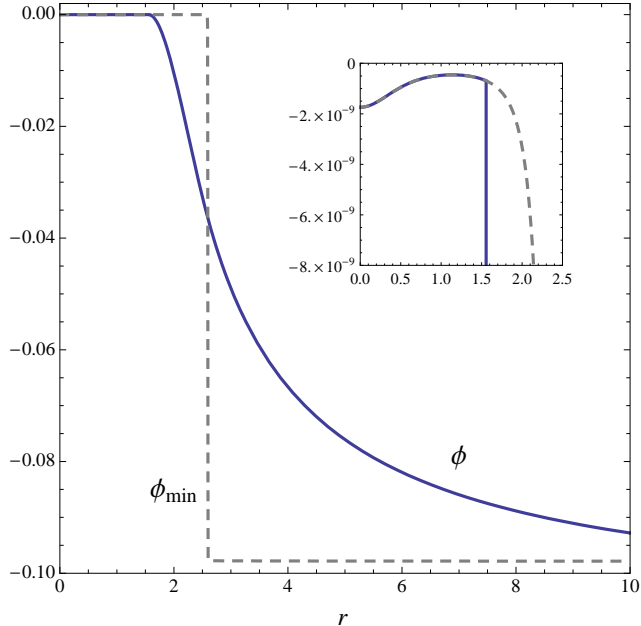}}
    \caption{The profile of the field $\phi=\sqrt{3/2}\,\ln\,F$ (in
      units of $M_{\mathrm{pl}}$) versus the radius $\tr$ (denoted as
      $r$ in the figure, in units of
      $M_{\mathrm{pl}}\rho_{\mathrm{center}}^{-1/2}$) for the
      model~(\ref{Bmodel}) with $n=1$, $R_{\infty}/R_c=3.6$, and
      $v_0=10^{-4}$ (shown as a solid line). The dashed line
      corresponds to the value $\phi_{\mathrm{min}}$ for the minimum
      of the effective potential. (Inset) The enlarged figure in the
      region $0<\tr<2.5$. From~\cite{Babi1}.}
    \label{phiv}
\end{figure}}

For the star with an equation of state $\tilde{\rho}_M<3\tilde{P}_M$, 
the effective potential of the field $\phi$ (in the presence of a matter coupling)
does not have an extremum, see Eq.~(\ref{be1}). 
In those cases the analytic results in Section \ref{constarsec}
are no longer valid. 
For the equation of state $\tilde{\rho}_M<3\tilde{P}_M$ 
there is a tachyonic instability that tends to 
prevent the existence of a static star configuration~\cite{Babi2}.
For realistic neutron stars, however, the equation of state proposed 
in the literature satisfies the condition $\tilde{\rho}_M>3\tilde{P}_M$
throughout the star. 

Babichev and Langlois~\cite{Babi1,Babi2} chose a polytropic equation 
of state for the energy density $\rho_M$ and the pressure $P_M$
in the Jordan frame:
\begin{equation}
\rho_M(n)=m_B \left( n+K\frac{n^2}{n_0} \right)\,,\qquad
P_M(n)=Km_B \frac{n^2}{n_0}\,,
\label{polytropic}
\end{equation}
where $m_B=1.66 \times 10^{-27}$\,kg, $n_0=0.1$\,fm$^{-1}$, 
and $K=0.1$. Solving the continuity equation $\nabla_\mu T^{\mu}_{\nu}=0$
coupled with Einstein equations, \cite{Babi1,Babi2} showed 
that $3\tilde{P}_M$  can remain smaller than $\tilde{\rho}_M$ for realistic 
neutron stars. Note that the energy density is a decreasing function with respect to 
the distance from the center of star.
Even for such a varying energy density, static star configurations
have been shown to exist~\cite{Babi1,Babi2}.

The ratio between the central density $\rho_{\mathrm{center}}$
and the cosmological density at infinity is parameterized by 
the quantity $v_0=M_{\mathrm{pl}}^2R_c/\rho_{\mathrm{center}}$.
Realistic values of $v_0$ are extremely small and it is
a challenging to perform precise numerical simulations 
in such cases. We also note that the field mass $m_A$ 
in the relativistic star is very much larger than its cosmological mass 
and hence a very high accuracy is required for 
solving the field equation numerically~\cite{Upadhye,Sami09}.
The authors in~\cite{Babi1,Upadhye,Babi2} carried out 
numerical simulations for the values of $v_0$ of 
the order of $10^{-3}\mbox{\,--\,}10^{-4}$.
Figure~\ref{phiv} illustrates an example of the thin-shell field 
profile for the polytropic equation of state~(\ref{polytropic})
in the model~(\ref{Bmodel}) with $n=1$ and 
$v_0=10^{-4}$~\cite{Babi1}.
In the regime $0<\tr<1.5$ the field is nearly frozen around 
the extremum of the effective potential, but it starts
to evolve toward its asymptotic value $\phi=\phi_B$ for $\tr>1.5$.

Although the above analysis is based on the \fR\ models
(\ref{Amodel}) and (\ref{Bmodel}) having a curvature singularity 
at $\phi=0$, such a singularity can be cured by adding the 
$R^2$ term~\cite{KM09}. The presence of the $R^2$ term
has an advantage of realizing inflation in the early universe.
However, the \fR\ models (\ref{Amodel}) and (\ref{Bmodel}) 
plus the $R^2$ term cannot relate the epoch of two
accelerations smoothly~\cite{Appleby09}.
An example of viable models that can allow a smooth transition 
without a curvature singularity is~\cite{Appleby09}
\begin{equation}
f(R)=(1-c)R+c \epsilon \ln \left[ \frac{\cosh (R/\epsilon-b)}
{\cosh b} \right]+\frac{R^2}{6M^2}\,,\qquad
\epsilon \equiv \frac{R_c}{b+\ln (2 \cosh b)}\,,
\label{stanew}
\end{equation}
where $b$, $c$ ($0<c<1/2$), $R_c$, and $M$ are constants.
In~\cite{Babi2} a static field profile 
was numerically obtained even for the model~(\ref{stanew}).

Although we have focused on the stellar configuration with 
$\Phi_c \lesssim 0.3$, there are also works of finding 
static or rotating black hole solutions in \fR\ 
gravity~\cite{blackhole,Psaltis}.
Cruz-Dombriz et al.~\cite{blackhole} 
derived static and spherically symmetric 
solutions by imposing that the curvature is constant.
They also used a perturbative approach around 
the Einstein--Hilbert action and found that only solutions of 
the Schwarzschild--Anti de~Sitter type are present up 
to second order in perturbations.
The existence of general black hole solutions in \fR\ gravity
certainly deserves for further detailed study.
It will be also of interest to study the transition from neutron stars 
to a strong-scalar-field state in \fR\ gravity~\cite{Novak}.
While such an analysis was carried out for a massless field
in scalar-tensor theory, we need to take into account the field mass
in the region of high density for realistic models of \fR\ gravity.

Pun et al.~\cite{Harko} studied physical properties of 
matter forming an accretion disk in the spherically symmetric metric 
in \fR\ models and found that specific signatures of modified 
gravity can appear in the electromagnetic spectrum.
In~\cite{Virial} the virial theorem for galaxy clustering 
in metric \fR\ gravity was derived by using the collisionless
Boltzmann equation. 
In~\cite{wormhole} the construction of traversable wormhole
geometries was discussed in metric \fR\ gravity.
It was found that the choice of specific shape functions and 
several equations of state gives rise to some exact solutions 
for \fR.

\newpage

\section{Gauss--Bonnet Gravity}
\label{GBsec}
\setcounter{equation}{0}

So far we have studied modification to the Einstein--Hilbert action via
the introduction of a general function of the Ricci scalar. Among the
possible modifications of gravity this may be indeed a very special
case. Indeed, one could think of a Lagrangian with all the infinite
and possible scalars made out of the Riemann tensor and its
derivatives. If one considers such a Lagrangian as a fundamental
action for gravity, one usually encounters serious problems in the
particle representations of such theories. It is well known that such
a modification would introduce extra tensor degrees of
freedom~\cite{Zwiebach:1985uq, Gross:1986mw, Gross:1986iv}. 
In fact, it is possible to show that these theories in
general introduce other particles and that some of them may lead to
problems.

For example, besides the graviton, another spin-2 particle typically
appears, which however, has a kinetic term opposite in sign with
respect to the standard one~\cite{ghost1, ghost1d, ghost1dd, ghost2,
  ghost2d, Hindawi:1995cu, Boulanger:2000rq}. The graviton does
interact with this new particle, and with all the other standard
particles too. The presence of ghosts, implies the existence of
particles propagating with negative energy. This, in turn, implies
that out of the vacuum a particle (or more than one) and a ghost (or
more than one) can appear at the same time without violating energy
conservation. This sort of vacuum decay makes each single background
unstable, unless one considers some explicit Lorentz-violating cutoff
in order to set a typical energy/time scale at which this phenomenon
occurs~\cite{Trodden03, Cline03}.

However, one can treat these higher-order gravity Lagrangians only as
effective theories, and consider the free propagating mode only coming
from the strongest contribution in the action, the Einstein--Hilbert
one, for which all the modes are well behaved.  The remaining
higher-derivative parts of the Lagrangian can be regarded as
corrections at energies below a certain fundamental scale. This scale
is usually set to be equal to the Planck scale, but it can be lower,
for example, in some models of extra dimensions.
This scale cannot be nonetheless equal to the dark energy density today, 
as otherwise, one would need to consider all these corrections for energies 
above this scale.
This means that one needs to re-sum all these contributions at
all times before the dark energy dominance. 
Another possible approach to dealing with the ghost degrees
of freedom consists of using the Euclidean-action path formalism, for
which, one can indeed introduce a notion of probability amplitude for
these spurious degrees of freedom~\cite{Herto,Sasak}.

The late-time modifications of gravity considered in this review
correspond to those in low energy scales.
Therefore we have a correction which begins to be important at very
low energy scales compared to the Planck mass. 
In general this means that somehow these
corrections cannot be treated any longer as corrections to the
background, but they become the dominant contribution. 
In this case the theory cannot be treated as an effective one, 
but we need to assume that the form of the Lagrangian 
is exact, and the theory becomes a fundamental theory for gravity. 
In this sense these theories are similar to quintessence, that is, 
a minimally coupled scalar field with a suitable potential. 
The potential is usually chosen such that its energy scale matches 
with the dark energy density today. However, for this theory as well, 
one needs to consider this potential as fundamental, i.e., it does not 
get quantum corrections that can spoil the form of the potential itself.
Still it may not be renormalizable, but so far we do not know any 4-dimensional
renormalizable theory of gravity.  In this case then, if we introduce
a general modification of gravity responsible for the late-time cosmic
acceleration, we should prevent this theory from introducing spurious
ghost degrees of freedom.


\subsection{Lovelock scalar invariants}

One may wonder whether it is possible to remove these spin-2 ghosts.  To
answer this point, one should first introduce the Lovelock scalars
\cite{Lovelo}.  These scalars are particular combinations/contractions
of the Riemann tensor which have a fundamental property: if present in
the Lagrangian, they only introduce second-order derivative
contributions to the equations of motion. 
Let us give an example of this property~\cite{Lovelo}. 
Soon after Einstein proposed General
Relativity~\cite{Einstein} and Hilbert found the Lagrangian to describe it~\cite{Hilbert},
Kretschmann~\cite{Kretto} pointed out that general covariance alone
cannot explain the form of the Lagrangian for gravity.
In the action he introduced, instead of the Ricci scalar, the scalar
which now has been named after him, 
the Kretschmann scalar:
\begin{equation}
\label{eq:kret}
S=\int \de^4x \sqrt{-g}\,R_{\alpha\beta\gamma\delta}
\,R^{\alpha\beta\gamma\delta}\,.
\end{equation}

At first glance this action looks well motivated. The Riemann tensor
$R_{\alpha\beta\gamma\delta}$ is a fundamental tensor for gravitation,
and the scalar quantity
$P_1 \equiv R_{\alpha\beta\gamma\delta}\,R^{\alpha\beta\gamma\delta}$ can be
constructed by just squaring it. Furthermore, it is a theory for which
Bianchi identities hold, as the equations of motion have both sides
covariantly conserved. However, in the equations of motion, there are
terms proportional to $\nabla_\mu\nabla_\nu
R^\mu{}_{\alpha\beta}{}^\nu$ together with its symmetric partner ($\alpha\leftrightarrow\beta$). 
This forces us to give in general at a particular
slice of spacetime, together with the metric elements $g_{\mu\nu}$,
their first, second, and third derivatives.
Hence the theory has many more degrees of
freedom with respect to GR.

In addition to the Kretschmann scalar there is another scalar 
$P_2 \equiv R_{\alpha \beta}R^{\alpha \beta}$
which is quadratic in the Riemann tensor $R_{\alpha \beta}$.
One can avoid the appearance of terms proportional to 
$\nabla_\mu\nabla_\nu R^\mu{}_{(\alpha\beta)}{}^\nu$ for the scalar quantity, 
\begin{equation}
\label{eq:gbonn}
\GB\equiv R^2-4R_{\alpha\beta}\,R^{\alpha\beta}
+R_{\alpha\beta\gamma\delta}\,R^{\alpha\beta\gamma\delta}\,,
\end{equation}
which is called the Gauss--Bonnet (GB) term~\cite{ghost1,ghost1d}.
If one uses this invariant in the action of $D$ dimensions, as 
\begin{equation}
\label{eq:actG}
S=\int \de^Dx\sqrt{-g}\,\GB\, ,
\end{equation}
then the equations of motion coming from this Lagrangian include only the terms 
up to second derivatives of the metric. The difference between this scalar and the Einstein--Hilbert term is that this tensor is not linear in the second derivatives of the metric itself. It seems then an interesting theory to study in detail. Nonetheless, it is a topological property of four-dimensional manifolds that $\sqrt{-g}\,\GB$ can be expressed in terms of a total 
derivative~\cite{Cherubini:2003nj}, as
\begin{equation}
\label{eq:capozzo}
\sqrt{-g}\,\GB=\partial_\alpha{\cal D}^\alpha\,,
\end{equation}
where
\begin{equation}
\label{eq:totallo}
{\cal D}^\alpha=\sqrt{-g}\,\epsilon^{\alpha\beta\gamma\delta}\epsilon_{\rho\sigma}{}^{\mu\nu}\Gamma^\rho{}_{\mu\beta}\left[ R^\sigma{}_{\nu\gamma\delta}/2+\Gamma^\sigma{}_{\lambda\gamma}\,\Gamma^\lambda{}_{\nu\sigma}/3 \right] .
\end{equation}
Then the contribution to the equations of motion disappears for any boundaryless manifold
in four dimensions. 

In order to see the contribution of the GB term to the equations of motion 
one way is to couple it with a scalar field $\phi$, i.e., $f(\phi)\GB$,
where $f(\phi)$ is a function of $\phi$.
More explicitly the action of such theories is in general given by 
\begin{equation}
\label{eq:actG1}
S=\int \de^4x\sqrt{-g} \left[ \frac12 F(\phi)R-\frac12 \omega(\phi)
(\nabla \phi)^2-V(\phi)-f(\phi)\GB \right]\,,
\end{equation}
where $F(\phi)$, $\omega(\phi)$, and $V(\phi)$ are functions of $\phi$.
The GB coupling of this form appears in the low energy effective action of 
string-theory~\cite{string,GBearly2}, 
due to the presence of dilaton-graviton mixing terms.

There is another class of general GB theories with 
a self-coupling of the form~\cite{fGO},
\begin{equation}
\label{eq:actG2}
S=\int \de^4x\sqrt{-g}\,\left[ \frac12 R+f(\GB) \right]\,,
\end{equation}
where $f(\GB)$ is a function in terms of the GB term (here we used the unit $\kappa^2=1$).
The equations of motion, besides the standard GR contribution, will get contributions proportional to $\nabla_\mu\nabla_\nu f_{,\GB}$~\cite{DeTsuji1,DeTsuji2}. 
This theory possesses more degrees of freedom than GR, but the extra information 
appears only in a scalar quantity $f_{,\GB}$ and its derivative. 
Hence it has less degrees of freedom compared to Kretschmann 
gravity, and in particular these extra degrees of freedom are not tensor-like. 
This property comes from the fact that the GB term is a Lovelock scalar.
Theories with the more general Lagrangian density $R/2+f (R, P_1, P_2)$
have been studied by many people in connection to the dark energy 
problem~\cite{CaDe,Cal05,Sami05,Mena,Tsujiana,Hervik,Cognola:2007vq,Sokolowski,DeHind}. 
These theories are plagued by the appearance of 
spurious spin-2 ghosts, unless the Gauss--Bonnet (GB) combination is chosen
as in the action~(\ref{eq:actG2})~\cite{ghost2,ghost2d,ghost2dd}
(see also~\cite{Cal05,DeFelice06,DeFelice06d}).

Let us go back to discuss the Lovelock scalars. How many are they? 
The answer is infinite (each of them consists of linear combinations of equal powers 
of the Riemann tensor). However, because of topological reasons, the only non-zero 
Lovelock scalars in four dimensions are the Ricci scalar $R$ and the GB term $\GB$. 
Therefore, for the same reasons as for the GB term, a general function of \fR\ will only introduce terms in the equations of motion of the form $\nabla_\mu\nabla_\nu F$, where 
$F\equiv \partial f/\partial R$. 
Once more, the new extra degrees of freedom introduced into the theory 
comes from a scalar quantity, $F$.

In summary, the Lovelock scalars in the Lagrangian prevent the equations of motion from 
getting extra tensor degrees of freedom. A more detailed analysis of perturbations on maximally symmetric spacetimes shows that, if non-Lovelock scalars are used in the action, then new extra tensor-like degrees of freedom begin to propagate~\cite{ghost1,ghost1d,ghost1dd,ghost2,ghost2d,Hindawi:1995cu,Boulanger:2000rq}. 
Effectively these theories, such as Kretschmann gravity, introduce two gravitons, 
which have kinetic operators with opposite sign. Hence one of the two gravitons is a ghost. 
In order to get rid of this ghost we need to use the Lovelock scalars. 
Therefore, in four dimensions, one can in principle study the following action
\begin{equation}
\label{eq:genact}
S=\int \de^4 x\sqrt{-g}\, f(R,\GB)\,.
\end{equation}
This theory will not introduce spin-2 ghosts. Even so, the scalar modes need 
to be considered more in detail: they may still become ghosts. 
Let us discuss more in detail what a ghost is and why we need to avoid it 
in a sensible theory of gravity.


\subsection{Ghosts}

What is a ghost for these theories? A ghost mode is a propagating degree of freedom 
with a kinetic term in the action with opposite sign. In order to see if a ghost is propagating 
on a given background, one needs to expand the action at second order around 
the background in terms of the perturbation fields. 
After integrating out all auxiliary fields, one is left with a minimal number of gauge-invariant fields $\vec\phi$. 
These are not unique, as we can always perform a field redefinition (e.g., a field rotation). 
However, no matter which fields are used, we typically need -- for non-singular 
Lagrangians -- to define the kinetic operator, the operator which in the Lagrangian 
appears as 
${\cal L}=\dot{\vec\phi}^t A \dot{\vec\phi}+\dots$~\cite{DeFelice:2009ak,DeFelice:2009wp}. 
Then the sign of the eigenvalues of the matrix $A$ defines whether a mode is 
a ghost or not. A negative eigenvalue would correspond to a ghost particle. 
On a FLRW background the matrix $A$ will be in general time-dependent and 
so does the sign of the eigenvalues.
Therefore one should make sure that the extra scalar modes introduced 
for these theories do not possess wrong signs in the kinetic term 
at any time during the evolution of the Universe, at least up to today.

An overall sign in the Lagrangian does not affect the classical equations of motion. 
However, at the quantum level, if we want to preserve causality by keeping 
the optical theorem to be valid, then the ghost can be interpreted as a particle 
which propagates with negative energy, as already stated above. 
In other words, in special relativity, the ghost would 
have a four-momentum $(E_g,\vec p_g)$ with $E_g<0$. 
However it would still be a timelike particle as $E_g^2-\vec p_g^{\,2}>0$, 
whether $E_g$ is negative or not. 
The problem arises when this particle is coupled to some other normal particle, 
because in this case the process $0=E_g+E_1+E_2+\dots$ with $E_g<0$
can be allowed.
This means in general that for such a theory one would expect 
the pair creation of ghost and normal particles out of the vacuum. 
Notice that the energy is still conserved, but the energy is 
pumped out of the ghost particle.

Since all the particles are coupled at least to gravity, one would think that out of the vacuum 
particles could be created via the decay of a couple of gravitons emitted in the vacuum 
into ghosts and non-ghosts particles.
This process does lead to an infinite contribution unless one introduces a cutoff 
for the theory~\cite{Trodden03,Cline03}, for which one can set observational constraints.

We have already seen that, for metric \fR\ gravity, the kinetic operator in the FLRW 
background reduces to $Q_s$ given in Eq.~(\ref{Qsf}) with the 
perturbed action~(\ref{eq:actPhid}).
Since the sign of $Q_s$ is determined by $F$, one needs to impose $F>0$ 
in order to avoid the propagation of a ghost mode.


\subsection{$f(\GB)$ gravity}

Let us consider the theory (\ref{eq:actG2}) in the presence of matter, i.e.
\begin{equation}
\label{GB}
S=\frac{1}{\kappa^2} \int \de^4x\sqrt{-g}\,
\left[ \frac{1}{2}R+f(\GB) \right]+S_M\,,
\end{equation}
where we have recovered $\kappa^2$. For the matter action $S_M$
we consider perfect fluids with an equation of state $w$. 
The variation of the action~(\ref{GB}) leads to the following 
field equations~\cite{Davis,LiMota}
\begin{eqnarray}
\label{geeq}
G_{\mu \nu}&+&8\bigl[ R_{\mu \rho \nu \sigma} +R_{\rho \nu} g_{\sigma \mu}
-R_{\rho \sigma} g_{\nu \mu} -R_{\mu \nu} g_{\sigma \rho} 
+R_{\mu \sigma} g_{\nu \rho}+(R/2)\,(g_{\mu \nu} g_{\sigma \rho}
-g_{\mu \sigma} g_{\nu \rho})\bigr] \nabla^{\rho} \nabla^{\sigma} f_{,\GB} 
\notag \\
&+&(\GB f_{,\GB}-f)\, g_{\mu \nu}=\kappa^2\, T_{\mu \nu}\,,
\end{eqnarray}
where $T_{\mu \nu}$ is the energy-momentum tensor of matter.
If $f\propto\GB$, then it is clear that the theory reduces to GR.

\subsubsection{Cosmology at the background level and viable $f(\GB)$ models}

In the flat FLRW background the $(00)$ component 
of Eq.~(\ref{geeq}) leads to 
\begin{equation}
3H^2=\GB f_{,\GB}-f-24H^3 \dot{f_{,\GB}}+\kappa^2
\left( \rho_m+\rho_r \right)\,,
\label{GBbackeq}
\end{equation}
where $\rho_m$ and $\rho_r$ are the energy densities of non-relativistic
matter and radiation, respectively.
The cosmological dynamics in $f(\GB)$ dark energy models have been 
discussed in~\cite{fGO,Cognola,LiMota,DeTsuji1,Zhou,Mohseni:2009ns}.
We can realize the late-time cosmic acceleration by the existence of 
a de~Sitter point satisfying the condition~\cite{fGO}
\begin{equation}
3H_1^2=\GB_1 f_{,\GB} (\GB_1)-f (\GB_1)\,,
\end{equation}
where $H_1$ and $\GB_1$ are the Hubble parameter and the GB term 
at the de~Sitter point, respectively. 
From the stability of the de~Sitter point we require 
the following condition~\cite{DeTsuji1}
\begin{equation}
0<H_1^6 f_{,\GB \GB} (H_1)<1/384\,.
\label{dSsta}
\end{equation}

The GB term is given by 
\begin{equation}
\label{GBexp}
\GB=24H^2 (H^2+\dot{H})=-12 H^4 (1+3w_{\mathrm{eff}})\,,
\end{equation}
where $w_{\mathrm{eff}}=-1-2\dot{H}/(3H^2)$ is the effective equation 
of state. We have $\GB<0$ and $\dot{\GB}>0$ during both radiation 
and matter domination.
The GB term changes its sign from negative to positive 
during the transition from the matter era 
($\GB=-12H^4$) to the de~Sitter epoch ($\GB=24H^4$). 
Perturbing Eq.~(\ref{GBbackeq}) about the background radiation and matter
dominated solutions, the perturbations in the Hubble parameter
involve the mass squared given by 
$M^2 \equiv 1/(96H^4f_{,\GB \GB})$~\cite{DeTsuji1}.
For the stability of background solutions we require that $M^2>0$, 
i.e., $f_{,\GB \GB}>0$. 
Since the term $24H^3\dot{f_\GB}$ in Eq.~(\ref{GBbackeq}) is of the order 
of $H^8 f_{,\GB \GB}$, this is suppressed relative to $3H^2$ for 
$H^6f_{,\GB \GB} \ll 1$ during the radiation and matter dominated epochs. 
In order to satisfy this condition we require 
that $f_{,\GB \GB}$ approaches $0$ in the limit $|\GB| \to \infty$.
The deviation from the $\Lambda$CDM model can be quantified
by the following quantity~\cite{Mota09}
\begin{equation}
\label{mudeffG}
\mu \equiv H \dot{f_{,\GB}}=H \dot{\GB}f_{,\GB \GB}=
72 H^6 f_{,\GB \GB} \left[ (1+w_{\mathrm{eff}})(1+3w_{\mathrm{eff}})
-w_{\mathrm{eff}}'/2 \right]\,,
\end{equation}
where a prime represents a derivative with respect to $N=\ln\,a$.
During the radiation and matter eras we have 
$\mu=192H^6 f_{,\GB \GB}$ and $\mu=72 H^6  f_{,\GB \GB}$, 
respectively, whereas at the de~Sitter attractor $\mu=0$.

The GB term inside and outside a spherically symmetric body 
(mass $M_{\odot}$ and radius $r_{\odot}$) with a homogeneous density
is given by $\GB=-48(G M_{\odot})^2/r_{\odot}^6$ 
and $\GB=48(G M_{\odot})^2/r^6$, respectively
($r$ is a distance from the center of symmetry).
In the vicinity of Sun or Earth,  $|\GB|$ is much larger than  
the present cosmological GB term, $\GB_0$.
As we move from the interior to the exterior of the star, 
the GB term crosses 0 from negative to positive.
This means that $f(\GB)$ and its derivatives with respect to $\GB$
need to be regular for both negative and positive values of $\GB$ 
whose amplitudes are much larger than $\GB_0$.

The above discussions show that viable $f(\GB)$ models need to
obey the following conditions: 
\begin{enumerate}
\item $f(\GB)$ and its derivatives $f_{,\GB}$, $f_{, \GB \GB}$, \dots
are regular. 
\item $f_{,\GB \GB}>0$ for all $\GB$ and $f_{,\GB \GB}$ 
approaches $+0$ in the limit $|\GB| \to \infty$.
\item $0<H_1^6 f_{,\GB \GB} (H_1)<1/384$ at the 
de~Sitter point.
\end{enumerate}
A couple of representative models that can satisfy these conditions 
are~\cite{DeTsuji1}
\begin{eqnarray}
\mathrm{(A)} \qquad f (\GB) &=& \lambda \frac{\GB}{\sqrt{\GB_*}}\, \arctan \! 
\left( \frac{\GB}{\GB_*} \right)
-\frac{1}{2}\lambda \sqrt{\GB_*}\, \ln \!
\left(1+\frac{\GB^2}{\GB_*^2} \right)-\alpha \lambda \sqrt{\GB_*}\,,
\label{modela}
\\
\mathrm{(B)} \qquad f (\GB) &=&
\lambda \frac{\GB}{\sqrt{\GB_*}}\, \arctan \!
\left( \frac{\GB}{\GB_*} \right)
-\alpha \lambda \sqrt{\GB_*}\,,
\label{model2}
\end{eqnarray}
where $\alpha$, $\lambda$ and $\GB_*\sim H_0^4$ are positive constants.  The
second derivatives of $f$ in terms of $\GB$ for the models (A) and (B)
are $f_{,\GB\GB}=\lambda/[\GB_*^{3/2} (1+\GB^2/\GB_*^2)]$ and
$f_{,\GB\GB}=2\lambda/[\GB_*^{3/2} (1+\GB^2/\GB_*^2)^2]$, respectively.
They are constructed to give rise to the positive $f_{,\GB \GB}$
for all $\GB$. Of course other models can be introduced by following the same prescription. 
These models can pass the constraint of successful expansion history that allows
the smooth transition from radiation and matter eras to the accelerated 
epoch~\cite{DeTsuji1,Zhou}.
Although it is possible to have a viable expansion history at the background level, 
the study of matter density perturbations places tight constraints on these models.
We shall address this issue in Section~\ref{fGpersec}.

\subsubsection{Numerical analysis}
\label{setioNumero}

In order to discuss cosmological solutions in the low-redshift regime 
numerically for the models (\ref{modela}) and (\ref{model2}),
it is convenient to introduce the following dimensionless quantities
\begin{equation}
x \equiv \frac{\dot{H}}{H^2}\,,\quad
y \equiv \frac{H}{H_*}\,, \quad
\Omega_m \equiv \frac{\kappa^2 \rho_m}{3H^2}\,,\quad
\Omega_r \equiv \frac{\kappa^2 \rho_r}{3H^2}\,,
\end{equation}
where $H_*=G_*^{1/4}$.
We then obtain the following equations of motion~\cite{DeTsuji1}
\begin{eqnarray}
\label{beeq1GG}
x'&=&-4x^2-4x+\frac{1}{24^2H^6 f_{,\GB \GB}}
\left[ \frac{\GB f_{,\GB}-f}{H^2}
-3(1-\Omega_m-\Omega_r) \right]\,, \\
y'&=&xy\,,\\
\Omega_m'&=&-(3+2x)\Omega_m\,,\\
\label{beeq4GG}
\Omega_r'&=&-(4+2x)\Omega_r\,,
\end{eqnarray}
where a prime represents a derivative with respect to $N= \ln\,a$. 
The quantities $H^6 f_{,\GB \GB}$ and $(\GB f_{,\GB}-f)/H^2$ can be 
expressed by $x$ and $y$ once the model is specified.

\epubtkImage{mu.png}{%
  \begin{figure}[hptb]
    \centerline{\includegraphics[width=2.9in]{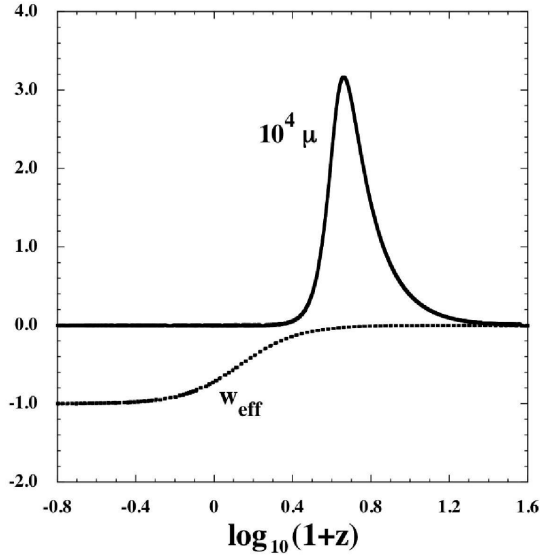}}
    \caption{The evolution of $\mu$ (multiplied by $10^4$) and
      $w_{\mathrm{eff}}$ versus the redshift $z=a_0/a-1$ for the
      model~(\ref{modela}) with parameters $\alpha=100$ and $\lambda=3
      \times 10^{-4}$. The initial conditions are chosen to be
      $x=-1.499985$, $y=20$, and $\Omega_m=0.99999$. We do not take
      into account radiation in this simulation. From~\cite{Mota09}.}
    \label{mufig} 
\end{figure}}

Figure~\ref{mufig} shows the evolution of $\mu$ and $w_{\mathrm{eff}}$ 
without radiation for the model~(\ref{modela}) with parameters 
$\alpha=100$ and $\lambda=3 \times 10^{-4}$.  
The quantity $\mu$ is much smaller than unity in the deep
matter era ($w_{\mathrm{eff}} \simeq 0$) and it reaches a 
maximum value prior to the accelerated epoch. 
This is followed by the decrease of $\mu$ toward 0, as the solution
approaches the de~Sitter attractor with $w_{\mathrm{eff}}=-1$. 
While the maximum value of $\mu$ in this case is of the order 
of $10^{-4}$, it is also possible to realize larger
maximum values of $\mu$ such as $\mu_{\mathrm{max}} \gtrsim 0.1$. 

For high redshifts the equations become too stiff to be integrated directly. 
This comes from the fact that, as we go back to the past, 
the quantity $f_{,\GB \GB}$ (or $\mu$) becomes smaller and smaller. 
In fact, this also occurs for viable \fR\ dark energy models in which $f_{,RR}$
decreases rapidly for higher $z$.
Here we show an iterative method 
(known as the ``fixed-point'' method)~\cite{Mena,DeTsuji1} 
that can be used in these cases, provided no singularity is 
present in the high redshift regime~\cite{DeTsuji1}.
We define $\bar H^2$ and $\bar \GB$ to be $\bar{H}^2\equiv H^2/H_0^2$ and $\bar{\GB} \equiv \GB/H_0^4$, where the subscript ``0'' represents present values. 
The models (A) and (B) can be written in the form 
\begin{equation}
 f(\GB)=\bar{f}(\GB)H_0^2-\bar{\Lambda}\,H_0^2\,, 
\end{equation}
where $\bar{\Lambda}=\alpha \lambda \sqrt{G_*}/H_0^2$ and $\bar{f}(\GB)$ is a function of $\GB$.
The modified Friedmann equation reduces to
\begin{equation}
\label{Hite}
\bar H^2- \bar H_{\Lambda}^2=\frac13\,
(\bar{f}_{,\bar{\GB}} \bar\GB-\bar f)-8\frac{\mathrm{d}
\bar{f}_{,\bar{\GB}}}
{\mathrm{d} N}\,\bar H^4\,,
\end{equation}
where $\bar H_\Lambda^2=\Omega_m^{(0)}/a^3+
\Omega_{r}^{(0)}/a^4+\bar{\Lambda}/3$ (which represents 
the Hubble parameter in the $\Lambda$CDM model).
In the following we omit the tilde for simplicity.

In Eq.~(\ref{Hite}) there are derivatives of $H$ in terms of $N$
up to second-order. Then we write Eq.~(\ref{Hite}) in the form
\begin{equation}
H^2-H^2_\Lambda=
C\!\left( H^2, {H^2}',{H^2}'' \right) \,,
\end{equation}
where $C=(f_{,\GB} \GB- f)/3-8H^4\, (\mathrm{d} f_{,\GB}/\mathrm{d} N)$.
At high redshifts ($a \lesssim 0.01$) the models (A) and (B)  
are close to the $\Lambda$CDM model, i.e., $H^2 \simeq H_\Lambda^2$.
As a starting guess we set the solution to be $H^2_{(0)}=H^2_\Lambda$. 
The first iteration is then $H^2_{(1)}=H^2_\Lambda+C_{(0)}$, where
$C_{(0)}\equiv C\bigl(H^2_{(0)}, {H^2_{(0)}}',{H^2_{(0)}}'' \bigr)$. 
The second iteration is $H^2_{(2)}=H^2_\Lambda+C_{(1)}$,
where $C_{(1)}\equiv C\bigl(H^2_{(1)}, {H^2_{(1)}}',{H^2_{(1)}}''\bigr)$.

\epubtkImage{abserr-relerr.png}{%
  \begin{figure}[hptb]
    \centerline{
      \includegraphics[width=2.9in]{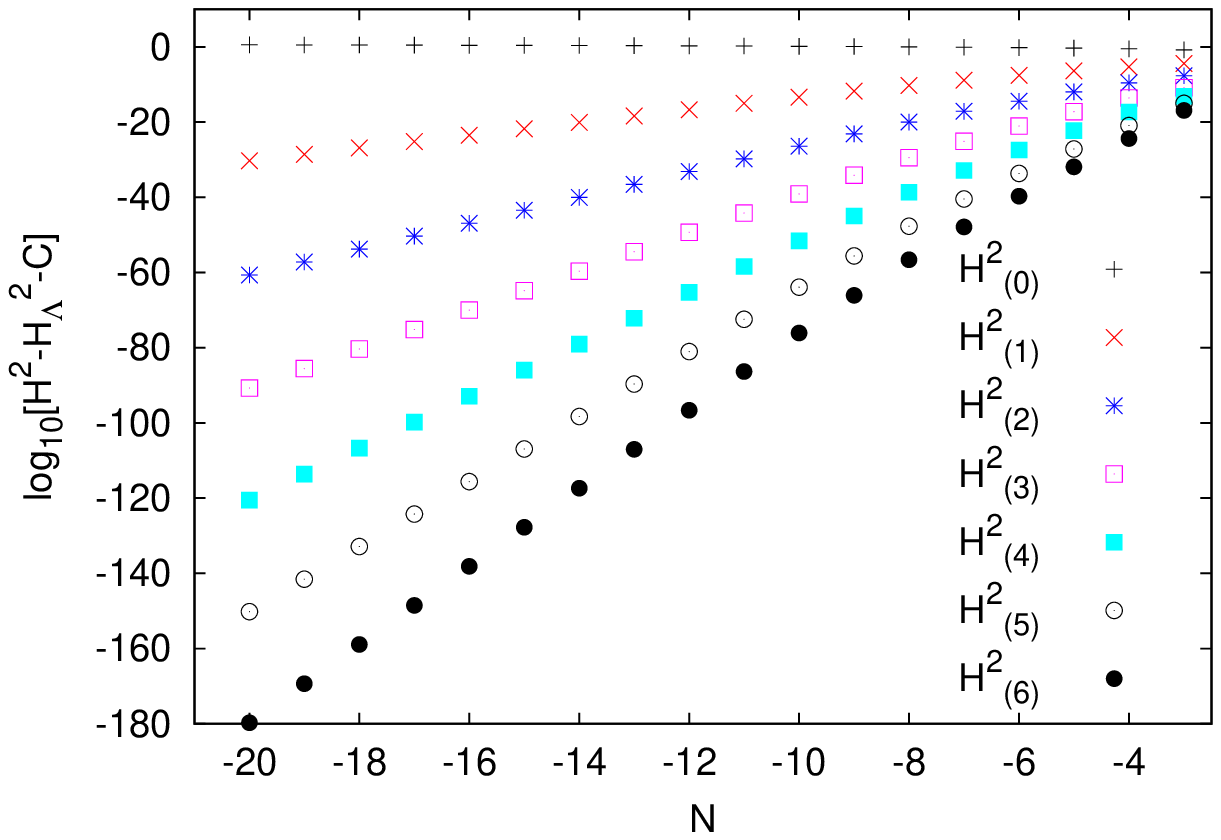}
      \includegraphics[width=2.9in]{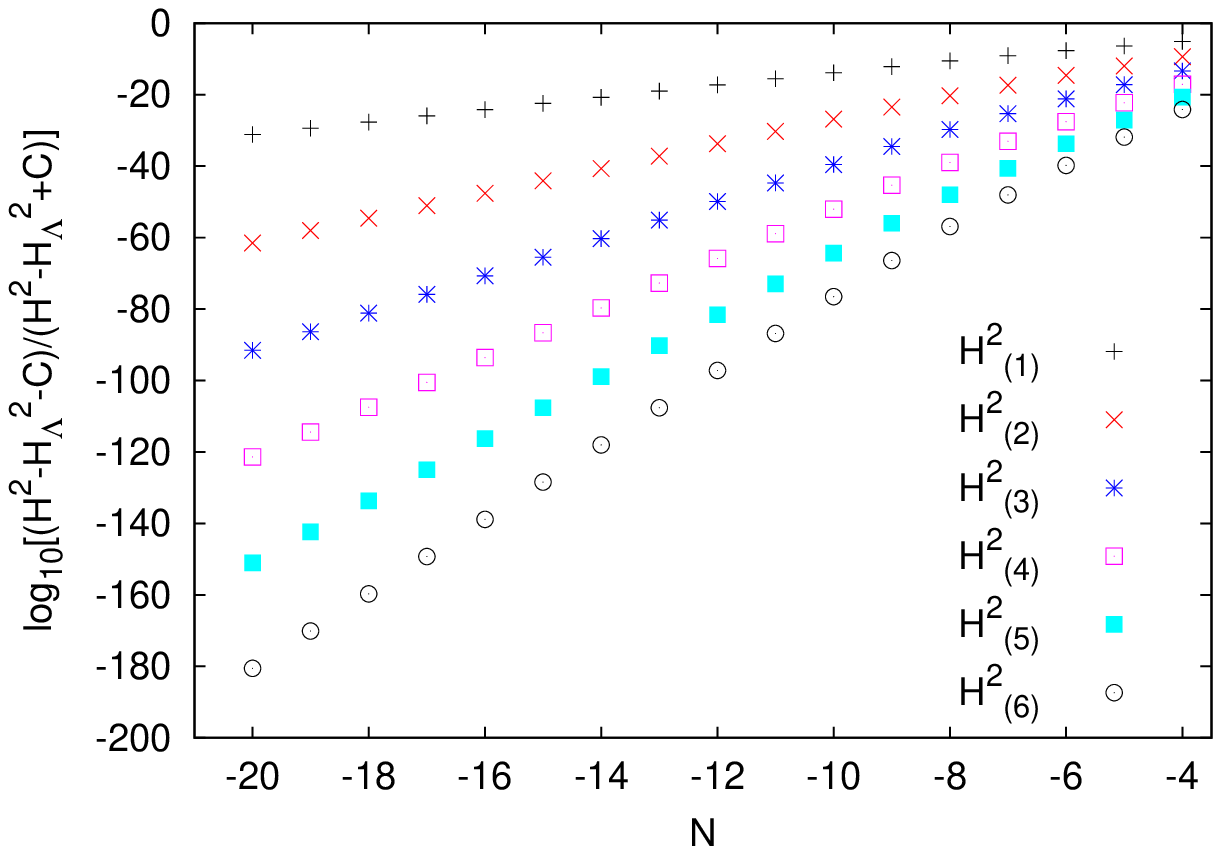}
    }
    \caption{Plot of the absolute errors
      $\log_{10}(|H_i^2-H_{\Lambda}^2-C_i|)$ (left) and
      $\log_{10}\left[\frac{|H_{i}^2-H_{\Lambda}^2-C_{i}|}{|H_i^2-H_{\Lambda}^2+C_i|}\right]$
      (right) versus $N=\ln a$ for the model~(\ref{modela}) with
      $i=0,1,\cdots,6$. The model parameters are $\alpha=10$ and
      $\lambda=0.075$. The iterative method provides the solutions
      with high accuracy in the regime $N \lesssim
      -4$. From~\cite{DeTsuji1}.}
    \label{figNUMfG4}
\end{figure}}

If the starting guess is in the basin of a fixed point, 
$H^2_{(i)}$ will converge to the solution of the equation 
after the $i$-th iteration. For the convergence 
we need the following condition
\begin{equation}
\frac{H^2_{i+1}-H^2_i}{H^2_{i+1}+H^2_i}
<\frac{H^2_{i}-H^2_{i-1}}{H^2_i+H^2_{i-1}}\,,
\end{equation}
which means that each correction decreases for larger $i$. 
The following relation is also required to be satisfied:
\begin{equation}
\frac{H^2_{i+1}-H^2_\Lambda-C_{i+1}}{H^2_{i+1}-H^2_\Lambda+C_{i+1}}
<\frac{H^2_{i}-H^2_\Lambda-C_i}{H^2_i-H^2_\Lambda+C_i}\,.
\end{equation}
Once the solution begins to converge, one can stop the iteration 
up to the required/available level of precision. 
In Figure~\ref{figNUMfG4} we plot absolute errors for the model~(\ref{modela}), 
which shows that the iterative method can produce solutions accurately 
in the high-redshift regime.
Typically this method stops working when the initial guess is outside 
the basin of convergence. 
This happen for low redshifts in which the modifications 
of gravity come into play. In this regime we just need to 
integrate Eqs.~(\ref{beeq1GG})\,--\,(\ref{beeq4GG}) directly.

\subsubsection{Solar system constraints}
\label{solarfG}

We study local gravity constraints on cosmologically viable $f(\GB)$ models.
First of all there is a big difference between $f(\GB)$ and \fR\ theories. 
The vacuum GR solution of a spherically symmetric manifold, the Schwarzschild metric, 
corresponds to a vanishing Ricci scalar ($R=0$) outside the star. 
In the presence of non-relativistic matter, 
$R$ approximately equals to the matter density 
$\kappa^2 \rho_m$ for viable \fR\ models. 

On the other hand, even for the vacuum exterior of  the Schwarzschild metric,
the GB term has a non-vanishing value 
$\GB=R_{\alpha\beta\gamma\delta}R^{\alpha\beta\gamma\delta}
=12\, r_s^2/r^6$~\cite{Davis,DeFelice:2009wp}, 
where $r_s=2GM_{\odot}/r_{\odot}$ is the Schwarzschild radius of the object. 
In the regime $|\GB| \gg \GB_*$ the models (A) and (B) have a correction term
of the order $\lambda \sqrt{\GB_*}\,\GB_*^2/\GB^2$ plus a cosmological 
constant term $-(\alpha+1)\lambda \sqrt{\GB_*}$.
Since $\GB$ does not vanish even in the vacuum, the correction term 
$\GB_*^2/\GB^2$ can be much smaller than 1 
even in the absence of non-relativistic matter.
If matter is present, this gives rise to the contribution of the order of 
$R^2 \approx (\kappa^2 \rho_m)^2$ to the GB term. 
The ratio of the matter contribution to the vacuum 
value $\GB^{(0)}=12\, r_s^2/r^6$ is estimated as
\begin{equation}
R_m \equiv \frac{R^2}{\GB^{(0)}} \approx
\frac{(8\pi)^2}{48} \frac{\rho_m^2 r^6}{M_{\odot}^2}\,.
\end{equation}
At the surface of Sun (radius $r_{\odot}=6.96 \times
10^{10}\mathrm{\ cm} =3.53\times 10^{24}\mathrm{\ GeV}^{-1}$ and mass
$M_{\odot}=1.99 \times 10^{33}\mathrm{\ g}=1.12 \times
10^{57}\mathrm{\ GeV}$), 
the density $\rho_m$ drops down rapidly from the order $\rho_m \approx 10^{-2}\mathrm{\ g/cm}^3$
to the order $\rho_m \approx 10^{-16}\mathrm{\ g/cm}^3$.
If we take the value $\rho_m = 10^{-2}\mathrm{\ g/cm}^3$ we have 
$R_m \approx 4 \times 10^{-5}$ 
(where we have used $1\mathrm{\ g/cm}^3=4.31 \times 10^{-18}\mathrm{\ GeV}^4$).
Taking the value $\rho_m = 10^{-16}\mathrm{\ g/cm}^3$ leads to 
a much smaller ratio: $R_m \approx 4 \times 10^{-33}$. The matter
density approaches a constant value $\rho_m \approx 10^{-24}\mathrm{\ g/cm}^3$
around the distance $r=10^3r_{\odot}$ from the center of Sun.
Even at this distance we have $R_m \approx 4 \times 10^{-31}$, 
which means that the matter contribution to the GB term can 
be neglected in the solar system we are interested in.

In order to discuss the effect of the correction term $\GB_*^2/\GB^2$
on the Schwarzschild metric, we introduce a dimensionless parameter
\begin{equation}
\label{eq:epsGG}
\varepsilon=\sqrt{\GB_*/\GB_s}\,,
\end{equation}
where $\GB_s=12/r_s^4$ is the scale of the GB term in the solar system. 
Since $\sqrt{\GB_*}$ is of the order of the Hubble parameter 
$H_0 \approx 70\mathrm{\ km\,sec}^{-1}\mathrm{\ Mpc}^{-1}$, the
parameter for the Sun is approximately given by $\varepsilon \approx
10^{-46}$. We can then decompose the vacuum equations in the form
\begin{equation}
\label{eq:GGsola}
G_{\mu \nu}+\varepsilon \Sigma_{\mu \nu}=0\,,
\end{equation}
where $G_{\mu \nu}$ is the Einstein tensor and 
\begin{eqnarray}
\label{eq:VC2}
\Sigma_{\mu\nu} &=& 8 \left[ R_{\mu \rho \nu \sigma} +R_{\rho \nu} g_{\sigma \mu}
-R_{\rho \sigma} g_{\nu \mu} -R_{\mu \nu} g_{\sigma \rho}+
R_{\mu \sigma} g_{\nu \rho}+R(g_{\mu \nu} g_{\sigma \rho}
-g_{\mu \sigma} g_{\nu \rho})/2 \right] \nabla^{\rho} \nabla^{\sigma} 
\tilde{f}_{,\GB} \nonumber \\
&& +(\GB \tilde{f}_{,\GB}-\tilde{f}) g_{\mu \nu}.
\end{eqnarray}
Here $\tilde{f}$ is defined by $f=\varepsilon \tilde{f}$.

We introduce the following ansatz for the metric
\begin{equation}
\label{eq:metGG}
\de s^2=-A(r,\varepsilon)\,\de t^2+B^{-1}(r,\varepsilon)\de r^2
+r^2(\de\theta^2+\sin^2\theta\de\phi^2)\,,
\end{equation}
where the functions $A$ and $B$ are expanded as power series 
in $\varepsilon$, as 
\begin{equation}
\label{eq:GGAB}
A=A_0(r)+A_1(r)\varepsilon+O(\varepsilon^2)\,,\qquad
B=B_0(r)+B_1(r)\varepsilon+O(\varepsilon^2)\,.
\end{equation}
Then we can solve Eq.\ (\ref{eq:GGsola}) as follows. 
At zero-th order the equations read
\begin{equation}
\label{eq:GG0}
G^\mu{}_\nu{}^{(0)}(A_0,B_0)=0\,,
\end{equation}
which leads to the usual Schwarzschild solution, 
$A_0=B_0=1-r_s/r$.
At linear order one has
\begin{equation}
\label{eq:GG1}
\varepsilon\,[G^\mu{}_\nu{}^{(1)}(A_1,B_1,A_0,B_0)
+\Sigma^\mu{}_\nu{}^{(0)}(A_0,B_0)]=0\, .
\end{equation}
Since $A_0$ and $B_0$ are known, one can solve the differential equations for $A_1$ and $B_1$. This process can be iterated order by order in $\varepsilon$. 

For the model (A) introduced in (\ref{modela}), we obtain the following differential
equations for $A_1$ and $B_1$~\cite{DeFelice:2009wp}:
\begin{eqnarray}
\rho \frac{\de B_1}{\de\rho} + B_1 &=& 32\sqrt{3}\lambda \rho^3
+12\sqrt{3}\lambda \rho^2\ln(\rho)
+ (4\ln\varepsilon-2\alpha-28)\sqrt{3}\lambda \rho^2\,, \\
(\rho-\rho^2)\frac{\de A_1}{\de\rho}+A_1
&=&8\sqrt{3}\lambda\rho^4
-2\sqrt{3} (10+6\ln\rho+2\ln\varepsilon-\alpha) \lambda\rho^3 \nonumber \\ 
&& -2\sqrt{3}( \alpha-6\ln\rho-2\ln\varepsilon-6)\lambda\rho^2+\rho B_1,
\end{eqnarray}
where $\rho\equiv r/r_s$. The solutions to these equations are
\begin{eqnarray}
B_1 &=& 8\sqrt{3}\lambda\rho^3
+4\sqrt{3}\lambda\rho^2\ln\rho
+\frac23\sqrt{3}\left( 2\ln\varepsilon-\alpha-16 \right )
\lambda\rho^2\,, 
\label{psi1d} \\
A_1&=& -\frac{16}3\sqrt{3}\lambda\rho^3
+\frac23\sqrt{3} \left( 4-\alpha+6\ln\rho+2\ln \varepsilon
\right)\lambda\rho^2\,.
\label{psi1}
\end{eqnarray}

Here we have neglected the contribution coming from the homogeneous
solution, as this would correspond to an order $\varepsilon$
renormalization contribution to the mass of the system. Although
$\varepsilon\ll1$, the term in $\ln\varepsilon$ only contributes by a
factor of order $10^2$. Since $\rho\gg1$ the largest contributions to
$B_1$ and $A_1$ correspond to those proportional to $\rho^3$, which
are different from the Schwarzschild--de~Sitter contribution (which
grows as $\rho^2$). Hence the model~(\ref{modela}) gives rise to the
corrections larger than that in the cosmological constant case by a
factor of $\rho$. Since $\varepsilon$ is very small, the contributions
to the solar-system experiments due to this modification are too weak
to be detected. The strongest bound comes from the shift of the
perihelion of Mercury, which gives the very mild bound $\lambda<2
\times 10^5$~\cite{DeFelice:2009wp}. For the model~(\ref{model2}) the
constraint is even weaker, $\lambda (1+\alpha)<10^{14}$. In other
words, the corrections look similar to the Schwarzschild--de~Sitter
metric on which only very weak bounds can be placed.

\subsubsection{Ghost conditions in the FLRW background}
\label{fGpersec}

In the following we shall discuss ghost conditions for the action~(\ref{GB}).
For simplicity let us consider the vacuum case ($S_M=0$) in the FLRW background.
The action~(\ref{GB}) can be expanded at second order in perturbations
for the perturbed metric (\ref{permetric}), as we have done for 
the action~(\ref{geneaction}) in Section~\ref{secperlag}.
Before doing so, we introduce the gauge-invariant perturbed quantity
\begin{equation}
{\cal R} = \psi-\frac{H}{\dot\xi}\delta\xi\,,\quad \mathrm{where}
\quad \xi \equiv f_{,\GB}\,.
\label{defphi2} 
\end{equation}
This quantity completely describes the dynamics of all the scalar perturbations. 
Note that for the gauge choice $\delta \xi=0$ one has ${\cal R}=\psi$.
Integrating out all the auxiliary fields, we obtain the second-order 
perturbed action~\cite{DeFelice:2009ak}
\begin{equation}
\label{eq:actPhi2d}
\delta S^{(2)}=\int \de t\,\de^3x\,a^3\,Q_s\left[\frac12\,\dot{\cal R}^2
-\frac12\,\frac{c_s^2}{a^2}(\nabla{\cal R})^2\right]\,,
\end{equation}
where we have defined
\begin{eqnarray}
Q_s &\equiv& \frac{24(1+4\mu)\,\mu^2}{\kappa^2(1+6\mu)^2}\,,
\label{QsfG}
\\
\qquad
c_s^2 &\equiv& 1+\frac{2\dot{H}}{H^2}=-2-3w_{\mathrm{eff}}\,.
\label{eq:speedG}
\end{eqnarray}
Recall that $\mu$ has been introduced in Eq.~(\ref{mudeffG}).

In order to avoid that the scalar mode becomes a ghost, one requires 
that $Q_s>0$, i.e.
\begin{equation}
\mu>-1/4\,.
\label{ghosto}
\end{equation}
This relation is dynamical, as one requires to know how $H$ and its derivatives change in time.
Therefore whatever $f(\GB)$ is, the propagating scalar mode can still become a ghost.
If $\dot{f_{,\GB}}>0$ and $H>0$, then $\mu>0$ and hence the ghost does not appear.
The quantity $c_s$ characterizes the speed of propagation for the scalar mode, which 
is again dependent on the dynamics.
For any GB theory, one can give initial conditions of $H$ and $\dot H$ such that $c_s^2$ becomes negative. 
This instability, if present, governs the high momentum modes in Fourier space, which
corresponds to an Ultra-Violet (UV) instability. 
In order to avoid this UV instability in the vacuum, we require that the effective equation of
state satisfies $w_{\mathrm{eff}} < -2/3$.
At the de~Sitter point ($w_{\mathrm{eff}}=-1$) the speed $c_s$ 
is time-independent 
and reduces to the speed of light ($c_s=1$).

Suppose that the scalar mode does not have a ghost mode, i.e., $Q_s>0$. 
Making the field redefinition $u=z_s {\cal R}$ and $z_s=a \sqrt{Q_s}$, 
the action~(\ref{eq:actPhi2d}) can be written as 
\begin{equation}
\label{eq:canactG}
\delta S^{(2)}=\int \de\eta\,\de^3x\left[\frac12\,u'^{2}
-\frac12\,c_s^2(\nabla u)^2-\frac12\,a^2\,M_s^2\,u^2\right]\,,
\end{equation}
where a prime represents a derivative with respect to 
$\eta=\int a^{-1} \mathrm{d}t$ and
$M_s^2\equiv -z_s''/(a^2 z_s)$. 
In order to realize the positive mass squared ($M_s^2>0$), the condition 
 $f_{,\GB\GB}>0$ needs to be satisfied in the regime $\mu \ll 1$
(analogous to the condition $f_{,RR}>0$ in metric \fR\ gravity).

\subsubsection{Viability of $f(\GB)$ gravity in the presence of matter}

In the presence of matter, other degrees of freedom appear in the action. 
Let us take into account a perfect fluid with the barotropic equation of state 
$w_M=P_M/\rho_M$. 
It can be proved that, for small scales (i.e., for large momenta $k$) 
in Fourier space, there are two different propagation speeds
given by~\cite{Mota09}
\begin{eqnarray}
c_1^2&=&w_M\, ,\\
c_2^2&=& 1+\frac{2\dot H}{H^2}+\frac{1+w_M}{1+4\mu}
\frac{\kappa^2\rho_M}{3H^2}\,.
\label{eq:eqcs2w}
\end{eqnarray}

The first result is expected, as it corresponds to the matter propagation 
speed. Meanwhile the presence of matter gives rise to a correction term 
to $c_2^2$ in Eq.~(\ref{eq:speedG}).
This latter result is due to the fact that the background 
equations of motion are different between the two cases. 
Recall that for viable $f(\GB)$ models one has $|\mu| \ll 1$ at high redshifts.
Since the background evolution is approximately given by 
$3H^2 \simeq 8\pi G \rho_M$ and $\dot H/H^2 \simeq -(3/2)(1+w_M)$, 
it follows that 
\begin{equation}
\label{eq:c2neg}
c_2^2 \simeq -1-2w_M\,.
\end{equation}
Hence the UV instability can be avoided for $w_M<-1/2$.
During the radiation era ($w_M=1/3$) and the matter era ($w_M=0$), 
the large momentum modes are unstable. 
In particular this leads to the violent growth of matter density perturbations
incompatible with the observations of large-scale structure~\cite{LiMota,Mota09}.
The onset of the negative instability can be characterized by the condition~\cite{Mota09}
\begin{equation}
\mu \approx (aH/k)^2\,.
\label{mucon}
\end{equation}
As long as $\mu \neq 0$ we can always find a wavenumber $k~(\gg aH)$
satisfying the condition (\ref{mucon}).
For those scales linear perturbation theory breaks down, and in principle one should look for 
all higher-order contributions. Hence the background solutions cannot be trusted 
any longer, at least for small scales, which makes the theory unpredictable. 
In the same regime, one can easily see that the scalar mode is not a ghost, 
as Eq.~(\ref{ghosto}) is satisfied (see Figure~\ref{mufig}).
Therefore the instability is purely classical.
This kind of UV instability sets serious problems for any theory, 
including $f(\GB)$ gravity.

\subsubsection{The speed of propagation in more general modifications of gravity}

We shall also discuss more general theories given by Eq.~(\ref{eq:genact}), i.e.
\begin{equation}
\label{eq:genact2}
S=\int \de^4 x\sqrt{-g}\, f(R,\GB)\,,
\end{equation}
where we do not take into account the matter term here.
It is clear that this function allows more freedom with respect to the background 
cosmological evolution\epubtkFootnote{There are several works about the background 
cosmological dynamics for some $f(R, \GB)$ models~\cite{Alimo08,Alimo09,Elizalde10}.}, 
as now one needs a two-parameter function to choose. 
However, once more the behavior of perturbations proves to be a strong tool 
in order to have a deep insight into the theory. 

The second-order action for perturbations is given by 
\begin{equation}
\label{eq:GENPP}
S=\int \de t\,\de^3x\,a^3\,Q_s \left[\frac12\,\dot{\cal R}^2
-\frac12\,\frac{B_1}{a^2}(\nabla{\cal R})^2
-\frac12\,\frac{B_2}{a^4}(\nabla^2{\cal R})^2\right]\, ,
\end{equation}
where we have introduced the gauge-invariant field
\begin{equation}
{\cal R} = \psi-\frac{H(\delta F+4H^2\delta\xi)}{\dot F+4H^2\dot\xi}\,,
\label{defGGphi2} 
\end{equation}
with $F \equiv f_{,R}$ and $\xi \equiv f_{,\GB}$.
The forms of $Q_s(t)$, $B_1(t)$ and $B_2(t)$ are given explicitly 
in~\cite{DeFelice:2009ak}. 

The quantity $B_2$ vanishes either on the de~Sitter solution or
for those theories satisfying 
\begin{equation}
\label{eq:GGPPQQ}
\Delta\equiv\frac{\partial^2 f}{\partial R^2}\frac{\partial^2 f}{\partial\GB^2}-\left(\frac{\partial^2 f}{\partial R\partial\GB}\right)^2=0\,.
\end{equation}
If $\Delta\neq0$, then the modes with high momenta $k$ 
have a very different propagation. 
Indeed the frequency $\omega$ becomes $k$-dependent, 
that is~\cite{DeFelice:2009ak}
\begin{equation}
\label{eq:KKGG}
\omega^2=B_2\,\frac{k^4}{a^4}\,.
\end{equation}
If $B_2<0$, then a violent instability arises. If $B_2>0$, then these modes propagate with a group velocity
\begin{equation}
\label{eq:GRUPGG}
v_g=2\sqrt{B_2}\,\frac{k}a\, .
\end{equation}
This result implies that the superluminal propagation is always
present in these theories, and the speed is scale-dependent. On the
other hand, when $\Delta=0$, this behavior is not present at
all. Therefore, there is a physical property by which different
modifications of gravity can be distinguished. The presence of an
extra matter scalar field does not change this regime at high
$k$~\cite{DeFelice:2009wp}, because the Laplacian of the gravitational
field is not modified by the field coupled to gravity in the form
$f(\phi,R,\GB)$.


\subsection{Gauss--Bonnet gravity coupled to a scalar field}

At the end of this section we shall briefly discuss theories with a GB
term coupled to a scalar field with the action given in
Eq.~(\ref{eq:actG1}). The scalar coupling with the GB term often
appears as higher-order corrections to low-energy, tree-level
effective string theory based on toroidal
compactifications~\cite{string,stringreview}. More explicitly the
low-energy string effective action in four dimensions is given by 
\begin{equation}
 \label{effactions}
S =\int \mathrm{d}^4x \sqrt{-g} e^{-\phi}
\left[ \frac12 R+\frac12 (\nabla \phi)^2+{\cal L}_M+
{\cal L}_c
\cdots \right]\,,
\end{equation}
where $\phi$ is a dilaton field that controls the string coupling
parameter, $g_{s}^2=e^{\phi}$. The above action is the string frame
action in which the dilaton is directly coupled to a scalar curvature,
$R$. The Lagrangian ${\cal L}_M$ is that of additional matter fields
(fluids, axion, modulus etc.). The Lagrangian ${\cal L}_c$ corresponds
to higher-order string corrections including the coupling between the
GB term and the dilaton. A possible set of corrections include terms
of the form~\cite{GBearly2,GBearly3,GBearly4}
\begin{equation}
 {\cal L}_c = -\frac12 \alpha' \lambda
  \zeta(\phi) \left[ c\,\GB+ d
 (\nabla \phi)^4 \right]\,,
\label{lagalpha}
\end{equation}
where $\alpha'$ is a string expansion parameter and 
$\zeta(\phi)$ is a general function of $\phi$.
The constant $\lambda$ is an additional parameter which depends on 
the types of string theories: $\lambda=-1/4, -1/8$, and 0 correspond to bosonic, heterotic, 
and superstrings, respectively.
If we require that the full action agrees with the three-graviton scattering amplitude, 
the coefficients $c$ and $d$ are fixed to be $c=-1$, $d=1$, and 
$\zeta (\phi)=-e^{-\phi}$~\cite{Metsaev}.

In the Pre-Big-Bang (PBB) scenario~\cite{string} 
the dilaton evolves from a weakly coupled regime ($g_s \ll 1$)
toward a strongly coupled region ($g_s \gtrsim 1$) during 
which the Hubble parameter grows in the string frame (superinflation).
This superinflation is driven by a kinetic energy of 
the dilaton field and it is called a PBB branch. 
There exists another Friedmann branch 
with a decreasing curvature. 
If ${\cal L}_c=0$ these branches are disconnected to 
each other with the appearance of a curvature singularity.
However the presence of the correction ${\cal L}_c$
allows the existence of non-singular solutions that 
connect two branches~\cite{GBearly2,GBearly3,GBearly4}.

The corrections ${\cal L}_c$ are the sum of the tree-level $\alpha'$
corrections and the quantum $n$-loop corrections ($n=1, 2, 3,\cdots$)
with the function $\zeta(\phi)$ given by $\zeta(\phi)=-\sum_{n=0} C_n
e^{(n-1)\phi}$, where $C_n$ ($n \ge 1$) are coefficients of $n$-loop
corrections (with $C_0=1$). In the context of the PBB cosmology it was
shown in~\cite{GBearly3} there exist regular cosmological solutions in
the presence of tree-level and one-loop corrections, but this is not
realistic in that the Hubble rate in Einstein frame continues to
increase after the bounce. Nonsingular solutions that connect to a
Friedmann branch can be obtained by accounting for the corrections up
to two-loop with a negative coefficient
($C_2<0$)~\cite{GBearly3,GBearly4}. In the context of Ekpyrotic
cosmology where a negative potential $V(\phi)$ is present in the
Einstein frame, it is possible to realize nonsingular solutions by
taking into account corrections similar to ${\cal L}_c$ given
above~\cite{GBearly5}. For a system in which a modulus field is
coupled to the GB term, one can also realize regular solutions even
without the higher-derivative term $(\nabla \phi)^4$ in
Eq.~(\ref{lagalpha})~\cite{GBearly1,Easther,Kanti,Kawai1,Kawai2,Yajima,Alexei1,Alexei2}. These
results show that the GB term can play a crucial role to eliminate the
curvature singularity.

In the context of dark energy there are some works which studied
the effect of the GB term on the late-time cosmic acceleration.
A simple model that can give rise to cosmic acceleration is
provided by the action~\cite{NO05}
\begin{equation}
S=\int \mathrm{d}^4 x \sqrt{-g} \left[ \frac12 R-\frac12 (\nabla \phi)^2
-V(\phi)-f(\phi)\,\GB \right]+S_M\,,
\label{GBeins}
\end{equation}
where $V(\phi)$ and $f(\phi)$ are functions of a scalar field $\phi$.
This can be viewed as the action in the Einstein frame corresponding to 
the Jordan frame action~(\ref{effactions}).
We note that the conformal transformation gives rise to 
a coupling between the field $\phi$ and non-relativistic matter 
in the Einstein frame.
Such a coupling is assumed to be negligibly small at low energy 
scales, as in the case of the runaway dilaton scenario~\cite{Piazza1,Piazza2}.  
For the exponential potential $V(\phi)=V_0e^{-\lambda \phi}$ and 
the coupling $f(\phi)=(f_0/\mu)e^{\mu \phi}$, cosmological dynamics 
has been extensively studied in~\cite{NO05,Koi,Koi2,TS}
(see also~\cite{Sanyal,Neupane,Neupaned,Neupane2}).
In particular it was found in~\cite{Koi,TS} that a scaling matter era 
can be followed by a late-time de~Sitter solution which appears due 
to the presence of the GB term.

Koivisto and Mota~\cite{Koi} placed observational constraints
on the above model using the Gold data  set of Supernovae Ia 
together with the CMB shift parameter data of WMAP.
The parameter $\lambda$ is constrained to be $3.5<\lambda<4.5$
at the 95\% confidence level. In the second paper~\cite{Koi2},
they included the constraints coming from the BBN, LSS,
BAO and solar system data and showed that these data strongly disfavor
the GB model discussed above.
Moreover, it was shown in~\cite{TS} that tensor perturbations
are subject to negative instabilities in the above model when the
GB term dominates the dynamics (see also~\cite{Guo07}). 
Amendola et al.~\cite{Amendola} studied local gravity constraints on 
the model~(\ref{GBeins}) and
showed that the energy contribution coming from the GB term needs
to be strongly suppressed for consistency with solar-system experiments.
This is typically of the order of $\Omega_{\mathrm{GB}} \lesssim 10^{-30}$
and hence the GB term of the coupling $f(\phi)\,\GB$ cannot
be responsible for the current accelerated expansion of the universe.

In summary the GB gravity with a scalar field coupling allows nonsingular 
solutions in the high curvature regime, but such a coupling is difficult to be 
compatible with the cosmic acceleration at low energy scales.
Recall that dark energy models based on $f(\GB)$ gravity also 
suffers from the UV instability problem.
This shows how the presence of the GB term makes it difficult 
to satisfy all experimental and observational constraints 
if such a modification is responsible for the late-time acceleration.
This property is different from metric \fR\ gravity in which 
viable dark energy models can be constructed.

\newpage

\section{Other Aspects of \fR\ Theories and Modified Gravity}
\label{othersec}
\setcounter{equation}{0}

In this section we discuss a number of topics related with \fR\
theories and modified gravity.
These include weak lensing, thermodynamics and horizon entropy, 
unified models of inflation and dark energy, \fR\ theories in 
the extra dimensions, Vainshtein mechanism, DGP model,
Noether and Galileon symmetries.


\subsection{Weak lensing}
\label{wlensingsec}

Weak gravitational lensing is sensitive to the growth of large scale 
structure as well as the relation between matter and gravitational potentials.
Since the evolution of matter perturbations and gravitational potentials
is different from that of GR, the observations of weak lensing 
can provide us an important test for probing modified 
gravity on galactic scales (see~\cite{Acq04,Uzanlensing,Sapone08,TsujiTate,Schmidt,SongKoyama}
for theoretical aspects and~\cite{Song05,Knox06,Ishak,Acq,Zhang07,Kunz07,Jain08,Daniel08,Bean09}
for observational aspects).
In particular a number of wide-field galaxy surveys are planned to 
measure galaxy counts and weak lensing shear with high accuracy, 
so these will be useful to distinguish between modified gravity and 
the $\Lambda$CDM model in future observations.

Let us consider BD theory with the action~(\ref{action2}), which includes 
\fR\ gravity as a specific case.
Note that the method explained below can be applied 
to other modified gravity models as well. 
The equations of matter perturbations
$\delta_m$ and gravitational potentials $\Phi, \Psi$ in BD theory have been already 
derived under the quasi-static approximation on sub-horizon scales ($k \gg aH$), 
see Eqs.~(\ref{PsiPhisca0}), (\ref{PsiPhisca}), and (\ref{mattereqsca2a}).
In order to discuss weak lensing observables,
we define the lensing deflecting potential 
\begin{equation}
\Phi_{\mathrm{wl}} \equiv \Phi+\Psi\,,
\label{Phiwl}
\end{equation}
and the effective density field 
\begin{equation}
\delta_{\mathrm{eff}} \equiv -\frac{a}{3H_0^2 \Omega_m^{(0)}}
k^2 \Phi_{\mathrm{wl}}\,,
\label{deleff}
\end{equation}
where the subscript ``0'' represents present values with $a_0=1$.
Using the relation $\rho_m=3F_0H_0^2 \Omega_m^{(0)}/a^3$ 
with Eqs.~(\ref{Phiwl}) and (\ref{deleff}), it follows that 
\begin{equation}
\Phi_{\mathrm{wl}}=-\frac{a^2}{k^2}
\frac{\rho_m}{F} \delta_m\,,\qquad
\delta_{\mathrm{eff}}=\frac{F_0}{F}\delta_m\,.
\label{Pdeltare}
\end{equation}

Writing the angular position of a source and 
the direction of weak lensing observation to be $\vec{\theta}_S$
and $\vec{\theta}_I$, respectively, 
the deformation of the shape of galaxies can be quantified
by the amplification matrix ${\cal A}=\rd \vec{\theta}_S/
\rd \vec{\theta}_I$.
The components of the matrix ${\cal A}$ 
are given by~\cite{Bartel}
\begin{equation}
\label{A}
{\cal A}_{\mu \nu}=I_{\mu \nu}
-\int_0^{\chi} \frac{\chi'(\chi-\chi')}{\chi}
\partial_{\mu \nu} \Phi_{\mathrm{wl}} 
[\chi' \vec{\theta}, \chi'] \mathrm{d} \chi'\,,
\end{equation}
where $\chi=\int_0^z \mathrm{d}z'/H(z')$ is 
a comoving radial distance ($z$ is a redshift).
The convergence $\kappa_{\mathrm{wl}}$ 
and the shear $\vec{\gamma}=(\gamma_1,\gamma_2)$
can be derived from the components of the 
$2 \times 2$ matrix ${\cal A}$, as
$\kappa_{\mathrm{wl}}=1-(1/2) \mathrm{Tr}\,{\cal A}$ and 
$\vec{\gamma}=\left([{\cal A}_{22}-{\cal A}_{11}]/2, 
{\cal A}_{12} \right)$.
For a redshift distribution $p(\chi)\mathrm{d}\chi$
of the source, the convergence can be expressed as
$\kappa_{\mathrm{wl}} (\vec{\theta})=\int p(\chi) 
\kappa_{\mathrm{wl}} (\vec{\theta},\chi)\mathrm{d}\chi$.
Using Eqs.~(\ref{deleff}) and (\ref{A}) it follows that  
\begin{equation}
\kappa_{\mathrm{wl}} (\vec{\theta})=\frac32 H_0^2 
\Omega_{m}^{(0)}
\int_0^{\chi_H} g(\chi)\chi 
\frac{\delta_{\mathrm{eff}}[\chi\,\vec{\theta},\chi]}{a} \rd \chi\,,
\end{equation}
where $\chi_H$ is the maximum distance to the source and 
$g(\chi) \equiv \int_{\chi}^{\chi_H}p(\chi')\,(\chi'-\chi)/\chi'
\rd \chi'$.

The convergence is a function on the 2-sphere and hence
it can be expanded in the form $\kappa_{\mathrm{wl}} (\vec{\theta})=
\int \hat{\kappa}_{\mathrm{wl}} (\vec{\ell}) e^{i \vec{\ell}\cdot \vec{\theta}}
\frac{\mathrm{d}^2 \vec{\ell}}{2\pi}$, where
$\vec{\ell}=(\ell_1, \ell_2)$ with 
$\ell_1$ and $\ell_2$ integers.
We define the power spectrum of the shear to be 
$\langle \hat{\kappa}_{\mathrm{wl}} (\vec{\ell}) \hat{\kappa}_{\mathrm{wl}}^* 
(\vec{\ell'})\rangle=P_{\kappa}(\ell) \delta^{(2)} 
(\vec{\ell}-\vec{\ell'})$.
Then the convergence has the same
power spectrum as $P_{\kappa}$, 
which is given by~\cite{Bartel,Uzan}
\begin{equation}
\label{Pkappa0}
P_{\kappa} (\ell)=\frac{9H_0^4 (\Omega_m^{(0)})^2}{4}
\int_0^{\chi_H} \left[ \frac{g(\chi)}{a(\chi)} \right]^2
P_{\delta_{\mathrm{eff}}} \left[ \frac{\ell}{\chi}, \chi 
\right] \rd \chi\,.
\end{equation}
We assume that the sources are located 
at the distance $\chi_s$ (corresponding to the redshift $z_s$), 
giving the relations $p(\chi)=\delta (\chi-\chi_s)$ and 
$g(\chi)=(\chi_s-\chi)/\chi_s$.
From Eq.~(\ref{Pdeltare}) $P_{\delta_{\mathrm{eff}}}$ can be 
expressed as $P_{\delta_{\mathrm{eff}}}=(F_0/F)^2 P_{\delta_m}$, 
where $P_{\delta_m}$ is the matter power spectrum.
Hence the convergence spectrum (\ref{Pkappa0}) reads
\begin{equation}
\label{Pkappa}
P_{\kappa} (\ell)=\frac{9H_0^4 (\Omega_{m}^{(0)})^2}
{4} \int_0^{\chi_s} \left( \frac{\chi_s-\chi}{\chi_s a} 
\frac{F_0}{F} \right)^2
P_{\delta_m} \left[ \frac{\ell}{\chi}, \chi 
\right] \rd \chi. 
\end{equation}

We recall that, during the matter era, the transition from 
the GR regime ($\delta_m \propto t^{2/3}$
and $\Phi_{\mathrm{wl}}= \mathrm{constant}$) to the scalar-tensor regime 
($\delta_m \propto t^{(\sqrt{25+48Q^2}-1)/6}$ and 
$\Phi_{\mathrm{wl}} \propto t^{(\sqrt{25+48Q^2}-5)/6}$)
occurs at the redshift $z_k$ characterized by the condition (\ref{zk2}).
Since the early evolution of perturbations is similar to that in the 
$\Lambda$CDM model, the weak lensing potential 
at late times is given by the formula~\cite{Dodelsonbook}
\begin{equation}
\label{Phikdef}
\Phi_{\mathrm{wl}} (k, a)=\frac{9}{10} \Phi_{\mathrm{wl}}(k, a_i)
T(k) \frac{D(k, a)}{a}\,,
\end{equation}
where $\Phi_{\mathrm{wl}}(k, a_i) \simeq 2\Phi (k, a_i)$ is 
the initial potential generated during inflation, 
$T(k)$ is a transfer function describing the epochs of 
horizon crossing and radiation/matter 
transition ($50 \lesssim z \lesssim 10^6$), 
and $D(k,a)$ is the growth function at late times defined by 
$D(k,a)/a=\Phi_{\mathrm{wl}}(a)/\Phi_{\mathrm{wl}} (a_I)$
($a_I$ corresponds to the scale factor at a redshift $1 \ll z_I<50$).
Our interest is the case in which the transition redshift 
$z_k$ is smaller than 50, so that we can use the standard transfer 
function of Bardeen--Bond--Kaiser--Szalay~\cite{BBKS}:
\begin{equation}
T(x)= 
\frac{\ln (1+0.171x)}{0.171 x}
\biggl [1.0+0.284x+(1.18x)^2+(0.399x)^3 
+(0.490x)^4 \biggr]^{-0.25},
\end{equation}
where $x \equiv k/k_{\mathrm{EQ}}$ and 
$k_{\mathrm{EQ}}=0.073\,\Omega_{m}^{(0)}h^{2}\mathrm{\ Mpc}^{-1}$.
In the $\Lambda$CDM model the growth function $D(k,a)$
during the matter dominance is scale-independent 
($D(a)=a$), but in BD theory with 
the action~(\ref{action2}) the growth of perturbations is 
generally scale-dependent.

From Eqs.~(\ref{deleff}) and (\ref{Phikdef}) we obtain 
the matter perturbation $\delta_m$ for $z<z_I$:
\begin{equation}
\delta_m (k, a)=-\frac{3}{10} \frac{F}{F_0}
\frac{k^2}{\Omega_m^{(0)} H_0^2}
\Phi_{\mathrm{wl}} (k, a_i) T(k) D(k, a)\,.
\end{equation}
The initial power spectrum generated 
during inflation is $P_{\Phi_{\mathrm{wl}}} \equiv 
4|\Phi|^2 =(200 \pi^2/9k^3)(k/H_0)^{n_{\Phi}-1} \delta_H^2$, 
where $n_{\Phi}$ is the spectral index and $\delta_H^2$ 
is the amplitude of $\Phi_{\mathrm{wl}}$~\cite{Bassett,Dodelsonbook}. 
Therefore we obtain the power spectrum of matter perturbations, as
\begin{equation}
\label{delm}
P_{\delta_m}(k,a) \equiv |\delta_m|^2=
2\pi^2 \left(\frac{F}{F_0}\right)^2 
\frac{k^{n_\Phi}}{(\Omega_{m}^{(0)})^2 H_0^{n_\Phi+3}}
\delta_H^2 T^2(k) D^2(k, a).
\end{equation}
Plugging Eq.~(\ref{delm}) into Eq.~(\ref{Pkappa}), we find
that the convergence spectrum is given by 
\begin{equation}
\label{Pkappa2}
P_{\kappa} (\ell) =
\frac{9\pi^2}{2}
\int_0^{z_s} \left( 1-\frac{X}{X_s} \right)^2
\frac{1}{E(z)} \delta_H^2 
\left( \frac{\ell}{X} \right)^{n_\Phi} T^2(x)
\left(\frac{\Phi_{\mathrm{wl}}(z)}
{\Phi_{\mathrm{wl}}(z_I)} \right)^2 \rd z\,,
\end{equation}
where 
\begin{equation}
E(z)=\frac{H(z)}{H_0}\,,\qquad 
X=H_0 \chi\,,\qquad
x=\frac{H_0}{k_{\mathrm{EQ}}}
\frac{\ell}{X}\,.
\end{equation}
Note that $X$ satisfies the 
differential equation ${\rd X}/{\rd z}=1/E(z)$.

\epubtkImage{weak.png}{%
  \begin{figure}[hptb]
    \centerline{\includegraphics[width=3.3in]{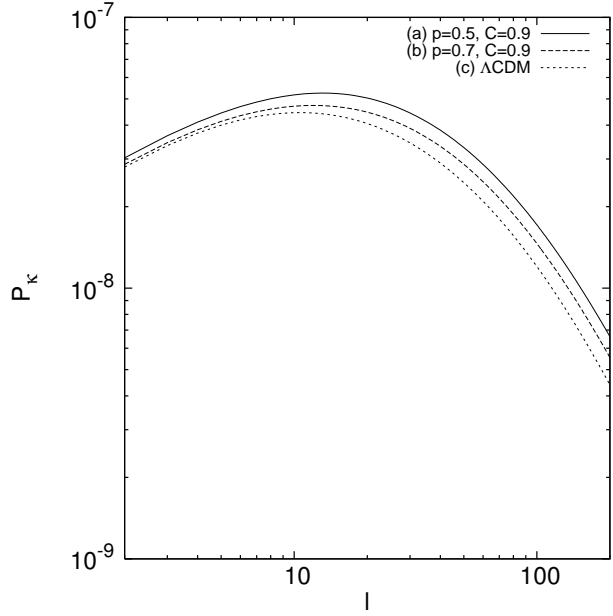}}
    \caption{The convergence power spectrum $P_{\kappa}(\ell)$ in
      \fR\ gravity ($Q=-1/\sqrt{6}$) for the
      model~(\ref{fRasy}). This model corresponds to the field
      potential (\ref{modelscalar}). Each case corresponds to (a)
      $p=0.5$, $C=0.9$, (b) $p=0.7$, $C=0.9$, and (c) the $\Lambda$CDM
      model. The model parameters are chosen to be
      $\Omega_m^{(0)}=0.28$, $n_{\Phi}=1$, and $\delta_H^2=3.2 \times
      10^{-10}$. From~\cite{TsujiTate}.}
    \label{weakfig} 
\end{figure}}

In Figure~\ref{weakfig} we plot the convergence spectrum
in \fR\ gravity with the potential (\ref{modelscalar})
for two different values of $p$
together with the $\Lambda$CDM spectrum.
Recall that this model corresponds to the \fR\ model
$f(R)=R-\mu R_c \left[1-(R/R_c)^{-2n} \right]$
with the correspondence $p=2n/(2n+1)$.
Figure~\ref{weakfig} shows the convergence spectrum in the 
linear regime characterized by $\ell \lesssim 200$.
The $\Lambda$CDM model corresponds to 
the limit $n \to \infty$, i.e., $p \to 1$.
The deviation from the $\Lambda$CDM model 
becomes more significant for smaller $p$ away from 1.
Since the evolution of $\Phi_{\mathrm{wl}}$ changes from 
$\Phi_{\mathrm{wl}}= \mathrm{constant}$ to 
$\Phi_{\mathrm{wl}} \propto t^{(\sqrt{25+48Q^2}-5)/6}$
at the transition time $t_{\ell}$ characterized by the condition 
$M^2/F=(\ell/\chi)^2/a^2$, this leads to a difference
of the spectral index of the convergence spectrum compared to 
that of the $\Lambda$CDM model~\cite{TsujiTate}:
\begin{equation}
\frac{P_{\kappa}(\ell)}{P_{\kappa}^{\Lambda\mathrm{CDM}}(\ell)}
\propto \ell^{\Delta n_s}\,,\quad \mathrm{where} \quad
\Delta n_s=\frac{(1-p)(\sqrt{25+48Q^2}-5)}{4-p}\,.
\label{Pkappa3}
\end{equation}
This estimation is reliable for the transition redshift $z_{\ell}$
much larger than 1.
In the simulation of Figure~\ref{weakfig} the numerical value of 
$\Delta n_s$ for $p=0.7$ at $\ell=200$ is 0.056 (with $z_\ell=3.26$), 
which is slightly smaller than the analytic value $\Delta n_s=0.068$
estimated by Eq.~(\ref{Pkappa3}).
The deviation of the spectral index of $P_{\kappa}$ from 
the $\Lambda$CDM model will be useful to probe modified
gravity in future high-precision observations.
Note that the galaxy-shear correlation spectrum will be also 
useful to constrain modified gravity models~\cite{Schmidt}.

Recent data analysis of the weak lensing shear field from the Hubble 
Space Telescope's COSMOS survey along with the ISW effect of CMB
and the cross-correlation between the ISW and galaxy distributions 
from 2MASS and SDSS surveys shows that the anisotropic 
parameter $\eta \equiv \Psi/\Phi$ is constrained to be 
$\eta<1$ at the 98\% confidence level~\cite{Bean09}.
For BD theory with the action~(\ref{action2}) the quasi-static 
results (\ref{PsiPhisca0}) and (\ref{PsiPhisca}) of 
the gravitational potentials give
\begin{equation}
\eta \simeq \frac{(k^2/a^2)(1-2Q^2)F+M^2}
{(k^2/a^2)(1+2Q^2)F+M^2}\,.
\end{equation}
Since $\eta \simeq (1-2Q^2)/(1+2Q^2)$
in the scalar-tensor regime ($k^2/a^2 \gg M^2/F$),
one can realize $\eta<1$ in BD theory.
Of course we need to wait for further observational data 
to reach the conclusion that modified gravity is favored 
over the $\Lambda$CDM model.

To conclude this session we would like to point out the possibility of using 
the method of gravitational lensing tomography~\cite{Takada}. 
This method consists of considering lensing on different redshift data-bins. 
In order to use this method, one needs to know the evolution of both 
the linear growth rate and the non-linear one (typically found through 
a standard linear-to-non-linear mapping). 
Afterward, from observational data, one can separate different bins 
in order to make fits to the models. 
Having good data sets, this procedure is strong 
enough to further constrain the models, 
especially together with other probes 
such as CMB~\cite{Ishak, ishakTomo, Zhao09, Guzik}.


\subsection{Thermodynamics and horizon entropy}

It is known that in Einstein gravity the gravitational entropy $S$
of stationary black holes is proportional to the horizon area $A$, 
such that $S=A/(4G)$, where $G$ is gravitational constant~\cite{Beken}.
A black hole with mass $M$ obeys the first law of
thermodynamics, $T \rd S=\rd M$~\cite{Bardeenthe}, 
where $T=\kappa_s/(2\pi)$ is a Hawking temperature 
determined by the surface gravity $\kappa_s$~\cite{Hawking}.
This shows a deep physical connection between gravity 
and thermodynamics.
In fact, Jacobson~\cite{Jacobson} showed that Einstein equations can 
be derived by using the Clausius relation $T\rd S=\rd Q$ on local 
horizons in the Rindler spacetime together with the relation $S \propto A$, 
where $\rd Q$ and $T$ are the energy flux across the horizon and
the Unruh temperature seen by an accelerating observer just inside 
the horizon respectively.

Unlike stationary black holes the expanding universe 
with a cosmic curvature $K$ has a dynamically changing 
apparent horizon with the radius $\bar{r}_A=(H^2+K/a^2)^{-1/2}$, 
where $K$ is a cosmic curvature~\cite{Cai07} (see also~\cite{Hayward}).
Even in the FLRW spacetime, however, the Friedmann 
equation can be written in the thermodynamical form 
$T\rd S=-\rd E+W \rd V$, where $W$ is the work density present
in the dynamical background~\cite{Akbar}. 
For matter contents of the universe with energy density $\rho$
and pressure $P$, the work density is given by 
$W=(\rho-P)/2$~\cite{Hayward2,Mukoh}.
This method is identical to the one established by 
Jacobson~\cite{Jacobson}, that is, $\rd Q=-\rd E+W \rd V$.

In metric \fR\ gravity Eling et al.~\cite{Eling} showed that a non-equilibrium 
treatment is required such that the Clausius relation is modified to 
$\rd S=\rd Q/T+\rd_i S$, where $S=FA/(4G)$ is the Wald horizon
entropy~\cite{Waldentropy} and $\rd_i S$ is the bulk viscosity entropy
production term. Note that $S$ corresponds to a Noether charge entropy.
Motivated by this work, the connections between thermodynamics and 
modified gravity have been extensively discussed -- including 
metric \fR\ gravity~\cite{Akbar2,Akbar3,Gong,Sadjadi,Wu1,Wu2,Emilio,Emilio2,Bamba09d,Bamba09,Liberati09}
and scalar-tensor theory~\cite{Gong,Wu1,Wu2,Cai07}.

Let us discuss the relation between thermodynamics and modified gravity
for the following general action~\cite{BGT}
\begin{equation}
I = \int \rd^4 x \sqrt{-g} \left[ \frac{f(R,\phi,X)}{16\pi G} +
{\mathcal{L}}_{M} \right]\,,
\label{eq:2.1}
\end{equation}
where $X \equiv -\left(1/2\right) g^{\mu\nu} \nabla_{\mu}\phi 
\nabla_{\nu}\phi$ 
is a kinetic term of a scalar field $\phi$. 
For the matter Lagrangian ${\mathcal{L}}_{M}$
we take into account perfect fluids (radiation and non-relativistic matter) 
with energy density $\rho_M$ and pressure $P_M$.
In the FLRW background with the metric 
$\rd s^2=h_{\alpha \beta} \rd x^{\alpha} \rd x^{\beta}
+\bar{r}^2 \rd \Omega^2$, 
where $\bar{r}=a(t)r$ and $x^0=t$, $x^1=r$ 
with the two dimensional metric $h_{\alpha \beta}= \mathrm{diag}(-1, a^2(t)/[1-Kr^2])$, 
the Friedmann equations are given by 
\begin{eqnarray}
H^2&+&\frac{K}{a^2}=\frac{8\pi G}{3F} 
\left( \rho_d+\rho_M \right)\,,
\label{frire1} \\
\dot{H}&-&\frac{K}{a^2}=-\frac{4\pi G}{F}
\left( \rho_d+P_d+\rho_M+P_M \right)\,,
\label{frire2} \\
\dot{\rho}_M&+&3H (\rho_M+P_M)=0\,,
\label{rhof}
\end{eqnarray} 
where $F \equiv \partial f/\partial R$ and
\begin{eqnarray}
\rho_d &\equiv& \frac{1}{8\pi G} \left[ f_{,X}X
+\frac12 (FR-f)-3H \dot{F} \right]\,,
\label{rodef1}
\\
P_d &\equiv& \frac{1}{8\pi G}
\left[ \ddot{F}+2H \dot{F}-\frac12 (FR-f) \right]\,.
\label{pdef1}
\end{eqnarray} 
Note that $\rho_d$ and $P_d$ originate from the 
energy-momentum tensor $T_{\mu \nu}^{(d)}$ defined by 
\begin{equation}
T_{\mu \nu}^{(d)} \equiv \frac{1}{8\pi G}
\biggl[ \frac12 g_{\mu \nu}
(f-RF)+\nabla_{\mu}\nabla_{\nu}F
-g_{\mu \nu} \Box F +\frac12 f_{,X}
\nabla_{\mu} \phi \nabla_{\nu} \phi  \biggr]\,,
\label{Tmunud}
\end{equation}
where the Einstein equation is given by 
\begin{equation}
G_{\mu \nu}=\frac{8\pi G}{F}
\left( T_{\mu \nu}^{(d)}+T_{\mu \nu}^{(M)}
\right)\,.
\label{Ein1}
\end{equation}
Defining the density $\rho_d$ and the pressure $P_d$
of ``dark'' components in this way, they obey the 
following equation 
\begin{equation}
\dot{\rho}_d+3H(\rho_d+P_d)
=\frac{3}{8\pi G} (H^2+K/a^2) \dot{F}\,.
\label{rhocon1}
\end{equation}
For the theories with $\dot{F} \neq 0$ (including
\fR\ gravity and scalar-tensor theory)
the standard continuity equation does not hold
because of the presence of the last term 
in Eq.~(\ref{rhocon1}).

In the following we discuss the thermodynamical property of 
the theories given above.
The apparent horizon is determined by the condition 
$h^{\alpha \beta} \partial_{\alpha} \bar{r} \partial_{\beta} \bar{r}=0$, 
which gives $\bar{r}_A=\left( H^2+K/a^2 \right)^{-1/2}$ in the 
FLRW spacetime. Taking the differentiation of this relation with respect to 
$t$ and using Eq.~(\ref{frire2}), we obtain 
\begin{equation}
\frac{F\rd \bar{r}_A}{4\pi G}=\bar{r}_A^3 H
\left( \rho_d+P_d+\rho_M+P_M \right)\rd t\,.
\label{Fdr}
\end{equation}

In Einstein gravity the horizon entropy is given by
the Bekenstein--Hawking entropy $S=A/(4G)$, 
where $A=4\pi \bar{r}_A^2$ is the
area of the apparent horizon~\cite{Bardeenthe,Beken,Hawking}.
In modified gravity theories one can introduce the Wald entropy 
associated with the Noether charge~\cite{Waldentropy}:
\begin{equation}
S=\frac{AF}{4G}\,.
\label{Sdef}
\end{equation}
Then, from Eqs.~(\ref{Fdr}) and (\ref{Sdef}), it follows that 
\begin{equation}
\frac{1}{2\pi \bar{r}_A} \rd S=4\pi \bar{r}_A^3 H
\left( \rho_d+P_d+\rho_M+P_M \right)\rd t +
\frac{\bar{r}_A}{2G} \rd F\,.
\label{dSre}
\end{equation}
The apparent horizon has the Hawking temperature 
$T=|\kappa_s|/(2\pi)$, where $\kappa_s$ is 
the surface gravity given by 
\begin{equation}
\kappa_s =-\frac{1}{\bar{r}_A}
\left( 1-\frac{\dot{\bar{r}}_A}{2H\bar{r}_A} \right)
=-\frac{\bar{r}_A}{2} \left( \dot{H}+2H^2
+\frac{K}{a^2} \right)=
-\frac{2\pi G}{3F} \bar{r}_A 
\left( \rho_T-3P_T \right)\,.
\end{equation}
Here we have defined $\rho_T \equiv \rho_d+\rho_M$ and 
$P_T \equiv P_d+P_M$. 
For the total equation of state $w_T=P_T/\rho_T$ less than 1/3, 
as is the case for standard cosmology, one has $\kappa_s \le 0$ 
so that the horizon temperature is given by 
\begin{equation}
T=\frac{1}{2\pi \bar{r}_A}
\left( 1-\frac{\dot{\bar{r}}_A}{2H\bar{r}_A} \right)\,.
\label{tempe}
\end{equation}
Multiplying the term $1-\dot{\bar{r}}_A/(2H\bar{r}_A)$ for 
Eq.~(\ref{dSre}), we obtain 
\begin{equation}
T \rd S = 4\pi \bar{r}_A^3 H (\rho_d+
P_d+\rho_M+P_M)\rd t
-2\pi \bar{r}_A^2 (\rho_d+P_d+\rho_M+P_M)
\rd \bar{r}_A+\frac{T}{G}\pi \bar{r}_A^2 \rd F. 
\label{TdS}
\end{equation}

In Einstein gravity the Misner-Sharp energy~\cite{Misner} is 
defined by $E=\bar{r}_A/(2G)$.
In \fR\ gravity and scalar-tensor theory this can be extended
to $E=\bar{r}_AF/(2G)$~\cite{Gong}.
Using this expression for $f(R, \phi, X)$ theory,  we have
\begin{equation}
E=\frac{\bar{r}_A F}{2G}=
V \frac{3F (H^2+K/a^2)}{8\pi G}=V(\rho_d+\rho_M)\,,
\label{Edef}
\end{equation}
where $V=4\pi \bar{r}_A^3/3$ is the volume inside
the apparent  horizon.
Using Eqs.~(\ref{rhof}) and (\ref{rhocon1}), we find 
\begin{equation}
\rd E= -4\pi \bar{r}_A^3 H (\rho_d
+P_d+\rho_M+P_M)\rd t
+4\pi \bar{r}_A^2 (\rho_d+\rho_M) 
\rd \bar{r}_A+\frac{\bar{r}_A }{2G} \rd F\,.
\label{dE}
\end{equation}
From Eqs.~(\ref{TdS}) and (\ref{dE}) it follows that 
\begin{equation}
T \rd S = -\rd E+2\pi \bar{r}_A^2 (\rho_d+\rho_M-P_d-P_M) 
\rd \bar{r}_A+\frac{\bar{r}_A}{2G} 
\left( 1+2\pi \bar{r}_A T \right) \rd F\,.
\label{TS}
\end{equation}
Following~\cite{Hayward2,Mukoh,Cai07} we introduce the work density  
$W=(\rho_d+\rho_M-P_d-P_M)/2$.
Then Eq.~(\ref{TS}) reduces to
\begin{equation}
T \rd S=-\rd E+W \rd V
+\frac{\bar{r}_A}{2G} \left( 1+2\pi \bar{r}_A T \right) \rd F\,,
\end{equation}
which can be written in the form~\cite{BGT}
\begin{equation}
T \rd S+T\rd_i S=-\rd E+W \rd V\,,
\label{noneqfirst}
\end{equation}
where 
\begin{equation}
\rd_i S=-\frac{1}{T} \frac{\bar{r}_A}{2G}
\left( 1+2\pi \bar{r}_A T \right) \rd F
=-\left( \frac{E}{T}+S \right) \frac{\rd F}{F}\,.
\label{diS}
\end{equation}
The modified first-law of thermodynamics~(\ref{noneqfirst}) suggests a deep connection 
between the horizon thermodynamics and Friedmann equations in modified gravity. 
The term $\rd_i S$ can be interpreted as a term of entropy production 
in the non-equilibrium thermodynamics~\cite{Eling}. 
The theories with $F= \mathrm{constant}$ lead to $\rd_i \hat{S}=0$, 
which means that the first-law of equilibrium thermodynamics holds. 
The theories with $\rd F \neq 0$, including \fR\ gravity and scalar-tensor theory, 
give rise to the additional non-equilibrium term~(\ref{diS})
\cite{Akbar2,Akbar3,Gong,Wu1,Wu2,Cai07,Bamba09,BGT}.

The main reason why the non-equilibrium term $\rd_i S$ appears is that
the energy density $\rho_d$ and the pressure $P_d$ defined in
Eqs.~(\ref{rodef1}) and (\ref{pdef1}) do not satisfy the standard
continuity equation for $\dot{F} \neq 0$.  On the other hand, if we
define the effective energy-momentum tensor $T_{\mu \nu}^{(D)}$ as
Eq.~(\ref{TmunuD}) in Section~\ref{fieldsec}, it satisfies the continuity
equation~(\ref{TmunuD2}). This correspond to rewriting the Einstein
equation in the form (\ref{Einmo}) instead of (\ref{Ein1}).  Using
this property, \cite{BGT} showed that equilibrium description of
thermodynamics can be possible by introducing the Bekenstein--Hawking
entropy $\hat{S}=A/(4G)$. In this case the horizon entropy $\hat{S}$
takes into account the contribution of both the Wald entropy $S$ in
the non-equilibrium thermodynamics and the entropy production term.


\subsection{Curing the curvature singularity in \fR\ dark energy
  models, unified models of inflation and dark energy}
\label{curesinsec}

In Sections~\ref{chameleonsec} and \ref{mattersec} we showed that there is 
a curvature singularity for viable \fR\ models such as (\ref{Amodel}) and (\ref{Bmodel}).
More precisely this singularity appears for the models having the asymptotic 
behavior (\ref{fRasy}) in the region of high density ($R \gg R_c$).
As we see in Figure~\ref{pofigfR}, the field potential $V(\phi)=(FR-f)/(2\kappa^2 F)$ 
has a finite value $\mu R_c/(2\kappa^2)$
in the limit $\phi=\sqrt{3/(16\pi)}m_{\mathrm{pl}}\,\ln \, F \to 0$.
In this limit one has $f_{,RR} \to 0$, so that the scalaron mass
$1/(3f_{,RR})$ goes to infinity.

This problem of the past singularity can be cured by adding 
the term $R^2/(6M^2)$ to the Lagrangian in \fR\
dark energy models~\cite{Appleby09}.
Let us then consider the modified version of the model~(\ref{Amodel}):
\begin{equation}
f(R)=R-\mu R_{c}\frac{(R/R_{c})^{2n}}{(R/R_{c})^{2n}+1}
+\frac{R^2}{6M^2}\,.
\label{fRHS}
\end{equation}
For this model one can easily show that the potential $V(\phi)=(FR-f)/(2\kappa^2 F)$ 
extends to the region $\phi>0$ and that the curvature singularity disappears
accordingly. Also the scalaron mass approaches the finite value $M$
in the limit $\phi \to \infty$. The perturbation $\delta R$ is bounded 
from above, which can evade the problem of the dominance of the
oscillation mode in the past.

Since the presence of the term $R^2/(6M^2)$ can drive inflation in the early universe, 
one may anticipate that both inflation and the late-time acceleration can be realized 
for the model of the type (\ref{fRHS}). 
This is like a modified gravity version of quintessential inflation
based on a single scalar field~\cite{PeebVil,DeFeTrod,DeFeTrod2,Liddle06}.
However, we have to caution that the transition between two accelerated 
epochs needs to occur smoothly for successful cosmology. 
In other words, after inflation, we require a mechanism in which the universe is 
reheated and then the radiation/matter dominated epochs follow.
However, for the model~(\ref{fRHS}), the Ricci scalar $R$ evolves to the 
point $f_{,RR}=0$ and it enters the region $f_{,RR}<0$.
Crossing the point $f_{,RR}=0$ implies the divergence of the scalaron mass.
Moreover, in the region $f_{,RR}<0$, the Minkowski space is not a stable
vacuum state. This is problematic for the particle creation from 
the vacuum during reheating. The similar problem arises for the models 
(\ref{Bmodel}) and (\ref{tanh}) in addition to the model
proposed by Appleby and Battye~\cite{Appleby}. 
Thus unified \fR\ models of inflation and dark energy cannot be 
constructed easily in general 
(unlike a number of related works~\cite{fRearly4,NOuni1,NOuni2}).
Brookfield et al.~\cite{Brookfield} studied the viability of the model 
$f(R)=R-\alpha/R^n+\beta R^m$ ($n, m>0$)
by using the constraints coming from Big Bang Nucleosynthesis and fifth-force experiments
and showed that it is difficult to find a unique parameter range for consistency of this model.

In order to cure the above mentioned problem, Appleby et al.~\cite{Appleby09} 
proposed the \fR\ model (\ref{stanew}).
Note that the case $c=0$ corresponds to the Starobinsky inflationary 
model $f(R)=R+R^2/(6M^2)$ \cite{Star80} and the case $c=1/2$ 
corresponds to the model of Appleby and Battye~\cite{Appleby} plus the $R^2/(6M^2)$ term.
Although the above mentioned problem can be evaded in this model, 
the reheating proceeds in a different way compared to that 
in the model $f(R)=R+R^2/(6M^2)$
[which we discussed in Section~\ref{reheatingsec}].
Since the Hubble parameter periodically evolves between $H=1/(2t)$ and $H=\epsilon/M$, 
the reheating mechanism does not occur very efficiently~\cite{Appleby09}.
The reheating temperature can be significantly lower than that in 
the model $f(R)=R+R^2/(6M^2)$.
It will be of interest to study observational signatures in such unified 
models of inflation and dark energy.


\subsection{\fR\ theories in extra dimensions}

Although \fR\ theories have been introduced mainly in four dimensions, 
the same models may appear in the context of braneworld~\cite{RS1,RS2} 
in which our universe is described by a brane embedded in extra dimensions
(see~\cite{Maartensliving} for a review).
This scenario implies a careful use of \fR\ theories, because a boundary (brane) appears. 
Before looking at the real working scenario in braneworld, it is necessary to focus on the 
mathematical description of \fR\ models through a sensible definition of boundary
conditions for the metric elements on the surface of the brane.

Some works appeared regarding the possibility of introducing
\fR\ theories in the context of braneworld scenarios~\cite{Rador2,
  Atazadeh08, Borzou, Saavedra, Mariam}. In doing so one requires a
surface term~\cite{Dyer, Parry, Barvinsky, Balce, Balce2, Guarnizo},
which is known as the Hawking--Luttrell term~\cite{Luttrell}
(analogous to the York--Gibbons--Hawking one for General
Relativity). The action we consider is given by 
\begin{equation}
\label{eq:yorko}
S=\int_\Omega \mathrm{d}^nx\sqrt{-g} f(R)+2\int_{\partial\Omega} 
\mathrm{d}^{n-1}x\sqrt{|\gamma|}\,FK\,,
\end{equation}
where $F \equiv \partial f/\partial R$, $\gamma$ is the determinant of the
induced metric on the $n-1$ dimensional 
boundary, and $K$ is the trace of the extrinsic curvature tensor.

In this case particular attention should be paid to boundary conditions on the brane,
that is, the Israel junction conditions~\cite{Israel}.
In order to have a well-defined geometry in five dimensions,
we require that the metric is continuous across
the brane located at $y=0$. 
However its derivatives with respect 
to $y$ can be discontinuous at $y=0$. 
The Ricci tensor $R_{\mu \nu}$ in Eq.~(\ref{fREin}) is made of the metric 
up to the second derivatives $g''$ with respect to $y$. 
This means that $g''$ have a 
delta-function dependence proportional to 
the energy-momentum tensor at a distributional source 
(i.e., with a Dirac's delta function centered on the 
brane) \cite{Bine1,Bine2,Shiromizu}.
In general this also leads to the discontinuity of the Ricci 
scalar $R$ across the brane.
Since the discontinuity of $R$ can lead to inconsistencies in \fR\ gravity,
one should add this extra-constraint as a junction condition. 
In other words, one needs to impose that, 
although the metric derivative is discontinuous, 
the Ricci scalar should still remain continuous on the brane.

This is tantamount to imposing that the extra scalar degree of freedom 
introduced is continuous on the brane. 
We use Gaussian normal coordinates with the metric
\begin{equation}
\label{eq:gaussCOD}
\de s^2=\de y^2+\gamma_{\mu\nu}\,\de x^\mu \de x^\nu\,.
\end{equation}
In terms of the extrinsic curvature tensor $K_{\mu\nu}=-\partial_y \gamma_{\mu\nu}/2$ 
for a brane, the l.h.s.\ of the equations of 
motion tensor [which is analogous to the l.h.s.\ of Eq.~(\ref{fREin}) in 4 dimensions]
is defined by
\begin{equation}
\Sigma_{AB} \equiv FR_{AB}-\frac{1}{2}fg_{AB}
-\nabla_{A}\nabla_{B}F(R)+g_{AB}\square F(R)\,.
\label{eq:SigmCOD}
\end{equation}
This has a delta function behavior for the $\mu$-$\nu$ components, 
leading to~\cite{SasakifRED}
\begin{equation}
\label{eq:sigmMunu}
D_{\mu\nu} \equiv [F\,(K_{\mu\nu}-K\,\gamma_{\mu\nu})+\gamma_{\mu\nu}\,F_{,R}\,
\partial_y R]^+_-=T_{\mu\nu}\,,
\end{equation}
where $T_{\mu\nu}$ is the matter stress-energy tensor on the brane. 
Hence $R$ is continuous, whereas its first derivative is not, in general. 
This imposes an extra condition on the metric crossing the brane at $y=0$, 
compared to General Relativity in which the condition
for the continuity of $R$ is not present.
However, it is not easy to find a solution for which the metric derivative 
is discontinuous but $R$ is not.
Therefore some authors considered matter on the brane which is not universally 
coupled with the induced metric. 
This approach leads to the relaxation of the condition that $R$ is continuous. 
Such a matter action can be found by analyzing the action in the Einstein frame
and introducing a scalar field $\psi$ coupled to the scalaron 
$\phi$ on the brane as follows~\cite{SasakifRED}
\begin{equation}
\label{eq:pipp}
S_M=\int \de^{n-1}x\sqrt{-\gamma}\exp[(n-1)\,C(\phi)]
\left[ 
-\frac12\exp[-2C(\phi)]\gamma^{\mu\nu}
\nabla_\mu\psi \nabla_\nu\psi-V(\psi)
\right]\,.
\end{equation}
The presence of the coupling $C(\phi)$ with the field $\phi$ modifies the Israel 
junction conditions. Indeed, if $C=0$, then $R$ must be continuous, 
but if $C\neq0$, $R$ can have a delta function profile. 
This method may help for finding a solution for the bulk 
that satisfies boundary conditions on the brane.


\subsection{Vainshtein mechanism}

Modifications of gravity in recent works have been introduced mostly in order to explain 
the late-time cosmic acceleration. This corresponds to the large-distance modification 
of gravity, but gravity at small distances is subject to change as well.
Modified gravity models of dark energy must pass local gravity tests in the 
solar system. The \fR\ models discussed in Section~\ref{denergysec} are designed 
to satisfy local gravity constraints by having a large scalar-field mass, while 
at the same time they are responsible for dark energy with a small mass 
compatible with the Hubble parameter today.

It is interesting to see which modified gravity theories have successful Newton limits. 
There are two known mechanisms for satisfying local gravity constraints, 
(i) the Vainshtein mechanism~\cite{Vaino}, and 
(ii) the chameleon mechanism~\cite{chame1,chame2}
(already discussed in Section~\ref{chameleonsec}). 
Both consist of using non-linearities in order to prevent any other fifth force from 
propagating freely. The chameleon mechanism uses the non-linearities coming from
matter couplings, whereas the Vainshtein mechanism uses the self-coupling 
of a scalar-field degree of freedom as a source for the non-linear effect.

There are several examples where the Vainshtein mechanism plays an important role.
One is the massive gravity in which a consistent free massive graviton is 
uniquely defined by Pauli--Fierz theory~\cite{Pauli,Pauli2}.
The massive gravity described by the Fierz--Pauli action cannot be studied 
through the linearization close to a point-like mass source, because of the crossing of the 
Vainshtein radius, that is the distance under which the linearization 
fails to study the metric properly~\cite{Vaino}. 
Then the theory is in the strong-coupling regime, and things become 
obscure as the theory cannot be understood well mathematically. 
A similar behavior also appears for the Dvali--Gabadadze--Porrati (DGP) model 
(we will discuss in the next section), in which the Vainshtein mechanism 
plays a key role for the small-scale behavior of this model.

Besides a standard massive term, other possible operators which could 
give rise to the Vainshtein mechanism come from non-linear self-interactions 
in the kinetic term of a matter field $\phi$.
One of such terms is given by 
\begin{equation}
\label{eq:vainsh}
\nabla_{\mu} \phi\,\nabla^{\mu}\phi\,\Box\phi\,, 
\end{equation}
which respects the Galilean invariance under which $\phi$'s 
gradient shifts by a constant~\cite{RattazziGal}
(treated in section \ref{Gallo}).
This allows a robust implementation of the Vainshtein mechanism
in that nonlinear self-interacting term can allow the decoupling of 
the field $\phi$ from matter in the gravitationally bounded system 
without introducing ghosts.

Another example of the Vainshtein mechanism may be seen in $f(\GB)$ gravity.
Recall that in this theory the contribution to the GB term from matter can be 
neglected relative to the vacuum value $\GB^{(0)}=12\, (2GM)^2/r^6$.
In Section~\ref{solarfG} we showed that on the Schwarzschild geometry 
the modification of gravity 
is very small for the models (\ref{modela}) and (\ref{model2}), because the GB term 
has a value much larger than its cosmological value today. 
The scalar-field degree of freedom acquires a large mass in the region of 
high density, so that it does not propagate freely.
For the model~(\ref{modela}) we already showed that 
at the linear level the coefficients $A$ and $B$ of 
the spherically symmetric metric (\ref{eq:metGG}) are 
\begin{equation}
\label{eq:GGAB2}
A=1-\frac1\rho+A_1(\rho)\,\varepsilon+O(\varepsilon^2)\,,\qquad
B=1-\frac1\rho+B_1(\rho)\,\varepsilon+O(\varepsilon^2)\,,
\end{equation}
where $\rho\equiv r/(2GM)$, $A_1 (\rho)$ and $B_1 (\rho)$ are given by
Eqs.~(\ref{psi1d}) and (\ref{psi1}), and $\varepsilon \approx10^{-46}$ 
for our solar system. Of course this result is trustable only in the region for 
which $A_1\varepsilon\ll1/\rho$. 
Outside this region non-linearities are important and one cannot rely 
on approximate methods any longer. 
Therefore, for this model, we can define the Vainshtein radius $r_V$
as
\begin{equation}
\label{eq:vannofG}
\lambda \varepsilon \rho_V^3 \sim \frac{1}{\rho_V} \quad
\to \quad r_V \sim 2GM (\lambda \varepsilon)^{-1/4}\,.
\end{equation}
For $\lambda\sim1$, this value is well outside the region in which 
solar-system experiments are carried out.
This example shows that the Vainshtein radius is 
generally model-dependent.

In metric \fR\ gravity a non-linear effect coming from 
the coupling to matter fields
(in the Einstein frame) is crucially important, 
because $R$ vanishes
in the vacuum Schwarzschild background.
The local gravity constraints can be satisfied under the chameleon 
mechanism rather than the non-linear self coupling of 
the  Vainshtein mechanism.


\subsection{DGP model}

The Dvali--Gabadadze--Porrati (DGP)~\cite{DGP} braneworld model has been 
considered as a model which could modify gravity 
because of the existence of the extra-dimensions. 
In the DGP model the 3-brane is embedded in a Minkowski bulk spacetime with 
infinitely large 5th extra dimensions.
The Newton's law can be recovered by adding a 4-dimensional (4D) 
Einstein--Hilbert action sourced by the brane curvature to the 5D action~\cite{DGP2}.
While the DGP model recovers the standard 4D gravity for small distances,
the effect from the 5D gravity manifests itself for large distances. 
Remarkably it is possible to realize the late-time cosmic
acceleration without introducing an exotic matter 
source~\cite{Deffayet1,Deffayet2}.

The DGP model is given by the action 
\begin{equation}
S=\frac{1}{2\kappa_{(5)}^{2}}\int\rd^{5}X\sqrt{-\tilde{g}}\,
\tilde{R}+\frac{1}{2\kappa_{(4)}^{2}}\int\rd^{4}x\sqrt{-g}\,R
+\int\rd^{4}x\sqrt{-g}\,{\cal L}_{M}^{\mathrm{brane}}\,,
\label{DGPaction}
\end{equation}
where $\tilde{g}_{AB}$ is the metric in the 5D bulk and 
$g_{\mu\nu}=\partial_{\mu}X^{A}\partial_{\nu}X^{B}\tilde{g}_{AB}$
is the induced metric on the brane with $X^{A}(x^{c})$ being the
coordinates of an event on the brane labeled by $x^{c}$.
The first and second terms in Eq.~(\ref{DGPaction}) correspond to
Einstein--Hilbert actions in the 5D bulk and on the brane, respectively.
Note that $\kappa_{(5)}^{2}$ and 
$\kappa_{(4)}^{2}$ are 5D and 4D gravitational constants, 
respectively, which are related with 5D and 4D Planck masses,
$M_{(5)}$ and $M_{(4)}$, via 
$\kappa_{(5)}^{2}=1/M_{(5)}^3$ and 
$\kappa_{(4)}^{2}=1/M_{(4)}^2$.
The Lagrangian ${\cal L}_{M}^{\mathrm{brane}}$ describes 
matter localized on the 3-brane.

The equations of motion read
\begin{equation}
G^{(5)}_{AB}=0\,,
\end{equation}
where $G^{(5)}_{AB}$ is the 5D Einstein tensor.
The Israel junction conditions on the brane, under which 
a $Z_2$ symmetry is imposed, read~\cite{Hinter}
\begin{equation}
\label{branejun}
G_{\mu\nu}-\frac{1}{r_c}(K_{\mu\nu}-g_{\mu\nu} K)
=\kappa_{(4)}^2T_{\mu\nu}\, ,
\end{equation}
where $K_{\mu\nu}$ is the extrinsic curvature~\cite{Wald} calculated on the brane
and $T_{\mu \nu}$ is the energy-momentum tensor of localized matter.
Since $\nabla^{\mu} (K_{\mu \nu}-g_{\mu \nu}K)=0$,
the continuity equation $\nabla^{\mu}T_{\mu \nu}=0$ follows from 
Eq.~(\ref{branejun}). The length scale $r_c$ is defined by 
\begin{equation}
\label{crossover}
r_{c}\equiv\frac{\kappa_{(5)}^{2}}{2\kappa_{(4)}^{2}}=
\frac{M_{(4)}^{2}}{2M_{(5)}^{3}}\,.
\end{equation}

If we consider the flat FLRW brane ($K=0$), we obtain 
the modified Friedmann equation~\cite{Deffayet1,Deffayet2}
\begin{equation}
H^{2}-\frac{\epsilon}{r_{c}}H=\frac{\kappa_{(4)}^{2}}{3}\rho_{M}\,,
\label{DGPK0eq}
\end{equation}
where $\epsilon=\pm 1$, $H$ and $\rho_M$ are the Hubble parameter and 
the matter energy density on the brane, respectively.
In the regime $r_c \gg H^{-1}$ the first term in Eq.~(\ref{DGPK0eq})
dominates over the second one and hence the standard Friedmann 
equation is recovered.
Meanwhile, in the regime $r_c \lesssim H^{-1}$, the second
term in Eq.~(\ref{DGPK0eq}) leads to a modification to the standard
Friedmann equation. 
If $\epsilon=1$, there is a de~Sitter solution characterized by 
\begin{equation}
\label{dSDGP}
H_{\mathrm{dS}}=1/r_c\,.
\end{equation}
One can realize the cosmic acceleration today if $r_c$ is of the
order of the present Hubble radius $H_0^{-1}$.
This self acceleration is the result of gravitational
leakage into extra dimensions at large distances.
In another branch ($\epsilon=-1$) such cosmic
acceleration is not realized.

In the DGP model the modification of gravity comes from 
a scalar-field degree of freedom, usually called $\pi$, which
is identified as a brane bending mode in the bulk. 
Then one may wonder if such a field mediates a fifth force
incompatible with local gravity constraints.
However, this is not the case, as the Vainshtein mechanism is
at work in the DGP model for the length scale smaller than 
the Vainshtein radius $r_*=(r_g r_c^2)^{1/3}$, where 
$r_g$ is the Schwarzschild radius of a source.
The model can evade solar system constraints
at least under some range of conditions on the energy-momentum 
tensor~\cite{Deffayet:2001uk,Gruzinov:2001hp,Porrati:2002cp}.
The Vainshtein mechanism in the 
DGP model originates from a non-linear self-interaction of the 
scalar-field degree of freedom.

Although the DGP model is appealing and elegant, it is also 
plagued by some problems.
The first one is that, although the model does not possess
ghosts on asymptotically flat manifolds, at the
quantum level, it does have the problem of strong coupling for 
typical distances smaller than 1000~km, 
so that the theory is not easily under control~\cite{Luty}. 
Besides the model typically possesses superluminal modes. 
This may not directly violate causality, but it implies a non-trivial
non-Lorentzian UV completion of the theory~\cite{Hinter}.
Also, on scales relevant for structure formation (between cluster scales and the Hubble radius), 
a quasi-static approximation to linear cosmological perturbations shows that the DGP model 
contains a ghost mode~\cite{Koyama06}.
This linear analysis is valid as long as the Vainshtein radius $r_*$
is smaller than the cluster scales. 

The original DGP model has been tested by using a number of observational data
at the background level~\cite{Sawicki05,Fair,Maje,Alambrane,Song06}.
The joint constraints from the data of  SN~Ia, BAO, and the CMB shift parameter
show that the flat DGP model is under strong observational pressure, while
the open DGP model gives a slightly better fit~\cite{Maje,Song06}.
Xia~\cite{Xia} showed that the parameter $\alpha$
in the modified Friedmann equation 
$H^2-H^{\alpha}/r_c^{2-\alpha}=\kappa_{(4)}^{2}\rho_M/3$~\cite{DvaliTurner}
is constrained to be $\alpha=0.254 \pm 0.153$ (68\% confidence level) by using 
the data of SN~Ia, BAO, CMB, gamma ray bursts, and the linear growth factor 
of matter perturbations. Hence the flat DGP model ($\alpha=1$) is not 
compatible with current observations.

On the sub-horizon scales larger than the Vainshtein radius, 
the equation for linear matter perturbations $\delta_m$ 
in the DGP model was derived in~\cite{Lue,Koyama06} 
under a quasi-static approximation:
\begin{equation}
\ddot{\delta}_m+2H\dot{\delta}_m-4\pi G_{\mathrm{eff}} \rho_m \delta_m \simeq 0\,,
\end{equation}
where $\rho_m$ is the non-relativistic matter density 
on the brane and 
\begin{equation}
G_{\mathrm{eff}}=\left( 1+\frac{1}{3\beta} \right)\,G\,,\qquad
\beta(t) \equiv
1-2Hr_{c}\left(1+\frac{\dot{H}}{3H^{2}}\right)\,.
\label{betadef}
\end{equation}
In the deep matter era one has $Hr_{c}\gg1$ and hence $\beta\simeq-Hr_{c}$,
so that $\beta$ is largely negative ($|\beta|\gg1$). In this regime
the evolution of $\delta_m$ is similar to that in GR 
($\delta_{m}\propto t^{2/3}$). 
Since the background solution finally approaches
the de~Sitter solution characterized by Eq.~(\ref{dSDGP}), it follows that 
$\beta\simeq1-2Hr_{c}\simeq-1$ asymptotically.
Since $1+1/(3\beta)\simeq 2/3$, the growth rate in this regime is
smaller than that in GR.

The index $\gamma$ of the growth rate $f_{\delta}=(\Omega_{m})^{\gamma}$ 
is approximated by $\gamma \approx 0.68$~\cite{Linder05}.
This is quite different from the value $\gamma\simeq0.55$
for the $\Lambda$CDM model. If the future imaging survey of galaxies
can constrain $\gamma$ within 20\%, it will be possible to distinguish
the $\Lambda$CDM model from the DGP model 
observationally~\cite{Yamamoto}.
We recall that in metric \fR\ gravity the growth index today can be 
as small as $\gamma=0.4$ because of the enhanced growth rate, 
which is very different from 
the value in the DGP model.

Comparing Eq.~(\ref{betadef}) with the effective gravitational constant (\ref{Geffmassless}) 
in BD theory with a massless limit (or the absence of the field potential), 
we find that the parameter $\omega_{\mathrm{BD}}$ has the following 
relation with $\beta$:
\begin{equation}
\omega_{\mathrm{BD}}=\frac32 (\beta-1)\,.
\end{equation}
Since $\beta<0$ for the self-accelerating DGP solution, it follows that 
$\omega_{\mathrm{BD}}<-3/2$.
This corresponds to the theory with ghosts, because the kinetic energy 
of a scalar-field degree of freedom becomes negative 
in the Einstein frame~\cite{Damour}.
There is a claim that the ghost may disappear
for the Vainshtein radius $r_*$ of the order of $H_0^{-1}$, because
the linear perturbation theory is no longer applicable~\cite{Dvalinew}.
In fact, a ghost does not appear in a Minkowski brane in the DGP model.
In~\cite{KoyamaSilva} it was shown that the Vainshtein radius in the early universe
is much smaller than the one in the Minkowski background, while in the self accelerating 
universe they agree with each other.
Hence the perturbative approach seems to be still possible 
for the weak gravity regime beyond the Vainshtein radius.

There have been studies regarding a possible regularization in order to avoid the ghost/strong coupling limit. 
Some of these studies have focused on smoothing out the delta profile of the Ricci scalar on the brane, 
by coupling the Ricci scalar to some other scalar field with 
a given profile~\cite{Kolano1, Kolano2}.
In~\cite{Varunbrane} the authors included the brane and bulk cosmological constants
in addition to the scalar curvature in the action for the brane and showed that 
the effective equation of state of dark energy can be smaller than $-1$.
A monopole in seven dimensions generated by a SO(3) invariant matter Lagrangian 
is able to change the gravitational law at its core, leading to a lower dimensional gravitational law.
This is a first approach to an explanation of trapping of gravitons, due to topological defects 
in classical field theory~\cite{Ringeval:2004ju,DeFeRing}.
Other studies have focused on re-using the delta function profile but in a higher-dimensional 
brane~\cite{Nemanja1,Nemanja2,CascoDGP}. 
There is also an interesting work about the possibility of self-acceleration in the normal DGP branch
[$\epsilon=-1$ in Eq.~(\ref{DGPK0eq})] by considering an \fR\ term on the brane 
action~\cite{Mariam} (see also~\cite{Petrov1}).
All these attempts indeed point to the direction that some mechanism, 
if not exactly DGP rather similar to it, may avoid 
a number of problems associated with the original DGP model.

\subsection{Special symmetries}

Since general covariance alone does not restrict the choice of the
Lagrangian function, e.g., for \fR\ theory, one can try to shrink
the set of allowed functions by imposing some extra symmetry. In
particular one can assume that the theory possesses a symmetry on some
special background. However, allowing some theories to be symmetrical
on some backgrounds does not imply these theories are viable by
default. Nevertheless, this symmetry helps to give stronger
constraints on them, as the allowed parameter space drastically
reduces. We will discuss here two of the symmetries studied in the
literature: Noether symmetries on a FLRW background and the Galileon
symmetry on a Minkowski background.

\subsubsection{Noether symmetry on FLRW}

The action for metric \fR\ gravity can be evaluated on a FLRW background, 
in terms of the fields $a(t)$ and $R(t)$,
see~\cite{Capozziello:1996bi, Capozziello:2008ch} 
(and also~\cite{Capozziello:2007iu, Capozziello:1999xs,
  Marmo:1985cg, Morandi:1990su, Modak,Vakili:2008ea, Vakili:2008uj,
  deSouza:2008az, deSouza:2008nj, Capozziello:2009te}). Then the
Lagrangian turns out to be non-singular, ${\cal L}(q_i,\dot q_i)$,
where $q_1=a$, and $q_2=R$. Its Euler--Lagrange equation is given by
$(\partial {\cal L}/\partial \dot{q}^i)^{\cdot}-\partial {\cal
  L}/\partial q^i=0$. Contracting these equations with a vector
function $\alpha^j(q_i)$, (where $\alpha^1=\alpha$ and
$\alpha^2=\beta$ are two unknown functions of the $q^i$), we obtain 
\begin{equation} 
\label{06a} 
\alpha^{i}\left(
\frac{\de}{\de t}\frac{\partial {\cal L}}{\partial \dot q^i}
-\frac{\partial {\cal L }}{\partial q^i}\right)=0\,. 
\quad \to \quad
\frac{\de}{\de t}\left(\alpha^{i}\frac{\partial {\cal L}}{\partial
\dot q^i} \right)=L_{{\bm X}}{\cal L}\,.
\end{equation} 
Here $L_{{{\bm X}}}{\cal L}$ is the Lie 
derivative of ${\cal L}$ with respect to the vector field
\begin{equation} 
\label{05} 
{\bm X}=\alpha^{i}({\bm q})\frac{\partial}{\partial q^{i}}+
\left(\frac{\de}{\de t}\alpha^{i}({\bm q})\right)\frac{\partial}{\partial
\dot q^i}\,.
\end{equation}
If $L_{\bm X}{\cal L}=0$, the \emph{Noether Theorem} states that the function $\Sigma_{0}=\alpha^{i} (\partial {\cal L }/\partial \dot q^i)$ is a constant of motion. 
The generator of the Noether symmetry in metric \fR\ gravity on the flat 
FLRW background is
\begin{equation}
\label{vec}{\bm X}=\alpha\frac{\partial}{\partial a}+\beta\frac{\partial}{\partial
R}+\dot{\alpha}\frac{\partial}{\partial
\dot{a}}+\dot{\beta}\frac{\partial}{\partial\dot{R}}\,.
\end{equation} 
A symmetry exists if the equation $L_{\bm X}{\cal L}=0$ has
non-trivial solutions. As a byproduct, the form of \fR, not
specified in the point-like Lagrangian ${\cal L}$, is determined in
correspondence to such a symmetry.

It can be proved that such $\alpha^i$ do exist~\cite{Capozziello:2008ch}, 
and they correspond to
\begin{equation}
\alpha=c_1\,a+\frac{c_2}a\,,\qquad
\beta=-\left[3\,c_1+\frac{c_2}{a^2}\right]
\frac{f_{,R}}{f_{,RR}}+\frac{c_3}{a\,f_{,RR}}\,,
\end{equation} 
where $c_1$, $c_2$, $c_3$ are constants. 
However, in order that $L_{\bm X}{\cal L}$ vanishes,
one also needs to set the constraint 
(provided that $c_2\,R \neq 0$)
\begin{equation}
f_{,R}=\frac{3(c_1\,a^2+c_2)\,f-c_3aR}{2c_2R}
+\frac{(c_1 a^2+c_2) \kappa^2 \rho_{r}^{(0)}}{a^4 c_2 R}\,,
\label{fRab} 
\end{equation}
where $\rho_{r}^{(0)}$ is the radiation density today.
Since now $L_{\bm X}{\cal}=0$, 
then $\alpha^{i}(\partial {\cal L}/\partial
\dot q^i)= \mathrm{constant}$. This constant of motion corresponds to
\begin{equation}
\alpha\,(6\,f_{,RR}\,a^2\,\dot R+12\,f_{,R}\,a\,\dot a)
+\beta\,(6\,f_{,RR}\,a^2\,\dot a)
=6\,\mu^3_0= \mathrm{constant}\,,
\label{abete} 
\end{equation}
where $\mu_0$ has a dimension of mass.

For a general $f$ it is not possible to solve the Euler--Lagrange
equation and the constraint equation~(\ref{fRab}) at the same
time. Hence, we have to use the Noether constraint in order to find the
subset of those $f$ which make this possible. Some partial solutions
(only when $\mu_0=0$) were found, but whether this symmetry helps
finding viable models of \fR\ is still not certain. However, the
\fR\ theories which possess Noether currents can be more easily
constrained, as now the original freedom for the function $f$ in the
Lagrangian reduced to the choice of the parameters $c_i$ and $\mu_0$.

\subsubsection{Galileon symmetry}
\label{Gallo}

Recently another symmetry, the Galileon symmetry, for a scalar field
Lagrangian was imposed on the Minkowski
background~\cite{RattazziGal}. This idea is interesting as it tries to
decouple light scalar fields from matter making use of
non-linearities, but without introducing new ghost degrees of
freedom~\cite{RattazziGal}. This symmetry was chosen so that the
theory could naturally implement the Vainshtein mechanism. However,
the same mechanism, at least in cosmology, seems to appear also in the
FLRW background for scalar fields which do not possess such a symmetry
(see~\cite{KazuyaGal, KobayashiGal, AT10}).

Keeping a universal coupling with matter (achieved through a pure
nonminimal coupling with $R$), Nicolis et al.~\cite{RattazziGal}
imposed a symmetry called the Galilean invariance on a scalar field
$\pi$ in the Minkowski background. If the equations of motion are
invariant under a constant gradient-shift on Minkowski spacetime,
that is
\begin{equation}
\label{eq:gal1}
\pi\to\pi+c+b_\mu x^\mu\,,
\end{equation}
where both $c$ and $b_\mu$ are constants, we call $\pi$ a Galileon field. 
This implies that the equations of motion fix the field up to such a transformation. 
The point is that the Lagrangian must implement the Vainshtein mechanism in order 
to pass solar-system constraints. This is achieved by introducing self-interacting 
non-linear terms in the Lagrangian. It should be noted that the Lagrangian is studied 
only at second order in the fields (having a nonminimal coupling with $R$) and 
the metric itself, whereas the non-linearities are fully kept by neglecting their backreaction 
on the metric (as the biggest contribution should come only from standard matter). 
The equations of motion respecting the Galileon symmetry contain terms such as a constant, $\Box\,\pi$ (up to fourth power), and other power contraction of the tensor $\nabla_\mu \nabla_\nu\pi$. It is due to these non-linear derivative terms by which the Vainshtein mechanism can be implemented, as it happens in the DGP model~\cite{Luty}.

Nicolis et al.~\cite{RattazziGal} found that there are only five terms ${\cal L}_i$ with $i=1,\dots,5$ which can be inserted into a Lagrangian, such that the equations of motion respect the Galileon symmetry in 4-dimensional Minkowski spacetime. The first three terms are given by 
\begin{eqnarray}
{\cal L}_1&=&\pi\, ,\\
{\cal L}_2&=&\nabla_\mu\pi\nabla^\mu\pi\,,\\
{\cal L}_3&=&\Box\pi\,\nabla_\mu\pi\nabla^\mu\pi\,.
\end{eqnarray}
All these terms generate second-order derivative terms only in the equations of motion. The approach in the Minkowski spacetime has motivated to try to find a fully covariant framework in the curved spacetime. In particular, Deffayet et al.~\cite{DeffaGal} found that all the previous 5-terms can be written in a fully covariant way. However, if we want to write down ${\cal L}_{4}$ and ${\cal L}_{5}$ covariantly in curved spacetime and keep the equations of motion free from higher-derivative terms, we need to introduce couplings between the field $\pi$ and the Riemann tensor~\cite{DeffaGal}. The following two terms keep the field equations to second-order,
\begin{eqnarray}
  {\cal L}_4&=&(\nabla_\mu\pi\nabla^\mu\pi)\,\bigl[
2(\Box\pi)^2
-2(\nabla_{\alpha\beta}\pi)\,(\nabla^{\alpha\beta}\pi)
-(1/2)\,R\,\nabla_\mu\pi\nabla^\mu\pi\bigr]\, ,
\label{L4} \\
{\cal L}_5&=&(\nabla_\lambda\pi\nabla^\lambda\pi)\,\bigl[
(\Box\pi)^3
-3\Box\pi\,(\nabla_{\alpha\beta}\pi)\,(\nabla^{\alpha\beta}\pi)
+2(\nabla_\mu\nabla^\nu\pi)\,(\nabla_\nu\nabla^\rho\pi)\,(\nabla_\rho\nabla^\mu\pi)\notag\\
&&\qquad\qquad\qquad\qquad -6(\nabla_\mu\pi)\,(\nabla^\mu\nabla^\nu\pi)\,(\nabla^\rho\pi)\, G_{\nu\rho}\bigr]\,,
\label{L5}
\end{eqnarray}
where the last terms in Eqs.~(\ref{L4}) and (\ref{L5}) are newly introduced in the curved spacetime. 
These terms possess the required symmetry in Minkowski spacetime, 
but mostly, they do not introduce derivatives higher than two into the equations of motion.
In this sense, although originated from an implementation of the DGP idea, the covariant Galileon field is closer to the approach of the modifications of gravity in $f(R,\GB)$, that is, a formalism which would introduce only second-order equations of motion.

This result can be extended to arbitrary $D$ dimensions~\cite{DeffaGalED}. One can find, analogously to the Lovelock action-terms, an infinite tower of terms that can be introduced with the same property of keeping the equations of motion at second order. In particular, let us consider the action 
\begin{equation}
\label{eq:lovGal}
S=\int \mathrm{d}^Dx\sqrt{-g}\sum_{p=0}^{p_{\mathrm{max}}}
{\cal C}_{(n+1,p)}{\cal L}_{(n+1,p)}\, ,
\end{equation}
where $p_{\mathrm{max}}$ is the integer part of $(n-1)/2$ ($n \le D$), 
\begin{equation}
\label{eq:lovGal2}
{\cal C}_{(n+1,p)}=\left(-\frac18\right)^{p}
\frac{(n-1)!}{(n-1-2p)!(p!)^2}\, ,
\end{equation}
and
\begin{eqnarray}
\label{eq:lovGal3}
{\cal L}_{(n+1,p)}&=&-\frac1{(D-n)!}
\varepsilon^{\mu_1\mu_3\dots\mu_{2n-1}\nu_1\dots\nu_{D-n}}\varepsilon^{\mu_2\mu_4\dots\mu_{2n}}{}_{\nu_1\dots\nu_{D-n}}\,
\pi_{;\mu_1}\pi_{;\mu_2}\,(\pi^{;\lambda}\pi_{;\lambda})^p\notag\\
&& \times\prod_{i=1}^{p}R_{\mu_{4i-1}\mu_{4i+1}\mu_{4i}\mu_{4i+2}}
\prod_{j=0}^{n-2-2p}\pi_{;\mu_{2n-1-2j}\mu_{2n-2j}}\,.
\end{eqnarray}
Here $\varepsilon^{1 \cdots n}$ is the Levi-Civita tensor. The first
product in Eq.~(\ref{eq:lovGal3}) is defined to be one when $p=0$ and $0$
for $p<0$, and the second product is one when $n=1+2p$, and $0$ when
$n<2+2p$. In ${\cal L}_{(n+1,p)}$ there will be $n+1$ powers of $\pi$,
and $p$ powers of the Riemann tensor. In four dimensions, for example,
${\cal L}_{(1,0)}$ and ${\cal L}_{(2,0)}$ are identical to ${\cal
  L}_1$ and ${\cal L}_2$ introduced before, respectively. Instead,
${\cal L}_{(3,0)}$, ${\cal L}_{(4,0)}-(1/4)\,{\cal L}_{(4,1)}$, and
${\cal L}_{(5,0)}-(3/4){\cal L}_{(5,1)}$ reduce to ${\cal L}_{3}$,
${\cal L}_{4}$, and ${\cal L}_{5}$, up to total derivatives,
respectively.

In general non-linear terms discussed above may introduce the
Vainshtein mechanism to decouple the scalar field from matter around a
star, so that solar-system constraints can be satisfied. However the
modes can have superluminal propagation, which is not surprising as
the kinetic terms get heavily modified in the covariant
formalism. Some studies have focused especially on the ${\cal L}_3$
term only, as this corresponds to the simplest case. For some models
the background cosmological evolution is similar to that in the DGP
model, although there are ghostlike modes depending on the sign of the
time-velocity of the field $\pi$~\cite{JustinGal}. There are some
works for cosmological dynamics in Brans--Dicke theory in the presence
of the non-linear term ${\cal L}_3$~\cite{KazuyaGal,
  KobayashiGal,AT10} (although the original Galileon symmetry is not
preserved in this scenario). Interestingly the ghost can disappear
even for the case in which the Brans--Dicke parameter
$\omega_{\mathrm{BD}}$ is smaller than $-2$. Moreover this theory
leaves a number of distinct observational signatures such as the
enhanced growth rate of matter perturbations and the significant ISW
effect in CMB anisotropies.

At the end of this section we should mention conformal gravity
in which the conformal invariance forces the gravitational action 
to be uniquely given by a Weyl action~\cite{Mann1,Mann2}.
Interestingly the conformal symmetry also forces the cosmological 
constant to be zero at the level of the action \cite{Mann3}.
It will be of interest to study the cosmological aspects of such theory, 
together with the possibility for the avoidance of ghosts and instabilities.

\newpage

\section{Conclusions}
\label{consec}
\setcounter{equation}{0}

We have reviewed many aspects of \fR\ theories studied extensively 
over the past decade. This burst of activities is strongly 
motivated by the observational discovery of dark energy.
The idea is that the gravitational law may be modified on cosmological 
scales to give rise to the late-time acceleration, while 
Newton's gravity needs to be recovered on solar-system scales.
In fact, \fR\ theories can be regarded as the simplest 
extension of General Relativity. 

The possibility of the late-time cosmic acceleration in metric \fR\
gravity was first suggested by Capozziello in 2002~\cite{fRearly1}.
Even if \fR\ gravity looks like a simple theory, 
successful \fR\ dark energy models need to satisfy 
a number of conditions for consistency with 
successful cosmological evolution
(a late-time accelerated epoch preceded by a matter era)
and with local gravity tests on solar-system scales.
We summarize the conditions under which metric \fR\ 
dark energy models are viable:
\begin{enumerate}
\item $f_{,R}>0$ for $R \ge R_0$, where $R_0$ is the 
Ricci scalar today. This is required to avoid a ghost state.
\item $f_{,RR}>0$ for $R \ge R_0$.
This is required to avoid the negative mass squared
of a scalar-field degree of freedom (tachyon).
\item $f(R) \to R-2\Lambda$ for $R \ge R_0$.
This is required for the presence of the matter era
and for consistency with local gravity constraints.
\item $0<\frac{Rf_{,RR}}{f_{,R}}(r=-2)<1$ at 
$r=-\frac{Rf_{,R}}{f}=-2$.
This is required for the stability and the presence 
of a late-time de~Sitter solution. Note that there is
another fixed point that can be responsible for
the cosmic acceleration (with an effective equation of 
state $w_{\mathrm{eff}}>-1$).
\end{enumerate}
We clarified why the above conditions are required
by providing detailed explanation about the background 
cosmological dynamics (Section~\ref{denergysec}), 
local gravity constraints (Section~\ref{lgcsec}), 
and cosmological perturbations
(Sections~\ref{persec}\,--\,\ref{cosmodark}).

After the first suggestion of dark energy scenarios based on metric
\fR\ gravity, it took almost five years to construct viable 
models that satisfy all the above conditions~\cite{AGPT,LiBarrow,AmenTsuji07,Hu07,Star07,Appleby,Tsuji08}.
In particular, the models (\ref{Amodel}), (\ref{Bmodel}), and (\ref{tanh})
allow appreciable deviation from the $\Lambda$CDM model during the late
cosmological evolution, while the early cosmological dynamics is similar to that of the 
$\Lambda$CDM. The modification of gravity manifests itself in the evolution 
of cosmological perturbations through the change of the effective gravitational
coupling.  As we discussed in Sections~\ref{cosmodark} and \ref{othersec}, 
this leaves a number of interesting observational signatures such as
the modification to the galaxy and CMB power spectra and the effect on 
weak lensing. This is very important to distinguish \fR\ dark energy 
models from the $\Lambda$CDM model in future high-precision observations.

As we showed in Section~\ref{fieldsec}, the action in metric \fR\ gravity 
can be transformed to that in the Einstein frame.
In the Einstein frame, non-relativistic matter couples to a scalar-field
degree of freedom (scalaron) with a coupling $Q$ of the order of unity
($Q=-1/\sqrt{6}$). 
For the consistency of metric \fR\ gravity with local gravity constraints,
we require that the chameleon mechanism~\cite{chame1,chame2} 
is at work to suppress such a large coupling.
This is a non-linear regime in which the linear expansion of the Ricci scalar
$R$ into the (cosmological) background value $R_0$ and the perturbation 
$\delta R$ is no longer valid, that is, the condition $\delta R \gg R_0$ 
holds in the region of high density. 
As long as a spherically symmetric 
body has a thin-shell, the effective matter coupling $Q_{\mathrm{eff}}$ is 
suppressed to avoid the propagation of the fifth force.
In Section~\ref{lgcsec} we provided detailed explanation about the chameleon
mechanism in \fR\ gravity and showed that the models (\ref{Amodel}) and (\ref{Bmodel}) 
are consistent with present experimental bounds of local gravity tests for $n>0.9$.
 
The construction of successful \fR\ dark energy models triggered 
the study of spherically symmetric solutions in those models.
Originally it was claimed that a curvature singularity present in the models 
(\ref{Amodel}) and (\ref{Bmodel}) may be accessed in the strong 
gravitational background like neutron stars~\cite{Frolov,KM08}.
Meanwhile, for the Schwarzschild interior and exterior background with 
a constant density star, one can approximately derive analytic thin-shell solutions 
in metric \fR\ and Brans--Dicke theory by taking into account the
backreaction of gravitational potentials~\cite{TTT09}.
In fact, as we discussed in Section~\ref{starsec}, a static star configuration 
in the \fR\ model~(\ref{Bmodel}) was numerically found both 
for the constant density star and the star with a polytropic equation 
of state~\cite{Babi1,Upadhye,Babi2}.
Since the relativistic pressure is strong  
around the center of the star, the choice of correct boundary 
conditions along the line of~\cite{TTT09} is important 
to obtain static solutions numerically.

The model $f(R)=R+R^2/(6M^2)$ proposed by Starobinsky in 1980
is the first model of inflation in the early universe.
Inflation occurs in the regime $R \gg M^2$, which is followed
by the reheating phase with an oscillating Ricci scalar.
In Section~\ref{inflationsec} we studied the dynamics of inflation and 
(p)reheating (with and without nonminimal couplings between 
a field $\chi$ and $R$) in detail.
As we showed in Section~\ref{cosmoinf}, this model is well consistent 
with the WMAP 5-year bounds of the spectral index $n_s$ 
of curvature perturbations and of the tensor-to-scalar ratio $r$.
It predicts the values of $r$ smaller than the order of 0.01, 
unlike the chaotic inflation model with $r={\cal O}(0.1)$.
It will be of interest to see whether this model continues to be 
favored in future observations. 

Besides metric \fR\ gravity, there is another formalism dubbed
the Palatini formalism in which the metric $g_{\mu \nu}$ 
and the connection $\Gamma^{\alpha}_{\beta \gamma}$
are treated as independent variables when we vary the action
(see Section~\ref{Palasec}).
The Palatini \fR\ gravity gives rise to the specific trace
equation~(\ref{Pala2}) that does not have a propagating degree of
freedom. Cosmologically we showed that even for the model $f(R)=R-\beta/R^n$
($\beta>0$, $n>-1$) it is possible to realize a sequence of
radiation, matter, and de~Sitter epochs (unlike the same model
in metric \fR\ gravity). However the Palatini \fR\ gravity is 
plagued by a number of shortcomings such as the inconsistency 
with observations of large-scale structure, the conflict 
with Standard Model of particle physics, and the divergent behavior of
the Ricci scalar at the surface of a static 
spherically symmetric star with a polytropic equation of 
state $P=c\rho_0^{\Gamma}$ with $3/2<\Gamma<2$.
The only way to avoid these problems is that the \fR\
models need to be extremely close to the $\Lambda$CDM 
model. This property is different from metric \fR\
gravity in which the deviation from the $\Lambda$CDM 
model can be significant for $R$ of the order of
the Ricci scalar today.

In Brans--Dicke (BD) theories with the action~(\ref{BDaction}), 
expressed in the Einstein frame, non-relativistic 
matter is coupled to a scalar field with a constant coupling $Q$.
As we showed in in Section~\ref{BDtheory}, this coupling $Q$ is related 
to the BD parameter $\omega_{\mathrm{BD}}$ with the relation 
$1/(2Q^2)=3+2\omega_{\mathrm{BD}}$.
These theories include metric and Palatini \fR\ gravity theories as special cases 
where the coupling is given by $Q=-1/\sqrt{6}$ (i.e., $\omega_{\mathrm{BD}}=0$)
and $Q=0$ (i.e., $\omega_{\mathrm{BD}}=-3/2$), respectively.
In BD theories with the coupling $Q$ of the order of unity
we constructed a scalar-field potential responsible 
for the late-time cosmic acceleration, while satisfying 
local gravity constraints through the chameleon mechanism.
This corresponds to the generalization of metric \fR\ gravity, 
which covers the models (\ref{Amodel}) and (\ref{Bmodel}) 
as specific cases.
We discussed a number of observational signatures in those models
such as the effects on the matter power spectrum and weak lensing.

Besides the Ricci scalar $R$, there are other scalar quantities 
such as $R_{\mu \nu}R^{\mu \nu}$ and $R_{\mu \nu \rho \sigma}R^{\mu
\nu \rho \sigma}$ constructed from the Ricci tensor $R_{\mu \nu}$
and the Riemann tensor $R_{\mu \nu \rho \sigma}$.
For the Gauss--Bonnet (GB) curvature invariant
$\GB \equiv R^2-4R_{\mu \nu}R^{\mu \nu}
+R_{\mu \nu \rho \sigma}R^{\mu \nu \rho \sigma}$
one can avoid the appearance of spurious spin-2 ghosts.
There are dark energy models in which the Lagrangian density is
given by ${\cal L}=R+f(\GB)$, 
where $f(\GB)$ is an arbitrary function in terms of $\GB$.
In fact, it is possible to explain the late-time cosmic acceleration 
for the models such as (\ref{modela}) and (\ref{model2}), while
at the same time local gravity constraints are satisfied.
However density perturbations in perfect fluids
exhibit violent negative instabilities during both the radiation and 
the matter domination, irrespective of the form of $f(\GB)$. 
The growth of perturbations gets stronger on smaller
scales, which is incompatible with the observed galaxy
spectrum unless the deviation from GR is very 
small. Hence these models are effectively ruled out from 
this Ultra-Violet instability. This implies that metric \fR\ 
gravity may correspond to the marginal theory 
that can avoid such instability problems.

In Section~\ref{othersec} we discussed other aspects of \fR\ gravity 
and modified gravity theories -- such as weak lensing, thermodynamics
and horizon entropy, Noether symmetry in \fR\ gravity, unified 
\fR\ models of inflation and dark energy, \fR\ theories in extra
dimensions, Vainshtein mechanism, DGP model, and Galileon field.
Up to early 2010 the number of papers that include the word ``\fR''
in the title is over 460, and more than 1050 papers including the words
``\fR'' or ``modified gravity'' or ``Gauss--Bonnet'' have been written 
so far. This shows how this field is rich and fruitful in application to many aspects
to gravity and cosmology. 

Although in this review we have focused on \fR\ gravity and some 
extended theories such as BD theory and Gauss--Bonnet gravity, there 
are other classes of modified gravity theories, e.g., 
Einstein--Aether theory~\cite{Jacobson2}, 
tensor-vector-scalar theory of gravity~\cite{TEVES}, ghost condensation~\cite{Arkani}, 
Lorentz violating theories~\cite{Lorentz1,Lorentz2,Lorentz3}, 
and Ho{\v{r}}ava--Lifshitz gravity~\cite{Horava}.
There are also attempts to study \fR\ gravity in the context of 
Ho{\v{r}}ava--Lifshitz gravity~\cite{Kluson1,Kluson2}.
We hope that future high-precision observations can distinguish between these modified 
gravity theories, in connection to solving the fundamental problems 
for the origin of inflation, dark matter, and dark energy.

\newpage

\section{Acknowledgements}

We thank Clifford Will for inviting us to write this article in \textit{Living
Reviews in Relativity}. We are grateful to Luca Amendola, Kazuharu
Bamba, Bruce A.\ Bassett, Gianluca Calcagni, Salvatore Capozziello,
Sean M.\ Carroll, Edmund J.\ Copleand, Beatriz de Carlos, Vikram
Duvvuri, Damien A.\ Easson, Stephane Fay, Radouane Gannouji,
Chao-Qiang Geng, Jean-Marc Gerard, Burin Gumjudpai, Zong-kuan Guo,
Mark Hindmarsh, Maxim Libanov, Roy Maartens, Kei-ichi Maeda, Gianpiero
Mangano, Shuntaro Mizuno, Bruno Moraes, David F.\ Mota, Pia Mukherjee,
Nobuyoshi Ohta, Eleftherios Papantonopoulos, David Polarski,
Christophe Ringeval, Valery Rubakov, M.\ Sami, Pasquale D.\ Serpico,
Alexei Starobinsky, Teruaki Suyama, Takashi Tamaki, Takayuki Tatekawa,
Reza Tavakol, Alexey Toporensky, Takashi Torii, Peter Tretjakov,
Michael S.\ Turner, Mark Trodden, Kotub Uddin, David Wands, Yun Wang,
and Jun'ichi Yokoyama for fruitful collaborations about \fR\ theory
and modified gravity. We also thank Eugeny Babichev, Nathalie
Deruelle, Kazuya Koyama, David Langlois, and Shinji Mukohyama for
useful discussions. The work of A.D.\ and S.T.\ was supported by the
Grant-in-Aid for Scientific Research Fund of the JSPS Nos.~09314 and
30318802. S.T.\ also thanks financial support for the Grant-in-Aid
for Scientific Research on Innovative Areas (No.~21111006).

\newpage

\bibliography{refs}

\end{document}